\def\fileversion{v0.02} \def\filedate{2005/06/03}
\documentclass[11pt,oneside]{book}
\usepackage{natbib}
\bibpunct{(}{)}{;}{a}{,}{,}
\usepackage{amsmath}
\usepackage{amssymb}
\usepackage{amscd}
\usepackage{deluxetable}
\usepackage{fullpage, graphicx, url}
\usepackage[fancy]{uiucthesis}
\usepackage{subfigure}
\usepackage{setspace}

\newcommand\bv{\mbox{\boldmath $v$}}
\renewcommand\>{{\rangle}}

\newcommand{\bld}[1]{\mbox{\boldmath $#1$}}
\newcommand\ex{\hat{\bld{x}}}
\newcommand\ey{\hat{\bld{y}}}
\newcommand\ez{\hat{\bld{z}}}
\newcommand\tco{{\tau_{co}}}
\newcommand\tc{{\tau_c}}
\newcommand\tce{{\<\< \tau_c \>\>}}
\newcommand\del{\partial}
\newcommand\<{{\langle}}

\newcommand\mdot{\dot{M}}
\newcommand\qq{\frac{3}{2}}
\newcommand\s{{\rm\,s}}

\newcommand\cm{{\rm\,cm}}

\newcommand\pc{{\rm\,pc}}

\newcommand\yr{{\rm\,yr}}
\newcommand\gm{{\rm\,g}}
\newcommand\kms{{\rm\,km\,s^{-1}}}
\newcommand\gcm{{\rm\,g\,cm^{-3}}}
\newcommand\erg{{\rm\,erg}}

\newcommand\msun{{M_\odot}}
\newcommand\lsun{{L_\odot}}

\newcommand\K{{\rm\,K}}
\newcommand\qe{{\tilde{q}}}
\newcommand\bB{\mbox{\boldmath $B$}}
\newcommand\bnabla{\mbox{\boldmath $\nabla$}}
\newcommand\bk{\mbox{\boldmath $k$}}

\newcommand\bx{\mbox{\boldmath $x$}}
\newcommand\bO{\mbox{\boldmath $\Omega$}}
\newcommand\bxi{\mbox{\boldmath $\xi$}}
\newcommand{\pdv}[2]{\frac{\partial#1}{\partial#2}}
\newcommand{\dv}[2]{\frac{d#1}{d#2}}
\newcommand{\be}{\begin{equation}}
\newcommand{\ee}{\end{equation}}
\newcommand{\Ri}{{\rm Ri}}

\begin{document}
\ifx\href\undefined\else\hypersetup{linktocpage=true}\fi 

\newcommand\aj{\rmfamily{AJ}}%
\newcommand\araa{\rmfamily{ARA\&A}}%
\newcommand\apj{\rmfamily{ApJ}}%
\newcommand\apjl{\rmfamily{ApJ}}%
\newcommand\apjs{\rmfamily{ApJS}}%
\newcommand\ao{\rmfamily{Appl.~Opt.}}%
\newcommand\apss{\rmfamily{Ap\&SS}}%
\newcommand\aap{\rmfamily{A\&A}}%
\newcommand\aapr{\rmfamily{A\&A~Rev.}}%
\newcommand\aaps{\rmfamily{A\&AS}}%
\newcommand\azh{\rmfamily{AZh}}%
\newcommand\baas{\rmfamily{BAAS}}%
\newcommand\jrasc{\rmfamily{JRASC}}%
\newcommand\memras{\rmfamily{MmRAS}}%
\newcommand\mnras{\rmfamily{MNRAS}}%
\newcommand\pra{\rmfamily{Phys.~Rev.~A}}%
\newcommand\prb{\rmfamily{Phys.~Rev.~B}}%
\newcommand\prc{\rmfamily{Phys.~Rev.~C}}%
\newcommand\prd{\rmfamily{Phys.~Rev.~D}}%
\newcommand\pre{\rmfamily{Phys.~Rev.~E}}%
\newcommand\prl{\rmfamily{Phys.~Rev.~Lett.}}%
\newcommand\pasp{\rmfamily{PASP}}%
\newcommand\pasj{\rmfamily{PASJ}}%
\newcommand\qjras{\rmfamily{QJRAS}}%
\newcommand\skytel{\rmfamily{S\&T}}%
\newcommand\solphys{\rmfamily{Sol.~Phys.}}%
\newcommand\sovast{\rmfamily{Soviet~Ast.}}%
\newcommand\ssr{\rmfamily{Space~Sci.~Rev.}}%
\newcommand\zap{\rmfamily{ZAp}}%
\newcommand\nat{\rmfamily{Nature}}%
\newcommand\iaucirc{\rmfamily{IAU~Circ.}}%
\newcommand\aplett{\rmfamily{Astrophys.~Lett.}}%
\newcommand\apspr{\rmfamily{Astrophys.~Space~Phys.~Res.}}%
\newcommand\bain{\rmfamily{Bull.~Astron.~Inst.~Netherlands}}%
\newcommand\fcp{\rmfamily{Fund.~Cosmic~Phys.}}%
\newcommand\gca{\rmfamily{Geochim.~Cosmochim.~Acta}}%
\newcommand\grl{\rmfamily{Geophys.~Res.~Lett.}}%
\newcommand\jcp{\rmfamily{J.~Chem.~Phys.}}%
\newcommand\jgr{\rmfamily{J.~Geophys.~Res.}}%
\newcommand\jqsrt{\rmfamily{J.~Quant.~Spec.~Radiat.~Transf.}}%
\newcommand\memsai{\rmfamily{Mem.~Soc.~Astron.~Italiana}}%
\newcommand\nphysa{\rmfamily{Nucl.~Phys.~A}}%
\newcommand\physrep{\rmfamily{Phys.~Rep.}}%
\newcommand\physscr{\rmfamily{Phys.~Scr}}%
\newcommand\planss{\rmfamily{Planet.~Space~Sci.}}%
\newcommand\procspie{\rmfamily{Proc.~SPIE}}%
\let\astap=\aap
\let\apjlett=\apjl
\let\apjsupp=\apjs
\let\applopt=\ao
\newcommand\phn{\phantom{0}}%
\newcommand\phd{\phantom{.}}%
\newcommand\phs{\phantom{$-$}}%
\newcommand\phm[1]{\phantom{#1}}%
\let\la=\lesssim            
\let\ga=\gtrsim
\newcommand\sq{\mbox{\rlap{$\sqcap$}$\sqcup$}}%
\newcommand\arcdeg{\mbox{$^\circ$}}%
\newcommand\arcmin{\mbox{$^\prime$}}%
\newcommand\arcsec{\mbox{$^{\prime\prime}$}}%
\newcommand\fd{\mbox{$.\!\!^{\mathrm d}$}}%
\newcommand\fh{\mbox{$.\!\!^{\mathrm h}$}}%
\newcommand\fm{\mbox{$.\!\!^{\mathrm m}$}}%
\newcommand\fs{\mbox{$.\!\!^{\mathrm s}$}}%
\newcommand\fdg{\mbox{$.\!\!^\circ$}}%
\newcommand\case[2]{\mbox{$\frac{#1}{#2}$}}%
\newcommand\slantfrac{\case}%
\newcommand\onehalf{\slantfrac{1}{2}}%
\newcommand\onethird{\slantfrac{1}{3}}%
\newcommand\twothirds{\slantfrac{2}{3}}%
\newcommand\onequarter{\slantfrac{1}{4}}%
\newcommand\threequarters{\slantfrac{3}{4}}%
\newcommand\ubvr{\mbox{$U\!BV\!R$}}
\newcommand\ub{\mbox{$U\!-\!B$}}
\newcommand\vr{\mbox{$V\!-\!R$}}
\newcommand\ur{\mbox{$U\!-\!R$}}
\newcommand\ion[2]{#1$\;${\small\rmfamily\@Roman{#2}}\relax}%
\newcommand\diameter{\ooalign{\hfil/\hfil\crcr\mathhexbox20D}}%
\newcommand\degr{\arcdeg}%
\newcommand\Sun{\sun}%
\newcommand\Sol{\sun}%
\newcommand\sun{\odot}%
\newcommand\Mercury{\astro{\char1}}
\newcommand\Venus{\astro{\char2}}
\newcommand\Earth{\earth}%
\newcommand\Terra{\earth}%
\newcommand\earth{\oplus}%
\newcommand\Mars{\astro{\char4}}
\newcommand\Jupiter{\astro{\char5}}
\newcommand\Saturn{\astro{\char6}}
\newcommand\Uranus{\astro{\char7}}
\newcommand\Neptune{\astro{\char8}}
\newcommand\Pluto{\astro{\char9}}
\newcommand\Moon{\astro{\char10}}
\newcommand\Luna{\Moon}%
\newcommand\Aries{\astro{\char11}}%
\newcommand\VEq{\Aries}
\newcommand\Taurus{\astro{\char12}}%
\newcommand\Gemini{\astro{\char13}}%
\newcommand\Cancer{\astro{\char14}}%
\newcommand\Leo{\astro{\char15}}%
\newcommand\Virgo{\astro{\char16}}%
\newcommand\Libra{\astro{\char17}}%
\newcommand\AEq{\Libra}
\newcommand\Scorpius{\astro{\char18}}%
\newcommand\Sagittarius{\astro{\char19}}%
\newcommand\Capricornus{\astro{\char20}}%
\newcommand\Aquarius{\astro{\char21}}%
\newcommand\Pisces{\astro{\char22}}%

\bibliographystyle{plainnat}

\title{Turbulent Angular Momentum Transport\\
in Weakly-Ionized Accretion Disks}
\author{Bryan Mark Johnson}
\department{Physics}
\schools{B.S., LeTourneau University, 1996}
\phdthesis

\degreeyear{2005}

\maketitle

\frontmatter

\chapter*{Abstract}

\begin{spacing}{1.5}

Accretion disks are ubiquitous in the universe.  Although difficult to observe directly,
their presence is often inferred from the unique signature they imprint on the spectra 
of the systems in which they are observed.  In addition, many properties of accretion-disk
systems that would be otherwise mysterious are easily accounted for by the presence of
matter accreting (accumulating) onto a central object.  Since the angular momentum of the
infalling material is conserved, a disk naturally forms as a repository of angular momentum.
Dissipation removes energy and angular momentum from the system and allows the disk to accrete.
It is the energy lost in this process and ultimately converted to radiation that we observe.

Understanding the mechanism that drives accretion has been the primary challenge in accretion
disk theory.  Turbulence provides a natural means of dissipation and the removal of angular
momentum, but firmly establishing its presence in disks proved for many years to be difficult.
The realization in the 1990s that a weak magnetic field will destabilize a
disk and result in a vigorous turbulent transport of angular momentum has revolutionized
the field.  Much of accretion disk research now focuses on understanding the implications
of this mechanism for astrophysical observations.  At the same time, the success of this
mechanism depends upon a sufficient ionization level in the disk for the flow to be well-coupled
to the magnetic field.  Many disks, such as disks around young stars and disks in binary systems
that are in quiescence, are too cold to be sufficiently ionized, and so efforts to establish the
presence of turbulence in these disks continues.

This dissertation focuses on several possible mechanisms for the turbulent transport of
angular momentum in weakly-ionized accretion disks: gravitational instability, radial convection
and vortices driving compressive motions.  It appears that none of these mechanisms are very
robust in driving accretion.  A discussion is given, based on these results, as to the most
promising directions to take in the search for a turbulent transport mechanism that does not
require magnetic fields.  Also discussed are the implications of assuming that no turbulent
transport mechanism exists for weakly-ionized disks.

\end{spacing}

\newpage
\leavevmode\vfill
\begin{center}
``Since we astronomers are priests of the highest God in regard to the book of nature,
it benefits us to be thoughtful, not of the glory of our minds, but rather, above all
else, of the glory of God.''

\vspace{0.2in}
-- Johannes Kepler

\vspace{0.2in}
{\it Soli Deo gloria}
\end{center}
\vfill

\chapter*{Acknowledgments}

\begin{spacing}{1.5}

There are many people without whose encouragement and assistance I would not have
begun, let alone completed, this dissertation.  I am grateful to my parents, Larry
and Joyce Johnson, for raising me in an environment of loving affection and
discipline and for providing me with every opportunity to pursue a love for learning.
I am grateful for the encouragement of Josh Coe and Jim Wolfe in making what seemed
at the time like an uncertain switch from engineering to physics, and for the
Department of Physics at the University of Illinois for giving me the chance to pursue
a doctorate even though I barely made the minimum required score on the Physics GRE.
I am also grateful to my teachers, both at the University of Illinois and at the other
institutions I have attended throughout my life, for their investment in my education.

The bulk of the research in this dissertation has been published in collaboration
with my adviser, Charles Gammie.  I have thoroughly enjoyed working with him and
have greatly appreciated his unselfishness and accessibility.  This dissertation
and my own progress in astrophysics have benefited immensely from his insights and guidance.
I am also grateful to Stu Shapiro, Alfred Hubler and Doug Beck for their willingness to serve on my defense committee and for their comments on improving the dissertation.  Additional improvements have come through critical reviews of an earlier version of the
manuscript by Po Kin Leung and Ruben Krasnopolsky.

The completion of this dissertation would not have proceeded as smoothly as it
has were it not for the constant and loyal support of my wife, Amy Banner Dau Johnson.
She has helped me in countless ways, and has admirably accepted the sacrifices
associated with living on a graduate-student salary.  The blessing of a family,
including my three children Luke, Emily and Oliver (who could care less about the
salary), is one of the primary things that makes my work worthwhile.

The funding for this work has come from the National Science Foundation
(grants AST 00-03091 and PHY 02-05155), the National Aeronautics and Space Administration
(grant NAG 5-9180) and a Drickamer Research Fellowship from the Department of Physics at
the University of Illinois.

Above all, I wish to acknowledge the Lord Jesus Christ as the Giver of all these
good things, and as the Creator and Sustainer of the heavens that declare His glory,
the fascinating study of which He has enabled me to pursue.

\end{spacing}


\tableofcontents




\chapter{List of Figures}\newpage
\addtocounter{page}{-1}
\listoffigures

\chapter{List of Tables}\newpage
\addtocounter{page}{-1}
\listoftables

\mainmatter

\chapter{Introduction}\label{intro}

\begin{spacing}{1.5}

Accretion disks form around gravitating objects because the angular momentum
of the infalling gas is conserved.  In order for the gas to accrete, however,
its angular momentum must be removed.  Understanding the mechanism underlying this
angular momentum transport is a long outstanding puzzle in accretion disk
theory.  Much progress has been made in recent decades through the realization
that a weak magnetic field will destabilize the flow in an ionized disk and result in a
turbulent transport of angular momentum.  One of the key features of this mechanism,
however, is that the gas in the disk must be sufficiently ionized to couple to the
magnetic field.  It is likely that there are portions of disks, and perhaps entire
classes of disks, in which the ionization is too low for the gas to destabilize.  The
search for a turbulent transport mechanism in weakly-ionized disks continues, therefore,
to this day, in an effort to place our understanding of the evolution of these disks on
as firm a theoretical footing as that for ionized disks.  My research, in collaboration
with Charles Gammie, has focused on several possible mechanisms: self-gravity, radial
convection and vortices driving compressive disturbances.  This dissertation summarizes
our main results and discusses their relevance to the question of what drives accretion
in weakly-ionized disks.

The purpose of this chapter is to describe all of these ideas in detail and put them
in their astrophysical context.  I begin in \S\ref{intros1} with an overview of
accretion disks: the systems in which they are observed and their general properties.
The importance of angular momentum transport for the evolution of disks is discussed in
\S\ref{intros2}, followed by a more detailed discussion of angular momentum transport
in both ionized and weakly-ionized disks (\S\S\ref{intros3} and \ref{intros4}).  I
give a brief overview in \S\ref{intros5} of the model that is used throughout the
dissertation for analytic and numerical studies.  \S\ref{intros6} looks ahead to
Chapter~\ref{conclusion} in which I summarize and chart a course for future 
work.\footnote{Portions of this chapter will be published in the proceedings of the
Workshop on Chondrites and the Protoplanetary Disk, November 8-11, 2004, Kauai, Hawaii.}

\section{Accretion Disks}\label{intros1}

An accretion disk is a roughly cylindrical distribution of matter in orbit around a
gravitating central object.  It is supported against the gravitational pull of its
host object primarily by the centrifugal forces arising from the angular momentum of
the orbiting material.  This support is slightly compromised, however, by the presence of
dissipation or the application of external torques.  As a result, angular momentum is
redistributed through the disk and some of the disk material falls onto the central
object, i.e., it accretes.  The gravitational potential energy lost during this process
is typically converted into radiation, which is the basis for our observations of
accretion disk systems.  Although disks are rarely observed directly (i.e., by being
resolved in a telescope image), they imprint a unique signature on the spectra of their
host systems.  In addition, the accretion-disk paradigm easily accounts for many properties
of astrophysical systems that would be difficult to explain otherwise.

The material in an accretion disk covers a wide range of density and temperature
scales \citep{bh98}.  In most cases, collisional mean free paths are extremely short
compared to the length scales of interest, and mean times between collisions are short
compared to the time scales of interest.  Disks are therefore usually modeled as a
continuous fluid, using the macroscopic equations of gas dynamics rather than the
microscopic equations of kinetic theory.  If the fluid is ionized, it is referred to
as a plasma.

\begin{figure}[h]
\centering
\includegraphics[width=6.0in,clip]{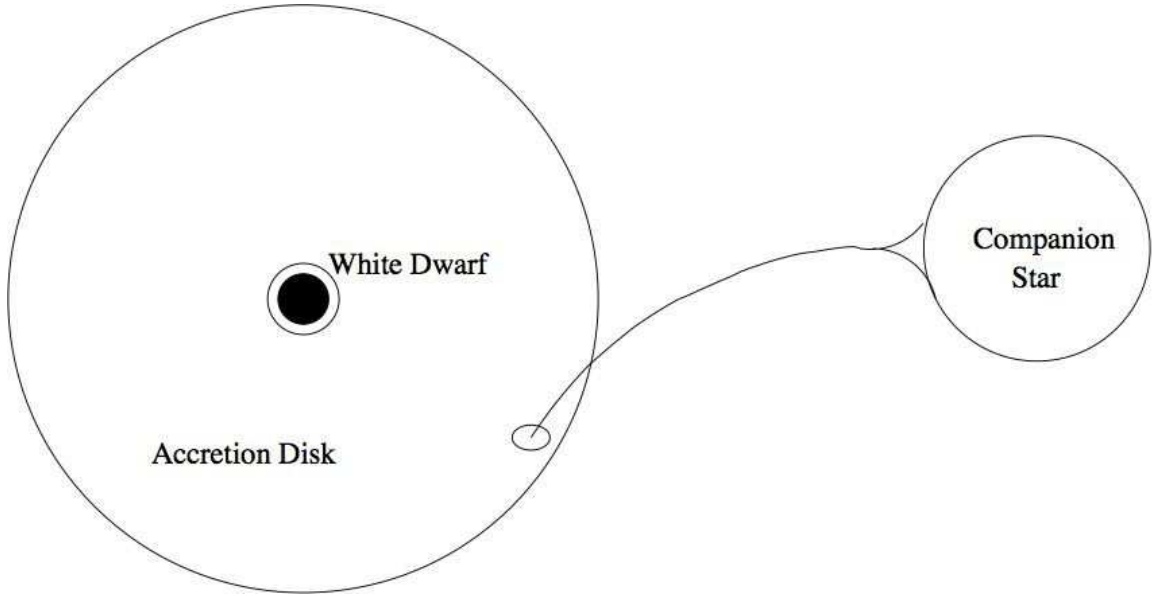}
\caption{Schematic of a cataclysmic variable system.}
\label{introf1}
\end{figure}

Accretion disks are found in a variety of astrophysical settings,
including compact binary systems (with a white dwarf, black
hole, or neutron star), active galactic nuclei (AGN), and young stars.
AGN consist of a supermassive black hole ($M \sim 10^8 \msun$) surrounded
by an accretion flow.  The presence of a disk in AGN is inferred
primarily from the large luminosities of these systems, luminosities
that cannot be accounted for by stellar nuclear burning
but are easily provided by the large gravitational energy
of the compact object.  A spectacular exception to this indirect verification
is the system NGC4258, an AGN in which the disk is directly observed via
maser emission \citep{ww94}.  Low Mass and High Mass X-Ray Binary (LMXRB and HMXRB)
systems consist of a neutron star (NS) or black hole (BH) accreting matter from its companion;
Cataclysmic Variables (CV) are binary systems in which the accreting object
is a white dwarf (see Figure~\ref{introf1}).  The variability observed in these
systems can be accounted for by models in which instabilities in a disk surrounding
the compact object give rise to episodic accretion.  Young Stellar Objects (YSO)
are pre-main-sequence stars with a circumstellar disk, also known as protoplanetary
disks since they are thought to be the sites of planet formation.  The presence of a
disk in these systems is confirmed in many cases by direct observation (e.g., 
\citealt{burr96}).

\begin{deluxetable}{lccccccccc}
\tablecolumns{10}
\tablewidth{0pc}
\tablecaption{Example Accretion-Disk Systems \label{introt1}}
\tablehead{System & Type & $\frac{M}{\msun}$ & $\frac{\mdot}{\msun
    \yr^{-1}}$ & $\frac{L_{acc}}{\lsun}$ & $R_c$ (cm) & $T$ (K)
    & $\frac{H}{R_c}$ }
\startdata
 NGC 4258 & AGN & $4 \times 10^7$ & $1 \times 10^{-2}$ & $1 \times 10^{10}$ &
    $1 \times 10^{18}$ & $7 \times 10^2$ & $2 \times 10^{-3}$ \\
 Sco X-1 & LMXRB-NS & $1$ & $1 \times 10^{-8}$ & $3 \times 10^{4}$ & $1 \times 10^{10}$
    & $3 \times 10^{5}$ & $5 \times 10^{-2}$ \\
 LMC X-3 & HMXRB-BH & $10$ & $1 \times 10^{-8}$ & $1 \times 10^{4}$ & $1 \times 10^{11}$
    & $7 \times 10^{4}$ & $3 \times 10^{-2}$ \\
 UX Uma & CV & $0.5$ & $1 \times 10^{-8}$ & $1 \times 10^1$ & $1 \times 10^{10}$ &
    $2 \times 10^{5}$ & $6 \times 10^{-2}$ \\
 TW Hydrae & YSO & $0.7$ & $5 \times 10^{-10}$ & $1 \times 10^{-2}$ & $1 \times 10^{13}$
    & $1 \times 10^2$ & $2 \times 10^{-2}$ \\
\enddata
\end{deluxetable}

Table~\ref{introt1} lists typical values for various physical properties of these systems.
The mass of the accreting object is denoted by $M$ in Table~\ref{introt1}, in
units of the solar mass ($\msun = 2.0 \times 10^{33}$ g), and the accretion rate
by $\mdot$.  The accretion luminosity
\be
L_{acc} \equiv \frac{GM\mdot}{R_{in}}
\ee
(where $R_{in}$
is the inner radius of the disk and where $G = 6.673 \times 10^{-8} \cm^3 \gm^{-1}
\s^{-2}$ is the gravitation constant) is given in units of the solar luminosity
($\lsun = 3.9 \times 10^{33} \erg \s^{-1}$).  $R_c$ is a characteristic radius for
the accretion disk, and $T$ is a characteristic temperature.  The final column in
Table~\ref{introt1} contains the ratio of the vertical scale height $H$ to the local radius.
Here
\be
H \equiv \frac{c_s}{\Omega_K},
\ee
where $c_s$ is the isothermal sound speed and
\be
\Omega_K = \sqrt{\frac{GM}{r^3}}
\ee
is the local Keplerian orbital frequency.  The values for $T$ and $H$ in
Table~\ref{introt1} are obtained from the standard $\alpha$-disk model, a
derivation of which is outlined in the following section, using $\alpha = 0.01$.
Values for the other quantities in Table~\ref{introt1} were obtained from the
literature (NGC4258: \citealt{miy95}, \citealt{gnb99}; Sco X-1: \citealt{vea91};
LMC X-3: \citealt{pac83}, \citealt{vanp96}; UX Uma: \citealt{fea81}; TW Hydrae:
\citealt{mea00}, \citealt{wea03}).  The standard disk model assumes that the internal
energy of the disk material is efficiently radiated from the surfaces of the disk.
One implication of this assumption is that the disk is typically quite thin, as can
be seen from the last column of Table~\ref{introt1}.  In addition, if the mass of the
disk is much less than the mass of the central star, the orbital frequency of the gas
$\Omega = \Omega_K + O(H/r)^2$, so thin, low-mass disks have a nearly-Keplerian
rotation profile.

\section{Angular Momentum Transport and Disk Evolution}\label{intros2}

The accretion process consists of a net inward transport of matter and a
net outward transport of angular momentum; a small fraction of the matter
carries angular momentum outward, enabling the bulk of the matter to accrete.
This angular momentum transport can take place 1) internally via a local exchange of
momentum between fluid elements at adjacent radii or 2) externally via
a global mechanism such as the removal of angular momentum by a wind off
the surface of the disk (e.g. \citealt{bp82}).  The focus of this dissertation
is on the former: the internal diffusion of angular momentum.  While global
mechanisms such as winds and jets are certain to play a role in many accretion
systems, their operation likely depends in a complicated manner upon the details
of each particular system.  Standard disk modeling typically ignores their effects 
and assumes that angular momentum is transported internally (e.g., \citealt{prin81,rp91,
sm94,nkh94,step97,gam99}).  Whether or not it is possible to isolate this aspect of
disk evolution and get meaningful results is a question that can only be answered 
as more comprehensive disk models are developed.

As an example of the importance of global effects, as well as some of the
difficulties involved in modeling them, consider the torques from
MHD winds (axisymmetric pressure-driven winds have zero torque).
Outflows are widely observed from YSO, and it is likely that the
outflows are magnetically driven.  Outflows are more common in young,
high-accretion-rate systems.  Highly uncertain estimates for the
mass loss rate suggest that about $10\%$ of the accreted mass goes
into the jet and associated outflow.  What is even less certain is the
amount of angular momentum in the outflows, and therefore the role that
they play in the evolution of disks on large scales (as opposed to the
disk at radii less than a few tenths of an AU).  Wind models exist
(e.g., \citealt{bp82,shu94,shu00,wk93}), but there are large gaps in our
understanding.  We do not know what the strength of the mean vertical
magnetic field, which organizes the wind, ought to be, nor how that mean
field is transported radially through the disk, nor how the wind evolves
in time.  A nice summary of this situation is given in \cite{kp00}.

Molecular shear viscosity is a natural mechanism for coupling fluid elements locally
and transporting angular momentum internally, but the large Reynolds
numbers of astrophysical flows (due to the large length scales involved) imply
molecular shear viscosities that are much too tiny to account for the observations.
The coupling due to molecular viscosity is simply too weak to explain the rapid
variability and accretion rates that are observed.  For example, the outburst duration
in CV ranges from $2-20$ days, with the interval between outbursts ranging from tens
of days to tens of years \citep{warn95}.  The timescale for viscous diffusion over a
distance $l$ due to molecular viscosity is $l^2/\nu_m$, where $\nu_m$ is the molecular
(or kinematic) shear viscosity.  Using a value $\nu_m = 10^5 \cm \s^{-1}$ and a
characteristic distance $l = 10^{10} \cm$ (see \citealt{bal03} and Table~\ref{introt1})
yields a viscous timescale of about $10^7$ years, which is orders of magnitude too large
to account for the timescales of CV outbursts.

Standard disk modeling circumvents this problem by assuming the presence of an
enhanced ``anomalous viscosity'' due to turbulence.  The large Reynolds numbers
are used to advantage, since our experience with laboratory flows indicates
that the onset of turbulence typically occurs above a critical Reynolds number.
The {\it assumption} of turbulent flow, along with the picture of turbulent eddies
exchanging momentum with one another to drive accretion, underlies the majority of the
phenomenological disk modeling that is currently used to explain observations.

The construction of a standard model for disk evolution proceeds as follows.  We
will use a cylindrical coordinate system centered on the accreting object, with $r$, $\phi$ and $z$ the radial,
azimuthal and vertical coordinates, respectively.  We start with mass conservation:
\begin{equation}
\frac{\partial \rho}{\partial t}
=
-\frac{1}{r} \partial_r (r \rho v_r)
- \frac{\partial}{\partial z} (\rho v_z)
\end{equation}
where $\rho$ is the mass density and we
assume axisymmetry ($\partial/\partial\phi = 0$), on average.  Integrating this
equation vertically through the disk gives
\begin{equation}\label{MASSCON}
\frac{\partial \Sigma}{\partial t}
=
-\frac{1}{r} \partial_r (r \Sigma \bar{v}_r) + \dot{\Sigma}_{ext}
\end{equation}
where $\Sigma = \int dz \rho$ is the surface density,
$\bar{v}_r$ is a vertical average of the radial velocity, and
$\dot{\Sigma}_{ext}$ is the difference of $-\rho v_z$ evaluated at the
upper and lower surface of the disk.  It includes infall onto the disk,
mass loss in winds, and mass loss through photoevaporation.  It is
positive when mass flows into the disk, and negative when mass flows
out.

The problem now is to find the radial velocity $\bar{v}_r$, which we can do
using angular momentum conservation:
\begin{equation}
\frac{\partial (\rho l)}{\partial t}
=
-\frac{1}{r} \frac{\partial}{\partial r} (r^2 \Pi_{r\phi})
- \frac{\partial}{\partial z} (r \Pi_{z \phi}).
\end{equation}
Here $\rho l$ is evidently the local density of angular momentum,
and the right hand side of the equation is the divergence of an angular
momentum flux density.  $\Pi_{r\phi}$ is a component of the stress tensor,
sometimes referred to as the shear stress, with dimensions of pressure;
it is the flux density of $\phi$ momentum in the $r$ direction.  Likewise
$\Pi_{z\phi}$ is the flux density of $\phi$ momentum in the $z$ direction.
Again integrating vertically,
\begin{equation}\label{AMCON}
\frac{\partial (\Sigma l)}{\partial t}
=
-\frac{1}{r} \frac{\partial}{\partial r} (r^2 W_{r\phi} + r \Sigma \bar{v}_r l)
+ \tau + l \dot{\Sigma}_{ext}.
\end{equation}
Here
\begin{equation}
W_{r\phi} \equiv \int dz \Pi_{r\phi} - r \Sigma \bar{v}_r l
\end{equation}
is the integrated shear stress, but with one piece of it, proportional to
the radial mass flux, peeled off.  In models which assume that angular momentum
transport is due to turbulence, $W_{r\phi}$ is referred to as the ``turbulent
shear stress.''  The external torque
\begin{equation}
\tau \equiv -r \Pi_{z\phi}|_{\rm lower}^{\rm upper} - l \dot{\Sigma}_{ext}
\end{equation}
is the angular momentum flux into the upper and lower surface of the
disk with one piece, proportional to the mass flux into the disk, peeled
off.  $\tau$ includes the effects of, e.g., MHD winds; it is positive
when angular momentum flows into the disk and negative when angular
momentum flows out.  In a steady state ($\partial/\partial t = 0$), the 
condition $W_{r\phi} > 0$ must be met for an outward transport of angular 
momentum and inward accretion.

The mass and angular momentum conservation equations (\ref{MASSCON}) and
(\ref{AMCON}) can be combined into a single equation governing the
evolution of thin, Keplerian disks (multiply the continuity equation by
$r v_\phi$, subtract the angular momentum equation, solve for $\bar{v}_r$,
substitute back into the continuity equation):
\begin{equation}\label{BASIC}
\frac{\partial \Sigma}{\partial t}
=
\frac{2}{r}
\frac{\partial}{\partial r} \left(
\frac{1}{r\Omega}
\frac{\partial}{\partial r} ( r^2 W_{r\phi} )
\, - \, \frac{\tau}{\Omega} \right)
\, + \, \dot{\Sigma}_{ext}.
\end{equation}

If we assume that the shear stress $W_{r\phi}$ is due to an
anomalous viscosity $\nu$, and that the external torques and mass
loss/infall are negligible, equation (\ref{BASIC}) becomes the
basic equation for standard disk modeling:
\begin{equation}\label{BASICPRIME}
\frac{\partial \Sigma}{\partial t} =
\frac{3}{r}
\frac{\partial}{\partial r} \left(
\frac{1}{r\Omega}
\frac{\partial}{\partial r} ( r^2 \Sigma \nu \Omega) \right).
\end{equation}
In a steady state one can show that the accretion rate (inward mass flux
$= -2 \pi \Sigma r \bar{v}_r$) is given by
\be
\dot{M} = 3 \pi \Sigma \nu.
\ee

In addition to setting $\tau$ and $\dot{\Sigma}_{ext}$ to zero, 
standard disk theory usually sets $\nu = \alpha c_s
H$, which parameterizes our ignorance of $W_{r\phi}$.  If one
reasonably assumes that the turbulent stress (an off-diagonal component
of the stress tensor) must be associated with a pressure (an isotropic,
diagonal component of the stress tensor), then $\alpha \lesssim 1$.
Most disk evolution models take $\alpha = const.$, or allow it to assume
a few discrete values.  For a disk around a solar-type star
with a temperature of $300{\rm\,K}$ at $1$ AU, this yields $\dot{M} =
9.9 \times 10^{-9} \alpha \Sigma {\rm\,M_\odot} {\rm\,yr}^{-1}$, or
$\sim 10^{-8} {\rm\,M_\odot} {\rm\,yr}^{-1}$ for $\alpha = 0.01$ and
$\Sigma = 10^2 {\rm\,g} {\rm\,cm}^{-2}$, roughly consistent with
observed accretion rates \citep[e.g.][]{ghbc98}.

\section{Angular Momentum Transport in Ionized Disks}\label{intros3}

While phenomenological disk modeling can proceed along its merry way
without a clear demonstration of the onset of turbulence in accretion
disk flows, recent progress towards a first-principles
understanding of disk turbulence has raised the possibility that more
physically-motivated disk models can be developed.  The primary
breakthrough in our understanding came with the realization that
the presence of a weak magnetic field destabilizes a disk on a dynamical
time scale, resulting in the onset of magnetohydrodynamic (MHD) turbulence and
a vigorous outward flux of angular momentum \citep{bh91,bh98,bal03}.

This section gives a summary of the essential physics of this instability,
generally termed the magneto-rotational instability, or MRI.  The importance
of ionization for the successful operation of the MRI in driving turbulence
is also discussed, which leads in the following section to an overview of the
main question addressed by this dissertation: what drives accretion in disks
that are too weakly ionized to be unstable to the MRI?

\subsection{Magneto-Rotational Instability}\label{MRI}

\begin{figure}[h]
\centering
\includegraphics[width=6.5in,clip]{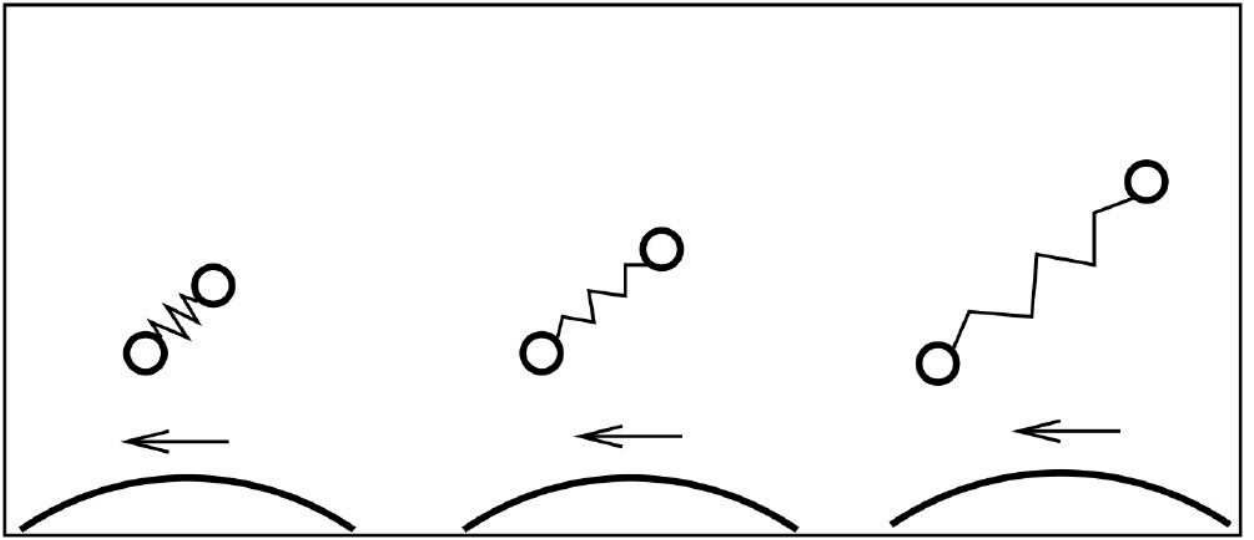}
\caption[Mechanical analogy for the MRI.]{
Mechanical analogy for the MRI.  Two masses, orbiting in the direction
indicated by the arrow about a massive body (bottom of the frame), are
connected by a spring.  The outer mass has higher angular momentum and
lower orbital frequency.  The lower mass is pulled back in its orbit by
the spring, reducing its angular momentum.  It sinks to a lower orbit,
where it orbits faster, stretching the string and increasing the torque.
A runaway ensues.  The masses may be thought of as fluid elements
connected by a magnetic field.}
\label{introf2}
\end{figure}

The MRI grows directly through exchange of angular momentum between
radially-separated fluid elements.  This can be understood with a simple
mechanical analogy introduced by \cite{bh92} and illustrated in
Figure~\ref{introf2}.

Imagine that two small masses orbit with frequency $\Omega(r)$ about a
third, massive body.  The masses are coupled by a spring; the natural
frequency of the spring-mass system is $\gamma$.  If $\gamma \gg
\Omega$, the bodies behave like a perturbed harmonic oscillator.  But if
$\gamma$ is lowered until $\gamma \sim \Omega$ the orbital motion of the
bodies begins to influence the dynamics, and something interesting
happens.  The outer mass has higher angular momentum but lower orbital
frequency.  It is pulled forward in its orbit by the spring; its angular
momentum increases, moves to a higher orbit, and lowers its orbital
frequency further.  This stretches the spring, increases the rate at
which the outer body gains angular momentum, and a runaway ensues.  The
outer body heads outward, acquiring angular momentum from the inner
body.  The inner body moves in a mirror image of the outer body as it
loses angular momentum and falls inward.

There is an exact correspondence between the modes of the spring-mass
model and the MRI.  One can think of the masses as fluid elements and the
spring as magnetic field.  The field has characteristic frequency
$\gamma \sim v_A/\lambda$, where $\lambda$ is the separation of the
masses and $v_A = B/\sqrt{4\pi\rho}$ is the Alfv\'en speed.
$v_A/\lambda \lesssim \Omega$ implies instability.

The simplicity of the dynamics shown in Figure~\ref{introf2} suggests that the MRI
is robust.  Instability requires the presence of a weak (subthermal,
i.e. $B^2/(8 \pi \rho) \lesssim c_s^2$) magnetic field.  A stronger
magnetic field seems unlikely (it would likely be ejected from the disk
by magnetic buoyancy), but if it were present it would likely be
associated with other, even more powerful instabilities.  The MRI also
requires an angular velocity that decreases outward ($d\Omega^2/d\ln r <
0$), which is always satisfied in Keplerian disks ($\Omega \propto
r^{-3/2}$).  The maximum growth rate is $\sim \Omega$, independent of
the field strength (unless diffusive effects are present).  This
surprising fact is easily understood once one realizes that the scale
$\lambda$ of the instability decreases with the field strength: $\lambda
\sim v_A/\Omega$.

What does the MRI tell us about $W_{r\phi}$, the key quantity in the
evolution equation?  Numerical integration of the compressible MHD
equations shows that the linear MRI initiates nonlinear turbulence.
In the turbulent state one can measure the average value of
\begin{equation}
W_{r\phi} = \int dz \left( \rho v_r \delta v_\phi
- \frac{B_r B_\phi}{4\pi} \right)
\end{equation}
where $\delta v_\phi$ is the noncircular azimuthal component of the
velocity.  There are two distinct contributors to the shear stress:
hydrodynamic velocity fluctuations, and magnetic field fluctuations,
sometimes referred to as the Reynolds and Maxwell stresses.  Then $\alpha
= const. \times W_{r\phi}/(\Sigma c_s^2)$ (the constant is a matter of
convention and takes several values in the literature).  Local
numerical models yield $\alpha \approx 0.01$ \citep{hgb95,mt95,hgb96,
mmts97}.  Global numerical models yield similar but slightly larger
values \citep[e.g.][]{arm98,haw00,mhm00,ar01}.

The presence of turbulence does not guarantee the $W_{r\phi} > 0$
necessary for outward angular momentum transport and accretion.  For
example, the turbulence associated with vertical convection can produce
$\alpha < 0$ \citep{sb96,cab96}.  Also, vortices in nearly incompressible
disks produce $\alpha \approx 0$ (see Chapter~\ref{paper3}).  The fact
that MRI-driven turbulence has $W_{r\phi} > 0$ is thus a nontrivial result,
although one that might have been anticipated because of the central
role of angular momentum exchange in driving the linear MRI.

\subsection{MRI in Low-Ionization Disks}\label{MRILowI}

MRI-generated MHD turbulence is the likely angular momentum transport 
mechanism in AGN and in binary systems during outbursts.  In portions of 
YSO disks, however, as well as in CV disks and X-Ray transients
in quiescence \citep{sgbh00,gm98,men00}, the plasma is cool
and nearly neutral.  The conductivity of the gas is small by
astrophysical standards and the field is no longer frozen into the gas.
In some regions the field may be completely decoupled from the fluid,
just as the Earth's lower atmosphere is decoupled from the Earth's
magnetic field.

Low ionization levels change the field evolution through three separate
effects: Ohmic diffusion, Hall drift, and ambipolar diffusion.
Following the beautiful discussion of \cite{bt01} \citep[see also][]
{war99,des04}, a measure of the correction to the field evolution
equation (\ref{INDUC}) due to Ohmic diffusion is given by the magnetic
Reynolds number $Re_M^{-1}
\equiv \eta/(v_A H)$, where $\eta = c^2/(4\pi\sigma_e)$ is the
resistivity, $c$ is the speed of light and the conductivity $\sigma_e$
is proportional to the collision timescale for electrons with neutrals.  Then
\begin{equation}
Re_M \simeq 2 \times 10^{19} B \left(\frac{r}{{\rm\,AU}}\right)^{3/2}
\left(\frac{n_e}{n}\right) \left(\frac{M_*}{{\rm\,M_\odot}}\right)^{-1/2}n^{-1/2}
\end{equation}
Here $n$ is the neutral number density and $n_e$ is the electron number
density.  Ohmic diffusion destroys flux (via reconnection) and converts
magnetic energy to thermal energy.

Hall drift can be thought of as arising from the relative mean motion of
the electrons and ions.  The associated correction to the field
evolution equation is, for conditions appropriate to circumstellar
disks, typically comparable to Ohmic diffusion.  Hall drift does not
change the magnetic energy.

Ambipolar diffusion arises from the relative mean motion of the
ions and neutrals.  Unlike Hall drift, it converts magnetic energy
to thermal energy.  The ratio of the ambipolar to Ohmic term $\sim 5
\times 10^{28} B^2 T^{-1/2} n^{-2}$ \citep[all cgs units; we have assumed
that the number densities of electrons and ions are equal, but see][]{des04}.
In low-density environments such as
the clouds from which young stars condense, ambipolar diffusion is the
dominant nonideal effect; one can then think of the field as being
locked into the ion-electron fluid, which gradually diffuses through the
neutrals.  At higher densities ambipolar diffusion becomes less dominant,
and at the highest densities found in disks Ohmic diffusion dominates.
Evidently the precise variation of the relative importance of these effects
with location depends on the variation of ionization fraction, temperature,
and field strength.  In YSO disks, ambipolar diffusion tends to dominate in the
disk atmosphere and at $r \gtrsim 10$ AU.

The linear theory of the MRI with Ohmic diffusion was first considered
by \cite{jin96}.  Stability is recovered when the Ohmic diffusion rate
$\eta/\lambda^2$ exceeds the growth rate of the instability
$v_A/\lambda$.  Since $\lambda \lesssim H$, this occurs when $\eta/(v_A
H) \sim 1$.  One can avoid expressing the stability condition in terms
of the unknown field strength $B$, which likely arises through dynamo
action induced by the MRI, by noting that for a subthermal field $c_s >
v_A$.  Then when $\eta/(c_s H) > 1$ the disk is stable, although the MRI
may be suppressed at even lower $\eta$.  Circumstellar disks have
$\eta/(c_s H) > 1$, and are therefore stable to the MRI, over a large
range in radius \citep[e.g.][]{gam96}.

The linear theory of the MRI with ambipolar diffusion was first treated
by \cite{bb94}, and reconsidered more recently by
\cite{des04,kb04,sw03,sal04}.  The natural expectation is that the MRI
develops even in the presence of ambipolar diffusion as long as a
neutral particle manages to collide with an ion (which can see the
magnetic field) at least once per orbit.  Assuming common values for the
collision strengths and ion mass \citep[e.g.][]{bt01}, this condition
becomes
\begin{equation}
A \equiv 0.01 \, n_e \, (r/AU)^{3/2} (M/{\rm\,M_\odot})^{-1/2}
\gtrsim 1.
\end{equation}
The more recent round of papers points out that even when $A < 1$ there
are unstable perturbations within a band of wavevectors
${\mbox{\boldmath $k$}}$ outside the purely axial wavevectors considered
by \cite{bb94}.  These perturbations evade ambipolar damping by
orienting their magnetic field perturbations perpendicular to both
${\mbox{\boldmath $B$}}$ and ${\mbox{\boldmath $k$}}$
\citep[see][]{des04}.  Instability can thus survive when $A < 1$, albeit
in a narrow band of wavevectors and at greatly reduced growth rates.

The linear theory of the MRI with Hall drift has been considered by
\cite{war99,bt01,des04}.  There are always perturbations that become
more unstable as Hall drift is turned on.  The maximum growth rate is
not affected.  The combination of all these nonideal effects, together
with a best guess for the disk ionization structure is discussed in
\cite{sw03,sal04,des04}.

Early numerical experiments by \cite{hgb95} using a scalar resistivity,
no explicit viscosity, no Hall drift and no ambipolar diffusion
suggested that the MRI dynamo fails when $c_s H/\eta \lesssim 10^4$.  A
similar but more thorough study by \cite{fsh00} found similar results.
\cite{ss02a,ss02b} considered models that incorporated Ohmic diffusion
and Hall drift, but not ambipolar diffusion (relevant in some regions of
the disk).  The most relevant of Sano \& Stone's models are probably
those with net toroidal field or zero net field.  Their results
\citep[see Figs. 14 and 19 of][]{ss02b} suggest that Ohmic diffusion is
the governing nonideal effect; $\alpha$ drops sharply when $c_s H/\eta <
3 \times 10^3$, and is only weakly dependent on the Hall parameter.

\section{Angular Momentum Transport in Weakly-Ionized Disks}\label{intros4}

Accretion rates inferred from observations of weakly-ionized disks indicate
that an enhanced viscosity or some other mechanism for angular momentum transport
is operating in these disks.  As discussed in the previous section, however, these same
disks are likely to be stable to the MRI.  This leaves open the question of what
generates turbulence in weakly-ionized disks.  As long as there was no firm
theoretical understanding of the onset of turbulence in disks (ionized or not),
it was reasonable to assume that all disks are turbulent due to their large
Reynolds numbers.  Laboratory shear flows are turbulent above a critical
Reynolds number even though they are stable to infinitesimal perturbations
(i.e., there is no linear instability to trigger the turbulence), and the
extrapolation to astrophysical shear flows was a natural one to make.  With the
establishment of a robust transport mechanism in MRI-induced MHD turbulence,
this assumption has come under critical scrutiny.  If a mechanism can be
established from first principles for ionized disks, it seems reasonable to
maintain the same standard for disks which are too weakly ionized to be
MHD-turbulent.  Although many attempts have been made, to date no robust transport
mechanism akin to the MRI has been established for low-ionization disks.  This
dissertation investigates three possible mechanisms in detail: gravitational
instability, convection, and vortices driving compressive motions.  Each of these
mechanisms is summarized briefly here and discussed in detail in the remainder of
the dissertation.  Since there are those who continue to argue for turbulence in
disks by way of analogy with laboratory shear flows, a brief overview is also given
of the current state of this controversy.

\subsection{Gravitational Instability}

Gravitational instability arises when self-gravity in the disk overcomes
the stabilizing influences of pressure and rotation.  The nonlinear
outcome of this instability is either a gravito-turbulent state of
marginal stability or fragmentation of the fluid into bound clumps.
If a sustained gravito-turbulent state can be established, then steady
outward angular momentum transport ensues.  Instability tends to form temporary
spiral enhancements in the density with a trailing orientation,
and gravity then carries angular momentum along the spiral
(there is a new term in $W_{r\phi}$, a ``Newton'' stress given by $\int
\, dz \, g_r g_\phi/(8\pi G)$, where $g_r$ and $g_\phi$ are components of the gravitational field).  Local numerical experiments exhibit a
gravitational $\alpha$ up to $\sim 1$ \citep{gam01}.  If this state can
be maintained steadily throughout the disk, it would provide an effective
turbulent transport mechanism in weakly-ionized disks \citep{pac78}.

A sustained gravito-turbulent state cannot be established, however, if
the cooling (due to radiation from the surfaces of the disk) is too strong.
Then the clumps of matter formed by the
instability cool before they can collide and heat each other via shocks.
Fragmentation-- the formation of small, bound clumps-- results.
The mean cooling time can be used to distinguish a gravito-turbulent disk
from a fragmenting disk:
\begin{equation}\label{COOLT}
\tau_c \equiv \frac{\langle U \rangle}{\langle \Lambda \rangle}.
\end{equation}
where $U = \int dz u$ and $u$ is the internal energy per unit volume,
and $\Lambda$ is the cooling function.  The brackets $\langle\rangle$
indicate an average over space and time, since the disk may have
nonuniform density and temperature.

Chapter~\ref{paper1} discusses in detail local numerical experiments
which show that fragmentation occurs when $\tau_c \Omega \lesssim 1$.\footnote{
This result has been been demonstrated in other numerical experiments as well, both
local and global \citep{gam01,rabb03}.}  These experiments also show
that fragmentation occurs for a wide range of parameters, indicating that a
gravito-turbulent state is difficult to sustain.  In addition, cooling
typically becomes more efficient with an increase in disk radius, making
an extended, marginally-stable region unlikely.  Gravitational instability thus
does not appear to be a likely candidate for a turbulent transport mechanism
in weakly-ionized disks.

\subsection{Convection}

As mentioned in \S\ref{MRI}, vertical convection appears to transport angular
momentum {\it inward}, opposite to what is usually required for accretion.
Indeed, arguments have been made that any incompressive disturbance or
incompressible turbulence (of which convection is just one example) will
drive angular momentum in the wrong direction \citep{bal00,bal03}.

The possible role of radial convection in driving angular momentum
transport has come to the fore recently with the work of \cite{kb03} and
\cite{klr04} on the ``Global Baroclinic Instability''.  One would expect
that the combination of weak radial gradients and strong Keplerian shear
in circumstellar disks would preclude any instabilities due to radial
convection, yet \cite{kb03} found turbulence and angular momentum
transport in global hydrodynamic simulations with a modest radial
equilibrium entropy gradient.  The claim in \cite{klr04} is that this
activity, which grows on a dynamical time scale, is the result of a {\it
local} hydrodynamic instability due to the presence of the global
entropy gradient.

Chapters~\ref{paper2} and \ref{paper4} describe analytic and numerical
work in a local model that attempts to confirm or refute these unexpected results.
Chapter~\ref{paper2} describes a local stability analysis in radially-stratified
disks, an analysis which uncovers no exponentially-growing instabilities for
disks with a Keplerian rotation profile.
Chapter~\ref{paper4} describes local numerical experiments which attempt
to uncover any nonlinear instabilities that may be present.  Disks with
Keplerian shear are again found to be stable.  It appears, therefore, that
the ``Global Baroclinic Instability'' claimed by \cite{kb03} is either
global or nonexistent.

\subsection{Vortices}

The absence of a robust instability mechanism for generating
hydrodynamic turbulence does not necessarily imply the absence of
internal angular momentum transport.  Chapter~\ref{paper3} describes
numerical experiments which show significant
shear stresses associated with finite-amplitude vortices that emit
compressive waves and shocks.  In these experiments, an initial field of random
velocity perturbations with Mach number $\sim 1$ forms anticyclonic
vortices that provide an outward flux of angular momentum corresponding
to an initial $\alpha \sim 0.001$ and decaying as $t^{-1/2}$.  These results were
obtained in a two-dimensional local model, and are likely to be modified
considerably by three-dimensional instabilities, which tend to destroy
two-dimensional vortices \citep{ker02,bm05}.  In addition, these results
leave open the key question of what generates the initial vorticity.
Both of these issues are discussed in more detail in Chapter~\ref{paper3}.

\subsection{Nonlinear Hydrodynamic Instability}

For some, the lack of a well-established mechanism for generating turbulence in
weakly-ionized disks does not necessarily imply the absence of turbulence (e.g.,
\citealt{rz99}).  In the first place, our understanding of the onset of turbulence
in simple laboratory shear flows is still incomplete, despite over a century of
theoretical effort.  Even when linear theory predicts stability of these flows at
all values of the Reynolds number, experiments consistently show the onset of
turbulence above a critical Reynolds number.  The failure of linear theory to predict
the outcome of experiments indicates that nonlinear instabilities (i.e., instabilities
due to finite-amplitude disturbances) are the likely source of turbulence in these
flows.  Perhaps an analogous mechanism operates in weakly-ionized disks.  In addition,
since nonlinear stability is extremely difficult to {\it prove} due to the complexity
of nonlinear dynamics, the question of stability in weakly-ionized disks remains, in
some sense, an open question.  As discussed in this section, however, no nonlinear
instability mechanism has yet been established for a Keplerian shear flow, despite its
apparent similarities with laboratory shear flows.  In addition, the two main features 
that distinguish an accretion disk flow from a laboratory shear flow-- rotational effects
and the absence of rigid boundaries-- seem to argue for the nonlinear {\it stability} of
the former.

The laboratory flow that most closely resembles an accretion disk flow is Couette-Taylor
flow, which is flow between two concentric cylinders.  Early theoretical and
experimental results for this flow were obtained by Rayleigh, Couette and Taylor
(see \citealt{dr81}).  Its {\it linear} stability is governed by the Rayleigh stability
criterion, which states that a necessary and sufficient criterion for stability
to axisymmetric disturbances is that
\be\label{RAY}
\kappa^2 = \frac{1}{r^3}\frac{d}{dr}\left(r^2\Omega(r)\right)^2 \geq 0,
\ee
where $\kappa^2$ is the square of the epicyclic frequency (also known as the Rayleigh 
discriminant).  For a Rayleigh-stable flow, fluid elements displaced from circular orbits
will undergo epicycles about their equilibrium velocity at a frequency $\kappa$.
The stability criterion (\ref{RAY}) is equivalent to the requirement that the specific
angular momentum of the mean flow decrease with radius.

Another laboratory flow that has been studied in depth is planar shear flow, also known
as plane Couette flow \citep{dr81}.  This is flow between two parallel walls, the laminar
state of which is a streamwise (parallel to the walls) velocity that varies
linearly with distance from the walls.  A necessary condition for the {\it linear}
instability of a parallel shear flow with an arbitrary shear profile (i.e., an
equilibrium velocity that is an arbitrary function of the coordinate perpendicular
to the direction of flow) is that there be an inflection point in the equilibrium
velocity profile.  Since a linear shear profile does not meet this condition, plane
Couette flow is also predicted to be stable based upon linear theory.

Both Couette-Taylor flow and plane Couette flow show the onset of turbulence above a critical
Reynolds number, against the predictions of linear theory.  Theoretical efforts to
explain this transition to turbulence have focused on 1) the transient amplification
of linear disturbances coupled with a nonlinear feedback mechanism to close the amplifier
loop (e.g., \citealt{bt97}); 2) self-sustaining nonlinear processes that are triggered at
finite amplitude and are therefore not treatable by a linear analysis (e.g., \citealt{wal97});
or 3) some combination of nonlinear mechanisms and secondary linear instabilities (e.g.,
\citealt{fi93}).  Reviews of these mechanisms can be found in \cite{bo88}, \cite{gross00}
and \cite{rem03}.  All of them include some aspects of the nonlinear dynamics and are
generically referred to as nonlinear instabilities.  While a discussion of their detailed
operation is not necessary for the purposes of this dissertation, it is important to note
that none of them has provided a complete understanding of the transition to turbulence
in laboratory shear flows.

The application of these ideas to accretion disks has continued since the discovery of
the MRI \citep{zahn91,dk92,dk93,dub93,ik01,rich01,rddz01,long02,cztl03,rich03,klr04,amn04,
man04,rd04,yeck04,ur04,hdh05,man05,unrs05}.  Much of this work has focused on the mechanism
of transient amplification of linear disturbances coupled with nonlinear feedback, since
there are local nonaxisymmetric vortical perturbations which can experience an arbitrary
amount of transient growth at infinite Reynolds number (a result that was recognized as early as 1907 by Orr; see \citealt{shep85}).  These solutions are discussed in
detail in Chapter~\ref{paper2}, where it is shown (\S\ref{pap2eis}) that an isotropic
superposition of these perturbations has an energy that is constant with time.  This seems
to indicate that any potential mechanism for the onset of hydrodynamic turbulence in disks
would be an entirely nonlinear process.  Only nonlinear simulations can fully answer this
question, however, and a full investigation of the effects of transient amplification over
a wide range of initial perturbation amplitudes and spectra has not yet been made.  To date,
however, no numerical simulations have demonstrated a transition to turbulence from
infinitesimal perturbations, and the results of \S\ref{pap2eis} indicate that such a
transition may not occur for a physically-realistic set of low-amplitude perturbations.

Early simulations of MHD turbulence in MRI-unstable disks \citep{hgb95,hgb96} found that
1) when the magnetic fields were turned off the turbulence decayed away and 2) when
rotational effects were removed, thereby converting the Keplerian flow into plane Couette
flow, the turbulence increased and the magnetic field decayed away.  While advocates of
nonlinear instabilities in disk flows will often attribute the absence of turbulence in
hydrodynamic simulations to numerical diffusion (e.g., \citealt{long02}), this latter
result confirms the ability of these simulations to identify a nonlinear instability.  In
addition, \cite{bal04} has argued that due to a nonlinear scale invariance of the equations
governing the local disk flow, any local instabilities that are present should be present
at all scales and therefore not require high resolutions for their manifestation in local
numerical simulations.

Two subsequent comprehensive studies of nonlinear instabilities in
local numerical simulations \citep{bhs96,hbw99} have confirmed these results.  Both
Keplerian and plane Couette flows were investigated, using codes with very different
diffusive properties, and rotational effects were cited as the key stabilizing factor in
disks.  One of the mechanisms for nonlinear instability in plane Couette flow is the
generation of streamwise vortices by the shear, resulting in a secondary instability due
to inflections in the spanwise (across the mean flow) direction.  The epicyclic motions
of fluid elements in a rotating flow prevent these streamwise vortices from developing.
When rotational effects are removed, the nonlinear instability of planar shear flow is
readily recovered.

Before the work of \cite{hbw99}, the only Couette-Taylor (rotating-flow) experiments that
showed the onset of turbulence had shear profiles that were sufficiently non-Keplerian to
more closely resemble plane Couette flow than a rotationally-dominated flow (the shear
profiles were near to linear instability, expression [\ref{RAY}]).  More recent experiments,
however, have shown the onset of turbulence in Couette-Taylor flow with a Keplerian
shear profile \citep{rich01,rddz01}, thus indicating the presence of a nonlinear instability.
These results, however, may simply highlight another key difference between laboratory
shear flows and disk flows, namely the presence or absence of rigid boundaries.  As noted in
\cite{go05}, early Couette-Taylor experiments revealed the importance of end effects in
disturbing the laminar flow.  Torque measurements in a system with an aspect ratio of $23$
(the ratio of the cylinder lengths to the width of the gap between the cylinders) and an end
plate corotating with the outer cylinder were $100\%$ larger than measurements with the end
plate stationary.  An aspect ratio $\gtrsim 40$ was required to minimize the end effects.
The experimental setup described in \cite{rddz01} has an aspect ratio of $25$.

\cite{go05} proposed a model for the nonlinear dynamics of turbulent shear flows and also used
their model to predict the onset of linear and nonlinear instability in shear flows both with
and without rotation.  The model accounts for many aspects of laboratory shear flow experiments.
For reasonable model parameters, the model predicts nonlinear {\it stability} for Keplerian
shear flows in the absence of boundaries and nonlinear {\it instability} for a wall-bounded
experiment with a Keplerian shear profile at sufficiently large Reynolds numbers.  This is 
another indication that the results observed by \cite{rddz01} may be due to boundary effects.

\section{Local Model}\label{intros5}

\begin{figure}[ht]
\centering
\includegraphics[width=2.8in,clip]{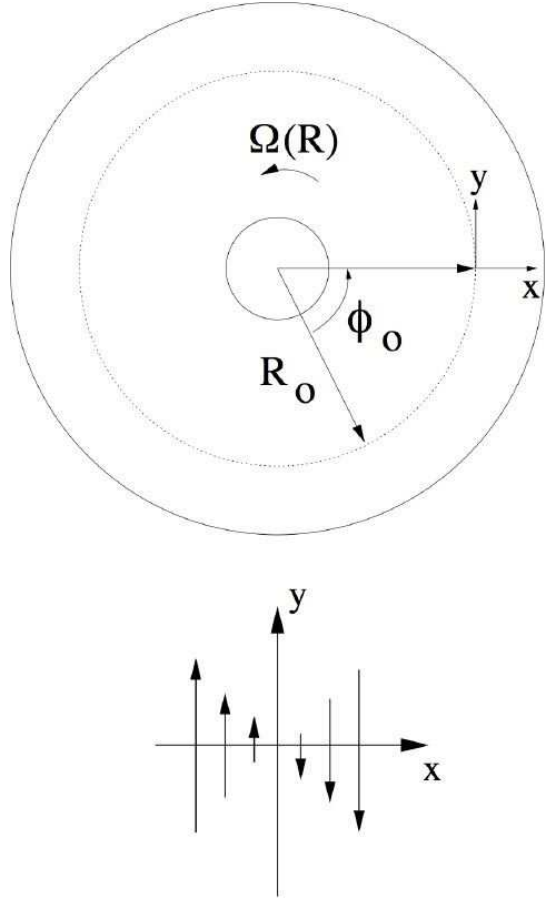}
\caption{Local coordinate system at $t = 0$.}
\label{introf3}
\end{figure}

Since the focus of this dissertation is on local mechanisms for angular momentum transport,
all the analytic and numerical results are obtained in a local model of an accretion disk.
Such a model can be obtained by a rigorous expansion of the fluid equations in $|{\bx}|/r$,
where ${\bx} = (x,y,z) \equiv (r-R_o, R_o(\phi - \phi_o - \Omega(R_o) t),z) \sim O(H)$ are
the local Cartesian coordinates of the fluid with respect to a fiducial radius $R_o$ and
fiducial angle $\phi_o$ (see Figure~\ref{introf3}).  Since the local coordinates are assumed
to vary on the order of the disk scale height $H$, the local model expansion is only valid for
thin disks with $H/r \ll 1$ (see Table~\ref{introt1}).  This local frame is corotating with
the fluid in the disk at a distance $R_o$ from the central object and at a frequency
$\Omega(R_o)$, the local rotation frequency of the disk.  Local curvature is neglected, but
centrifugal and Coriolis forces are retained.  The additional simplifying assumption of an
infinitesimally thin disk is made, which implies a vertical integration of the fluid variables.

The resulting equations of motion for a fluid in the local model (including self-gravity) are
\begin{equation}\label{EQM}
\partial_t \bv + (\bv \cdot \bnabla) \bv = -\frac{1}{\Sigma}\bnabla \left(P + \frac{B^2}{8\pi}
\right) + \frac{({\bB}\cdot \bnabla){\bB}}{4\pi \rho} - \bnabla\phi - 2\bO{\bf \times} \bv +
3 \Omega^2 x \ex,
\end{equation}
where $\bv$ is the fluid velocity with respect to the rotating frame, $B$ is the magnetic field,
$P$ is the two-dimensional pressure, $\Sigma$ is the column density and $\phi$ is the disk
potential with the time-steady axisymmetric component removed. The last two terms on the
right-hand side of equation (\ref{EQM}) incorporate the effects of Coriolis and centrifugal forces as well as the gravitational
acceleration due to the central point mass and the time-steady axisymmetric component of the
disk.  These equations of motions are valid for a disk system in which the gravitational
potential is dominated by the central object; the fluid in such a disk follows a Keplerian
rotation curve, $v_\phi \sim r^{-1/2}$.

The continuity, internal energy and induction equations retain their usual form:
\begin{equation}
\partial_t \Sigma + (\bv \cdot \bnabla) \Sigma + \Sigma \bnabla \cdot \bv = 0,
\end{equation}
\begin{equation}\label{EQE}
\partial_t U + (\bv \cdot \bnabla) U + (U + P) \bnabla \cdot \bv + \Lambda = 0,
\end{equation}
(where $U$ is the internal energy per unit area and $\Lambda$ is the cooling function) and
\begin{equation}\label{INDUC}
\partial_t {\bB} - \bnabla {\bf \times} (\bv {\bf \times} {\bB}) = 0.
\end{equation}
Equations (\ref{EQM}) through (\ref{INDUC}) are the equations of compressible ideal
magnetohydrodynamics (MHD) in the local model.  Magnetic fields are assumed to be dynamically
unimportant for most of research described in this dissertation, in which case equations
(\ref{EQM}) through (\ref{EQE}) reduce to the equations of compressible hydrodynamics.
The disk self-gravity ($\phi$ in equation [\ref{EQM}]) and explicit cooling ($\Lambda$ in
equation [\ref{EQE}]) are neglected except in the work discussed in Chapter~\ref{paper1}.
In Chapters~\ref{paper1} - \ref{paper4}, the fluid is assumed to obey an ideal-gas equation of
state, $P = (\gamma - 1) U$, where $\gamma$ is the adiabatic index.  Chapter~\ref{paper3}
assumes an isothermal equation of state $P = c_s^2 \Sigma$.

With constant density and pressure in equilibrium, an exact steady-state solution to
equations (\ref{EQM}) through (\ref{EQE}) is $\bv_o = -\frac{3}{2} \Omega x \ey$.  This uniform shear
velocity is a manifestation of differential rotation of the fluid in the disk.  As a result, 
the (two-dimensional) local model is referred to as the ``shearing sheet''.  The numerical 
implementation of the shearing sheet requires a careful treatment of the boundary conditions
in the radial direction.  These boundary conditions are described in detail in \cite{hgb95}.
In brief, one uses strictly periodic boundary conditions in $y$  and shearing-periodic boundary
conditions in $x$.  The latter is done by enforcing periodic boundary conditions in the radial
direction followed by an advection of the boundary fluid due to the shear.  This assumes that
the shearing sheet is surrounded by identical boxes that are strictly periodic initially, with
a large-scale shear flow present across all the boxes.

The shearing-sheet equations are evolved using a ZEUS-based scheme \citep{sn92}: a time-explicit,
operator-split, finite-difference method on a staggered grid which uses an artificial viscosity
to capture shocks.  An important modification of the standard shearing sheet, introduced by
\cite{mass00}, is the splitting of the overall shear velocity from the rest of the flow.
This overcomes a practical limitation of the standard shearing sheet, which is the small
Courant-limited time step imposed by the large shear velocities at the edges of the sheet;
for numerical stability of grid-based schemes the Courant condition requires time steps to
be lower than the grid spacing divided by the maximum velocity on the grid.  The larger the
box, the more severe this limitation becomes.  Separating out the shear removes this limitation
and allows one to increase the size of the shearing sheet arbitrarily.  This separation is done
by replacing $v_y$ by $v_o + \delta v_y$ in the fluid equations, and then evolving $\delta v_y$;
this can be done because there is no evolution of $v_o$ directly ($\partial_t v_o = 0$ and
$\bnabla v_o = {\rm{constant}}$).  Advection of other fluid variables by $v_o$ is done by
splitting the distance over which the fluid is sheared into an integral and fractional number
of grid zones: the fluid variables are simply shifted an integral number of zones and then
advected in the usual manner for the remaining fractional part (which does not require a
higher effective velocity than any other part of the flow).

\section{Discussion}\label{intros6}

While a rigorous proof of the stability of weakly-ionized disks may well be impossible,
the results of this dissertation add to the already strong evidence against a turbulent
angular momentum transport mechanism in weakly-ionized disks.  Gravitational instability likely
results in fragmentation, radial convection is suppressed by differential rotation and
two-dimensional vortices, which provide a decaying flux of angular momentum, are likely to be
unstable in three dimensions.  Evidence against a local, nonlinear, purely hydrodynamic
instability is mounting.

Accretion may be driven globally by a magneto-centrifugal wind \citep{bp82} or tidally-induced
spiral waves \citep{lars89,ls91}, or locally via spiral waves excited by planets embedded in
the disk \citep{gr01,sg04}.  There also exist global instabilities (e.g., \citealt{pp84,pp85})
that result in a small amount of turbulence and angular momentum transport (e.g.,
\citealt{haw87}).  In addition, there are instabilities associated with the dust layer in
YSO disks (e.g., \citealt{gl04}) that will generate some amount of turbulence in those
systems.  While one or more of these mechanisms may play a role in transporting angular
momentum in certain systems, their dependence upon global structure or other special features
in order to operate makes their broad application to weakly-ionized disks doubtful.  
Alternatively, weakly-ionized disks may simply be inactive except in ionized surface layers 
\citep{gam96}.

A detailed discussion of these possibilities is beyond the scope of this dissertation, but a
brief discussion of layered accretion is given in Chapter~\ref{conclusion}, along with some 
proposals for future modeling based upon that idea.  Chapter~\ref{conclusion} also summarizes
the main results and implications of this work, and provides some direction as to where to go
from here in the search for a turbulent angular momentum transport mechanism in weakly-ionized
accretion disks.  In addition, proposals are made for future investigations of the properties
of turbulent stresses in ionized disks, with a view towards incorporating these properties in
advanced, physically-motivated disk models.

\end{spacing}

\chapter{Nonlinear Outcome of Gravitational Instability in
Disks with Realistic Cooling}\label{paper1}

\begin{spacing}{1.5}

\section{Chapter Overview}

We consider the nonlinear outcome of gravitational instability in
optically-thick disks with a realistic cooling function.  We use a
numerical model that is local, razor-thin, and unmagnetized.  External
illumination is ignored.  Cooling is calculated from a one-zone model
using analytic fits to low temperature Rosseland mean opacities.  The
model has two parameters: the initial surface density $\Sigma_o$ and
the rotation frequency $\Omega$.  We survey the parameter space and find: (1)
The disk fragments when $\tce \Omega \sim 1$, where $\tce$ is an
effective cooling time defined as the average internal energy of the model
divided by the average cooling rate.  This is consistent with earlier
results that used a simplified cooling function.  (2) The initial
cooling time $\tco$ for a uniform disk with Toomre stability parameter $Q = 1$ can differ by orders
of magnitude from $\tce$ in the nonlinear outcome.  The difference is
caused by sharp variations in the opacity with temperature.  The
condition $\tco \Omega \sim 1$ therefore does not necessarily indicate
where fragmentation will occur.  (3) The largest difference
between $\tce$ and $\tco$ is near the opacity gap, where dust is absent
and hydrogen is largely molecular.  (4) In the limit of strong
illumination the disk is isothermal; we find that an isothermal version
of our model fragments for $Q \lesssim 1.4$.  Finally, we discuss some physical
processes not included in our model, and find that most are likely to
make disks more susceptible to fragmentation.  We conclude that disks
with $\tce\Omega \lesssim 1$ do not exist.\footnote{Published in ApJ
Volume 597, Issue 1, pp. 131-141. Reproduction for this dissertation is 
authorized by the copyright holder.}

\section{Introduction}

The outer regions of accretion disks in both active galactic nuclei
(AGN) and young stellar objects (YSO) are close to gravitational
instability (for a review see, for AGN: \citealt{sbf90}; YSOs:
\citealt{al93}).  Gravitational instability can be of central importance in
disk evolution.  In some disks, it leads to the efficient redistribution
of mass and angular momentum (e.g.  \citealt{lar84,lr96,gam01}).  In
other disks, gravitational instability leads to fragmentation and the
formation of bound objects.  This may cause the truncation of
circumnuclear disks \citep{good03}, or the formation of planets (e.g.
\citealt{boss97}, and references therein).

We will restrict attention to disks whose potential is dominated by the
central object, and whose rotation curve is therefore approximately
Keplerian.  Gravitational instability to axisymmetric perturbations sets in when the sound
speed $c_s$, the rotation frequency $\Omega$, and the surface density
$\Sigma$ satisfy
\begin{equation}\label{QDEF}
Q \equiv {\frac{c_s\Omega}{\pi G\Sigma}} < Q_{crit} \simeq 1
\end{equation}
\citep{toom64,glb65}.  Here $Q_{crit} = 1$ for a ``razor-thin'' (two-
dimensional) fluid disk model of the sort we will consider below, and
$Q_{crit} = 0.676$ for a finite-thickness isothermal disk \citep{glb65}. \footnote{For global 
models with radial structure, nonaxisymmetric instabilities typically set in for
slightly larger values of $Q$ (see \citealt{boss98} and references therein).}
The instability condition (\ref{QDEF}) can be rewritten, for a disk with
scale height $H \simeq c_s/\Omega$, around a central object of mass
$M_*$,
\begin{equation}\label{THICKCRIT}
M_{disk} \gtrsim \frac{H}{r} M_*,
\end{equation}
where $M_{disk} = \pi r^2 \Sigma$.  For YSO disks $H/r \sim 0.1$ and
thus a massive disk is required for instability.  AGN disks are expected
to be much thinner.  The instability condition can be rewritten in a
third, useful form if we assume that the disk is in a steady state and its
evolution is controlled by internal (``viscous'') transport of angular
momentum.  Then the accretion rate $\dot{M} = 3\pi \alpha c_s^2
\Sigma/\Omega$, where $\alpha \lesssim 1$ is the usual dimensionless
viscosity of \cite{ss73}, and
\begin{equation}\label{INSTCRIT}
\dot{M} \gtrsim \frac{3 \alpha c_s^3}{G}
= 7.1 \times 10^{-4}\, \alpha
\left( \frac{c_s}{1\kms}\right)^3 \,\msun \yr^{-1}
\end{equation}
implies gravitational instability (e.g. \cite{sbf90}).  Disks
dominated by external torques (e.g. a magnetohydrodynamic [MHD] wind)
can have higher accretion rates (but not arbitrarily higher; see
\citealt{good03}) while avoiding gravitational instability.

For a young, solar-mass star accreting from a disk with $\alpha =
10^{-2} $ at $10^{-6} \msun \yr^{-1}$, equation (\ref{INSTCRIT}) implies
that instability occurs where the temperature drops below $17 \K$.
Disks may not be this cold if the star is located in a warm molecular
cloud where the ambient temperature is greater than $17 \K$, or if the
disk is bathed in scattered infrared light from the central star
(although there is some evidence for such low temperatures in the solar
nebula, e.g. \citealt{owen99}).  If the vertically-averaged value of
$\alpha$ is small and internal dissipation is confined to surface
layers, as in the layered accretion model of \cite{gam96}, then
instability can occur at higher temperatures, although equation
(\ref{THICKCRIT}) still requires that the disk be massive.

AGN disk heating is typically dominated by illumination from a central
source.  The temperature then depends on the shape of the disk.  If the
disk is flat or shadowed, however, and transport is dominated by
internal torques, one can apply equation (\ref{INSTCRIT}).  For example,
in the nucleus of NGC 4258 \citep{miy95} the accretion rate may be as
large as $10^{-2} \msun \yr^{-1}$ \citep{las96,gnb99}.  Equation
(\ref{INSTCRIT}) then implies that instability sets in where $T < 10^4
(\alpha/10^{-2}) \K$.  If the disk is illumination-dominated then $Q$
fluctuates with the luminosity of the central source.

In a previous paper \citep{gam01}, one of us investigated the effect of
gravitational instability in cooling, gaseous disks in a local model.  A
simplified cooling function $\Lambda$ was employed in these simulations,
with a fixed cooling time $\tco$:
\begin{equation}
\Lambda = -\frac{U}{\tco},
\end{equation}
where $U \equiv$ the internal energy per unit area.  Disk fragmentation
was observed for $\tco\Omega \lesssim 3$.  The purpose of this
paper is to investigate gravitational instability in a local model with
more realistic cooling.  

Several recent numerical experiments have included cooling, as opposed
to isothermal or adiabatic evolution, and we can ask whether these
results are consistent with \cite{gam01}.  \cite{nbr00} studied a global
two-dimensional (thin) SPH model in which the vertical density and
temperature structure is calculated self-consistently and each particle
radiates as a blackbody at the surface of the disk.  The initial
conditions at a radius corresponding to the minimum initial value of Q
($\sim 1.5$) for these simulations were $\Sigma_o \approx 50
{\rm\,g\,cm^{-2}}, \Omega \approx 8 \times 10^{-10} {\rm \,s^{-1}}$; the
initial cooling time under these circumstances is $\tco \approx 250 \,
\Omega^{-1}$, so fragmentation is not expected and is not observed.

\cite{dmph01} consider a global three dimensional (3D) Eulerian
hydrodynamics model in which the volumetric cooling rate varies with
height above the midplane so as to preserve an isentropic vertical
structure.  The cooling time is fixed at each radius.  Their cooling
time $\gtrsim 10 \Omega^{-1}$ at all radii, so fragmentation is not expected
based on the criterion of \cite{gam01}.  The simulations show structure
formation due to gravitational instabilities but not fragmentation.

\cite{rabb03} consider a global 3D SPH model with a cooling time that is
a fixed multiple of $\Omega^{-1}(r)$.  They find that their disk
fragments when $\tco \approx 3 \Omega^{-1}$ and $M_{disk} = 0.1 M_*$.
For a more massive disk ($M_{disk} = 0.25 M_*$), fragmentation occurred
at somewhat higher cooling times ($\tco \approx 10 \Omega^{-1}$).  This
is effectively a global generalization of the local model problem
considered by \cite{gam01}.  The fact that the results are so consistent
suggests that the local, thin approximation used in \cite{gam01} and
here give a reasonable approximation to a global outcome.

\cite{mqws02} consider a global three dimensional SPH model of a
circumstellar disk.  Explicit cooling is not included, but the equation
of state switches from isothermal to adiabatic when gravitational
instability begins to set in.  This is designed to account for the
inefficient cooling of dense, optically thick regions.  Fragmentation is
observed.  Realistic cooling can have a complex influence on disk
evolution, and it is not clear that switching between isothermal and
adiabatic behavior ``brackets'' the outcomes that might be obtained when
full cooling is used.

Other notable recent work, such as that by \cite{boss02}, includes strong
radiative heating in the sense that the effective temperature of the
external radiation field $T_{irr}$ is comparable to or larger than the
disk midplane temperature $T_c$.  In the limit that $T_{irr} \ll T_c$ we
recover the limit considered here and in \cite{gam01}; in the limit that
$T_{irr} \gg T_c$ the disk is effectively isothermal.

The plan of this paper is as follows.  In \S\ref{pap1s2} we describe the model,
with a detailed description of the cooling function given in \S\ref{pap1s3}. The
results of numerical experiments are described in \S\ref{pap1s4}.  Conclusions are
given in \S\ref{pap1s5}.

\section{Model}\label{pap1s2}

The model we use here is identical to that used in \cite{gam01} in every
respect except that we use a more complicated cooling function.  To make
the description more self-contained, we summarize the basic equations of
the model here.  The model is local, in the sense that it considers a
region of size $L$ where $L/r_o \ll 1$ and $r_o$ is a fiducial radius.  We
use a {\it local Cartesian} coordinate system $x \equiv r - r_o$ and $y
\equiv (\phi - \Omega t) r_o$, where $r,\phi$ are the usual cylindrical
coordinates and $\Omega$ is the orbital frequency at $r_o$.  The model is also
thin in the sense that matter is confined entirely to the plane of the disk.

Using the local approximation one can perform a formal expansion of the
equations of motion in the small parameter $L/r_o$.  The resulting equations of
motion read, where $\bv$ is the velocity, $P$ is the (two-dimensional)
pressure, and $\phi$ is the gravitational potential with the time-steady
axisymmetric component removed:
\begin{equation}
\frac{D\bv}{D t} = -\frac{\bnabla P}{\Sigma} - 2\bO\times\bv
        + 3\Omega^2 x \ex - \bnabla\phi.
\end{equation}
For constant pressure and surface density, $\bv = -\frac{3}{2}\Omega x
\ey$ is an equilibrium solution to the equations of motion.  This linear
shear flow is the manifestation of differential rotation in the local
model.

The equation of state is
\begin{equation}
P = (\gamma - 1) U,
\end{equation}
where $P$ is the two-dimensional pressure and $U$ the two-dimensional
internal energy.  The two-dimensional (2D) adiabatic index $\gamma$ can
be mapped to a 3D adiabatic index $\Gamma$ in the low-frequency (static)
limit.  For a non-self-gravitating disk $\gamma = (3\Gamma - 1)/(\Gamma +
1)$ (e.g.  \citealt{ggn86,osa92}).  For
a strongly self-gravitating disk, one can show that $\gamma = 3 - 2/\Gamma$.  We
adopt $\Gamma = 7/5$ throughout, which yields $\gamma = 11/7$.

The internal energy equation is
\begin{equation}
\frac{\del U}{\del t} + \nabla \cdot (U \bv) =
	-P\bnabla\cdot\bv - \Lambda,
\end{equation}
where $\Lambda = \Lambda(\Sigma,U,\Omega)$ is the cooling function,
fully described below.  Notice that there is no heating term; heating is
due solely to shocks.  Numerically, entropy is increased by artificial
viscosity in shocks.

The gravitational potential is determined by the razor-thin disk Poisson
equation:
\begin{equation}
\nabla^2\phi = 4\pi G \Sigma \, \delta(z).
\end{equation}
For a single Fourier component of the surface density $\Sigma_{\bk}$
this has the solution
\begin{equation}
\phi = -\frac{2\pi G}{|\bk|} \Sigma_{\bk} e^{i \bk\cdot{\bf x}
	- |k z|}.
\end{equation}
A finite-thickness disk has weaker self-gravity, but this does not
qualitatively change the dynamics of the disk in linear theory
\citep{glb65}.

We integrate the governing equations using a self-gravitating
hydrodynamics code based on ZEUS \citep{sn92}. ZEUS is a
time-explicit, operator-split, finite-difference method on a staggered
mesh.  It uses an artificial viscosity to capture shocks.  Our
implementation has been tested on standard linear and nonlinear
problems, such as sound waves and shock tubes.  We use the ``shearing
box'' boundary conditions, described in detail by \cite{hgb95}, and solve
the Poisson equation using the Fourier transform method, modified for
the shearing box boundary conditions.  See \cite{gam01}  for further
details on boundary conditions, numerical methods and tests.

The numerical model is always integrated in a region of size $L \times
L$ at a numerical resolution of $N \times N$.  In linear theory the disk
is most responsive at the critical wavelength $2c_s^2/G\Sigma_o$.\footnote{The
wavelength corresponding to the minimum in the dispersion
relation for axisymmetric waves.}
We have checked the dependence of the outcome on $L$ and
found that as long as $L \gtrsim 2c_s^2/G\Sigma_o$ the outcome does
not depend on $L$.  We have also checked the dependence of the outcome
on $N$ and found that the outcome is
insensitive to $N$, at least for the models with $N \geq 256$ that we use.

\section{Cooling Function}\label{pap1s3}

Our cooling function is determined from a one-zone model for the
vertical structure of the disk.  The disk cools at a rate per unit area
\begin{equation}
\Lambda \equiv 2 \sigma T_e^4,
\end{equation}
which defines the effective temperature $T_e$.  The cooling function
depends on the heat content of the disk and how that content is
transported from the disk interior to the surface: by radiation,
convection, or perhaps some more exotic form of turbulent transport such
as MHD waves.  Low temperature disks are expected to be convectively
unstable (e.g.  \citealt{cam78,lp80}).  \cite{cass93} has argued,
however, that the radiative heat flux in an adiabatically-stratified
disk is comparable to the heat dissipated by turbulence (in an
$\alpha$-disk model), suggesting that convection is incapable of
radically altering the vertical structure of the disk.  We will consider
only radiative transport.

If the disk is optically thick in the Rosseland mean sense, so that
radiative transport can be treated in the diffusion approximation, then
\citep{hub90}
\begin{equation}\label{TEFF}
T_e^4 = \frac{8}{3}\frac{T_c^4}{\tau} 
\end{equation}
where $\tau$ is the Rosseland mean optical depth and $T_c$ is the
central temperature.  We will assume that $T_c \approx T$, where
\begin{equation}\label{TEMP}
T = \frac{\mu m_p c_s^2}{\gamma k_B},
\end{equation}
and
\begin{equation}\label{CS2}
c_s^2 = \gamma(\gamma - 1)\frac{U}{\Sigma},
\end{equation}
which follows from the equation of state and the assumption that the
radiation pressure is small (we have verified that this is never
seriously violated).  Here $k_B$ is Boltzmann's constant, $m_p$ is the
proton mass, and $\mu$ is the mean mass per particle, which we have set
to $2.4$ in models with initial temperature below the boundary between
the grain-evaporation opacity and molecular opacity and $\mu = 0.6$ in
models with initial temperature above the boundary.

The optical depth is
\begin{equation}
\tau \equiv \int_0^\infty \, dz \, \kappa(\rho_z, T_z) \rho_z
\end{equation}
where $\kappa$ is the Rosseland mean opacity, $\rho_z$ and $T_z$ are
local density and temperature, and $z$ is the height above the midplane.
Following the usual one-zone approximation,
\begin{equation}
\int_0^\infty \, dz \, \kappa(\rho_z, T_z) \rho_z \approx
H \kappa(\bar{\rho},\bar{T}) \bar{\rho}
\end{equation}
where the overbar indicates a suitable average and $H \approx
c_s(T)/\Omega$ is the disk scale height (we ignore the effects of
self-gravity on the disk scale height, which is valid when locally $Q
\gtrsim 1$).  Taking $\bar{T} \approx T$ and $\bar{\rho} \approx
\Sigma/(2 H)$ then gives a final, closed expression for $\Lambda$.

We have adopted the analytic approximations to the opacities provided by
\cite{bl94}.  These opacities are dominated by, in order of increasing
temperature: grains with ice mantles, grains without ice mantles,
molecules, H$^-$ scattering, bound-free/free-free absorption and
electron scattering. The molecular opacity regime is commonly called the
{\it opacity gap}; it is too hot for dust, but too cold for H$^-$
scattering to contribute much opacity.  The opacity can be as much as
$4$ orders of magnitude smaller than the $\sim 5 \gm \cm^{-2}$ typical
of the dust-dominated opacity regime.  It turns out that this feature
plays a significant role in the evolution of gravitationally-unstable
disks.

To sum up, the cooling function is
\begin{equation}\label{COOL}
\Lambda(\Sigma,U,\Omega) = \frac{16}{3} \frac{\sigma T^4}{\tau}.
\end{equation}
For a power-law opacity of the form $\kappa = \kappa_0 \rho^a T^b$, this
implies that
\begin{equation}
\Lambda \sim \Sigma^{-5 - 3 a/2 + b} U^{4 + a/2 - b}.
\end{equation}
From this it follows that the cooling time $\tc \equiv U/\Lambda$
scales as
\begin{equation}
\tc \sim \Sigma^{5 + 3 a/2 - b} U^{-3 - a/2 + b}.
\end{equation}
If the disk evolves quasi-adiabatically (as it does if the cooling time
is long compared to the dynamical time) then $U \sim \Sigma^\gamma$ and
\begin{equation}
\tc \sim \Sigma^{5 - 3 \gamma + (a/2) (3 - \gamma) + b (\gamma - 1)}.
\end{equation}

\begin{deluxetable}{lccc}
\tablecolumns{4}
\tablewidth{0pc}
\tabcolsep 0.5truecm
\tablecaption{Scaling Exponent for Cooling Time as a Function of Surface Density \label{pap1t1}}
\tablehead{Opacity Regime & a & b & Exponent}
\startdata
Ice grains & 0 & 2 & 10/7 \\
Evaporation of ice grains & 0 & -7  & -26/7 \\
Metal grains & 0 & 1/2  & 4/7 \\
Evaporation of metal grains & 1 & -24  & -89/7 \\
Molecules & 2/3 & 3  & 52/21 \\
H$^-$ scattering & 1/3 & 10  & 131/21 \\
Bound-free and free-free & 1 & -5/2  & -3/7 \\
Electron scattering & 0 & 0  & 2/7 \\
\enddata
\end{deluxetable}

\noindent
Table~\ref{pap1t1} gives a list of values for this scaling exponent for our nominal
value of $\gamma = 11/7$.  Notice that, when ice grains or metal grains
are evaporating, and in the bound-free/free-free opacity regime, cooling
time {\it decreases} as surface density {\it increases}.

Our cooling function is valid in the limit of large optical depth ($\tau
\gg 1$).  Since the disk becomes optically thin at some locations in the
course of a typical run, we must modify this result so that the cooling
rate does not diverge at small optical depth.  A modification that
produces the correct asymptotic behavior is
\begin{equation}\label{COOLNEW}
\Lambda = \frac{16}{3} \sigma T^4 \frac{\tau}{1 + \tau^2}.
\end{equation}
This interpolates smoothly between the optically-thick and
optically-thin regimes and is proportional to the (Rosseland mean)
optical depth in the optically-thin limit.  While it would be more
physically sensible to use a Planck mean opacity in the optically-thin
limit, usually the optically-thin regions contain little mass so their
cooling is not energetically significant.  An exception is in the
opacity gap, where even high density regions become optically thin.

Our simulations begin with $\Sigma$ and $U$ constant.  The velocity field is
perturbed from the equilibrium solution to initiate the gravitational
instability.  The initial velocities are $v_x = \delta v_x$, $v_y = -
\frac{3}{2}\Omega x + \delta v_y$, where $\delta \bv$ is a Gaussian
random field of amplitude $\< \delta v^2 \>/c_s^2 = 0.1$. The
power spectrum of perturbations is white noise ($v_k^2 \sim k^0$) in a band
in wavenumber $k_{crit}/4 < |k| < 4 k_{crit}$ surrounding the minimum
$k_{crit} = 1/(\pi Q^2)$ (with $G = \Sigma_o = \Omega = 1$) in the
density-wave dispersion relation.  We have checked in particular cases
that for $10^{-3} < \< \delta v^2 \>/c_s^2 < 10$ the outcome is
qualitatively unchanged.  This is expected because disk
perturbations (unlike cosmological perturbations) grow exponentially and
the initial conditions are soon forgotten.

Excluding the initial velocity field, the initial conditions for a
spatially-uniform disk consist of three parameters: $\Sigma_o, U_o$, and $\Omega$.
We fix $Q = 1$, leaving two degrees of freedom.  In models with simple,
scale-free cooling functions such as that considered by \cite{gam01},
these degrees of freedom remain and can be scaled away by setting $G =
\Sigma_o = \Omega = 1$.  That is, there is a two-dimensional continuum
of disks (with varying values of $\Sigma_o$ and $\Omega$, but the same
value of $Q$) that are described by a single numerical
model.

The opacity contains definite physical scales in density and
temperature.  The realistic cooling function considered here therefore removes our freedom to
rescale the disk surface density and rotation frequency.  That is, there
is now a one-to-one correspondence between disks with fixed $\Sigma$ and
$\Omega$ and our numerical models.

The choice of $\Sigma_o$ and $\Omega$ as labels for the parameter space
is not unique.  Internally in the code we fix the initial
volume density (in $\gcm$) and the initial temperature (in
Kelvins).  These choices are difficult to interpret, however, since they
are tied to quantities that change over the course of the simulation;
$\Omega$ and the mean value of $\Sigma$ do not.

The cooling is integrated explicitly using a first-order scheme.  The timestep is
modified to satisfy the Courant condition and to be less than a fixed fraction of
the shortest cooling time on the grid. We have varied this fraction and shown that
the results are insensitive to it, provided that it is sufficiently small.

\section{Nonlinear Outcome}\label{pap1s4}

\subsection{Standard Run}\label{STDRUN}

Consider the evolution of a single ``standard'' run, with $\Sigma_o =
1.4 \times 10^5 \gm \cm^{-2}$ and $\Omega = 1.1 \times 10^{-7}
\sec^{-1}$.  This corresponds to $T_o = 1200$ and $\tco = 9.0
\times 10^4 \Omega^{-1}$.  The model size is $L = 320
G\Sigma_o/\Omega^2$ and numerical resolution $1024^2$.  The model
initially lies at the lower edge of the opacity gap.

\begin{figure}[h]
\centering
\includegraphics[width=4.5in,clip]{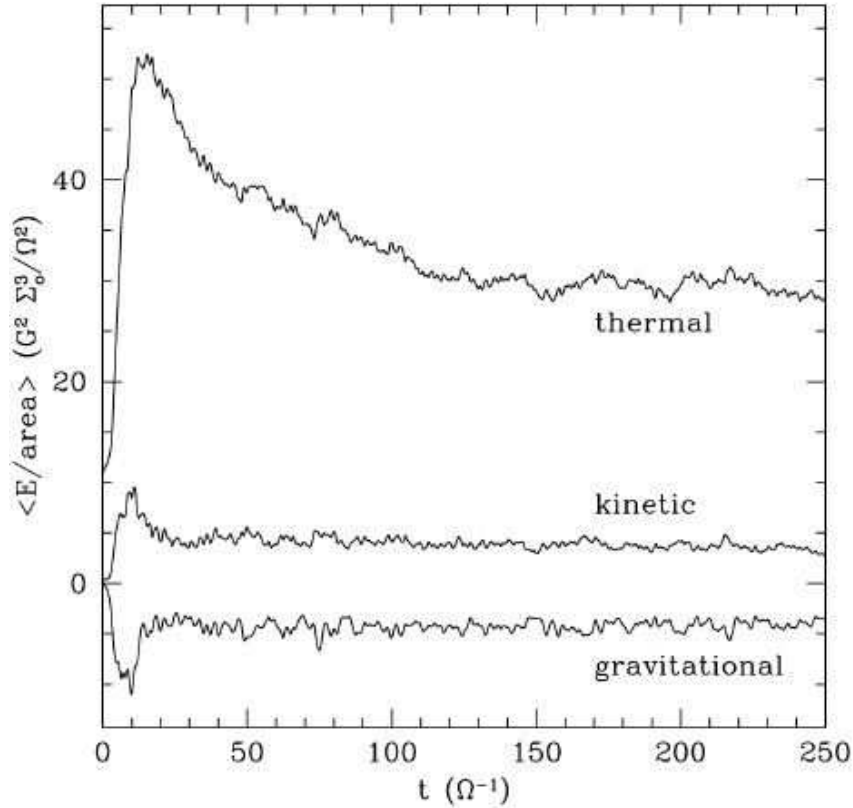}
\caption[Evolution of the kinetic, gravitational, and thermal energy per unit
area in the standard run.]
{Evolution of the kinetic, gravitational, and thermal energy per unit
area, normalized to $G^2 \Sigma_o^3/\Omega^2$, in the standard run, which
has $L = 320 G\Sigma_o/\Omega^2$, resolution $1024^2$, and $\tco = 9.0
\times 10^4 \Omega^{-1}$.}
\label{pap1f1}
\end{figure}

The evolution of the kinetic, gravitational and thermal energy per unit
area ($\<E_k\>$, $\<E_g\>$ and $\<E_{th}\>$ respectively) normalized to
$G^2\Sigma_o^3/\Omega^2$,\footnote{The natural unit that can be formed
from $G$, $\Sigma$ and $\Omega$.} are shown in Figure~\ref{pap1f1}.  After
the initial phase of gravitational instability the model settles into a
statistically-steady, gravito-turbulent state.  It does not fragment.
Cooling is balanced by shock heating.  Energy for driving the shocks is
extracted from the shear flow, and the mean shear flow is enforced by the
boundary conditions.

\begin{figure}[h]
\centering
\includegraphics[width=5.in,clip]{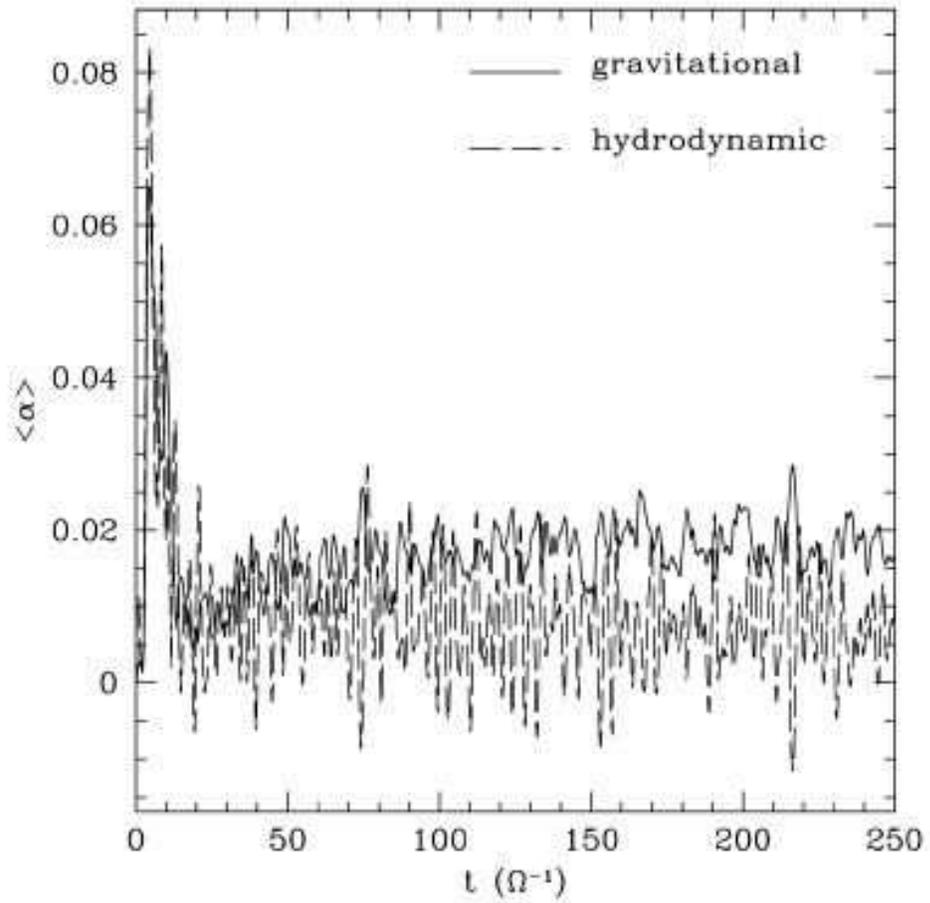}
\caption{Evolution of the gravitational and hydrodynamic 
pieces of $\< \alpha \>$ in the standard run.}
\label{pap1f2}
\end{figure}

The turbulent state transports angular momentum outward via hydrodynamic
and gravitational shear stresses.  The dimensionless gravitational shear stress
is
\begin{equation}
\alpha_{grav} = \frac{1}{\<\frac{3}{2} \Sigma c_s^2\>}
		\int_{-\infty}^{\infty} dz \frac{g_x g_y}{4 \pi G}
\end{equation}
where ${\bf g}$ is the gravitational acceleration, and
the dimensionless hydrodynamic shear stress is
\begin{equation}
\alpha_{hyd} = \frac{\Sigma v_x \delta v_y}{\<\frac{3}{2} \Sigma c_s^2\>}
\end{equation}
where $\<\>$ denote a spatial average. Figure~\ref{pap1f2} shows the evolution of $\< \alpha_{grav} \>$
and $\< \alpha_{hyd} \>$ in the standard run.  Averaged over the last $230 \Omega^{-1}$
of the run, $\<\< \alpha_{hyd} \>\> = 0.0079$, $\<\< \alpha_{grav} \>\> = 0.017$, and so the
total dimensionless shear stress is $\<\< \alpha \>\> = 0.025$, where $\<\<\>\>$ denote
a space and time average.

The mean stability parameter $\<Q\> \equiv \langle c_s
\rangle\Omega/\pi G\langle \Sigma\rangle$ averages $1.86$ over the
last $230 \Omega^{-1}$ of the run.  Because the temperature and
surface density vary strongly, other methods of averaging $Q$ will
give different results.

\begin{figure}[h]
\centering
\includegraphics[width=4.in,clip]{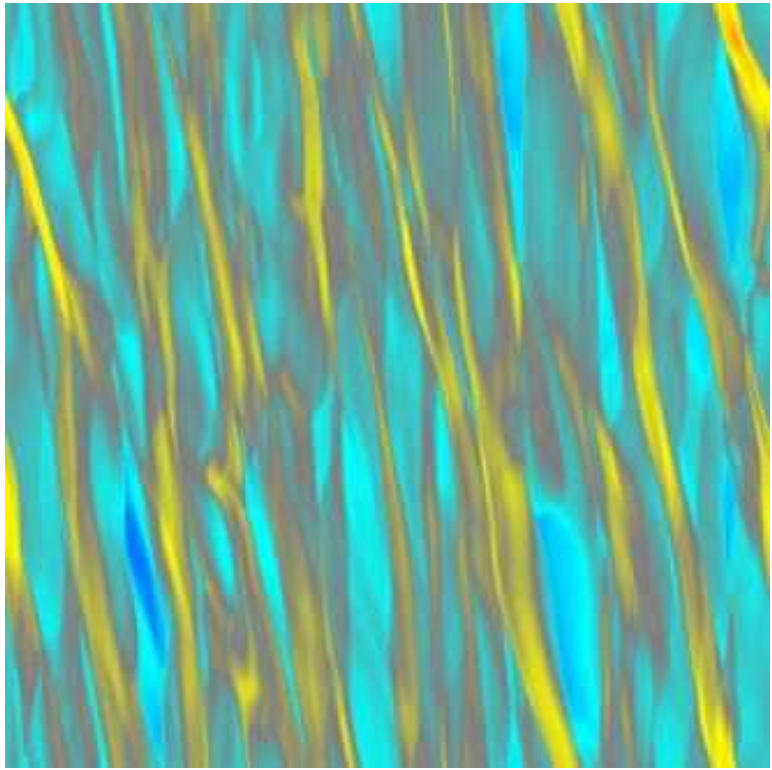}
\caption[Map of surface density in the standard run.]
{Map of surface density at $t = 50 \Omega^{-1}$ in the standard run.
Dark shades (blue in color version) indicate low density ($0.2 \Sigma_o$) 
and light shades (yellow in color version) indicate high density ($3
\Sigma_o$).}
\label{pap1f3}
\end{figure}

Figure~\ref{pap1f3} shows a snapshot of the surface density at $t = 50
\Omega^{-1}$.  The structure is similar to that observed in
\cite{gam01}, with trailing density structures.  The density
structures are stretched into a trailing configuration by the
prevailing shear flow.  Their scale is determined by the disk
temperature and surface density rather than the size of the box (see
\citealt{gam01}).

\subsection{Varying $\Sigma_o$ and $\Omega$}

We now turn to exploring the two-dimensional parameter space of
models.  First consider a series of models with the same initial
central temperature, but with varying $\tco$.  As $\tco$ is lowered
the time-averaged gravitational potential energy per unit area $\<\<E_g\>\>$ increases
monotonically in magnitude.  The gravito-turbulent state becomes more
extreme, with larger $\<\< \alpha \>\>$, larger perturbed velocities, and larger
density contrasts.  Eventually a threshold is crossed and the disk
fragments.

\begin{figure}[h]
  \hfill
  \begin{minipage}[t]{.45\textwidth}
    \begin{center}
      \includegraphics[width=0.9\textwidth,clip]{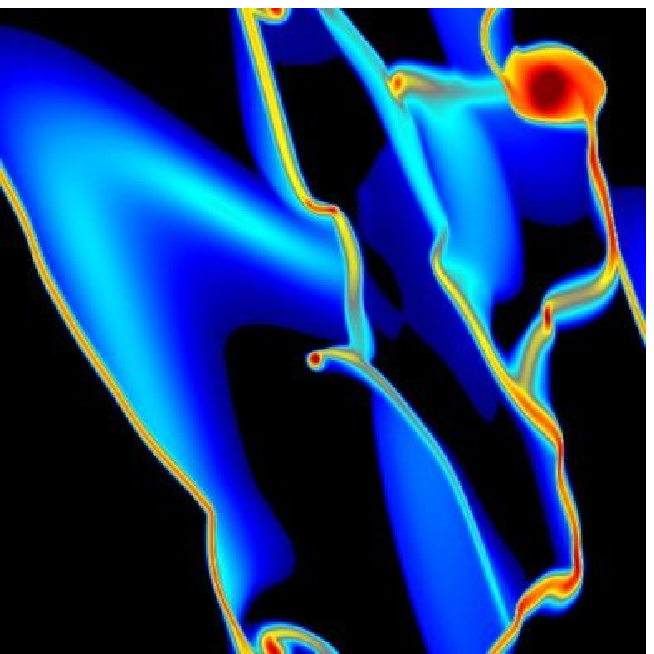}
      \caption[Map of surface density in a run with $\tco = 0.025\Omega^{-1}$.]
      {Map of surface density in a run with $\tco = 0.025\Omega^{-1}$.
	Dark shades indicate both low density ($10^{-2} \Sigma_0$, black in 
	color version) and high density ($10^2 \Sigma_0$, near the centers 
	of bound objects, red in color version).}
      \label{pap1f4}
    \end{center}
  \end{minipage}
  \hfill
  \begin{minipage}[t]{.45\textwidth}
    \begin{center}
      \includegraphics[width=0.9\textwidth,clip]{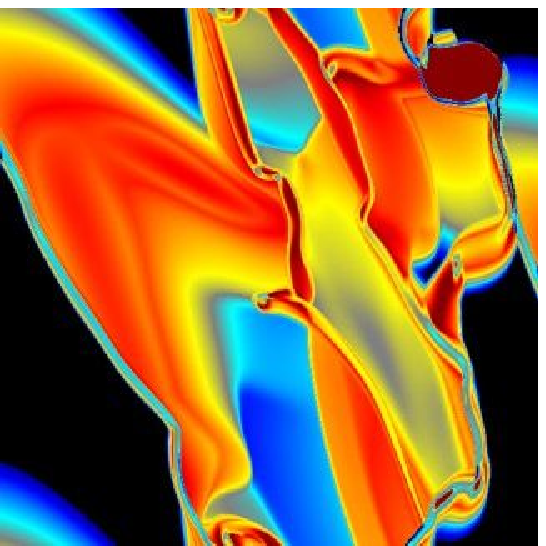}
      \caption[Map of optical depth in a run with $\tco = 0.025 \Omega^{-1}$.]
      {Map of optical depth $\tau$ in a run with $\tco = 0.025 \Omega^{-1}$. 
        Dark shades indicate both low $\tau$ ($10^{-2}$, black in
        color version) and high $\tau$ ($10^4$, near the centers
        of bound objects, red in color version).}
      \label{pap1f5}
    \end{center}
  \end{minipage}
  \hfill
\end{figure}

Fragmentation is illustrated in Figure~\ref{pap1f4}, which shows a snapshot from a
run with $\Sigma_o = 6.6 \times 10^3 \gm \cm^{-2}$, $\Omega = 5.4 \times
10^{-9} \sec^{-1}$.  This corresponds to $T_o = 1200$,  $\tco =
0.025\Omega^{-1}$.  The run has numerical resolution $256^2$ and $L = 80
G\Sigma_o/\Omega^2$.  The largest bound object in the center of the
figure was formed from the collision and coalescence of several smaller
bound objects.  A snapshot of the optical depth at the same point in the
simulation is given in Figure~\ref{pap1f5}.  For each snapshot, red indicates high
values of the mapped variable and blue indicates low values.  Much of the disk is
optically thick, but most of the low density regions are optically thin in the
Rosseland mean sense.

\begin{figure}[hp]
  \hfill
  \begin{minipage}[t]{1.\textwidth}
    \begin{center}
      \includegraphics[width=3.5in,clip]{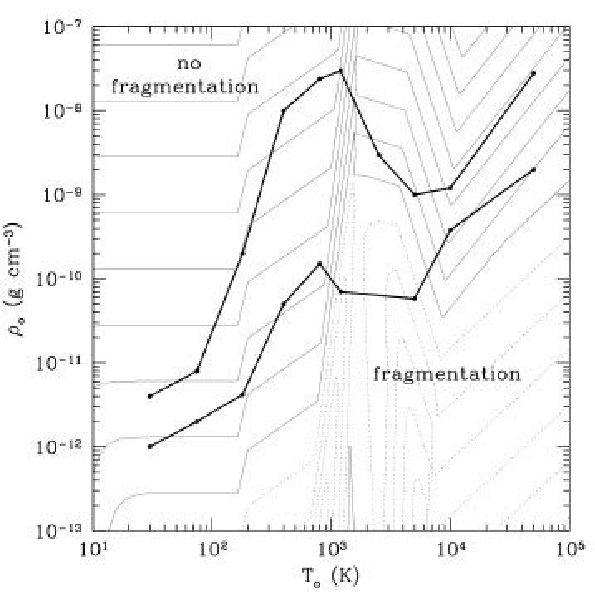}
      \caption[Location of the critical curves as a function of initial volume
      density and temperature.]
      {Location of the critical curves as a function of initial volume density and
      temperature (in cgs units). Each contour line is an order of magnitude
      change in $\tco$, solid/dotted lines indicating positive/negative
      integer values of log($\tco$).}
      \label{pap1f6}
    \end{center}
  \end{minipage}
  \hfill
  \begin{minipage}[b]{1.\textwidth}
    \begin{center}
      \includegraphics[width=3.5in,clip]{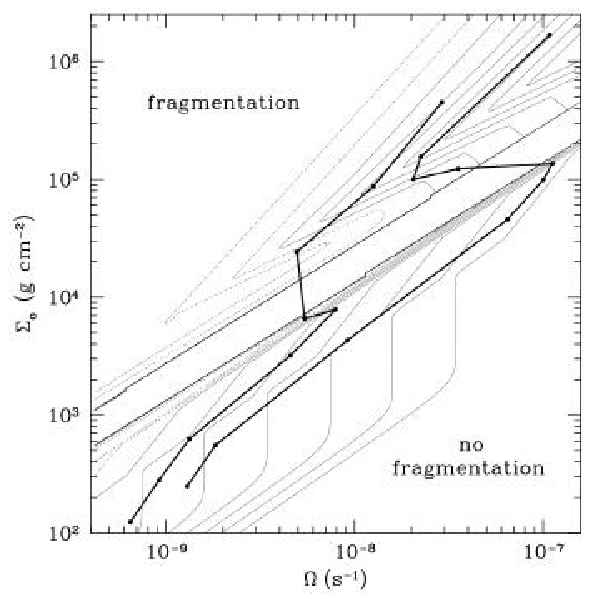}
      \caption[Location of the critical curves as a function of initial surface
      density and rotation frequency.]
      {Location of the critical curves as a function of initial surface density
      and rotation frequency (in cgs units).  Each contour line is an order of
      magnitude change in $\tco$, solid/dotted lines indicating
      positive/negative integer values of log($\tco$).  The gap in the center
      of the plot is due to the discontinuous jump in the value of $\mu$.}
      \label{pap1f7}
    \end{center}
  \end{minipage}
  \hfill
\end{figure}

Lowering $\tco$ sufficiently always leads to fragmentation.  We have
surveyed the parameter space of $\Omega$ and $\Sigma_o$ to
determine where the disk begins to fragment.  Each model was run to $100
\Omega^{-1}$.\footnote{In four cases we had to run the simulation longer
to get converged results.}  Figures~\ref{pap1f6} and \ref{pap1f7} summarize
the results.  Two heavy solid lines are shown on each diagram.  The upper line
shows the most rapidly cooling simulations that show no signs of
gravitational fragmentation ({\it nonfragmentation point}).  Quantitatively, we
define this as the point at which the time-averaged gravitational potential energy
per unit area is equal to $-3 G^2 \Sigma_o^3/\Omega^2$.\footnote{$-3$ is the
potential energy per unit area of a wave at the critical wavelength in a $Q = 1$
disk with $\delta \Sigma/\Sigma = \sqrt{3}/\pi$. No bound objects are
observed throughout the duration of these runs.} The lower line shows
the most slowly cooling simulations to show definite fragmentation ({\it
fragmentation point}).  Quantitatively, we define this as the point at
which the gravitational potential energy per unit area

\noindent
is equal to $-300
G^2 \Sigma_o^3/\Omega^2$ {\it at some point during the run}.\footnote{These
runs exhibit bound objects that persist for the duration of the run.}
Figure~\ref{pap1f6} shows the data in the $\rho_o, T_o$ plane, while
Figure~\ref{pap1f7} shows the results in the $\Sigma_o,\Omega$ plane.
Light contours are lines of constant $\tco$.

\begin{figure}[h]
\centering
\includegraphics[width=4.75in,clip]{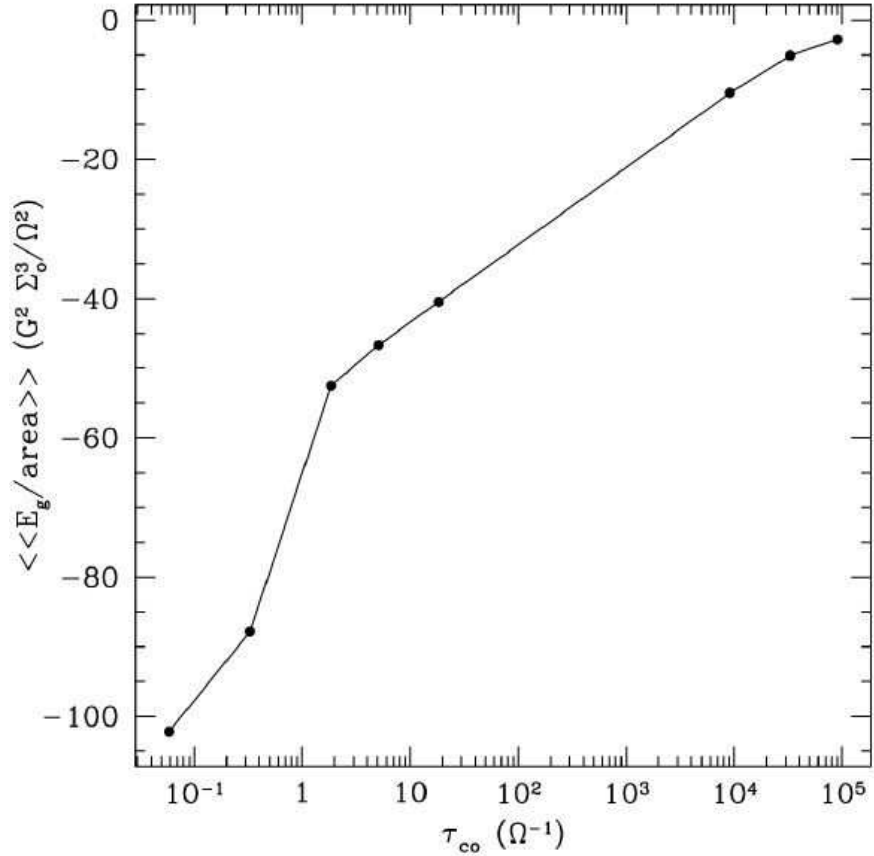}
\caption{Mean gravitational potential energy as a function of initial cooling
time for a series of models with varying initial cooling time and $T_o = 1200$.}
\label{pap1f8}
\end{figure}

The transition from persistent, gravito-turbulent outcomes to
fragmentation is gradual and statistical in nature.  Figure~\ref{pap1f8} shows the
gravitational potential energy per unit area in the transition region for a series
of runs with $T_o = 1200\K$.  The abscissa is labeled with the initial
cooling time $\tco\Omega$.  There is a gradual, approximately
logarithmic increase in the magnitude of $\<\<E_g\>\>$ as $\tco$ decreases.
Runs in this region exhibit the transient formation of small bound
objects which might collapse if additional physics (e.g.  the effects of
MHD turbulence) were included in the model.  Eventually $-\<\<E_g\>\>$ begins to
increase dramatically, and we define the {\it transition point} as the
beginning of this steep increase in gravitational binding energy.

\begin{figure}[h]
\centering
\includegraphics[width=5.5in,clip]{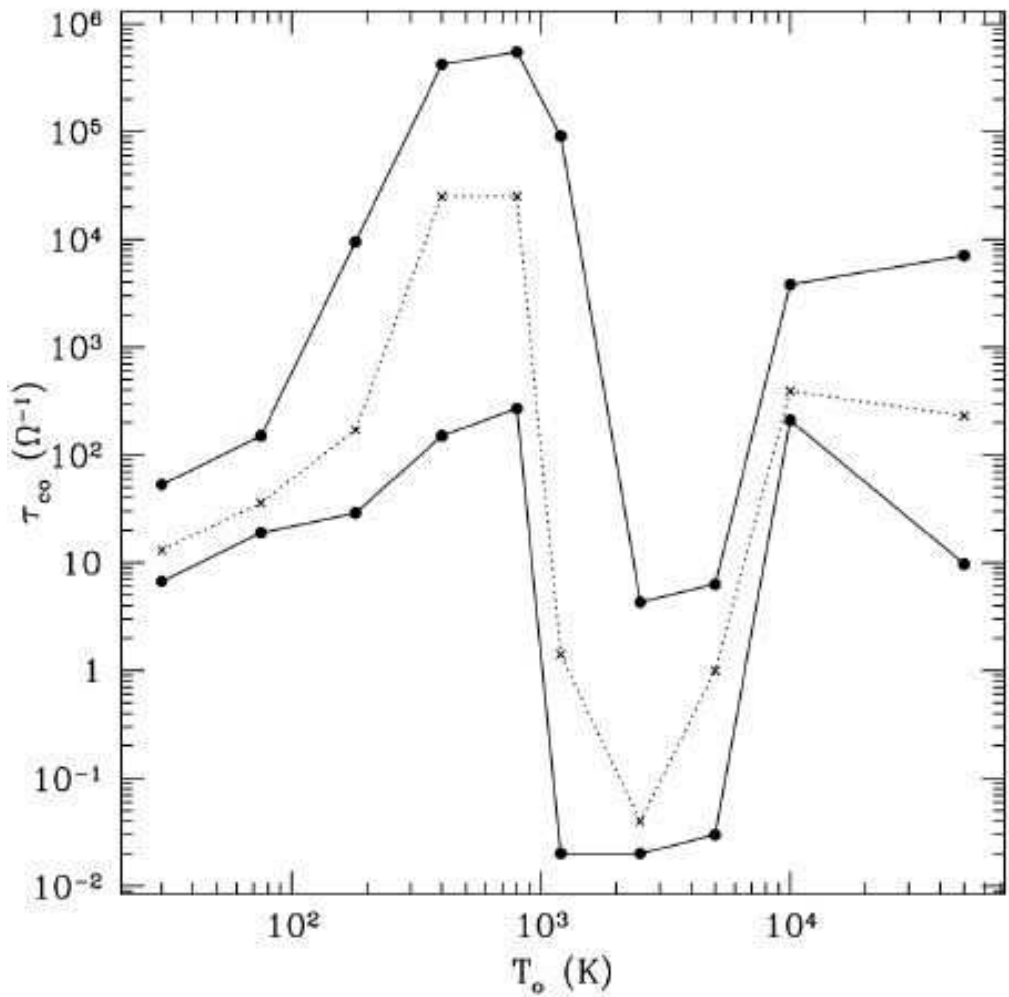}
\caption{
Initial cooling times at the points of non-fragmentation, fragmentation
and transition.
}
\label{pap1f9}
\end{figure}

Figure~\ref{pap1f9} shows the run of $\tco\Omega$ for the fragmentation point,
transition point, and nonfragmentation point as a function of $T_o$.
It is surprising that a disk can begin to exhibit signs of
gravitational collapse for $\tco\Omega$ as large as $10^6$, and evade
collapse for $\tco\Omega$ as small as $0.02$.  A naive application of
the results of \cite{gam01} would suggest that fragmentation should
occur for $\tco\Omega \lesssim 3$.  Evidently this estimate can be off by
orders of magnitude, with the largest error for $T_o \approx 10^3\K$,
just below the opacity gap.

\begin{figure}[h]
  \hfill
  \begin{minipage}[t]{.45\textwidth}
    \begin{center}
      \includegraphics[width=1.\textwidth,clip]{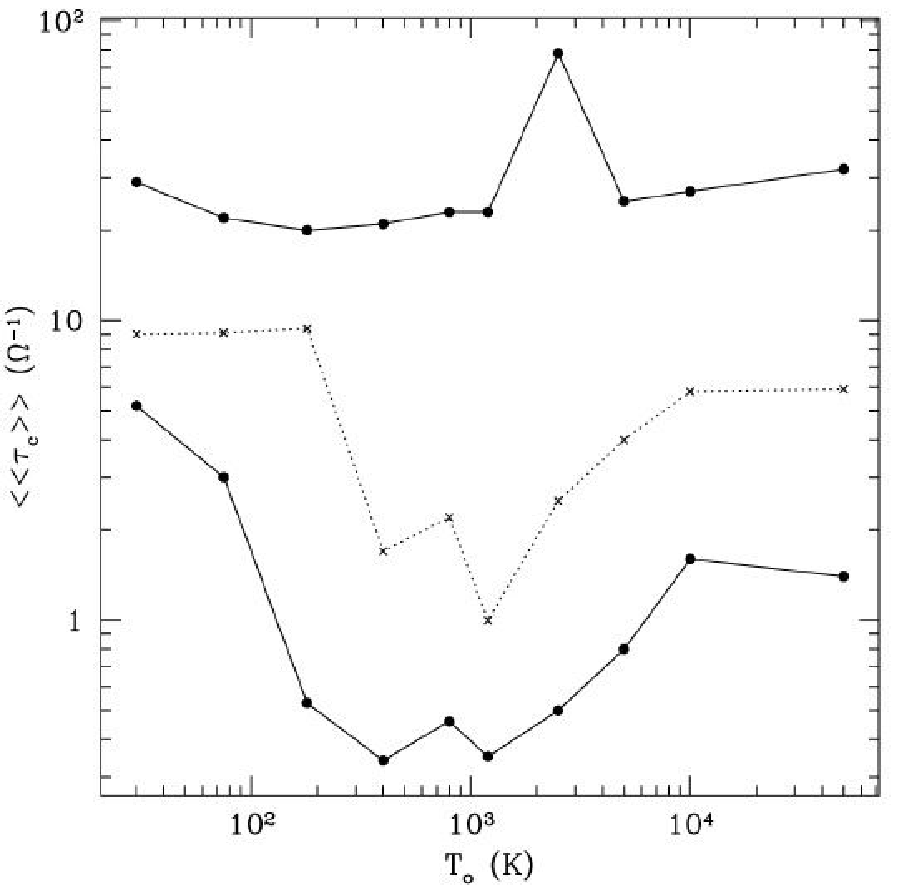}
      \caption{Effective cooling times at the points of non-fragmentation,
      fragmentation and transition.}
      \label{pap1f10}
    \end{center}
  \end{minipage}
  \hfill
  \begin{minipage}[t]{.45\textwidth}
    \begin{center}
      \includegraphics[width=1.\textwidth,clip]{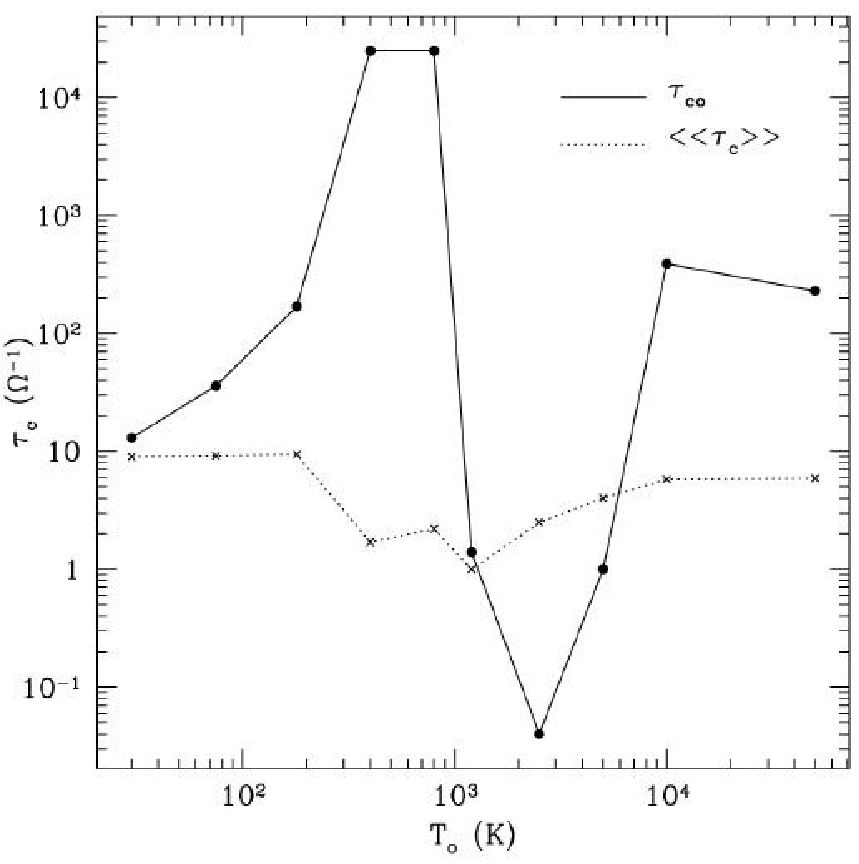}
      \caption{Initial and effective cooling times at the transition between
       non-fragmentation and fragmentation.}
      \label{pap1f11}
    \end{center}
  \end{minipage}
  \hfill
\end{figure}

The physical argument for fragmentation at short cooling times is as
follows (e.g. \citealt{sbf90}).  Thermal energy is supplied to the
disk via shocks.  Strong shocks occur when dense clumps collide with one
another; this occurs on a dynamical timescale $\sim \Omega^{-1}$.  If
the disk cools itself more rapidly then shock heating cannot match
cooling and fragmentation results.  This argument is apparently
contradicted by Figure~\ref{pap1f9}.  The resolution lies in finding an appropriate
definition of cooling time.  The disk loses thermal energy on the
effective cooling timescale
\begin{equation}
\tce^{-1} \equiv \frac{\<\< \Lambda \>\>}{\<\< U \>\>}.
\end{equation}
Figure~\ref{pap1f10} shows the run of
$\tce$ at the fragmentation, transition, and non-fragmentation points.
Evidently $\tce$ at transition lies between $\Omega^{-1}$ and $10
\Omega^{-1}$.  Figure~\ref{pap1f11} shows the run of $\tco$ and $\tce$ on the
transition line.  Just below the opacity gap they differ by as much as
four orders of magnitude.

Why do $\tco$ and $\tce$ differ by such a large factor?  The answer is
related to the existence of sharp variations in opacity with
temperature.  Consider a disk near the lower edge of the opacity gap.
Once gravitational instability sets in, fluctuations in temperature move
parts of the disk into the opacity gap.  There, the opacity is reduced
by orders of magnitude.  Since the cooling rate for an optically thick
disk is proportional to $\kappa^{-1}$, the cooling time drops by a
similar factor.   Relatively small variations in temperature can thus
produce large variations in cooling rate.

\begin{figure}[h]
\centering
\includegraphics[width=5.in,clip]{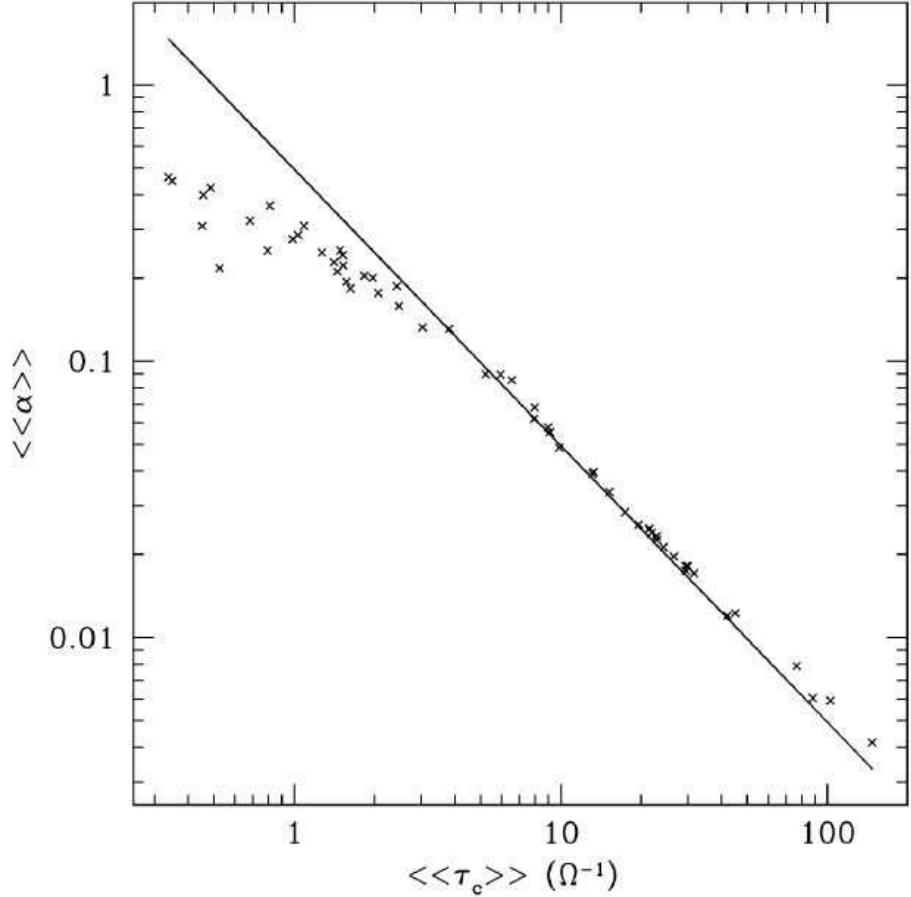}
\caption[Time-averaged shear stress vs. effective cooling time for a series of runs.]
{Time-averaged shear stress vs. effective cooling time for a series of
runs. The solid line shows the analytic result, based on energy
conservation, from equation (\ref{ANALPHA}).}
\label{pap1f12}
\end{figure}

As in \cite{gam01}, the result $\tce\Omega \gtrsim 1$ also implies a
constraint on $\<\< \alpha \>\>$.  Energy conservation implies that
\begin{equation}\label{AVGSOL}
\qq \Omega \<\< W_{xy} \>\> = \<\< \Lambda \>\>,
\end{equation}
where $W_{xy}$ is the total shear stress (hydrodynamic plus gravitational). Equivalently, stress by rate-of-strain
is equal to the dissipation rate. Using the definition of $\tce$, this implies
\begin{equation}\label{ANALPHA}
\<\< \alpha \>\> = \left(\gamma (\gamma - 1) \frac{9}{4} \Omega \tce
	\right)^{-1}.
\end{equation}
Hence $\tce\Omega \gtrsim 1$ implies $\<\< \alpha \>\> \lesssim 1$.  Figure~\ref{pap1f12}
shows $\<\< \alpha \>\>$ vs $\tce$ for a large number of runs plotted against
equation (\ref{ANALPHA}).  For small values of $\tce$ the numerical
values lie below the line.  These models are not in equilibrium (i.e., not in a
statistically-steady gravito-turbulent state), so the
time average used in equation (\ref{AVGSOL}) is not well defined.  For
larger values of $\tce$ numerical results typically (there is noise in
the measurement of both $\<\< \alpha \>\>$ and $\tce$ because the time
average is taken over a finite time interval) lie slightly above the
analytic result.

The bias toward points lying slightly above the line reflects the fact
that $\<\< \alpha \>\>$ measures the rate of energy extraction from the shear
while $\tce$ measures the rate at which that energy is transformed into
thermal energy.  If energy is lost, perhaps to numerical averaging at
the grid scale, then more energy must be extracted from the shear flow
to make up the difference.  Overall, however, the agreement with the
analytic result is good and demonstrates good energy conservation
in the code.

The relationship between $\tce$ and $\<\< \alpha \>\>$ is interesting but not
particularly useful because $\tce$ is no more readily calculated than
$\<\< \alpha \>\>$; it depends on a complicated moment of the surface density and
temperature.  Only for constant cooling time have we been able to evaluate this moment
analytically.

\subsection{Isothermal Disks}

We have assumed that external illumination of the disk is negligible.
This approximation is valid when the effective temperature $T_{irr}$ of
the external irradiation is small compared to the central temperature of
the disk.  In the opposite limit, illumination controls the energetics
of the disk and it is isothermal (if it is illuminated directly so that
shadowing effects, such as those considered by \cite{js03} are
negligible).

It is therefore worth studying the outcome of gravitational instability
in an isothermal disk.  The isothermal disk model has a single
parameter: the initial value of $Q$.  We ran models with varying values
of $Q$ and with $\<\delta v^2\>/c_s^2 = 0.1$.  We find that models with
$Q \lesssim 1.4$ fragment.

It is likely that the mass of the fragments, etc., depends on how an
isothermal disk becomes unstable.  Rapid fluctuation of the external
radiation field is likely to produce a different outcome than dimming on a
timescale long compared to the dynamical time.

\section{Discussion}\label{pap1s5}

Using numerical experiments, we have identified
those disks that are likely to fragment absent external
heating.  Disks with effective cooling times $\tce \lesssim
\Omega^{-1}$ are susceptible to fragmentation.  This is what one might
expect based on the simple argument of \cite{sbf90}: if the disk cools
more quickly than the self-gravitating condensations can collide with
one another, then those collisions (which occur on a timescale $\sim
\Omega^{-1}$) cannot reheat the disk and fragmentation is inevitable.  But our
results are at the same time surprising.

The effective cooling time depends on the nonlinear outcome of
gravitational instability.  It depends on the cooling function, which in
turn depends sensitively on $\Sigma$ and $U$.  Since $\Sigma$ and $U$
vary strongly over the disk once gravitational instability has set in,
it is difficult to estimate $\tce$ directly.  One might be tempted to
estimate $\tce(\Sigma,\Omega) \simeq \tco(\Sigma_o,\Omega,Q = 1)$, but
our experiments show that this estimate can be off by as much as four
orders of magnitude.  The effect is particularly pronounced near sharp
features in the opacity.  For example, consider a model initially
located just below the opacity gap with $\tco\Omega \gg 1$.
Gravitational instability creates dense regions with higher
temperatures, where dust is destroyed.  The result is rather like having
to shed one's blanket on a cold winter morning: the disk loses its
thermal energy suddenly.  Pressure support is lost and gravitational
collapse ensues.

The difference between $\tce$ and $\tco (Q = 1)$ implies that a much
larger region of the disk is susceptible to fragmentation than naive
estimates based on the approximation $\tce \approx \tco$ might suggest.
For example, consider an equilibrium disk model with $Q \gg 1$ at small
$r$.  As $r$ increases, $Q$ declines.  Eventually $Q\sim 1$ and
gravitational instability sets in.  There is then a range of radii where
$Q \sim 1$, $\tce\Omega \gtrsim 1$ and recurrent gravitational
instability can transport angular momentum and prevent collapse.
Generally speaking, however, the cooling time decreases with increasing
radius.  Eventually  $\tce\Omega \sim 1$ and fragmentation cannot be
avoided.  By lowering our estimate of $\tce$, we narrow the range of
radii over which recurrent gravitational instability can occur.

The general sense of our result is that it is extremely difficult to
prevent a marginally-stable, $Q \sim 1$, optically-thick disk from
fragmenting and forming planets (in circumstellar disks) or stars (in
circumstellar and circumnuclear disks).  This is particularly true for
disks with $T \sim 10^3\K$, whose opacity is dominated by dust grains,
i.e. disks whose temperature lies within a factor of several of the
opacity gap.

Our numerical model uses a number of approximations.  First, our
treatment is razor-thin, i.e. all the matter is in a thin slice at
$z = 0$.  The effect of finite thickness on linear stability has been
understood since \cite{glb65}: it is stabilizing because gravitational
attraction of neighboring columns of disk is diluted by finite
thickness.  The size of the effect may be judged by the fact that $Q =
0.676$ is required for marginal stability of a finite-thickness,
isothermal disk.  

The behavior of a finite-thickness disk in the nonlinear regime is more
difficult to predict.  Shocks will evidently deposit some of their
energy away from the midplane, where it can be radiated away more
quickly (because the energy is deposited at smaller optical depth - see \citealt{pcdl00}).
Radiative diffusion parallel to the disk plane (not included here) may
enhance cooling of dense, hot regions.  Both these effects are
destabilizing.  Ultimately, however, a numerical study is required.
This is numerically expensive:  one
must resolve the disk vertically, on the scale height $H$, and
horizontally, at the critical wavelength $2 \pi Q H$.

Second, we have ignored magnetic fields.  While there may be
astrophysical situations where cool disks have such low ionization that
they are unmagnetized, most disks are likely to contain dynamically
important magnetic fields that give rise to a dimensionless shear stress
$\<\< \alpha \>\> \gtrsim 0.01$ (e.g. \citealt{hgb95}).  These fields are likely to
remove spin angular momentum from partially collapsed objects,
destabilizing them.  Numerical experiments including both gravitational
fields and magnetohydrodynamics are necessarily three dimensional (the
instability of \cite{bh91} requires $\partial_z \ne 0$), and are thus
numerically expensive.

Third, we have fixed $\gamma$ and $\mu$ for the duration of each
simulation.  This eliminates the soft spots in the equation of state
associated with ionization of atomic hydrogen and dissociation of
molecular hydrogen.  In these locations the three dimensional $\gamma$
dips below $4/3$, which is destabilizing.

Fourth, we have treated the physics of grain destruction and formation
very simply.  In using the \cite{bl94} opacities we implicitly assume
that grains reform in cooling gas on much less than a dynamical time.
It is likely that grain re-formation will take some time (e.g.
\citealt{hess91}) and this will further reduce the disk opacity and
enhance fragmentation.

Fifth, we have neglected the effects of illumination.  In the limit of
strong external illumination, i.e. when the effective temperature of the
irradiation $T_{irr}$ is large compared to the disk central temperature
$T_c$, the disk is isothermal (here $T_c$ is the temperature of a dense
condensation).  We have carried out isothermal experiments and shown
that, for initial velocity perturbations with $\< \delta v^2 \>/c_s^2 = 0.1$,
disks with $Q \lesssim 1.4$ fragment.  Weaker illumination produces a more
complicated situation that we have not explored here.
Illumination-dominated disks that become unstable presumably do so
because the external illumination declines, and the rate at which the
external illumination changes may govern the nonlinear outcome.

We conclude that disks with $\tce\Omega \lesssim 1$ do not exist.
Cooling in this case is so effective that fragmentation into condensed
objects-- stars, planets, or smaller accretion disks-- is inevitable.

As an example application of this result, consider the model for the
nucleus of NGC 1068 recently proposed by \cite{lb03}.  Their model is an
extended marginally-stable self-gravitating disk of the type
investigated here and originally proposed by \cite{glb65} for galactic
disks and \cite{pac78} for accretion disks, although their disk is
sufficiently massive that it modifies the rotation curve as well.  Based
on their Figure 3, at a typical radius of $0.5\pc$, $\Sigma_o \simeq
10^4$ and $\Omega \simeq 10^{-9}$.  According to our Figure 7 this disk is
about 2 orders of magnitude too dense to avoid fragmentation.  While it
may be possible to avoid this conclusion by invoking strong external
heating, the energy requirements are severe, as outlined in
\cite{good03}.  The disk proposed by \cite{lb03} would therefore fragment
into stars on a short timescale.


\end{spacing}

\chapter{Linear Theory of Thin, Radially-Stratified Disks}\label{paper2}

\begin{spacing}{1.5}

\section{Chapter Overview}

We consider the nonaxisymmetric linear theory of radially-stratified
disks.  We work in a shearing-sheet-like approximation, where the
vertical structure of the disk is neglected, and develop equations for
the evolution of a plane-wave perturbation comoving with the shear flow
(a shearing wave, or ``shwave''). We calculate a complete solution set
for compressive and incompressive short-wavelength perturbations in both
the stratified and unstratified shearing-sheet models. We develop
expressions for the late-time asymptotic evolution of an individual
shwave as well as for the expectation value of the energy for an
ensemble of shwaves that are initially distributed isotropically in
$k$-space. We find that: (i) incompressive, short-wavelength
perturbations in the unstratified shearing sheet exhibit transient
growth and asymptotic decay, but the energy of an ensemble of such
shwaves is constant with time; (ii)
short-wavelength compressive shwaves grow asymptotically in the
unstratified shearing sheet, as does the energy of an ensemble of such
shwaves; (iii) incompressive shwaves in the stratified shearing sheet
have density and azimuthal velocity perturbations $\delta \Sigma$,
$\delta v_y \sim t^{-{\rm Ri}}$ (for $|{\rm Ri}| \ll 1$), where ${\rm
Ri} \equiv N_x^2/ (\qe \Omega)^2$ is the Richardson number, $N_x^2$ is
the square of the radial
Brunt-V$\ddot{\rm{a}}$is$\ddot{\rm{a}}$l$\ddot{\rm{a}}$ frequency and
$\qe \Omega$ is the effective shear rate; (iv) the energy of an ensemble
of incompressive shwaves in the stratified shearing sheet behaves
asymptotically as $\Ri \, t^{1-4{\rm Ri}}$ for $|{\rm Ri}| \ll 1$. For
Keplerian disks with modest radial gradients, $|{\rm Ri}|$ is expected
to be $\ll 1$, and there will therefore be weak growth in a single
shwave for ${\rm Ri} < 0$ and near-linear growth in the energy of an
ensemble of shwaves, independent of the sign of Ri.\footnote{To be published in ApJ
Volume 626, Issue 2. Reproduction for this dissertation is authorized by the
copyright holder.}

\section{Introduction}

Angular momentum transport is central to the evolution of astrophysical
disks.  In many disks angular momentum is likely redistributed
internally by magnetohydrodynamic (MHD) turbulence driven by the
magnetorotational instability (MRI; see \citealt{bh98}).  But in portions of
disks around young, low-mass stars, in cataclysmic-variable disks in
quiescence, and in X-ray transients in quiescence \citep{sgbh00,gm98,men00}, 
disks may be composed of gas that is so neutral that the MRI fails.  It is 
therefore of interest to understand if there are purely hydrodynamic 
mechanisms for driving turbulence {\it and} angular momentum transport 
in disks.

The case for hydrodynamic angular momentum transport is not promising.
Numerical experiments carried out under conditions similar to those
under which the MRI produces ample angular momentum fluxes-- local
shearing-box models-- show small or negative angular momentum fluxes
when the magnetic field is turned off \citep{hgb95,hgb96}.  Unstratified 
shearing-sheet models show decaying angular momentum flux and kinetic 
energy when nonlinearly perturbed, yet recover the well known, high 
Reynolds number nonlinear instability of plane Couette flow when the 
parameters of the model are set appropriately (\citealt{bhs96}; see, however, 
the recent results by \citealt{ur04}). Local models 
with unstable vertical stratification show overturning and the development 
of convective turbulence, but the mean angular momentum flux is small and 
of the wrong sign \citep{sb96}.

Linear theory of global disk models has long indicated the presence of
instabilities associated with reflecting boundaries or features in the
flow (see e.g., \citealt{pp84,pp85,pp87,ggn86,gng87,ngg87,llcn99,lflc00}).
Numerical simulations of the nonlinear outcome of these instabilities
suggest that they saturate at low levels and are turned off by modest 
accretion \citep{bla87,haw91}.  One might guess that in the nonlinear
outcome these instabilities will attempt to smooth out the features that 
give rise to them, much as convection tends to erase its parent entropy 
gradient.  There are some suggestions, however, that such instabilities saturate
into long-lived vortices, which may serve as obstructions in the flow that
give rise to angular momentum transport \citep{lcwl01}.  We will consider
this possibility in a later publication.

Linear theory has yet to uncover a {\it local} instability of
hydrodynamic disks that produces astrophysically-relevant angular
momentum fluxes.  Because of the absence of a complete set of modes in
the shearing-sheet model, however, local linear stability is difficult
to prove.  Local nonlinear stability may be impossible to prove.
Comparison with laboratory Couette flow experiments is complicated by
several factors, not least of which is the inevitable presence of solid
radial boundaries in the laboratory that have no analogue in
astrophysical disks.

Recently, however, \cite{kb03} (hereafter KB03) have claimed to find a local 
hydrodynamic instability in global numerical simulations: the ``Global 
Baroclinic Instability.'' The 
instability arises in a model with scale-free initial conditions (an 
equilibrium entropy profile that varies as a power-law in 
radius) and thus does not depend on sharp features in the flow. 
\cite{klr04} has performed a local linear stability analysis of a
radially-stratified accretion disk in an effort to explain the numerical
results obtained by KB03. The instability mechanism invoked is the
phenomenon of transient amplification as a shearing wave goes from
leading to trailing. This is the mechanism that operates for
nonaxisymmetric shearing waves in a disk that is nearly unstable to the
axisymmetric gravitational instability \citep{glb65,jt66,gt78}. It is the 
purpose of this work to clarify and extend the linear analysis of
\cite{klr04}. If this instability exists it could be important for the
evolution of low-ionization disks.

To isolate the cause for instabilities originally observed in global 3D
simulations, KB03 perform both local and global 2D calculations
in the ($R,\phi$)-plane. The local simulations use a new set of boundary
conditions termed the shearing-disk boundary conditions.  The model is
designed to simulate a local portion of the disk without neglecting
global effects such as curvature and horizontal flow gradients. The
boundary conditions, which are described in more detail in KB03,
require the assumption of a power-law scaling for the mean values of
each of the variables, as well as the assumption that the fluctuations
in each variable are proportional to their mean values. The radial
velocity component in the inner and outer four grid cells is damped by
$5\%$ each time step in order to remove artificial radial oscillations
produced by the model.\footnote{It is not surprising that shearing disk
boundary conditions as implemented in KB03 produce features on the
radial boundary, because the Coriolis parameter is discontinuous across
the radial boundary.}

The equilibrium profile for KB03's 2D runs was a constant surface density
$\Sigma$ with either a constant temperature $T$ or a temperature profile
$T \propto R^{-1}$. The constant-$T$ runs showed no instability while
those with varying $T$ (and thus varying entropy) sustained turbulence
and positive Reynolds stresses.\footnote{Notice that with a constant
$\Sigma$, the constant-$T$ runs have no variation in any of the
equilibrium variables, so it is not clear that the effects being
observed in the 2D calculations are due to the presence of an entropy
gradient rather than due simply to the presence of a pressure gradient.}
The fiducial local simulations were run at a resolution of $64^2$, with
a spatial domain of $R = 4$ to $6$ AU and $\Delta \phi = 30\,^{\circ}$.
The unstable run was repeated at a resolution of $128^2$, along with a
run at twice the physical size of the fiducial runs. One global model
(with nonreflecting outflow boundary conditions) was run at a
resolution of $128^2$ with a spatial domain of $R = 1$ to $10$ AU and
$\Delta \phi = 360\,^{\circ}$. All the runs yielded similar results,
with the larger simulations producing vortices and power on large
scales.

KB03 have chosen the term ``baroclinic instability'' by way of
analogy with the baroclinic instability that gives rise to weather
patterns in the atmosphere of the Earth and other planets (see e.g. 
\citealt{ped87}).\footnote{A baroclinic flow is one in which surfaces 
of constant density are inclined with respect to surfaces of constant
pressure. If these surfaces coincide, the flow is termed barotropic.} 
The analogy is somewhat misleading, however, since the baroclinic 
instability that arises in planetary contexts is due to a baroclinic 
equilibrium. In a planetary atmosphere, a baroclinically-unstable 
situation requires stratification in both the vertical and latitudinal 
directions.\footnote{Contrary to the claim in \cite{klr04}, the 
two-layer model \citep{ped87} does not ignore the vertical structure; 
it simply considers the lowest-order vertical mode.} The stratification 
in KB03 is only in the radial direction, and as a result the
equilibrium is barotropic. It is the perturbations that are baroclinic; 
i.e., the disk is only baroclinic at linear order in the amplitude of a 
disturbance.

\cite{cab84} and \cite{ks86} have analyzed a thin disk with a baroclinic
equilibrium state (with both vertical and radial gradients). The latter
find that due to the dominant effect of the Keplerian shear, the
instability only occurs if the radial scale height is comparable to the
vertical scale height, a condition which is unlikely to be
astrophysically relevant.  As pointed out in KB03, the salient feature
that is common to their analysis and the classical baroclinic
instability is an equilibrium entropy gradient in the horizontal
direction. As we show in \S 2, however, an entropy gradient is not
required in order for two-dimensional perturbations to be baroclinic;
any horizontal stratification will do.

The ``Global Baroclinic Instability'' claimed by KB03 is thus
analogous to the classical baroclinic instability in the sense that both
have the potential to give rise to convection.\footnote{The classical
baroclinic instability gives rise to a form of ``sloping convection''
\citep{hou02} since the latitudinal entropy gradient is inclined with
respect to the vertical buoyancy force.} When neglecting vertical
structure, however, the situation in an accretion disk is more closely
analogous to a shearing, stratified atmosphere, the stability of which
is governed by the classical Richardson criterion \citep{jwm61,chi70}. 
The only additional physics in a disk is the presence of the Coriolis force.
Most analyses of a shearing, stratified atmosphere, however, only 
consider stratification profiles that are stable to convection. The primary
question that \cite{klr04} and this work are addressing, then, is
whether or not the presence of shear stabilizes a stratified equilibrium 
that would be unstable in its absence.

We begin in \S\ref{pap2s2} by outlining the basic equations for a local model of a
thin disk.  \S\S\ref{pap2s3} and \ref{pap2s4} describe the local linear theory for
nonaxisymmetric sinusoidal perturbations in unstratified and
radially-stratified disks, respectively.  We summarize and discuss the
implications of our findings in \S\ref{pap2s5}.

\section{Basic Equations}\label{pap2s2}

The effect of radial gradients on the local stability of a thin disk can
be analyzed most simply in the two-dimensional shearing-sheet
approximation\footnote{See \cite{rg92} for a discussion of why this
approximation is appropriate for an analysis of local stability.  See 
also \cite{mp77}, who use a similar approach to demonstrate the
stability of unbounded viscous plane Couette flow.}.
This is obtained by a rigorous expansion of the equations
of motion in the ratio of the vertical scale height $H$ to the local radius $R$,
followed by a vertical integration of the fluid equations. The basic
equations that one obtains (e.g., \citealt{gt78}) are
\be\label{EQ1}
\dv{\Sigma}{t} + \Sigma \bnabla \cdot \bv = 0,
\ee
\be\label{EQ2}
\dv{\bv}{t} + \frac{\bnabla P}{\Sigma} + 2\bO\times\bv - 2q\Omega^2 x \ex = 0,
\ee
\be\label{EQ3}
\dv{\,{\rm{ln}} S}{t} = 0,
\ee
where $\Sigma$ and $P$ are the two-dimensional density and pressure,
$S \equiv P \Sigma^{-\gamma}$ is monotonically related to the fluid
entropy,\footnote{With the assumptions of vertical hydrostatic equilibrium 
and negligible self-gravity, the effective two-dimensional adiabatic index
can be shown to be $\gamma = (3\gamma_{3D} - 1)/(\gamma_{3D} + 1)$ 
(e.g. \citealt{ggn86}).} $\bv$ is the fluid velocity and $d/dt$ is the
Lagrangian derivative. The third and fourth terms in equation
(\ref{EQ2}) represent the Coriolis and centrifugal forces in the local
model expansion, where $\Omega$ is the local rotation frequency, $x$ is
the radial Cartesian coordinate and $q$ is the shear parameter (equal to
$1.5$ for a disk with a Keplerian rotation profile). The gravitational
potential of the central object is included as part of the centrifugal
force term in the local-model expansion, and we ignore the self-gravity
of the disk.

It is worth emphasizing at this point that we have integrated out the
vertical degrees of freedom in the model.  We will later focus on
perturbations with planar wavelengths that are small compared to a scale
height, and these perturbations will be strongly influenced by the
vertical structure of the disk.  

Equations (\ref{EQ1}) through (\ref{EQ3}) can be combined into a single
equation governing the evolution of the potential vorticity:
\be\label{PVEV}
\dv{}{t}\left(\frac{\bnabla \times \bv + 2\bO}{\Sigma}\right) \equiv
\dv{\bxi}{t} = \frac{\bnabla \Sigma \times \bnabla P}{\Sigma^3}.
\ee
In two dimensions, $\bxi$ has only one nonzero component and can
therefore be regarded as a scalar. Equation (\ref{PVEV}) demonstrates 
that for $P \equiv P(\Sigma)$ (as in the case of a strictly adiabatic evolution 
with isentropic initial conditions), the potential vorticity of fluid elements 
is conserved. For $P \neq P(\Sigma)$, however, the potential vorticity 
evolves with time. A barotropic equilibrium stratification can result in 
baroclinic perturbations that cause the potential vorticity to evolve at linear 
order. This can be seen by linearizing the scalar version of equation 
(\ref{PVEV}):
\be\label{PVEVLIN}
\pdv{\delta \xi}{t} + \bv_0 \cdot \bnabla \delta \xi + \delta \bv \cdot \bnabla \xi_0 = \frac{\ez \cdot(\bnabla 
\Sigma_0 \times \bnabla \delta P - \bnabla P_0 \times \bnabla \delta \Sigma)}{\Sigma_0^3},
\ee
where we have dropped the term $\propto \bnabla \Sigma_0 \times \bnabla
P_0$. Notice that an entropy gradient is not required for the evolution of 
the perturbed potential vorticity. For $S_0 = P_0 \Sigma_0^\gamma = 
constant$, equation (\ref{PVEVLIN}) reduces to
\be\label{PVEVLIN2}
\pdv{\delta \xi}{t} + \bv_0 \cdot \bnabla \delta \xi + \delta \bv \cdot \bnabla \xi_0 = \frac{\ez \cdot(\bnabla 
P_0 \times \bnabla \delta S)}{\gamma \Sigma_0^2 S_0}.
\ee
Potential vorticity is conserved only in the limit of zero stratification ($P_0 
= constant$) or adiabatic perturbations ($\delta S = 0$).

\section{Unstratified Shearing Sheet}\label{pap2s3}

Our goal is to understand the effects of radial stratification, but we
begin by developing the linear theory of the standard (unstratified)
shearing sheet, in which the equilibrium density and pressure are
assumed to be spatially constant.  This will serve to establish notation
and method of analysis and to highlight the changes introduced by
radial stratification in the next section.

Our analysis follows that of \cite{gt78} except for our neglect of
self-gravity. The equilibrium consists of a uniform sheet with $\Sigma =
\Sigma_0 = constant$, $P = P_0 = constant$, and $\bv_0 = -q\Omega x
\ey$. We consider nonaxisymmetric Eulerian perturbations about this
equilibrium with space-time dependence $\delta(t){\rm exp} (ik_x(t) x +
ik_y y)$, where
\be\label{KX}
k_x(t) \equiv k_{x0} + q\Omega k_y t
\ee
(with $k_{x0}$ and $k_y > 0$ constant) is required to allow for a spatial Fourier 
decomposition of the perturbation. We will refer to these perturbations 
as shearing waves, or with some trepidation, but more compactly, as 
``shwaves''.

\subsection{Linearized Equations}

To linear order in the perturbation amplitudes, the dynamical equations
reduce to
\be\label{LIN1s}
\frac{\dot{\delta \Sigma}}{\Sigma_0} + i k_x \delta v_x 
	+ i k_y \delta v_y = 0,
\ee
\be\label{LIN2s}
\dot{\delta v}_x - 2\Omega \delta v_y 
	+ ik_x \frac{\delta P}{\Sigma_0} = 0,
\ee
\be\label{LIN3s}
\dot{\delta v}_y + (2- q)\Omega \delta v_x 
	+ ik_y \frac{\delta P}{\Sigma_0} = 0,
\ee
\be\label{LIN4s}
\frac{\dot{\delta P}}{\Sigma_0} + c_s^2 (i k_x \delta v_x 
	+ i k_y \delta v_y)  = 0,
\ee
where $c_s^2 = \gamma P_0/\Sigma_0$ is the square of the equilibrium
sound speed and an over-dot denotes a time derivative.

The above system of equations admits four linearly-independent
solutions. Two of these are the nonvortical shwaves (solutions for 
which the perturbed potential vorticity is zero), which in the absence 
of self-gravity can be solved for exactly. The remaining two solutions are 
the vortical shwaves. When $k_y \rightarrow 0$ the latter reduce to the 
zero-frequency modes of the axisymmetric version of equations 
(\ref{LIN1s}) through (\ref{LIN4s}). One of these (the entropy mode) 
remains unchanged in nonaxisymmetry (in a frame comoving with the 
shear). There is thus only one nontrivial vortical shwave in the 
unstratified shearing sheet.

In the limit of tightly-wound shwaves ($|k_x| \gg k_y$), the nonvortical 
and vortical shwaves are compressive and incompressive, respectively. 
In the short-wavelength limit ($H k_y \gg 1$, where $H \equiv c_s/\Omega$ 
is the vertical scale height), the compressive and incompressive solutions  
remain well separated at all times, but for $H k_y \lesssim O(1)$ there is mixing 
between them near $k_x = 0$ as an incompressive shwave shears from leading 
to trailing.\citep[][; also Goodman 2005, private communication]{chag97}
With the understanding that the distinction between compressive shwaves 
and incompressive shwaves as separate solutions is not valid for all time 
when $H k_y \lesssim O(1)$, we generally choose to employ these terms over 
the more general but less intuitive terms ``nonvortical'' and ``vortical.''

Based upon the above considerations, it is convenient to study the 
vortical shwave in the short-wavelength, low-frequency ($\partial_t \ll 
c_s k_y$) limit. This is equivalent to working in the Boussinesq 
approximation,\footnote{We demonstrate this equivalence in the Appendix.} 
which in the unstratified shearing sheet amounts to assuming incompressible 
flow. In this limit, equation (\ref{LIN1s}) is replaced with
\be\label{LIN1b}
k_x \delta v_x + k_y \delta v_y = 0.
\ee
This demonstrates the incompressive nature of the vortical shwave in 
the short-wavelength limit.

\subsection{Solutions}

In the unstratified shearing sheet, equation (\ref{PVEVLIN}) for the 
perturbed potential vorticity can be integrated to give:
\be\label{POTVORT}
\delta \xi_u = \frac{i k_x \delta v_y - i k_y \delta v_x}{\Sigma_0} - 
\xi_0 \frac{\delta \Sigma}{\Sigma_0} = constant,
\ee
where $\xi_0 = (2 - q)\Omega/\Sigma_0$ is the equilibrium potential
vorticity and we have employed the subscript $u$ to highlight the fact 
that the perturbed potential vorticity is only constant in the unstratified 
shearing sheet. To obtain the compressive-shwave solutions, we set the 
constant $\delta \xi_u$ to zero. Combining equations (\ref{LIN3s}) and
(\ref{POTVORT}) with $\delta \xi_u = 0$, one obtains an expression for
$\delta v_{xc}$ in terms of $\delta v_{yc}$ and its derivative:
\be\label{CVX}
\delta v_{xc} = \frac{c_s^2 k_x k_y \delta v_{yc} - \xi_0 \Sigma_0
\dot{\delta v}_{yc}}{\xi_0^2 \Sigma_0^2 + c_s^2 k_y^2},
\ee
where the subscript $c$ indicates a compressive shwave.  The associated 
density and pressure perturbations are
\be\label{CS}
\delta \Sigma_c = \frac{\delta P_c}{c_s^2} = i \frac{\xi_0 \Sigma_0 k_x 
\delta v_{yc} + k_y \dot{\delta v}_{yc}}{\xi_0^2 \Sigma_0^2 + c_s^2 k_y^2}
\ee
via equation (\ref{POTVORT}). Reinserting equation (\ref{CVX}) into
equation (\ref{LIN3s}), taking one time derivative and replacing
$\dot{\delta P}$ via equation (\ref{LIN4s}), we obtain the following
remarkably simple equation:
\be\label{SOUND}
\ddot{\delta v}_{yc} + \left(c_s^2 k^2 + \kappa^2\right) \delta v_{yc} = 0,
\ee
where $k^2 = k_x^2 + k_y^2$ and $\kappa^2 = (2 - q)\Omega^2$ is the
epicyclic frequency. Changing to the dimensionless dependent variable
\be\label{TTAU}
T \equiv i \sqrt{\frac{2 c_s k_y}{q\Omega}} \left(q \Omega t + \frac{k_{x0}}
{k_y}\right) \equiv i \sqrt{\frac{2 c_s k_y}{q\Omega}} \, \tau
\ee
and defining
\be
C \equiv \frac{c_s^2 k_y^2 + \kappa^2}{2 q\Omega c_s k_y},
\ee
the equation governing $\delta v_{yc}$ becomes
\be\label{VYT}
\dv{^2\delta v_{yc}}{T^2} + \left(\frac{1}{4}T^2 - C\right) \delta v_{yc} = 0.
\ee
This is the parabolic cylinder equation (e.g. \citealt{as72}), 
the solutions of which are parabolic cylinder functions. One 
representation of the general solution is
\be\label{CVY}
\delta v_{yc} = e^{-\frac{i}{2}T^2}\left[c_1 \, M\left(\frac{1}{4} - 
\frac{i}{2}C,\frac{1}{2},\frac{i}{2}T^2\right) + c_2 \, T \, M\left(
\frac{3}{4} - \frac{i}{2}C,\frac{3}{2},\frac{i}{2}T^2\right)\right],
\ee
where $c_1$ and $c_2$ are constants of integration and $M$ is a 
confluent hypergeometric function. This completely specifies the 
compressive solutions for the unstratified shearing sheet, for any 
value of $k_y$.

Equation (\ref{VYT}) has been analyzed in detail by \cite{ngg87}; 
their modal analysis yields the analogue of equation (\ref{VYT}) in 
radial-position space rather than in the radial-wavenumber ($k_x = 
k_y \tau$) space that forms the natural basis for our shwave 
analysis. One way of seeing the correspondence between the 
modes and shwaves is to take the Fourier transform of the asymptotic 
form of the solution. Appropriate linear combinations of the solutions 
given in equation (\ref{CVY}) have the following asymptotic time 
dependence for $\tau \gg 1$:
\be
\delta v_{yc} \propto \sqrt{\frac{2}{T}}\exp\left(\pm \frac{i}{4}T^2 
\right) \propto \frac{1}{\sqrt{k_x}}\exp\left(\pm i \int c_s k_x 
\, dt \right).
\ee
The Fourier transform of the above expression, evaluated by the method of 
stationary phase for $H k_y \gg 1$, yields
\be
\delta v_{yc}(X) \propto \sqrt{\frac{2}{X}}\exp\left(\pm \frac{i}{4}X^2\right),
\ee
which is equivalent to the expressions given for the modes analyzed by 
\cite{ngg87}, in which the dimensionless spatial variable (with zero frequency, 
so that corotation is at $x = 0$) is defined as
\be
X \equiv \sqrt{\frac{2 q \Omega k_y}{c_s}} x.
\ee

To obtain the incompressive shwave, we use the condition of incompressibility
(equation (\ref{LIN1b})) to write $\delta v_y$ in terms of $\delta v_x$,
and then combine the dynamical equations (\ref{LIN2s}) and (\ref{LIN3s})
to eliminate $\delta P$. The incompressive shwave is given by:
\be\label{IVX}
\delta v_{xi} = \delta v_{xi0}\frac{k_0^2}{k^2},
\ee
\be\label{IVY}
\delta v_{yi} = -\frac{k_x}{k_y} \delta v_{xi},
\ee
\be\label{IS}
\frac{\delta \Sigma_i}{\Sigma_0} = \frac{\delta P_i}{\gamma P_0} = 
\frac{1}{i c_s k_y}\left(\frac{k_x}{k_y} \frac{\dot{\delta v}_{xi}}{c_s} 
+ 2(q - 1) \Omega \frac{\delta v_{xi}}{c_s}\right),
\ee
where the subscript $i$ indicates an incompressive shwave, $k_0^2 =
k_{x0}^2 + k_y^2$ and $\delta v_{xi0}$ is the value of $\delta v_{xi}$
at $t=0$.\footnote{\cite{cztl03} obtained this solution by starting with
the assumption of incompressibility.  In the incompressible limit, it is 
an exact nonlinear solution to the fluid equations.} This solution is uniformly valid  
for all time to leading order in $(H k_y)^{-1} \ll 1$.

\subsection{Energetics of the Incompressive Shwaves}\label{pap2eis}

We define the kinetic energy in a single incompressive shwave as
\be
E_{ki} \equiv \frac{1}{2}\Sigma_0 (\delta v_{xi}^2 + \delta v_{yi}^2) =
\frac{1}{2}\Sigma_0 \delta v_{xi}^2 \frac{k^2}{k_y^2} =
\frac{1}{2}\Sigma_0 \delta v_{xi0}^2 \frac{k_0^4}{k_y^2 k^2},
\ee
which peaks at $k_x = 0$. This is not the only possible
definition for the energy associated with a shear-flow disturbance;
see Appendix A of \cite{ngg87} for a discussion of the subtleties
involved in defining a perturbation energy in a differentially-rotating
system. The energy defined above can simply be regarded as a convenient
scalar measure of the shwave amplitude.

One can also define an amplification factor for an individual shwave,
\be
{\cal A } \equiv \frac{E_{ki}(k_x = 0)}{E_{ki}(t = 0)} =
1 + \frac{k_{x0}^2}{k_y^2},
\ee
which indicates that an arbitrary amount of transient amplification in 
kinetic energy can be obtained as one increases the amount of swing for 
a leading shwave ($k_{x0} \ll -k_y$). This is essentially the mechanism
invoked by \cite{cztl03}, \cite{ur04} and \cite{amn04} to argue for the onset 
of turbulence in unmagnetized Keplerian disks.

Because only a small subset of all Fourier components achieve large
amplification (those with initial wavevector very nearly aligned with
the radius vector), one must ask what amplification is achieved for an
astrophysically relevant set of initial conditions containing a
superposition of Fourier components.  It is natural to draw such a set
of Fourier components from a distribution that is isotropic, or nearly
so, when $k_0$ is large.

Consider, then, perturbing a disk with a random set of incompressive
perturbations (initial velocities perpendicular to $\bk_0$) drawn from an
isotropic, Gaussian random field and asking how the expectation value for the
kinetic energy associated with the perturbations evolves with time.  The
evolution of the expected energy density is given by the following
integral:
\be
\<E_i\> = L^2 \int d^2k_0 \<E_{ki}\> 
= L^2 \int d^2k_0 \frac{1}{2}\Sigma_0
\< \delta v_{xi0}^2\> \frac{k_0^4}{k_y^2 k^2}.
\ee
where $\<\>$ indicates an average over an ensemble of initial
conditions, the first equality follows from Parseval's theorem, the
second equality follows from the incompressive shwave solution
(\ref{IVX})-(\ref{IS}) and therefore applies only for $k_0 H \gg 1$,
and $L^2$ is a normalizing factor with units of length squared.

For initial conditions that are isotropic in $\bk_0$ ($\delta v_{xi0} =
\delta v_\perp (k_0,\theta) \sin \theta$, where $\<\delta v_\perp^2
(k_0)\>$ is the expectation value for the initial incompressive
perturbation as a function of $k_0$ and $\tan\theta=k_y/k_{x0}$), the
integral becomes
\be\label{EKI}
\< E_i\> = \frac{1}{2}\Sigma_0 L^2 \int k_0 dk_0 \< \delta v_\perp^2 (k_0)\>
\int_0^{2\pi} d\theta \, \frac{1}{\sin^2\theta + (q\Omega t \sin\theta 
+ \cos\theta)^2}.
\ee
Changing integration variables to $\tau = q\Omega t + \cot\theta$, the angular 
integral becomes
\be
\int_{-\infty}^{\infty} d\tau \, \frac{2}{1 + \tau^2} = 2\pi,
\ee
which is independent of time; hence
\be
\< E_i\> = \< E_i (t = 0)\>
\ee
and we do not expect the total energy in incompressive shwaves to
evolve.\footnote{Notice that while the energy of each individual shwave
decays asymptotically, the energy of an ensemble does not.  This is
due to the spread of amplification factors in the spectrum of shwaves;
some are amplified by very large factors while others are amplified
very little.}  This same calculation has been performed in the context of 
plane Couette flow by \cite{shep85}, who also points out that the 
amplification factor due to a distribution of wavevectors in an angular 
wedge $\Delta \theta$ has an upper bound of $2\pi/(\Delta \theta)$.  
This indicates that the amplification will be modest unless the initial 
disturbance is narrowly concentrated around a single wavevector.

Although this result may appear to depend in detail on the assumption of
isotropy, one can show that it really only depends on $\< E_{ki} (t =
0)\>$ being smooth near $\sin \theta = 0$, i.e. that there should not be
a concentration of power in nearly radial wavevectors. This can be seen
from the following argument. If we relax the assumption of isotropy, the
angular integral becomes
\be
\int_0^{2\pi} d\theta \, \frac{\< \delta v_\perp^2(k_0,\theta) \>}
{\sin^2\theta + (q\Omega t \sin\theta + \cos\theta)^2}.
\ee
For $q\Omega t \gg 1$ the above integrand is sharply peaked in the
narrow regions around $\tan \theta = -1/(q\Omega t) \ll 1$ (i.e., $\sin
\theta \simeq 0$). One can perform a Taylor-series expansion of $\<
\delta v_\perp^2 (k_0,\theta) \>$ in these regions, and as long as $\<
\delta v_\perp^2(k_0,\theta) \>$ itself is not sharply peaked it is well
approximated as a constant. A modest relaxation of the assumption of
isotropy, then, will result in an asymptotically constant value for the 
energy integral.

Based upon this analysis, large amplification in an individual shwave
does not in itself argue for a transition to turbulence due to transient
growth.  One must also demonstrate that a ``natural'' set of
perturbations can extract energy from the background shear flow.  In the
case of the unstratified shearing sheet, the energy of a random set of
incompressive perturbations remains constant with time.  This is
consistent with the results of \cite{ur04}, who see asymptotic decay in
linear theory, because they work with a finite set of wavevectors, each
of which must decay asymptotically.

\subsection{Energetics of the Compressive Shwaves}

Here we calculate the energy evolution of the compressive shwaves
for comparison purposes. We will consider the evolution of 
short-wavelength compressive shwaves in which only the initial velocity is 
perturbed, both for simplicity and for consistency with our calculation 
of the short-wavelength incompressive shwaves.  As before, we
will assume that the initial kinetic energy is distributed isotropically.

We use the WKB solutions to equation (\ref{SOUND}) with $H k_y \gg 
1$.\footnote{These solutions are the short-wavelength, {\it high}-frequency 
($\partial_t \sim O(c_s k_y)$) limit of the full set of linear equations in 
the shearing sheet; see the Appendix.}  With the initial density perturbation 
set to zero (consistent with our assumption of only initial velocity 
perturbations), the uniformly-valid asymptotic solution to leading order 
in $(H k_y)^{-1}$ is given by
\be\label{VYWKB}
\delta v_{yc} = \delta v_{yc0} \sqrt{\frac{k_0}{k}} \cos(W-W_0),
\ee
\be\label{VXWKB}
\delta v_{xc} = \frac{k_x}{k_y} \delta v_{yc},
\ee
\be\label{SWKB}
\delta \Sigma_c = \frac{i}{c_s^2 k_y}\dot{\delta v}_{yc},
\ee
where the WKB eikonal is given by
\be
W \equiv \int c_s k \, dt = \frac{H k_y}{q} \int \sqrt{1 + \tau^2} \, 
d\tau = \frac{H k_y}{2 q} \left(\tau \sqrt{1 + \tau^2} + \ln\left(\tau + 
\sqrt{1 + \tau^2}\right) \right),
\ee
with $W_0$ being the value of $W$ at $t=0$.\footnote{This is
not the same WKB solution that is calculated in the tight-winding
approximation by \cite{gt78}; in that case $c_s k_y/\kappa \ll 1$, the
opposite limit to that which we are considering here. The two WKB 
solutions match for $\tau \gg 1$ in the absence of self-gravity. We have 
verified the accuracy of this solution by comparing it to the exact 
solution with acceptable results, and it is valid to leading order for all time.}

Using equation (\ref{VXWKB}), the energy integral for the compressive 
shwaves in the short-wavelength limit is
\be
\< E_{c}\> = L^2 \int d^2k_0 \<E_{kc}\> = L^2  
\int d^2k_0 \frac{1}{2}\Sigma_0 \< \delta v_{yc}^2 \>\frac{k^2}{k_y^2}.
\ee
With initial velocities now parallel to ${\bk}_0$ (and again isotropic), 
this becomes
\be
\< E_{c}\> = 
\frac{1}{2}\Sigma_0 L^2
\int k_0 dk_0 \< \delta v_\|^2(k_0) \>
\int_0^{2\pi} d\theta \,
\sqrt{\sin^2\theta + (q\Omega t \sin\theta + \cos\theta)^2}\cos^2(W-W_0).
\ee
For $q \Omega t \gg 1$, the angular integral is approximated by
\be
\int_0^{2\pi} d\theta \, |\sin\theta| \left(1 + \cos(2W-2W_0)\right) \simeq 
2 q \Omega t + \sqrt{\frac{2\pi q \Omega}{c_s k_0}} \cos(c_s k_0 q 
\Omega t^2 - \pi/4),
\ee
where the second approximation comes from employing the method of
stationary phase.\footnote{The first approximation breaks down near
$\sin \theta = 0$, but the contribution of these regions to the integral
is negligible for $q \Omega t \gg 1$, in contrast to the situation for
incompressive shwaves.} In the short-wavelength limit, then,
\be
\< E_{c} (q \Omega t \gg 1) \> =  2 q \Omega t \, \< E_{c}(t = 0) \>.
\ee
Thus the kinetic energy of an initially isotropic distribution of compressive
shwaves grows, presumably at the expense of the background shear flow.  

The fate of a single compressive shwave is to steepen into a weak shock
train and then decay.  The fate of the field of weak shocks generated by
an ensemble of compressive shwaves is less clear, but the mere presence
of weak shocks does not indicate a transition to turbulence.

\section{Radially-Stratified Shearing Sheet}\label{pap2s4}

We now generalize our analysis to include the possibility that the
background density and pressure varies with $x$; this stratification is
required for the manifestation of a convective instability.  In order to
use the shwave formalism we must assume that the background
varies on a scale $L \sim H \ll R$ so that the local model expansion
(e.g., the neglect of curvature terms in the equations of motion) is
still valid.

With this assumption the equilibrium condition becomes
\be\label{V0}
\bv_0 = \left(-q\Omega x + \frac{P_0^\prime(x)}{2\Omega 
\Sigma_0(x)}\right)\ey,
\ee
where a prime denotes an $x$-derivative. 
One can regard the background flow as providing an effective shear rate
\be
\qe \Omega \equiv -v_0^\prime
\ee
that varies with $x$, in which case $\bv_0 = -\int^x \qe(s) ds \, \Omega
\ey$. 

Localized on this background flow we will consider a shearing
wave with $k_y L \gg 1$.  That is, we will consider nonaxisymmetric
short-wavelength Eulerian perturbations with spacetime dependence
$\delta(t) \exp(i\int^x \tilde{k}_x(t,s) ds + ik_y y + ik_z z)$,
where $k_y$ and $k_z$ are constants and
\be
\tilde{k}_x(t,x) \equiv k_{x0} + \qe(x)\Omega k_y t.
\ee  

It may not be immediately obvious that this is a valid expansion since 
the shwaves sit on top of a radially-varying background (see \citealt{toom69} 
for a discussion of waves in a slowly-varying background).  But
this is an ordinary WKB expansion in disguise.  To see this, one
need only transform to ``comoving'' coordinates $x' = x$, $y'
= y + \int^x \qe(s) ds \Omega t$, $t' = t$ (this procedure may be more familiar
in a cosmological context; as \cite{bal88} has pointed out, this
is possible for any flow in which the velocities depend linearly
on the spatial coordinates).  In this frame the time-dependent
wavevector given above is transformed to a time-independent wavevector.
The price paid for this is that $\partial_x \rightarrow
\partial_{x'} + \qe \Omega t \partial_{y'}$, so new explicit time
dependences appear on the right hand side of the perturbed equations
of motion, and the perturbed variables no longer have time dependence
$\exp(i\omega t')$.  Instead, we must solve an ODE for $\delta(t')$.
The $y'$ dependence can be decomposed as $\exp(ik_y y')$. The 
$x'$ dependence can be treated via WKB, since the perturbation may 
be assumed to have the form $W(\epsilon x', \epsilon t')\, \exp(i\bk' 
\cdot \bx')$. This ``nearly diagonalizes'' the operator $\partial_{x'}$.
Thus we are considering the evolution of a wavepacket in comoving
coordinates--- a ``shwavepacket''.

For this procedure to be valid two conditions must be met.  First the
usual WKB condition must apply, $k_y L \gg 1$.  Second, the parameters
of the flow that are ``seen'' by the shwavepacket must change little on
the characteristic timescale for variation of $\delta(t)$, which is
$\Omega^{-1}$ for the incompressive shwaves.  For solid body rotation
($\tilde{q} = 0$) the group velocity (derivable from equation [\ref{DRQ0}],
below) is $|v_g| < N_x/k$ (for positive squared Brunt-V\"ais\"al\"a
frequency $N_x^2$, defined below; for $N_x^2 < 0$ the waves grow in place),
so the timescale for change of wave packet parameters in this case is
$L/|v_g| > k L/N_x \gg \Omega^{-1}$.  It seems reasonable to anticipate
similarly long timescales when shear is present.  As a final check, we
have verified directly, using a code based on the ZEUS code of
\cite{sn92}, that a vortical shwavepacket in the stratified shearing
sheet remains localized as it swings from leading to trailing.

\subsection{Linearized Equations}

To linear order in the perturbation amplitudes, the dynamical equations
reduce to
\be\label{LIN1a}
\frac{\dot{\delta \Sigma}}{\Sigma_0} + \frac{\delta v_x}{L_{\Sigma}}  
+ i \tilde{k}_x \delta v_x + i k_y \delta v_y + i k_z \delta v_z = 0,
\ee
\be\label{LIN2}
\dot{\delta v}_x - 2\Omega \delta v_y + i\tilde{k}_x \frac{\delta P}{\Sigma_0}  -
\frac{c_s^2}{L_P} \frac{\delta \Sigma}{\Sigma_0} = 0,
\ee
\be\label{LIN3}
\dot{\delta v}_y + (2- \qe)\Omega \delta v_x + ik_y \frac{\delta P}{\Sigma_0} = 0,
\ee
\be\label{LIN4}
\dot{\delta v}_z + ik_z \frac{\delta P}{\Sigma_0} = 0,
\ee
\be\label{LIN5a}
\frac{\dot{\delta P}}{\Sigma_0} - c_s^2 \frac{\dot{\delta \Sigma}}
{\Sigma_0} + c_s^2 \frac{\delta v_x}{L_S} = 0,
\ee
where
\be
\frac{1}{L_P} \equiv \frac{P_0^\prime}{\gamma P_0} = 
\frac{1}{L_{\Sigma}} + \frac{1}{L_S} \equiv
\frac{\Sigma_0^\prime}{\Sigma_0} +
\frac{S_0^\prime}{\gamma S_0}
\ee
define the equilibrium pressure, density and entropy length scales in
the radial direction.
We have included the vertical component of the velocity in
order to make contact with an axisymmetric convective instability that
is present in two dimensions, after which we will set $k_z$ to zero. 

We will be mainly interested in the incompressive shwaves because the
short-wavelength compressive shwaves are unchanged at leading order by
stratification. We will therefore work solely in the Boussinesq
approximation.\footnote{We also drop the subscripts distinguishing
between the compressive and incompressive shwaves.} In addition to the
assumption of incompressibility, this approximation considers $\delta P$
to be negligible in the entropy equation; pressure changes are
determined by whatever is required to maintain nearly incompressible
flow.  The original Boussinesq approximation applies only to
incompressible fluids. It was extended to compressible fluids by
\cite{jeff30} and \cite{sv60}. We show in the Appendix that it is
formally equivalent to taking the short-wavelength, low-frequency limit
of the full set of linear equations. From this viewpoint, assuming that
$H k_y \delta P/P_0$ is of the same order as the other terms in the
dynamical equations implies that $\delta P/P_0 \sim (H k_y)^{-1} \delta
\Sigma/\Sigma_0$, thus justifying its neglect in the entropy equation.
We therefore replace equations (\ref{LIN1a}) and (\ref{LIN5a}) with
\be\label{LIN1}
\tilde{k}_x \delta v_x + k_y \delta v_y + k_z \delta v_z = 0
\ee
and
\be\label{LIN5}
\frac{\dot{\delta \Sigma}}{\Sigma_0} - \frac{\delta v_x}{L_S} = 0.
\ee

Using equations (\ref{LIN3}) and (\ref{LIN5}) and the time derivative of
equation (\ref{LIN1}), one can express $\dot{\delta v}_y$ and $\delta P$
in terms of $\delta v_x$ and $\dot{\delta v}_x$:
\be\label{DH}
\frac{\delta P}{\Sigma_0} = -i \frac{\tilde{k}_x \dot{\delta v}_x + 2(\qe - 1) 
\Omega k_y \delta v_x}{k_y^2 + k_z^2},
\ee
\be\label{DVY}
\dot{\delta v}_y = \frac{(-\qe k_y^2 + (\qe - 2)k_z^2) \Omega \delta v_x - 
\tilde{k}_x k_y \dot{\delta v}_x}{k_y^2 + k_z^2}.
\ee
Eliminating $\delta P$ in equation (\ref{LIN2}) via equation (\ref{DH}) gives
\be
\tilde{k}^2 \dot{\delta v}_x + 2(\qe - 1) \Omega \tilde{k}_x k_y \delta v_x = (k_y^2 + k_z^2) (2 \Omega \delta v_y + (c_s^2/L_P) \delta \Sigma/\Sigma_0),
\ee
where $\tilde{k}^2 = \tilde{k}_x^2 + k_y^2 + k_z^2$. Taking the time
derivative of this equation and eliminating $\dot{\delta \Sigma}$ and
$\dot{\delta v}_y$ via equations (\ref{LIN5}) and (\ref{DVY}), we obtain
the following differential equation for $\delta v_x$:
\be\label{BOUSSVX}
\tilde{k}^2 \ddot{\delta v}_x + 4 \qe \Omega \tilde{k}_x k_y \dot{\delta v}_x 
+ \left[k_y^2\left(N_x^2 + 2\qe^2 \Omega^2\right) + k_z^2\left(N_x^2 +
 \tilde{\kappa}^2\right)\right]\delta v_x = 0,
\ee
where $\tilde{\kappa}^2 = 2(2-\qe)\Omega^2$ is the square of the
effective epicyclic frequency and 
\be
N_x^2 \equiv -\frac{c_s^2}{L_S L_P}
\ee
is the square of the
Brunt-V$\ddot{\rm{a}}$is$\ddot{\rm{a}}$l$\ddot{\rm{a}}$ frequency in the
radial direction.\footnote{Notice that $N_x^2$, $\qe$ and
$\tilde{\kappa}^2$ are all functions of $x$ and vary on a scale $L \sim
H$.}

\subsection{Comparison with Known Results}

Setting $k_y = 0$ in equation (\ref{BOUSSVX}) yields the axisymmetric
modes with the following dispersion relation (for $\delta(t) \propto
e^{-i\omega t}$):
\be
\omega^2 = \frac{k_z^2}{k_{x0}^2 + k_z^2}\left(N_x^2 + \tilde{\kappa}^2\right).
\ee
This is the origin of the H{\o}iland stability criterion: the axisymmetric
modes are stable for $N_x^2 + \tilde{\kappa}^2 > 0$. In the absence of
rotation this reduces to the Schwarzschild stability criterion: $N_x^2 >
0$ is the necessary condition for stability. The effect of rotation is strongly 
stabilizing: if $N_x^2 < -\tilde{\kappa}^2$, as required for instability, then 
$L_S L_P \sim H^2$; pressure and entropy must vary on radial scales of 
order the scale height for the disk to be H{\o}iland unstable.

Notice that effective epicyclic frequency $\tilde{\kappa}^2$ only stabilizes 
modes with nonzero $k_z$. The stability of nonaxisymmetric
shwaves with $k_z = 0$ (as in the mid-plane of a thin disk) is the open 
question that this work is addressing. In this limit and in
the absence of shear the Schwarzschild stability criterion is again
recovered: with $k_z = 0$ and $\qe = 0$ in equation (\ref{BOUSSVX}) the
dispersion relation becomes
\be\label{DRQ0}
\omega^2 = \frac{k_y^2}{k_{x0}^2 + k_y^2}N_x^2,
\ee
If there is a region of the disk where the effective shear is zero, a
WKB normal-mode analysis will yield the above dispersion relation and
there will be convective instability for $N_x^2 < 0$. It appears from
equation (\ref{BOUSSVX}) that differential rotation provides a
stabilizing influence for nonaxisymmetric shwaves just as rotation
does for the axisymmetric modes. Things are not as simple in
nonaxisymmetry, however. The time dependence is no longer exponential,
nor is it the same for all the perturbation variables. There is no clear
cutoff between exponential and oscillatory behavior, so the question of
flow stability becomes more subtle. 

As discussed in the introduction, the Boussinesq system of equations in 
the shearing-sheet model of a radially-stratified disk bear a close resemblance 
to the system of equations employed in analyses of a shearing, stratified 
atmosphere. A sufficient condition for stability in the latter case is that 
\be\label{RICH}
{\rm Ri} \equiv \frac{N_x^2}{(v_0^\prime)^2} \geq \frac{1}{4}
\ee
everywhere in the flow, where Ri is the Richardson number, a measure of 
the relative importance of buoyancy and shear. This stability criterion was 
originally proved by \cite{jwm61} and \cite{how61} for incompressible 
fluids, and its extension to compressible fluids was demonstrated by 
\cite{chi70}. The stability criterion is based on a normal-mode analysis 
with rigid boundary conditions. Other than differences in notation (e.g., 
our radial coordinate corresponds to the vertical coordinate in a stratified 
atmosphere), the key differences in our system are: (i) the equilibrium 
pressure gradient in a disk is balanced by centrifugal forces rather than 
by gravity; (ii) the disk equations contain Coriolis force terms; (iii) most 
atmospheric analyses only consider an equilibrium that is convectively 
stable, whereas we are interested in an unstable stratification; (iv) we
do not employ boundary conditions in our analytic model since we are 
only interested in the possibility of a local instability. 

The lack of boundary conditions in our model makes the applicability of 
the standard Richardson stability criterion in determining local stability 
somewhat dubious, since the lack of boundary conditions precludes 
the decomposition of linear disturbances into normal modes. 
The natural procedure for performing a local linear analysis in disks 
is to decompose the perturbations into shwaves, as we have done.

\cite{ehr53} consider both stable and unstable atmospheres and analyze 
an initial-value problem by decomposing the perturbations in time via
Laplace transforms.  For flow between 
two parallel walls, they find that an arbitrary initial disturbance behaves 
asymptotically as $t^{(\alpha - 1)/2}$ for $-3/4 < {\rm Ri} < 1/4$, where
\be\label{ALPHA}
\alpha \equiv \sqrt{1 - 4 \, {\rm{Ri}}},
\ee
which grows algebraically for ${\rm Ri} < 0$. The disturbance grows 
exponentially only for ${\rm Ri} < -3/4$. For a semi-infinite flow, the 
power-law behavior in time holds for $-2 < {\rm Ri} < 1/4$, with 
exponential growth for ${\rm Ri} < -2$. These results illustrate the 
importance of boundary conditions in determining stability.

In the $k_z = 0$ limit that we are concerned with here, the correspondence 
between the disk and atmospheric models turns out to be exact in the 
shwave formalism. This is because the Coriolis force only appears in
equation (\ref{BOUSSVX}) via $\tilde{\kappa}^2$, which disappears
when $k_z = 0$. The equation describing the time evolution of shwaves
in both a radially-stratified disk and a shearing, stratified atmosphere 
is thus
\be\label{BOUSSVX2D}
\tilde{k}^2 \ddot{\delta v}_x + 4 \qe \Omega \tilde{k}_x k_y \dot{\delta v}_x
+ k_y^2\left(N_x^2 + 2\qe^2 \Omega^2\right)\delta v_x = 0.
\ee
We analyze the solutions to this equation in the following 
section.\footnote{This equation is also obtained in a shwave analysis
of interchange instability in a disk with a poloidal magnetic field
\citep{ssp95}, with $N_x^2$ replaced by a magnetic buoyancy frequency.}

\subsection{Solutions}

Changing time variables in equation (\ref{BOUSSVX2D}) to 
$\tilde{\tau} \equiv \tilde{k}_x/k_y$, the
differential equation governing $\delta v_x$ becomes
\be\label{ODE}
(1 + \tilde{\tau}^2)\dv{^2\delta v_x}{\tilde{\tau}^2} + 4 \tilde{\tau} 
\dv{\delta v_x}{\tilde{\tau}} + ({\rm{Ri}} + 2)\delta v_x = 0.
\ee
The solutions to equation (\ref{ODE}) are
hypergeometric functions. With the change of variables $z \equiv
-\tilde{\tau}^2$, equation (\ref{ODE}) becomes
\be\label{ODEZ}
z(1-z)\dv{^2\delta v_x}{z^2} + \frac{1- 5z}{2}\dv{\delta v_x}{z} - 
\frac{{\rm{Ri}} + 2}{4}\delta v_x = 0.
\ee
The hypergeometric equation \citep{as72}
\be\label{HGEQ}
z(1-z)\dv{^2\delta v_x}{z^2} + \left[c - (a + b + 1)z\right]\dv{\delta v_x}{z} 
- ab\delta v_x = 0
\ee
has as its two linearly independent solutions $F(a,b;c;z)$ and $z^{1-c}
F(a-c+1,b-c+1;2-c;z)$. Comparison of equations (\ref{ODEZ}) and 
(\ref{HGEQ}) shows that $a = (3 - \alpha)/4$, $b = (3 + \alpha)/4$ and 
$c = 1/2$, where $\alpha$ is defined in equation (\ref{ALPHA}).

The general solution for $\delta v_x$ is thus given by
\be\label{SOLVX}
\delta v_x = C_1 \,F\left(\frac{3 - \alpha}{4},\frac{3 + \alpha}{4};
\frac{1}{2};-\tilde{\tau}^2\right) + C_2 \, \tilde{\tau}\,F\left(\frac{
5 - \alpha}{4},\frac{5 + \alpha}{4};\frac{3}{2};-\tilde{\tau}^2\right),
\ee 
where $C_1$ and $C_2$ are constants of integration representing
the two degrees of freedom in our reduced system. These two 
degrees of freedom can be represented physically by the initial velocity 
and displacement of a perturbed fluid particle in the radial direction. 
The radial Lagrangian displacement $\xi_x$ is obtained from 
equation (\ref{SOLVX}) by direct integration,\footnote{In our notation, 
a subscript $x$ or $y$ on the symbol $\xi$ indicates a Lagrangian displacement, not 
a component of the potential vorticity, which is a scalar.}
\be\label{SOLX}
\xi_x = \int \delta v_x \, dt = -\frac{C_2}
{\qe \Omega {\rm Ri}} \,F\left(\frac{1 - \alpha}{4},\frac{1 + \alpha}{4};
\frac{1}{2};-\tilde{\tau}^2\right) + \frac{C_1}{\qe \Omega}\,\tilde{\tau}\,
F\left(\frac{3 - \alpha}{4},\frac{3 + \alpha}{4};\frac{3}{2};
-\tilde{\tau}^2\right).
\ee
The solutions for the other perturbation variables can be obtained from 
equations (\ref{LIN1}), (\ref{LIN5}) and (\ref{DH}) with $k_z = 0$:
\be
\delta v_y = -\tilde{\tau} \delta v_x,
\ee
\be\label{SOX}
\frac{\delta \Sigma}{\Sigma_0} = \frac{\xi_x}{L_S}
\ee
and
\be\label{SOLDH}
\frac{\delta P}{P_0} = \frac{\gamma \Omega}{i c_s k_y} \left[\qe 
\tilde{\tau} \dv{}{\tilde{\tau}}\left(\frac{\delta v_x}{c_s}\right) + 
2(\qe - 1) \frac{\delta v_x}{c_s}\right].
\ee
It can be seen from the latter equation and the solution for $\delta
v_x$ that $\delta P/P_0$ remains small compared to $\delta v_x/c_s$ in the
short-wavelength limit. This demonstrates the consistency of the Boussinesq approximation. 

The hypergeometric functions can be transformed to a form valid for
large $\tilde{\tau}$ (see \citealt{as72} equations 15.3.7 and 15.1.1).
An equivalent form of the solution for $|\tilde{\tau}| \gg 1$ is
\begin{eqnarray}
\delta v_x = (C_1 V_1 + {\rm{sgn}}(\tilde{\tau}) C_2 V_2) \, |\tilde{\tau}|^
{\frac{\alpha - 3}{2}} F\left(\frac{3 - \alpha}{4}, \frac{5 - \alpha}{4}; 1 -
 \frac{\alpha}{2}; -\frac{1}{\tilde{\tau}^2}\right) + \nonumber \\ (C_1 V_3 +
 {\rm{sgn}}(\tilde{\tau}) C_2 V_4) \, |\tilde{\tau}|^{-\frac{\alpha+3}{2}} 
F\left(\frac{3 + \alpha}{4}, \frac{5 + \alpha}{4}; 1 + \frac{\alpha}{2}; 
-\frac{1}{\tilde{\tau}^2}\right), \; \; \;
\end{eqnarray}
where ${\rm{sgn}}(\tilde{\tau})$ is the arithmetic sign of
$\tilde{\tau}$ and the constants $V_i$ are given by
\begin{eqnarray}
V_1 \equiv \frac{\Gamma\left(\frac{1}{2}\right) \Gamma\left(\frac{\alpha}
{2}\right)} {\Gamma\left(\frac{3 + \alpha}{4}\right)\Gamma\left(-\frac{1 -
 \alpha}{4}\right)} \;\; , \;\; V_2 \equiv \frac{\Gamma\left(\frac{3}{2}\right)
 \Gamma\left(\frac{\alpha}{2}\right)} {\Gamma\left(\frac{5 + \alpha
 }{4}\right)\Gamma\left(\frac{1 + \alpha}{4}\right)} , \;\; \nonumber \\
V_3 \equiv \frac{\Gamma\left(\frac{1}{2}\right) \Gamma\left(-\frac{\alpha}
{2}\right)} {\Gamma\left(\frac{3 - \alpha}{4}\right)\Gamma\left(-\frac{1 +
 \alpha}{4}\right)} \;\; , \;\; V_4 \equiv \frac{\Gamma\left(\frac{3}{2}\right)
 \Gamma\left(-\frac{\alpha}{2}\right)} {\Gamma\left(\frac{5 - \alpha}
{4}\right)\Gamma\left(\frac{1 - \alpha}{4}\right)}.
\end{eqnarray}
Expanding the above form of the solution for $|\tilde{\tau}| \gg 1$, we obtain
\be\label{DVAS}
\delta v_x = \left(C_1 V_1 + {\rm{sgn}}(\tilde{\tau}) C_2 V_2\right)\,
 |\tilde{\tau}|^{\frac{\alpha - 3}{2}} + \left(C_1 V_3 + {\rm{sgn}}
(\tilde{\tau}) C_2 V_4\right) \, |\tilde{\tau}|^{-\frac{\alpha + 3}{2}} + 
O(\tilde{\tau}^{-2}).
\ee

An equivalent form of $\xi_x$ for $|\tilde{\tau}| \gg 1$ is
\begin{eqnarray}
\xi_x = \left(-\frac{C_2 X_1}{\qe \Omega {\rm Ri}} + {\rm{sgn}}
(\tilde{\tau}) \frac{C_1 X_2}{\qe \Omega}\right) \, |\tilde{\tau}|^{\frac{
\alpha - 1}{2}} F\left(\frac{3 - \alpha}{4}, \frac{1 - \alpha}{4}; 1 - 
\frac{\alpha}{2}; -\frac{1}{\tilde{\tau}^2}\right) + \nonumber \\ \left(
-\frac{C_2 X_3}{\qe \Omega {\rm Ri}} + {\rm{sgn}}(\tilde{\tau}) 
\frac{C_1 X_4}{\qe \Omega}\right) \, |\tilde{\tau}|^{-\frac{\alpha+1}{2}} 
F\left(\frac{3 + \alpha}{4}, \frac{1 + \alpha}{4}; 1 + \frac{\alpha}{2};
-\frac{1}{\tilde{\tau}^2}\right), \; \; \;
\end{eqnarray}
where the constants $X_i$ are given by
\begin{eqnarray}
X_1 \equiv \frac{\Gamma\left(\frac{1}{2}\right) \Gamma\left(\frac{\alpha}
{2}\right)} {\Gamma\left(\frac{1 + \alpha}{4}\right)\Gamma\left(\frac{1 + 
\alpha}{4}\right)} \;\; , \;\; X_2 \equiv \frac{\Gamma\left(\frac{3}{2}\right)
 \Gamma\left(\frac{\alpha}{2}\right)} {\Gamma\left(\frac{3 + \alpha}
{4}\right)\Gamma\left(\frac{3 + \alpha}{4}\right)} , \;\; \nonumber \\
X_3 \equiv \frac{\Gamma\left(\frac{1}{2}\right) \Gamma\left(-\frac{\alpha}
{2}\right)} {\Gamma\left(\frac{1 - \alpha}{4}\right)\Gamma\left(\frac{1 - 
\alpha}{4}\right)} \;\; , \;\; X_4 \equiv \frac{\Gamma\left(\frac{3}{2}\right)
 \Gamma\left(-\frac{\alpha}{2}\right)} {\Gamma\left(\frac{3 - \alpha}
{4}\right)\Gamma\left(\frac{3 - \alpha}{4}\right)}.
\end{eqnarray}
Expanding $\xi_x$ for $|\tilde{\tau}| \gg 1$ yields
\be\label{DSAS}
\xi_x = \left(-\frac{C_2 X_1}{\qe \Omega {\rm Ri}} + {\rm{sgn}}
(\tilde{\tau}) \frac{C_1 X_2}{\qe \Omega}\right) \, |\tilde{\tau}|^
{\frac{\alpha - 1}{2}}  + \left(-\frac{C_2 X_3}{\qe \Omega {\rm Ri}} + 
{\rm{sgn}}(\tilde{\tau}) \frac{C_1 X_4}{\qe \Omega}\right) \, |\tilde{\tau}|^
{-\frac{\alpha+1}{2}} + O(\tilde{\tau}^{-2}). 
\ee

The dominant contribution for each perturbation variable at late times is thus
\be
\delta P \propto \delta v_x \sim t^{\frac{\alpha - 3}{2}},
\ee
\be
\delta \Sigma \propto \xi_x \sim t^{\frac{\alpha - 1}{2}},
\ee
and
\be
\delta v_y \propto t \delta v_x \sim t^{\frac{\alpha - 1}{2}}.
\ee
This leads to one of our main conclusions: the density and $y$-velocity
perturbations will grow asymptotically for $\alpha > 1$, i.e.
${\rm{Ri}} \propto N_x^2 < 0$.\footnote{Notice that this is the same time 
dependence obtained by \cite{ehr53} in a modal analysis; see the discussion 
surrounding equation (\ref{ALPHA}). These power law time-dependences 
can be obtained more efficiently by solving the large-$\tilde{\tau}$ limit 
of equation (\ref{SOLVX}).} For small Richardson number, 
however (as is expected for a Keplerian disk with modest radial gradients), 
$\alpha \sim 1 - 2{\rm Ri}$ and the asymptotic growth is extremely slow:
\be
\delta \Sigma \sim \delta v_y \sim t^{-{\rm Ri}}.
\ee

In the stratified shearing sheet, the right-hand side of equation 
(\ref{PVEVLIN}) governing the evolution of the perturbed potential 
vorticity is no longer zero. The form of this equation for the incompressive 
shwaves is
\be\label{DPVEV}
\dot{\delta \xi} = \frac{d}{d t}\left(\frac{i\tilde{k}_x \delta v_y - i k_y \delta v_x}
{\Sigma_0}\right) = \frac{c_s^2 k_y}{i L_P \Sigma_0^2}\delta \Sigma.
\ee
The asymptotic time dependence of the perturbed potential vorticity can
be obtained by integrating equation (\ref{DPVEV}):
\be
\delta \xi \sim t^{\frac{\alpha + 1}{2}} \sim t^{1 - {\rm Ri}}
\ee
for $\tilde{\tau} \gg 1$ and $|{\rm Ri}| \ll 1$. As noted in \S 2, an entropy 
gradient is not required to generate vorticity. For $N_x^2 = 0$, $\alpha = 1$ 
and the perturbed potential vorticity grows linearly with time. The unstratified 
shearing sheet is recovered in the limit of zero stratification ($1/L_P \rightarrow 
0$), since in this limit equation (\ref{DPVEV}) reduces to $\xi = constant$.

\subsection{Energetics of the Incompressive Shwaves}

For a physical interpretation of the incompressive shwaves in the stratified 
shearing sheet, we repeat the analysis of section 3.3 for the solution 
given in the previous section. For a complete description of the energy in
this case, however, we must include the potential energy of a fluid element 
displaced in the radial direction. Following \cite{jwm61}, an  
expression for the energy in the Boussinesq approximation is obtained by
summing equation (\ref{LIN2}) multiplied by $\delta v_x$ 
and equation (\ref{LIN3}) multiplied by $\delta v_y$. Replacing 
$\delta \Sigma/\Sigma_0$ by $\xi_x/L_S$ via equation (\ref{SOX}) 
results in the following expression for the energy evolution:
\be\label{ENERGY}
\dv{E_k}{\tilde{\tau}} \equiv \dv{}{\tilde{\tau}} \left(\frac{1}{2}\Sigma_0 
\delta v^2 + \frac{1}{2} \Sigma_0 N_x^2 \xi_x^2 \right) = \Sigma_0 \delta 
v_x \delta v_y,
\ee
where $\delta v^2 = \delta v_x^2 + \delta v_y^2$. The three terms in equation 
(\ref{ENERGY}) can be identified as the kinetic energy, potential energy 
and Reynolds stress associated with an individual shwave. One may readily 
verify that the vortical shwaves (see equations (\ref{IVX})-(\ref{IS})) in the 
unstratified shearing sheet ($N_x^2 = 0$) satisfy equation (\ref{ENERGY}).

The right hand side of equation (\ref{ENERGY}) can be rewritten $-\tilde{\tau}
\delta v_x^2$ and individual trailing shwaves ($\tilde{\tau} > 0$) are therefore
associated with a negative angular momentum flux.  If the energy were
positive definite this would require that individual shwaves always
decay.  But when $N_x^2 < 0$ (${\rm Ri} < 0$) the potential energy
associated with a displacement is negative, so the energy $E_k$ can be
negative and a negative angular momentum flux is not enough to halt
shwave growth.  

Our next step is to write the constants of integration $C_1$ and $C_2$ 
in terms of the initial radial velocity and displacement of the shearing 
wave, $\delta v_{x0}$ and $\xi_{x0}$:
\be
C_1 = \frac{\qe \Omega {\rm Ri}\, \delta v_{x2}(\tilde{\tau}_0) \, 
\xi_{x0} + \xi_{x1}(\tilde{\tau}_0) \, \delta v_{x0}}{\delta v_{x1}
(\tilde{\tau}_0) \, \xi_{x1}(\tilde{\tau}_0) + {\rm Ri}\,\delta v_{x2}
(\tilde{\tau}_0) \, \xi_{x2}(\tilde{\tau}_0)} \;\; , \;\;
C_2 = \frac{-\qe \Omega {\rm Ri}\, \delta v_{x1}(\tilde{\tau}_0) \,
\xi_{x0} + {\rm Ri}\,\xi_{x2}(\tilde{\tau}_0) \, \delta v_{x0}}{\delta 
v_{x1}(\tilde{\tau}_0) \, \xi_{x1}(\tilde{\tau}_0) + {\rm Ri}\,\delta 
v_{x2}(\tilde{\tau}_0) \, \xi_{x2}(\tilde{\tau}_0)},
\ee
where $\tilde{\tau}_0 = k_{x0}/k_y$, $\delta v_{x1}$ is the 
hypergeometric function given by equation (\ref{SOLVX}) with $C_1 
= 1$ and $C_2 = 0$, and the other functions are similarly defined. These 
expressions can be simplified by noticing that the denominator of $C_1$ 
and $C_2$ is the Wronskian of the differential equation for $\xi_x$:\footnote{
Based upon the relationship between a hypergeometric function 
and its derivatives, $\delta v_{x1} = d(\xi_{x2})/d\tilde{\tau}$ and 
${\rm Ri} \delta v_{x2} = -d(\xi_{x1})/d\tilde{\tau}$.}
\be\label{ODEX}
(1 + \tilde{\tau}^2)\dv{^2\xi_x}{\tilde{\tau}^2} + 2 \tilde{\tau} 
\dv{\xi _x}{\tilde{\tau}} + {\rm Ri}\xi_x = 0.
\ee
The Wronskian of this equation is
\be
{\cal W} \equiv \dv{\xi_{x2}}{\tilde{\tau}} \, \xi_{x1}
-  \dv{\xi_{x1}}{\tilde{\tau}} \, \xi_{x2}
= \exp \left(-\int^{\tilde{\tau}} \frac{2\tau^2}
{1 + \tau^2} \, d\tau \right) =  \frac{1}{1 + \tilde{\tau}^2}.
\ee
We further simplify the analysis by setting the initial displacement $\xi_{x0}$
to zero.

With these simplifications, the solution given by equations 
(\ref{SOLVX}) and (\ref{SOLX}) becomes
\begin{eqnarray}
\frac{\delta v_x}{\delta v_{x0}} = \left(1 + \tilde{\tau}_0^2 \right)
\left[F\left(\frac{1 - \alpha}{4},\frac{1 + \alpha}{4};\frac{1}{2};
-\tilde{\tau}_0^2\right) \,F\left(\frac{3 - \alpha}{4},\frac{3 + \alpha}{4};
\frac{1}{2};-\tilde{\tau}^2\right) + \right. \nonumber \\ \left. {\rm Ri}\,
\tilde{\tau}_0\,F\left(\frac{3 - \alpha}{4},\frac{3 + \alpha}{4};
\frac{3}{2};-\tilde{\tau}_0^2\right) \, \tilde{\tau}\,F\left(\frac{5 - \alpha}
{4},\frac{5 + \alpha}{4};\frac{3}{2};-\tilde{\tau}^2\right)\right],
\end{eqnarray}
\begin{eqnarray}
\frac{\xi_x}{\delta v_{x0}} = \left(1 + \tilde{\tau}_0^2\right)\left[
-\frac{1}{\qe \Omega} \tilde{\tau}_0\,F\left(\frac{3 - \alpha}{4},
\frac{3 + \alpha}{4};\frac{3}{2};-\tilde{\tau}_0^2\right) \,F\left(
\frac{1 - \alpha}{4},\frac{1 + \alpha}{4};\frac{1}{2};-\tilde{\tau}^2
\right) + \right. \nonumber \\ \left. \frac{1}{\qe \Omega} F\left(\frac{1 
- \alpha}{4},\frac{1 + \alpha}{4};\frac{1}{2};-\tilde{\tau}_0^2\right) 
\,\tilde{\tau}\, F\left(\frac{3 - \alpha}{4},\frac{3 + \alpha}{4};
\frac{3}{2};-\tilde{\tau}^2\right)\right].
\end{eqnarray}

As in section 3.3, the energy integral for the incompressive perturbations 
is given by
\be
\< E_i \> = \frac{1}{2}\Sigma_0 L^2 \int k_0 dk_0 \< \delta v_\perp^2(k_0)\>
\int_0^{2\pi} d\theta \sin^2\theta \left[ \left(1 + \tilde{\tau}^2\right) 
\left(\frac{\delta v_x}{\delta v_{x0}}\right)^2 + N_x^2 \left(
\frac{\xi_x}{\delta v_{x0}}\right)^2\right],
\ee
for initial perturbations perpendicular to and isotropic in $\bk_0$. 
Changing integration variables to $\tilde{\tau} = \qe\Omega t + 
\cot\theta$, the angular integral becomes
\begin{eqnarray}\label{EINT}
2 \int_{-\infty}^{\infty} d\tilde{\tau} \, 
\left[ \left(1 + \tilde{\tau}^2\right) \left\{\xi_{x1}
(\tilde{\tau} - \qe \Omega t)\delta v_{x1}(\tilde{\tau}) + {\rm{Ri}}
 \, \xi_{x2}(\tilde{\tau} - \qe \Omega t) \delta v_{x2}(\tilde{\tau})
\right\}^2 + \right. \nonumber \\ \left. {\rm Ri} \, \left\{\xi_{x2}
(\tilde{\tau} - \qe \Omega t)\xi_{x1} (\tilde{\tau}) - \xi_{x1}
(\tilde{\tau} - \qe \Omega t) \xi_{x2}(\tilde{\tau})\right\}^2\right],
\end{eqnarray}
where we have used the relation $\sin\theta = (1 + \tilde{\tau}_0^2)
^{-1}$. In the limit of large $\qe \Omega t$, the dominant 
contribution to the angular integral comes from the region $0 \lesssim 
\tilde{\tau} \lesssim \qe \Omega t$. This can be seen from the 
following argument. Using the expansions given by equations 
(\ref{DVAS}) and (\ref{DSAS}), we find the angular integrand is
\be
2 |\tilde{\tau}(\tilde{\tau} - \qe \Omega t)|^{\alpha-1}\left[(V_1 
X_1  + {\rm sgn}(\tilde{\tau}){\rm sgn}(\tilde{\tau} - \qe \Omega 
t){\rm{Ri}} \, V_2 X_2)^2 + {\rm Ri} X_1^2 X_2^2 ({\rm sgn}
(\tilde{\tau})  - {\rm sgn}(\tilde{\tau} - \qe \Omega t))^2\right]
\ee
for $|\tilde{\tau}| \gg 1$ and $|\tilde{\tau} - \qe \Omega t| \gg 1$. 
Using the relation $\Gamma(n+1) = n\Gamma(n)$, one can easily
show that
\be
X_2 = \frac{2}{\alpha - 1}V_1 \;\;\;\; {\rm and} \;\;\;\; V_2 = \frac{2}
{\alpha + 1}X_1.
\ee
The integrand therefore simplifies to
\be
|\tilde{\tau}(\tilde{\tau} - \qe \Omega t)|^{\alpha-1}V_1^2 
X_1^2\frac{2}{1 - \alpha}\left[{\rm sgn}
(\tilde{\tau}) - {\rm sgn}(\tilde{\tau} - \qe \Omega t)\right]^2,
\ee
which is zero unless $0 < \tilde{\tau} < \qe \Omega t$ (for $t > 0$). 
For large $\qe \Omega t$, therefore, the angular integral is 
approximately given by
\be
\frac{16 V_1^2 X_1^2}{1 - \alpha} \int_\nu^{\qe \Omega 
t - \nu} d\tilde{\tau} \, \left[\tilde{\tau}\left(\tilde{\tau}-\qe 
\Omega t\right)\right]^{\alpha-1} = \frac{16 V_1^2 X_1^2}{\alpha 
(1 - \alpha)} \left. \frac{(\tilde{\tau} \qe \Omega t)^{\alpha}}
{\qe \Omega t} F\left(\alpha,1-\alpha;1+\alpha;\frac{\tilde
{\tau}}{\qe \Omega t}\right) \right | ^{\qe \Omega t - \nu}_{\nu},
\ee
where $1 \ll \nu \ll  \qe \Omega t$. For $\qe \Omega t \gg \nu$, the 
above expression can be approximated by evaluating it at 
$\tilde{\tau} = \qe \Omega t$, giving
\be
\< E_i (\qe \Omega t \gg 1) \>  \simeq \, 16 V_1^2 X_1^2
\frac{\Gamma(1+\alpha) \Gamma(\alpha)}{\alpha (1 - \alpha) 
\Gamma(2\alpha)} (\qe \Omega t)^{2\alpha-1} \, \< E_i(t = 0) \>,
\ee
where we have used equation 15.3.7 in \cite{as72} to evaluate 
$F(a,b;c;1)$.\footnote{We have numerically integrated the angular 
integral (\ref{EINT}) and found this to be an excellent approximation 
at late times.}

Notice that there is no power-law growth in the perturbation energy for
${\rm{Ri}} > 1/4$,\footnote{For ${\rm{Ri}} > 1/4$, $\alpha$ is 
imaginary and ${\rm{Re}}[t^{2\alpha-1}] = t^{-1}\cos(2|\alpha| \ln t)$.}
consistent with the classical Richardson criterion (\ref{RICH}). In 
our analysis the energy decays with time for $2\alpha-1<0$, or 
${\rm{Ri}} > 3/16$.  Thus the energy of an initial isotropic set of 
incompressive perturbations in a radially-stratified shearing sheet-model 
grows asymptotically (for ${\rm{Ri}} < 3/16$), just like the compressive 
shwaves and {\it unlike} the incompressive shwaves in an unstratified 
shearing sheet, for which the energy is constant in time.

The growth of an ensemble of incompressive shwaves in a stratified disk is {\it not}
due to a Rayleigh-Taylor or convective type instability.  There is 
asymptotic growth for $0 < {\rm{Ri}} < 3/16$, and convective 
instability requires ${\rm{Ri}} < 0$.  One can also see 
this by examining the asymptotic energy for small values of
$|{\rm{Ri}}|$, such as would be expected for a Keplerian disk with
modest radial gradients:
\be
\< E_i (\qe \Omega t \gg 1) \>  \simeq \left[2 \pi^2 {\rm Ri} + O({\rm Ri}^2)\right]\qe 
\Omega t^{1-4{\rm Ri}+O({\rm Ri}^2)} \, \< E_i(t = 0) \>.
\ee
Evidently for small values of Ri the near-linear growth in time of the
energy is independent of the sign of Ri and therefore $N_x^2$.\footnote{
This asymptotic expression assumes $\Ri \neq 0$. Notice that the energy 
at late times can have the opposite sign to the initial energy because the 
potential energy is negative for $N_x^2 < 0$.}

\section{Implications}\label{pap2s5}

We have studied the nonaxisymmetric linear theory of a thin, 
radially-stratified disk.  Our findings are: (i) incompressive,
short-wavelength perturbations in the unstratified shearing sheet
exhibit transient growth and asymptotic decay, but the energy of an
ensemble of such shwaves is constant with time; 
(ii) short-wavelength compressive shwaves grow asymptotically
in the unstratified shearing sheet, as does the energy of an ensemble of
such shwaves, which in the absence of any other dissipative effects
(e.g., radiative damping) will result in a compressive shwave steepening
into a train of weak shocks; (iii) incompressive shwaves in the
stratified shearing sheet have density and azimuthal velocity
perturbations $\delta \Sigma$, $\delta v_y \sim t^{-{\rm Ri}}$ (for
$|{\rm Ri}| \ll 1$); (iv) incompressive shwaves in the stratified
shearing sheet are associated with an angular momentum flux proportional
to $-\tilde{k}_x/k_y$; leading shwaves therefore have positive angular
momentum flux and trailing shwaves have negative angular momentum
flux\footnote{This is consistent with the asymptotic result one obtains
from a WKB analysis of incompressive waves \citep{bal03}.}; (v) the
energy of an ensemble of incompressive shwaves in the stratified
shearing sheet behaves asymptotically as $t^{1-4{\rm Ri}}$ for $|{\rm
Ri}| \ll 1$.  For Keplerian disks with modest radial gradients, $|{\rm
Ri}|$ is expected to be $\ll 1$, and there will therefore be weak growth
in a single shwave for ${\rm Ri} < 0$ and near-linear growth in the
energy of an ensemble of shwaves, independent of the sign of Ri.

Along the way we have found the following solutions: (i) an exact solution 
for nonvortical shwaves in the unstratified shearing sheet, equations 
(\ref{CVX}), (\ref{CS}) and (\ref{CVY}); (ii) a WKB-solution for the 
nonvortical,
compressive shwaves in the short-wavelength, high-frequency limit,
equations (\ref{VYWKB})-(\ref{SWKB}); (iii)
a solution for incompressive shwaves in the unstratified shearing sheet
valid in the short-wavelength, low-frequency limit, equations
(\ref{IVX})-(\ref{IS}); (iv) a solution for incompressive shwaves in
the radially-stratified shearing sheet (also valid in the short-wavelength,
low-frequency limit), equations (\ref{SOLVX})-(\ref{SOLDH}).

\begin{figure}[h]
\centering
\includegraphics[width=6.in,clip]{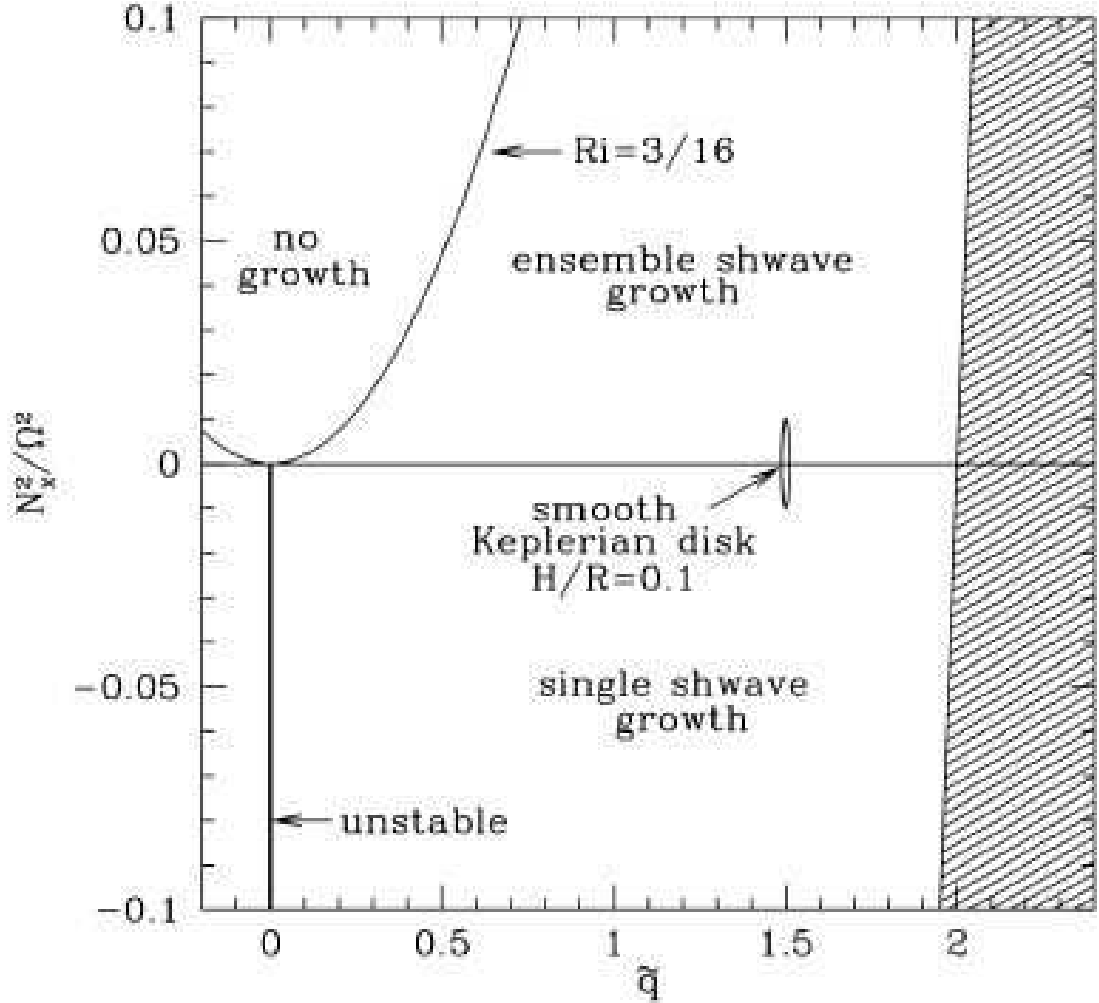}
\caption[A summary of analytic results for shwaves (shearing waves) in a
stratified disk.]
{A summary of analytic results for shwaves (shearing waves) in a
stratified disk.  The relevant parameters are the local dimensionless
shear rate
$\tilde{q} = -\frac{1}{2}d\ln\Omega^2/d\ln r$ and the dimensionless
Brunt-V\"ais\"al\"a frequency $N_x^2/\Omega^2$.  The expected location
of a thin, smooth disk is shown as a vertically extended ellipse near
$\tilde{q} = 1.5$, $N_x^2/\Omega^2 = 0$.  The far right region (shaded in
the figure) is forbidden by the H{\o}iland criterion.  When $\tilde{q} =
0$ shear is absent and a modal analysis is possible; instability is
present for $N_x^2 < 0$.  Solitary shwaves with $\Ri =
N_x^2/(\tilde{q}^2\Omega^2) < 0$ experience asymptotic power-law growth
($\propto t^{-\Ri}$ for small $\Ri$); since each shwave grows the energy
of an ensemble of shwaves does as well.  For $0 < \Ri < 3/16$ solitary
shwaves decay but the energy of an ensemble of shwaves grows as a
power-law in time.  For $\Ri > 3/16$ both solitary shwaves and the energy
of an ensemble of shwaves asymptotically decay.}
\label{pap2f1}
\end{figure}

Our results are summarized in Figure~\ref{pap2f1}, which shows the regions of
amplification and decay for shwaves in a stratified disk in the
$N_x^2/\Omega, \tilde{q}$ plane.

The presence of power-law growth of incompressive shwaves in stratified
disks opens the possibility of a transition to turbulence as amplified
shwaves enter the nonlinear regime.  Any such transition would depend,
however, on the nonlinear behavior of the disk after the shwaves break.
It is far from clear that they would continue to grow.  We will evaluate
the nonlinear behavior of the disk in subsequent work.

Our results are essentially in agreement with the numerical results
presented by \cite{klr04},
that is, we find that arbitrarily large amplification factors can be
obtained by starting with appropriate initial conditions.  Our results,
however, clarify the nature and asymptotic time dependence of the
growth.  Our results on the unstratified shearing sheet are also
consistent with the results of \cite{shep85} and \cite{amn04}, who find that an isotropic
ensemble of incompressive shwaves have fixed energy.


\end{spacing}

\chapter{Nonlinear Stability of Thin, Radially-Stratified Disks}\label{paper4}

\begin{spacing}{1.5}

\section{Chapter Overview}

We perform local numerical experiments to investigate the nonlinear
stability of thin, radially-stratified disks.  We demonstrate the
presence of radial convective instability when the disk is nearly in
uniform rotation, and show that the net angular momentum transport
is slightly inwards, consistent with previous investigations of
vertical convection.  We then show that a convectively-unstable
equilibrium is stabilized by differential rotation.  Convective instability
(corresponding to ${\rm Ri} \rightarrow -\infty$, where Ri is the radial
Richardson number) is suppressed when ${\rm Ri} \gtrsim -1$, i.e. when
the shear rate becomes greater than the growth rate.  Disks
with a nearly-Keplerian rotation profile and radial gradients on the
order of the disk radius have ${\rm Ri} \gtrsim -0.01$ and are
therefore stable to local nonaxisymmetric disturbances.  One
implication of our results is that the ``Global Baroclinic
Instability'' claimed by \cite{kb03} is either global or nonexistent.

\section{Introduction}

In order for astrophysical disks to accrete, angular momentum must be
removed from the disk material and transported outwards.  In many disks,
this outward angular momentum transport is likely mediated internally by
magnetohydrodynamic (MHD) turbulence driven by the magnetorotational
instability (MRI; see \citealt{bh98}).  A key feature of this transport
mechanism is that it arises from a local shear instability and is therefore 
very robust.  In addition,
MHD turbulence transports angular momentum {\it outwards}; some other forms
of turbulence, such as convective turbulence, appear to transport angular
momentum inwards \citep{sb96}.  The mechanism is only effective, however, if
the plasma in the disk is sufficiently ionized to be well-coupled to the
magnetic field (see \S\ref{MRILowI}).  In portions of disks around young,
low-mass stars, in cataclysmic-variable disks in quiescence, and in X-ray
transients in quiescence \citep{sgbh00,gm98,men00}, the plasma may be too
neutral for the MRI to operate.  This presents some difficulties for
understanding the evolution of these systems, since no robust transport mechanism
akin to MRI-induced turbulence has been established for purely-hydrodynamic
Keplerian shear flows.

Such a mechanism has been claimed recently by \cite{kb03},
who find vortices and an outward transport of angular momentum in the
nonlinear outcome of their global simulations.  The claim is that this
nonlinear outcome is due to a {\it local} instability (the ``Global
Baroclinic Instability'') resulting from the presence of an equilibrium
entropy gradient in the radial direction.  The instability mechanism
invoked \citep{klr04} is an interplay between transient amplification of
linear disturbances and nonlinear effects.  The existence of such a
mechanism would have profound implications for understanding the
evolution of weakly-ionized disks.

In Chapter~\ref{paper2}, we have
performed a linear stability analysis for local nonaxisymmetric
disturbances in the shearing-wave formalism.  While the linear theory
uncovers no exponentially-growing instability (except for convective
instability in the absence of shear), interpretation of the results
is somewhat difficult due to the nonnormal nature of the linear
differential operators\footnote{A nonnormal operator is one that is
not self-adjoint, i.e. it does not have orthogonal eigenfunctions.}:
one has a coupled set of differential
equations in time rather than a dispersion relation, which results
in a nontrivial time dependence for the perturbation amplitudes
$\delta(t)$.  In addition, transient amplification does occur for
a subset of initial perturbations, and linear theory cannot tell us
what effect this will have on the nonlinear outcome.  For these
reasons, and in order to test for the presence of local nonlinear
instabilities, we here supplement our linear analysis with local
numerical experiments.

We begin in \S2 by outlining the basic equations for a local model of
a thin disk. In \S3 we summarize the linear theory results from Chapter~\ref{paper2}.
We describe our numerical model and nonlinear results in \S\S4 and 5, and
discuss the implications of our findings in \S6.

\section{Basic Equations}

The simulations of \cite{kb03} are two-dimensional (without vertical
structure), since the salient feature supposedly giving rise to the
instability is a radial entropy gradient.  The simplest model to use
for a local verification of their global results is the two-dimensional
shearing sheet (see, e.g., \citealt{gt78}).  This local approximation
is made by expanding the equations of motion in the ratio of the disk
scale height $H$ to the local radius $R$, and is therefore only valid
for thin disks ($H/R \ll 1$).  The vertical structure is removed by
using vertically-integrated quantities for the fluid variables\footnote{This
vertical integration is not rigorous; we are assuming that important
vertical structure does not develop to affect our results.}.  The
basic equations that one obtains are
\be\label{EQN1}
\dv{\Sigma}{t} + \Sigma \bnabla \cdot \bv = 0,
\ee
\be\label{EQN2}
\dv{\bv}{t} + \frac{\bnabla P}{\Sigma} + 2\bO\times\bv - 2q\Omega^2 x \ex = 0,
\ee
\be\label{EQN3}
\dv{\,{\rm{ln}} S}{t} = 0,
\ee
where $\Sigma$ and $P$ are the two-dimensional density and pressure, $S
\equiv P \Sigma^{-\gamma}$ is the fluid entropy,\footnote{For a
non-self-gravitating disk the two-dimensional adiabatic index $\gamma =
(3\gamma_{3D} - 1)/(\gamma_{3D} + 1)$ (e.g. \citealt{ggn86}).} $\bv$ is
the fluid velocity and $d/dt$ is the Lagrangian derivative. The third and
fourth terms in equation (\ref{EQN2}) represent the Coriolis and centrifugal
forces in the local model expansion, where $\Omega$ is the local rotation
frequency, $x$ is the radial Cartesian coordinate and $q$ is the shear
parameter (equal to $1.5$ for a disk with a Keplerian rotation profile).
The gravitational potential of the central object is included as part of
the centrifugal force term in the local-model expansion, and we ignore the
self-gravity of the disk.

\section{Summary of Linear Theory Results}

An equilibrium solution to equations (\ref{EQN1}) through (\ref{EQN3}) is
\be\label{P0}
P = P_0(x),
\ee
\be
\Sigma = \Sigma_0(x),
\ee
\be\label{VEQ}
\bv \equiv \bv_0 = \left(-q\Omega x + \frac{P_0^\prime}{2\Omega
\Sigma_0}\right)\ey,
\ee
where a prime denotes an $x$ derivative.
One can regard the background flow as providing an effective shear rate
\be
\qe \Omega \equiv -v_0^\prime
\ee
that varies with $x$, in which case $\bv_0 = -\int^x \qe(s) ds \, \Omega
\ey$.  Due to this background shear, localized disturbances can be
decomposed in terms of ``shwaves'', Fourier modes in a frame comoving with
the shear.  These have a time-dependent radial wavenumber given by
\be\label{KXEFF}
\tilde{k}_x(t,x) \equiv k_{x0} + \qe(x)\Omega k_y t.
\ee
where $k_{x0}$ and $k_y$ are constants.  Here $k_y$ is the azimuthal
wave number of the shwave.

In the limit of zero stratification,
\be
P_0(x) \rightarrow constant,
\ee
\be
\Sigma_0(x) \rightarrow constant,
\ee
\be\label{V0U}
\bv_0 \rightarrow -q\Omega x \ey,
\ee
\be
\qe \rightarrow q,
\ee
and
\be
\tilde{k}_x \rightarrow k_x \equiv k_{x0} + q\Omega k_y t.
\ee
In Chapter~\ref{paper2}, we analyze the time dependence of the shwave amplitudes for both
an unstratified equilibrium and a radially-stratified equilibrium.  As
discussed in more detail in Chapter~\ref{paper2}, applying the shwave formalism to a
radially-stratified shearing sheet effectively uses a short-wavelength
WKB approximation, and is therefore only valid in the limit $k_y L \gg 1$,
where the background varies on a scale $L \sim H \ll R$.  The disk scale
height $H \equiv c_s \Omega$, where $c_s = \sqrt{\gamma P_0/\Sigma_0}$.

There are three nontrivial shwave solutions in the unstratified shearing
sheet, two nonvortical and one vortical.  The radial stratification
gives rise to an additional vortical shwave.  In the limit of 
tightly-wound shwaves ($|k_x| \gg k_y$), the nonvortical and vortical
shwaves are compressive and incompressive, respectively.  The former 
are the extension of acoustic modes to nonaxisymmetry, and to leading 
order in $(k_y L)^{-1}$ they are the same both with and without 
stratification.  Since the focus of our investigation is on convective 
instability and the generation of vorticity, we repeat here only the
solutions for the incompressive vortical shwaves and refer the
reader to Chapter~\ref{paper2} for further details on the nonvortical shwaves.

In the unstratified shearing sheet, the solution for the incompressive shwave is given
by:
\be
\delta v_x = \delta v_{x0}\frac{k_0^2}{k^2},
\ee
\be
\delta v_y = -\frac{k_x}{k_y} \delta v_x
\ee
and
\be\label{IS4}
\frac{\delta \Sigma}{\Sigma_0} = \frac{\delta P}{\gamma P_0} =
\frac{1}{i c_s k_y}\left(\frac{k_x}{k_y} \frac{\dot{\delta v}_x}{c_s}
+ 2(q - 1) \Omega \frac{\delta v_x}{c_s}\right),
\ee
where $k^2 = k_x^2 + k_y^2$, ($k_0,\delta v_{x0}$) are the values of
($k, \delta v_x$) at $t=0$ and an overdot denotes a time 
derivative.\footnote{As
discussed in Chapter~\ref{paper2}, this solution is valid for all time
only in the short-wavelength limit ($k_y H \gg 1$); for $H k_y \lesssim 
O(1)$, an initially-leading incompressive shwave will turn into a 
compressive shwave near $k_x = 0$.}

The kinetic energy for a single incompressive shwave can be defined as
\be
E_k \equiv \frac{1}{2}\Sigma_0 (\delta v_x^2 + \delta v_y^2) =
\frac{1}{2}\Sigma_0 \delta v_{x0}^2 \frac{k_0^4}{k_y^2 k^2},
\ee
an expression which varies with time and peaks at $k_x = 0$.
If one defines an amplification factor for an individual shwave,
\be\label{AMP}
{\cal A } \equiv \frac{E_k(k_x = 0)}{E_k(t = 0)} =
1 + \frac{k_{x0}^2}{k_y^2},
\ee
it is apparent that an arbitrary amount of transient amplification in
the kinetic energy of an individual shwave can be obtained as one 
increases the amount of swing for a leading shwave ($k_{x0} \ll -k_y$).

This transient amplification of local nonaxisymmetric disturbances is
reminiscent of the ``swing amplification'' mechanism that occurs in 
disks that are marginally-stable to the axisymmetric gravitational 
instability \citep{glb65,jt66,gt78}.  In that context, nonaxisymmetric 
shwaves experience a short period of exponential growth near $k_x = 0$ 
as they swing from leading to trailing.  In order for this mechanism 
to be effective in destabilizing a disk, however, a feedback mechanism 
is required to convert trailing shwaves into leading shwaves
\citep{bt87}.  The arbitrarily-large amplification implied by equation
(\ref{AMP}) has led some authors to argue for a bypass transition to
turbulence in hydrodynamic Keplerian shear flows \citep{cztl03,ur04,amn04}.
The reasoning is that nonlinear effects somehow provide the necessary
feedback.  We show in Chapter~\ref{paper2} that a ensemble of incompressive shwaves drawn
from an isotropic, Gaussian random field has a kinetic energy that is a
constant, independent of time.  This indicates that a random set
of vortical perturbations will not extract energy from the mean shear.  It
is clear, however, that the validity of this mechanism as a transition to
turbulence can only be fully explored via numerical experiments.  No
numerical experiments to date have demonstrated a {\it transition} to
turbulence in Keplerian shear flows.

In the presence of radial stratification, there are two linearly-independent
incompressive shwaves.  The radial-velocity perturbation satisfies the
following equation (we use a subscript $s$ to distinguish the stratified
from the unstratified case):
\be\label{BOUSS2D}
(1 + \tilde{\tau}^2)\dv{^2\delta v_{xs}}{\tilde{\tau}^2} + 4 \tilde{\tau}
\dv{\delta v_{xs}}{\tilde{\tau}} + ({\rm{Ri}} + 2)\delta v_{xs} = 0,
\ee
where
\be
\tilde{\tau} \equiv \tilde{k}_x/k_y = \qe\Omega t + k_{x0}/ky
\ee
is the time variable and
\be
{\rm Ri} \equiv \frac{N_x^2}{(\qe\Omega)^2}
\ee
is the Richardson number, a measure of the relative importance of buoyancy
and shear \citep{jwm61,how61,chi70}\footnote{As discussed in Chapter~\ref{paper2}, 
equation (\ref{BOUSS2D}) is the same equation that one obtains for the
incompressive shwaves in a shearing, stratified atmosphere.}.  Here
\be
N_x^2 \equiv -\frac{c_s^2}{L_S L_P}
\ee
is the square of the Brunt-V$\ddot{\rm{a}}$is$\ddot{\rm{a}}$l$\ddot{\rm{a}}$
frequency in the radial direction, where $L_P \equiv \gamma P_0/P_0^\prime$
and $L_S \equiv \gamma S_0/S_0^\prime$ are the equilibrium pressure and 
entropy length scales in the radial direction.  The solutions for the other
perturbation variables are related to $\delta v_{xs}$ by
\be\label{XIY}
\delta v_{ys} = -\tilde{\tau} \delta v_{xs},
\ee
\be
\frac{\delta \Sigma_s}{\Sigma_0} = \frac{1}{L_S} \int \delta v_{xs}\,dt
\ee
and
\be\label{SOLH}
\frac{\delta P_s}{P_0} = \frac{\gamma \Omega}{i c_s k_y} \left[\qe
\tilde{\tau} \dv{}{\tilde{\tau}}\left(\frac{\delta v_{xs}}{c_s}\right) +
2(\qe - 1) \frac{\delta v_{xs}}{c_s}\right].
\ee

Since the solutions to equation (\ref{BOUSS2D}) are hypergeometric
functions, which have a power-law time dependence, it cannot in general be
accurately treated with a WKB analysis; there is no asymptotic region in
time where equation (\ref{BOUSS2D}) can be reduced to a dispersion
relation.  If, however, there  is a region of the disk where the effective
shear is zero, $\tilde{\tau} \rightarrow constant$ and equation
(\ref{BOUSS2D}) can be expressed as a WKB dispersion relation:
\be\label{DRQ}
\omega^2 = \frac{k_y^2}{k_{x0}^2 + k_y^2}N_x^2,
\ee
with $\delta(t) \propto \exp(-i\omega t)$.
For $\qe \simeq 0$ and $N_x^2 < 0$, then, there is convective instability.
For disks with nearly-Keplerian rotation profiles and modest radial
gradients, $\qe \simeq 1.5$ and one would expect that the instability is
suppressed by the strong shear.  Due to the lack of a dispersion relation, 
however, there is no clear cutoff between exponential and oscillatory time 
dependence, and establishing a rigorous analytic stability criterion is difficult.

For $\qe \neq 0$, the asymptotic time dependence for each perturbation variable
at late times is
\be
\delta P_s \propto \delta v_{xs} \sim t^{\frac{\alpha - 3}{2}},
\ee
\be
\delta \Sigma_s \propto \delta v_{ys} \sim t^{\frac{\alpha - 1}{2}},
\ee
where
\be
\alpha \equiv \sqrt{1 - 4 \, {\rm{Ri}}}.
\ee
The density and $y$-velocity perturbations therefore grow asymptotically for
$\alpha > 1$, i.e. ${\rm{Ri}} < 0$.  For small Richardson number, as is
expected for a nearly-Keplerian disk with modest radial gradients,
$\alpha \simeq 1 - 2{\rm Ri}$ and the asymptotic growth is extremely slow:
\be\label{ASYMP}
\delta \Sigma_s \sim \delta v_{ys} \sim t^{-{\rm Ri}}.
\ee
The energy of an ensemble of shwaves grows asymptotically as
$t^{2\alpha-1}$, or $t^{1-4{\rm Ri}}$ for small Ri.  The ensemble energy
growth is thus nearly linear in time for small Ri, independent of the sign
of Ri.\footnote{We show in Chapter~\ref{paper2} that there is also linear growth in the
energy of an ensemble of compressive shwaves.}

The velocity perturbations are changed very little by a weak radial
gradient.  One would therefore expect that, at least in the linear
regime, transient amplification of the kinetic energy for an individual
shwave is relatively unaffected by the presence of stratification.
There is, however, an associated density perturbation in the stratified 
shearing sheet that is not present in the unstratifed sheet.\footnote{
The amplitude of the density perturbation in the unstratified sheet is 
an order-of-magnitude lower than the velocity perturbations in the 
short-wavelength limit; see equation(\ref{IS4}).}  This results in 
transient amplification of the {\it potential} energy of an individual 
shwave.  We do not derive in Chapter~\ref{paper2} I a general closed-form expression
for the energy of an ensemble of incompressive shwaves in the stratified 
shearing sheet, so it is not entirely clear what effect this 
qualitatively new piece of the energy will have on an ensemble of 
shwaves in the linear regime.

In any case, the question of whether or not radial stratification can play 
a role in generating turbulence by interacting with the transient 
amplification of linear disturbances or by some other nonlinear mechanism 
can only be fully answered with a nonlinear study.  For this reason, and 
due to the subtleties involved in the linear analysis, we now turn to the
main focus of this paper, which is a series of local numerical experiments
in a radially-stratified shearing sheet.

\section{Numerical Model}

To investigate local nonlinear effects in a radially-stratified thin
disk, we integrate the governing equations (\ref{EQN1}) through (\ref{EQN3})
with a hydrodynamics code based on ZEUS \citep{sn92}.  This is a
time-explicit, operator-split, finite-difference method on a staggered
mesh.  It uses an artificial viscosity to capture shocks.  The
computational grid is $L_x \times L_y$ in physical size with $N_x \times
N_y$ grid cells, where $x$ is the radial coordinate and $y$ is the
azimuthal coordinate.  The boundary conditions are periodic in the
$y$-direction and shearing-periodic in the $x$-direction.  The shearing-box
boundary conditions are described in detail in \cite{hgb95}.  As described 
in \cite{mass00} and \cite{gam01}, advection by the linear shear flow can be
done by interpolation.  Rather than using a linear interpolation scheme as
in \cite{gam01}, we now do the shear transport with the same upwind 
advection algorithm used in the rest of the code.  This is less diffusive 
than linear interpolation, and the separation of the shear from the bulk 
fluid velocity means that one is not Courant-limited by large shear 
velocities at the edges of the computational domain.

We use the following equilibrium profile, which in general gives rise to 
an entropy that varies with radius:
\be
h_0(x) = h_a\left[1 - \epsilon \cos\left(\frac{2\pi x}{L_x}\right)\right] \;
\; , \; \; \Sigma_0(x) =
\left[\frac{h_0(\Gamma-1)}{\Gamma K}\right]^{\frac{1}{\Gamma-1}} \; \; , \;
\; P_0(x) = K \Sigma_0^\Gamma,
\ee
where $h_a$, $\epsilon$,
$\Gamma$ and $K$ are model parameters. The flow is maintained in
equilibrium by setting the initial velocity according to equation
(\ref{VEQ}).  This equilibrium yields a
Brunt-V$\ddot{\rm{a}}$is$\ddot{\rm{a}}$l$\ddot{\rm{a}}$ frequency
\be\label{N2}
N_x^2(x) = \frac{(\gamma-\Gamma) {h_0^\prime}^2}{\gamma (\Gamma-1) h_0},
\ee
which can be made positive, negative or zero by varying $\gamma-\Gamma$.

We fix some of the model parameters to yield an equilibrium profile that
is appropriate for a thin disk.  In particular, we want $H/L_P \sim H/R \ll
1$ in order to be consistent with our use of a razor-thin (two-dimensional)
disk model.  In addition, we want the equilibrium values for each fluid
variable to be of the same order to ensure the applicability of our linear
analysis.  These requirements can be met by choosing $K = 1$, $\epsilon =
0.1$, $L_x = 12$ and $h_a = \bar{c}_s^2 \Gamma/(\Gamma-1)$, where
$\bar{c}_s \equiv \sqrt{\< P_0/\Sigma_0 \>} \equiv 1$ is (to within a factor of
$\sqrt{\gamma}$) the $x$ average of the sound speed.  Since the equilibrium
profile changes with $\Gamma$, we choose a fixed value of $\Gamma = 4/3$,
which for $\gamma = \Gamma$ corresponds to a three-dimensional adiabatic
index of $7/5$.  These numbers yield $|H/L_P| \leq 0.2$.  Our unfixed model 
parameters are thus $L_y$, $q$ and $\gamma$.

The sinusoidal equilibrium profile we are using generates radial
oscillations in the shearing sheet due to truncation error. We apply an
exponential-damping term to the governing equations in order to reduce the
spurious oscillations and therefore get cleaner growth-rate measurements.
We damp the oscillations until their amplitude is equal to that of
machine-level noise, and subsequently apply low-level random perturbations
to trigger any instabilities that may be present.

\begin{figure}[hp]
  \hfill
  \begin{minipage}[t]{1.\textwidth}
    \begin{center}
      \includegraphics[width=3.5in,clip]{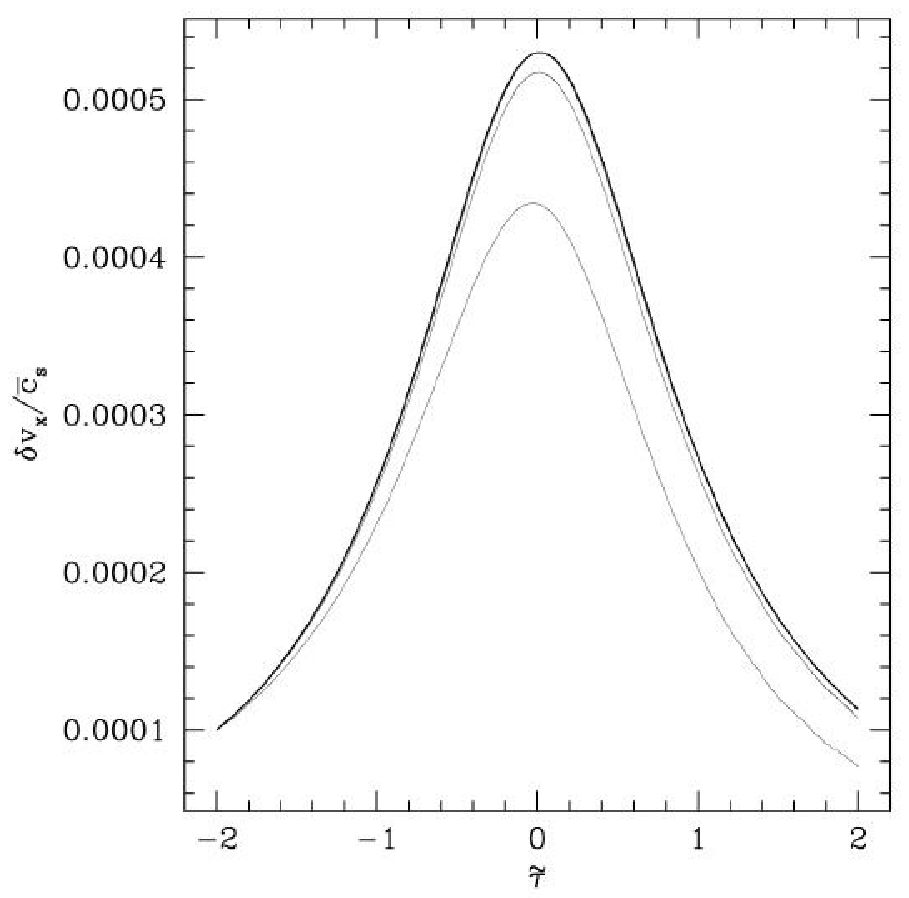}
      \caption[Evolution of the radial velocity amplitude for a vortical shwave in the
      radially-stratified shearing sheet.]
      {Evolution of the radial velocity amplitude for a vortical shwave in the
      radially-stratified shearing sheet (Run 1).  The heavy line is the analytic result,
      and the light lines are runs with a numerical resolution of (in order of increasing
      accuracy) $N_x \times N_y = 1024 \times 16$, $2048 \times 32$ and $4096 \times 64$.
      The number of grid cells are chosen so that the shwave initially has the same number
      of grid cells per wavelength in both the $x$ and $y$ directions.  The results are
      shown for a test point at the minimum in $N_x^2$.}
      \label{pap4f1a}
    \end{center}
  \end{minipage}
  \hfill
  \begin{minipage}[b]{1.\textwidth}
    \begin{center}
      \includegraphics[width=3.5in,clip]{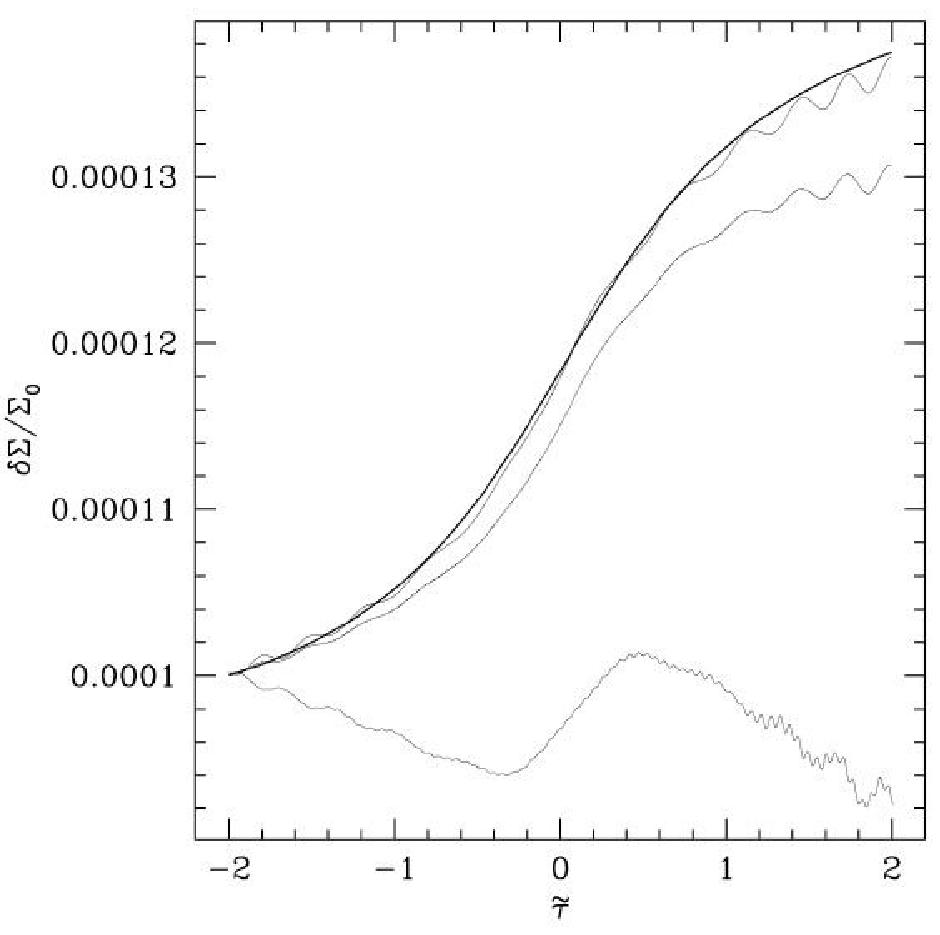}
      \caption[Evolution of the density amplitude for a vortical shwave in the
      radially-stratified shearing sheet.]
      {Evolution of the density amplitude for a vortical shwave in the
      radially-stratified shearing sheet (Run 1).  The heavy line is the analytic result,
      and the light lines are runs with a numerical resolution of (in order of increasing
      accuracy) $N_x \times N_y = 1024 \times 16$, $2048 \times 32$ and $4096 \times 64$.}
      \label{pap4f1b}
    \end{center}
  \end{minipage}
  \hfill
\end{figure}

As a test for our code, we evolve a particular solution for the
incompressive shwaves in the radially-stratified shearing sheet (equations
[\ref{BOUSS2D}] and [\ref{XIY}]-[\ref{SOLH}]).  The initial conditions
are $\delta v_x/\bar{c}_s = \delta \Sigma/\Sigma_0 = 1 \times 10^{-4}$ and
$k_{x0} = -128\pi/L_x$.  We set $L_y = 0.375$ and $k_y = 2\pi/L_y$ in
order to operate in the short-wavelength regime, and the other model
parameters are $q = 1.5$ and $\gamma - \Gamma = -0.3102$.  The latter
value yields a minimum value for $N_x^2(x)$ of $-0.01$.  The results of
the linear theory test are shown in Figures~\ref{pap4f1a} and \ref{pap4f1b}.

\section{Nonlinear Results}

\begin{deluxetable}{lll}
\tablecolumns{3}
\tablewidth{0pc}
\tabcolsep 0.5truecm
\tablecaption{Summary of Code Runs \label{pap4t1}}
\tablehead{Run & Description & Figure(s)}
\startdata
1 & Linear theory test & \ref{pap4f1a}-\ref{pap4f1b} \\
2 & External potential, $\Omega = 0$ & \ref{pap4f2}-\ref{pap4f4} \\
3 & External potential, $\Omega = 1$ & \ref{pap4f5}-\ref{pap4f6} \\
4 & Uniform rotation ($q = 0$) & \ref{pap4f7}-\ref{alpharun4} \\
5 & External potential, $\Omega = 1$, boost & \ref{pap4f9} \\
6 & Small shear ($-\infty \lesssim {\rm Ri} \lesssim -1$) & \ref{pap4f10} \\
7 & Small shear ($q = 0.2, {\rm Ri} \gtrsim -1$) & \ref{pap4f11} \\
8 & Aliasing ($q = 0.2, {\rm Ri} \gtrsim -1$) & \ref{pap4f12} \\
9 & Parameter survey & \ref{pap4f13} \\
10 & Keplerian disk ($q = 1.5, {\rm Ri} \simeq -0.004$) & \ref{pap4f14} \\
\enddata
\end{deluxetable}

Table~\ref{pap4t1} gives a summary of the runs that we have performed.  A
detailed description of the setup and results for each is given in the
following subsections.  Our primary diagnostic is a measurement of growth
rates, and the probe that we use for these measurements is an average over
azimuth of the absolute value of $v_x = \delta v_x$ at the minimum in
$N_x^2$.  Measuring $v_x$ allows us to demonstrate the damping of the
initial radial oscillations, and the average over azimuth masks the
interactions between multiple WKB modes with different growth rates in our
measurements. We will reference this probe with the following definition:
\be
v_t \equiv \langle |v_x(x_{min},y)| \rangle,
\ee
where here angle brackets denote an average over $y$ and $x_{min}$ is the
$x$-value at which $N_x^2(x)$ is a minimum.

\subsection{External Potential in Non-Rotating Frame}

As a starting problem, we investigate a stratified flow with $\bv_0 = 0$.
Such a flow can be maintained in equilibrium by replacing the tidal force
in equation (\ref{EQN2}) with an external potential $\Phi = -h_0$.  This can
be done in either a rotating or non-rotating frame.  It is a particularly
simple way of validating our study of convective instability in the shearing
sheet.  The condition $\bv_0 = 0$ implies $\qe = 0$, and therefore equation
(\ref{DRQ}) should apply in the WKB limit, with the expected growth rate
obtained by evaluating equation (\ref{N2}) locally.\footnote{The fastest growing
WKB modes will be the ones with a growth rate corresponding to the minimum
in $N_x^2$.}  We have performed a fiducial run with an imposed external
potential in a non-rotating frame ($\Omega = 0$ in equation [\ref{EQN2}]) to
compare with the outcome expected from the Schwarzschild stability criterion
implied by equation (\ref{DRQ}).  We set $\gamma - \Gamma = -0.3102$,
corresponding to $N_{x,min}^2 = -0.01$, and $L_y = L_x$.  The expected growth
rate for this Schwarzschild-unstable equilibrium is $0.1$ (in units of the
average radial sound-crossing time).  The numerical resolution for the fiducial
run is $512 \times 512$, and all the variables are randomly perturbed at an
amplitude of $1.0 \times 10^{-12}$.

\begin{figure}[hp]
\centering
\includegraphics[width=3.5in,clip]{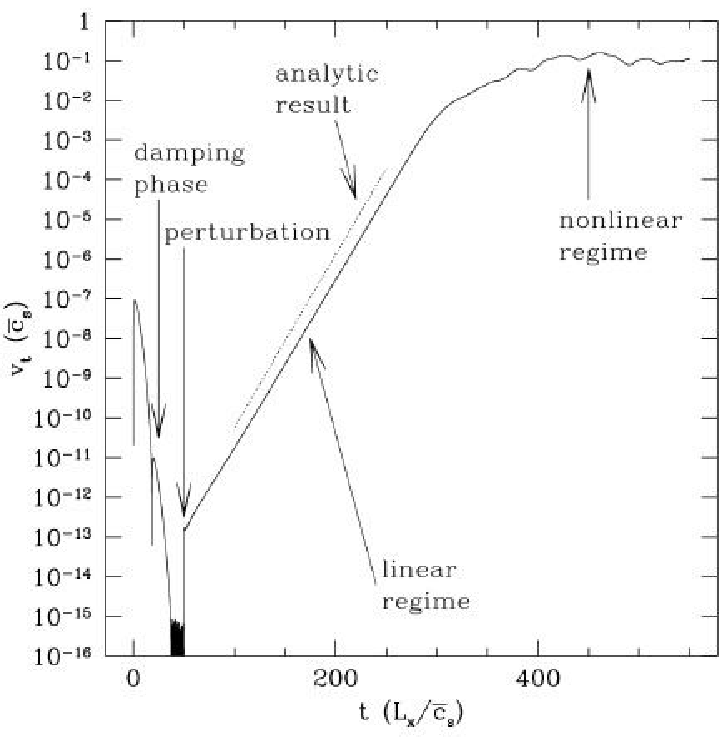}
\caption{Evolution of $v_t$ as a function of time for Run 2 (external potential,
non-rotating frame).}
\label{pap4f2}
\end{figure}

\begin{figure}[hp]
  \hfill
  \begin{minipage}[t]{1.\textwidth}
    \begin{center}
      \includegraphics[width=3.5in,clip]{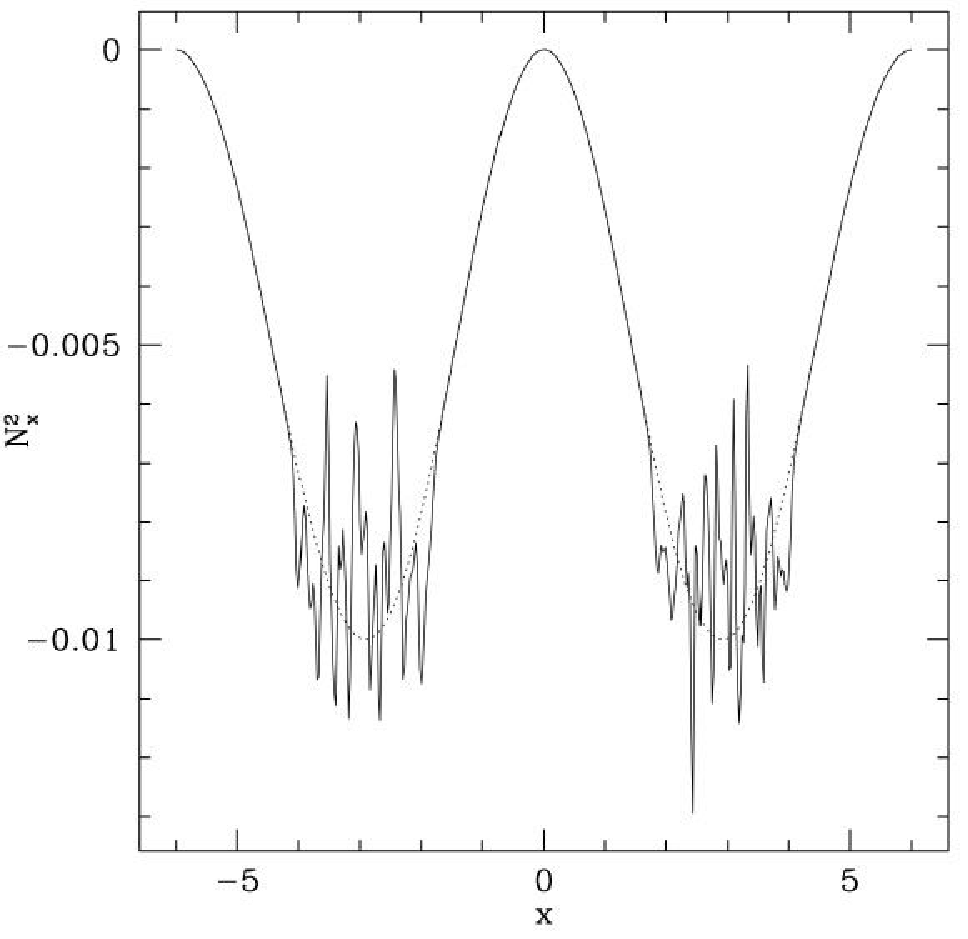}
      \caption[Plot of $N_x^2$ as a function of $x$ for Run 2.]
      {Plot of $N_x^2$ (averaged over $y$) as a function of $x$ for Run 2.  The dotted line
      shows the equilibrium profile, and the solid line shows a snapshot during the nonlinear
      regime.  Growth initially occurs at the minimum in $N_x^2$.}
      \label{pap4f3a}
    \end{center}
  \end{minipage}
  \hfill
  \begin{minipage}[b]{1.\textwidth}
    \begin{center}
      \includegraphics[width=3.5in,clip]{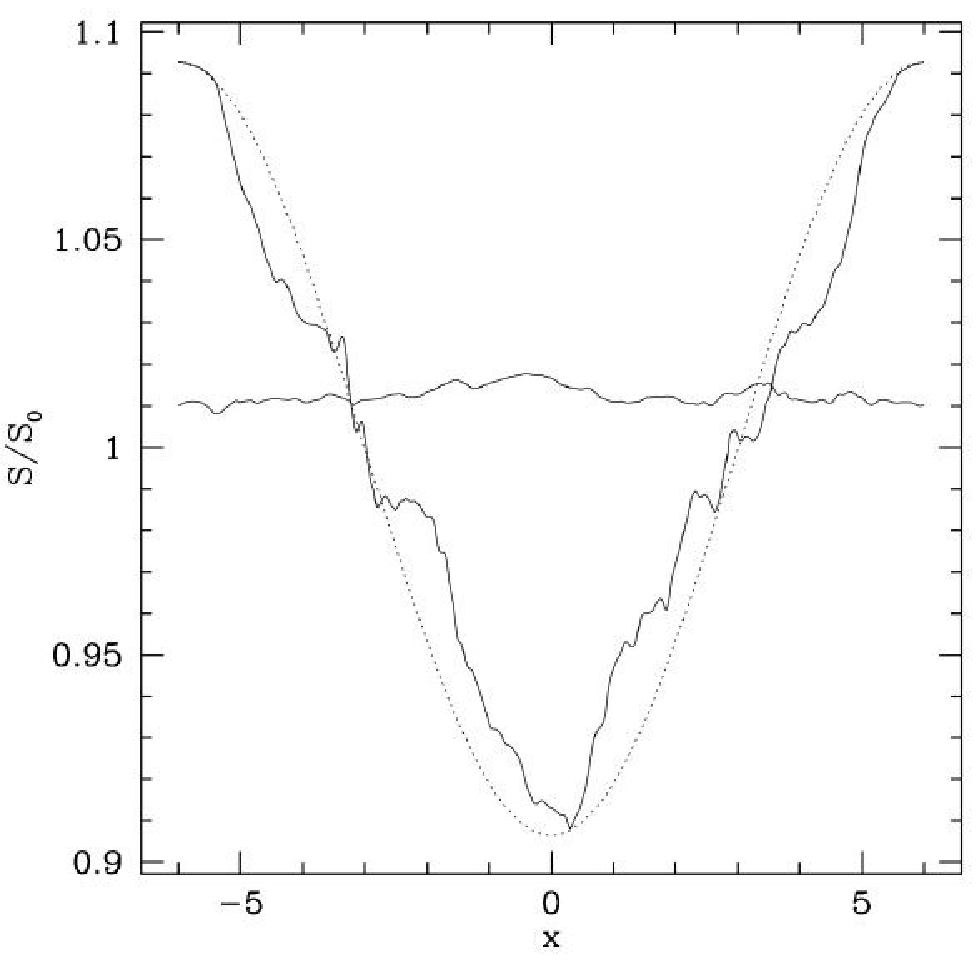}
      \caption[Plot of the entropy as a function of $x$ for Run 2.]
      {Plot of the entropy (averaged over $y$) as a function of $x$ for Run 2.  The dotted line
      shows the equilibrium profile, and the solid lines show snapshots during the nonlinear
      regime.  The entropy eventually settles to a nearly-constant value.}
      \label{pap4f3b}
    \end{center}
  \end{minipage}
  \hfill
\end{figure}

\begin{figure}[hp]
\centering
\includegraphics[width=6.in,clip]{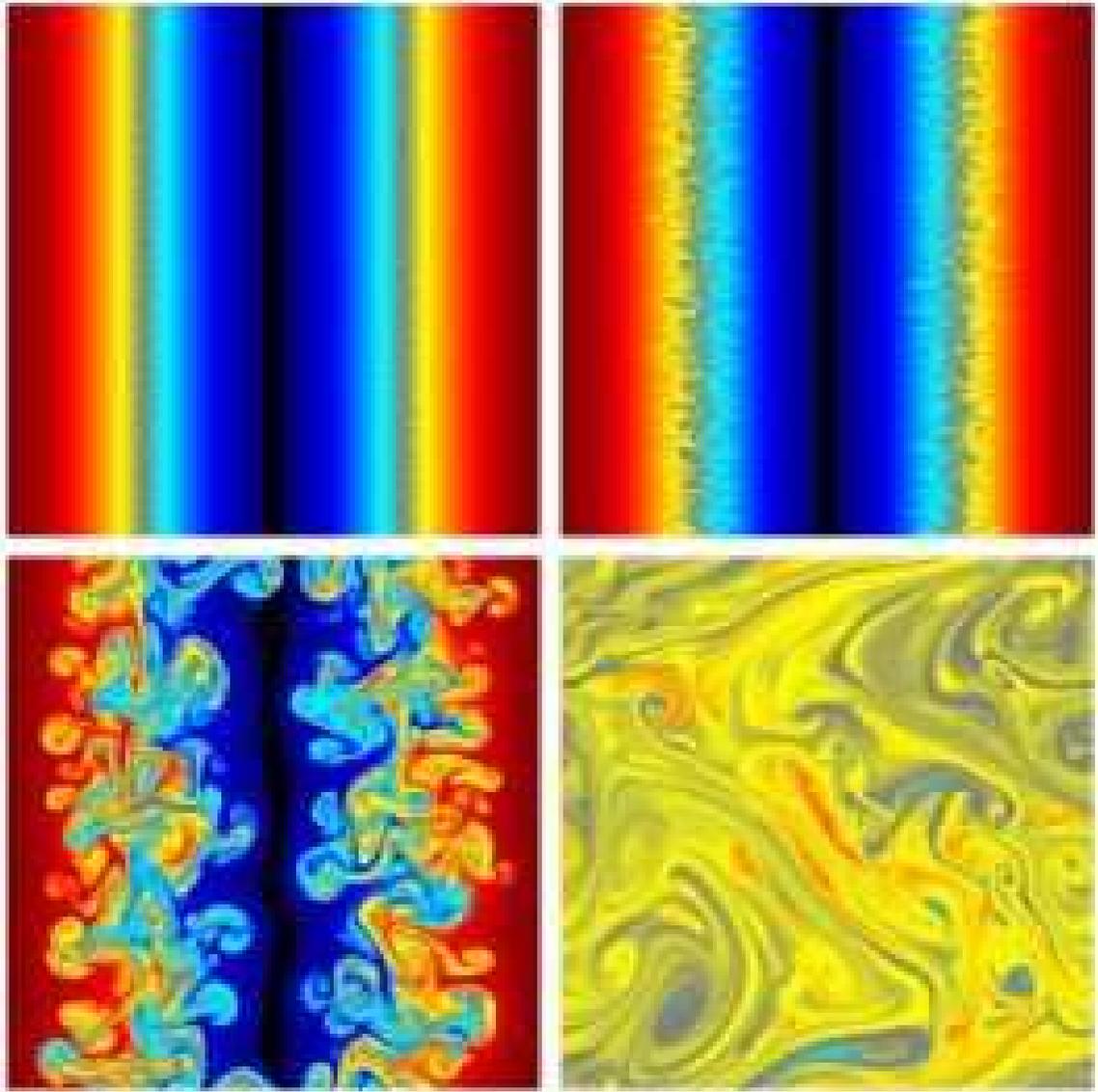}
\caption[Snapshots of the entropy in the nonlinear regime for Run 2.]
{Snapshots of the entropy in the nonlinear regime for Run 2, indicating
maximum growth for modes near the grid scale and the eventual turnover of
the equilibrium entropy profile to its average value.  Dark shades indicate 
values above (red in the color version) and below (blue in the color version) 
the average value (yellow in the color version).}
\label{pap4f4}
\end{figure}

A plot of $v_t$ as a function of time is given in Figure~\ref{pap4f2}, showing
the initial damping followed by exponential growth in the linear regime. The
analytic growth rate is shown on the plot for comparison.  A least-squares
fit of the data in the range $100 \leq t \leq 250$ yields a measured growth
rate of $0.0978$.\footnote{Measurements of the growth rate earlier in the
linear regime or over a larger range of data yield results that differ from
this value by at most $5\%$.}  Figure~\ref{pap4f3a} shows a cross section of $N_x^2$
as a function of $x$ after the instability has begun to set in, and Figure~\ref{pap4f3b}
shows cross
sections of the entropy early and late in the nonlinear regime.  The growth is
initially concentrated near the minimum points in $N_x^2(x)$.  Eventually the entropy
turns over completely and settles to a nearly constant value.   Figure~\ref{pap4f4}
shows two-dimensional snapshots of the entropy in the nonlinear regime.  Runs with
the same equilibrium profile except $\gamma - \Gamma \geq 0$ are stable.  There
is also a long-wavelength axisymmetric instability that is present for
$\gamma - \Gamma < 0$ even in the absence of the small-scale nonaxisymmetric
modes.  We measure its growth rate to be $0.07$.  Due to the long-wavelength
nature of these modes, they are not treatable by a local linear analysis.

\subsection{External Potential in Rotating Frame}

\begin{figure}[hp]
\centering
\includegraphics[width=5.5in,clip]{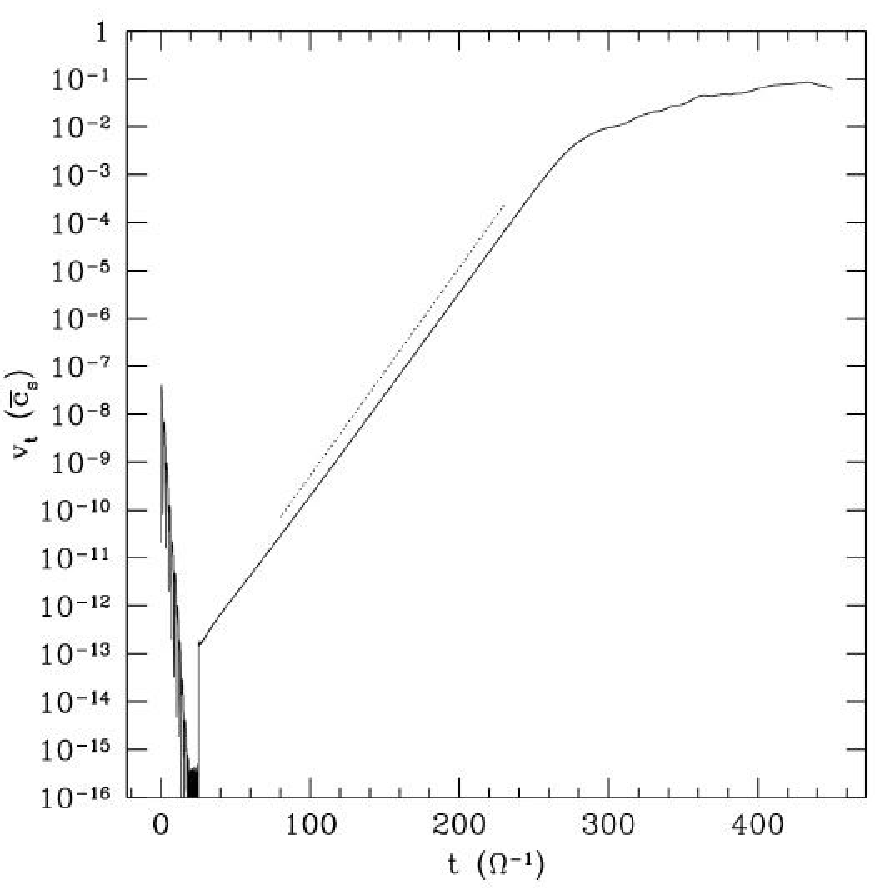}
\caption[Evolution of $v_t$ as a function of time for Run 3 (external potential,
rotating frame).]
{Evolution of $v_t$ as a function of time for Run 3 (external potential,
rotating frame).  The dotted line shows the expected growth rate.}
\label{pap4f5}
\end{figure}

\begin{figure}[hp]
\centering
\includegraphics[width=6.in,clip]{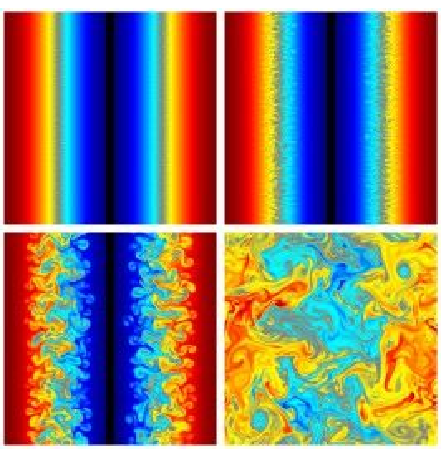}
\caption[Snapshots of the entropy in the nonlinear regime for Run 3.]
{Snapshots of the entropy in the nonlinear regime for Run 3. Dark shades 
indicate values above (red in the color version) and below (blue in the 
color version) the average value (yellow in the color version).}
\label{pap4f6}
\end{figure}

We have performed the same test as described in \S5.1 in a rotating frame
($\Omega = 1$ in equation [\ref{EQN2}]).  Figure~\ref{pap4f5} shows the
exponential growth in the linear regime for this run, with a measured growth
rate of $0.0977$.  Figure~\ref{pap4f6} shows snapshots of the entropy in the
nonlinear regime.  The results are similar to the nonrotating case, except that
1) rotation suppresses the long-wavelength axisymmetric instability; 2) the
nonlinear outcome exhibits more coherent structures in the rotating case
including transient vortices; and 3) these coherent structures eventually
become unstable to a Kelvin-Helmholz-type instability.

\subsection{Uniform Rotation}

\begin{figure}[hp]
\centering
\includegraphics[width=5.5in,clip]{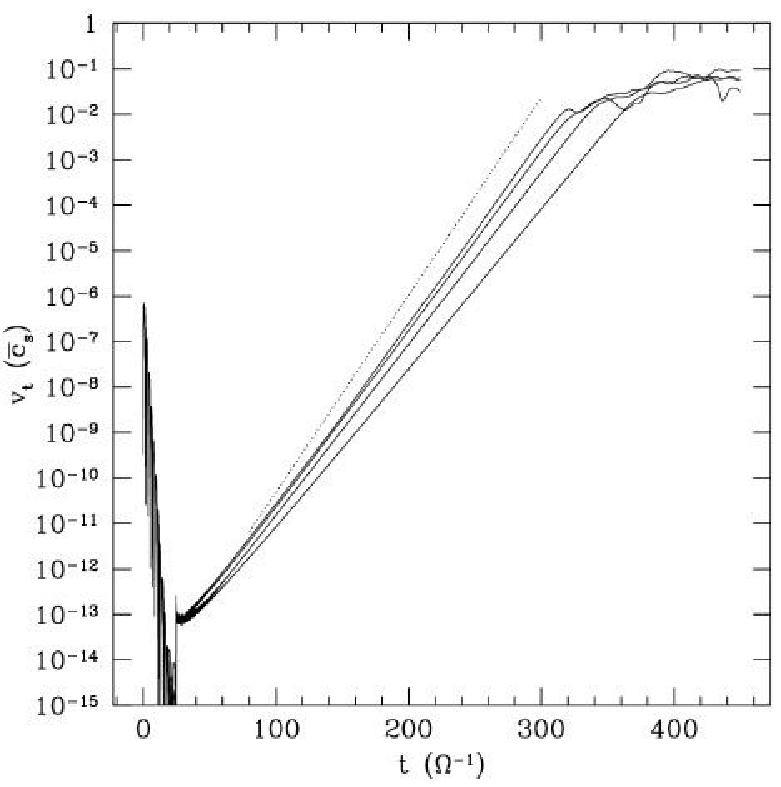}
\caption[Evolution of $v_t$ as a function of time for Run 4 ($q = 0$,
rotating frame).]
{Evolution of $v_t$ as a function of time for Run 4 ($q = 0$).  The
dotted line shows the expected growth rate, and the solid lines are runs
with (in order of increasing growth) $L_y = 12$, $6$, $3$ and $1.5$.}
\label{pap4f7}
\end{figure}

\begin{figure}[hp]
\centering
\includegraphics[width=6.in,clip]{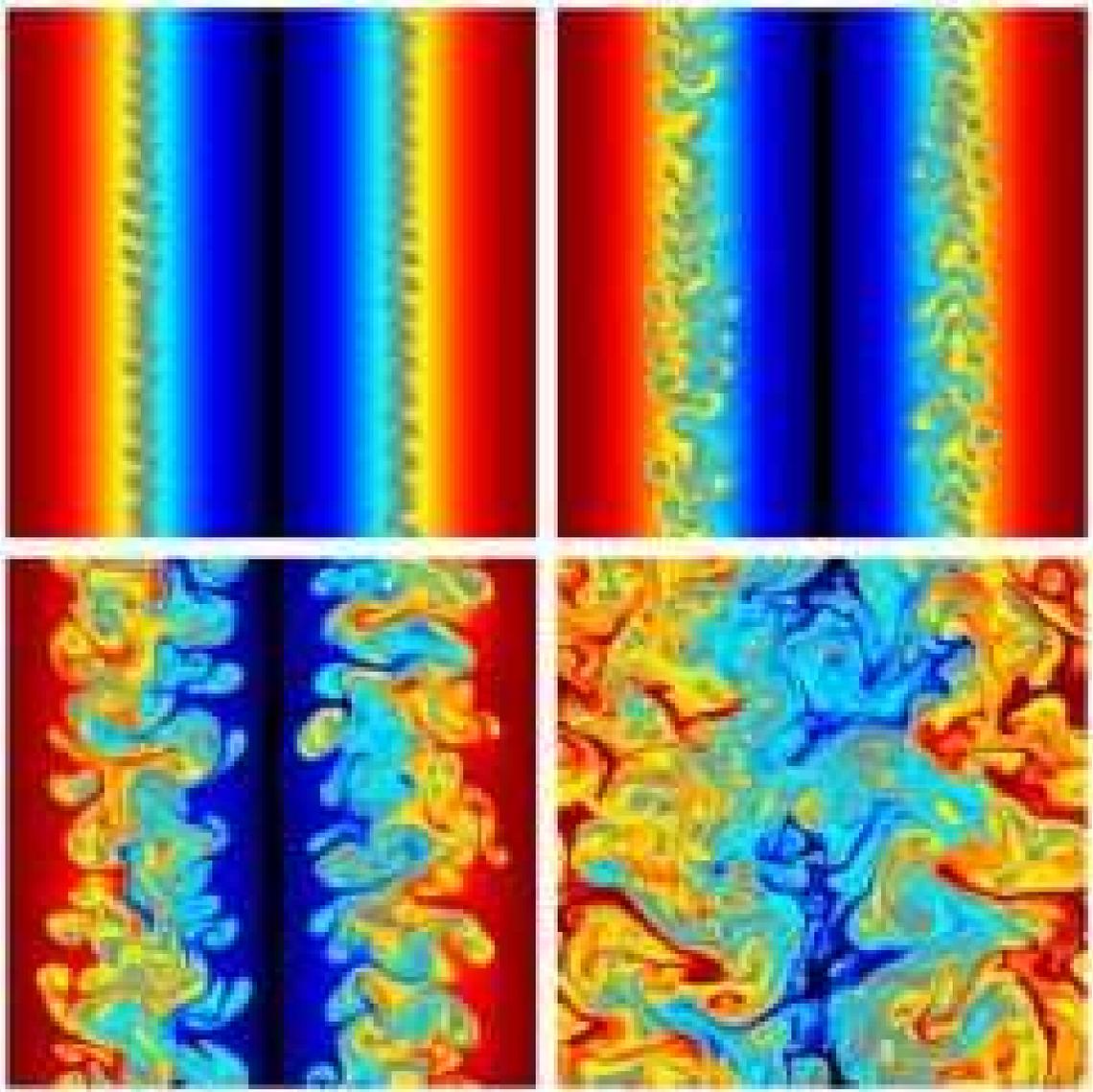}
\caption[Snapshots of the entropy in the nonlinear regime for Run 4.]
{Snapshots of the entropy in the nonlinear regime for Run 4.  Notice that
the maximum growth does not occur for modes at the grid scale.}
\label{pap4f8}
\end{figure}

Having demonstrated the viability of simulating convective instability in
the local model, we now turn to the physically-realistic equilibrium
described in \S4.  We begin by setting the shear parameter $q$ to zero
in order to make contact with the results of \S\S5.1 and 5.2.  This is
analogous to a disk in uniform rotation.  The other model parameters are
the same as for the previous runs.  While there is still an effective
shear $-0.05 \lesssim \qe \lesssim 0.05$, near $\qe = 0$ one expects the
modes to obey equation (\ref{DRQ}) in the WKB limit.  Figures~\ref{pap4f7}
and \ref{pap4f8} give the linear and nonlinear results for this run.  The
measured growth rate in the linear regime is $0.0809$.

\begin{figure}[hp]
\centering
\includegraphics[width=6.in,clip]{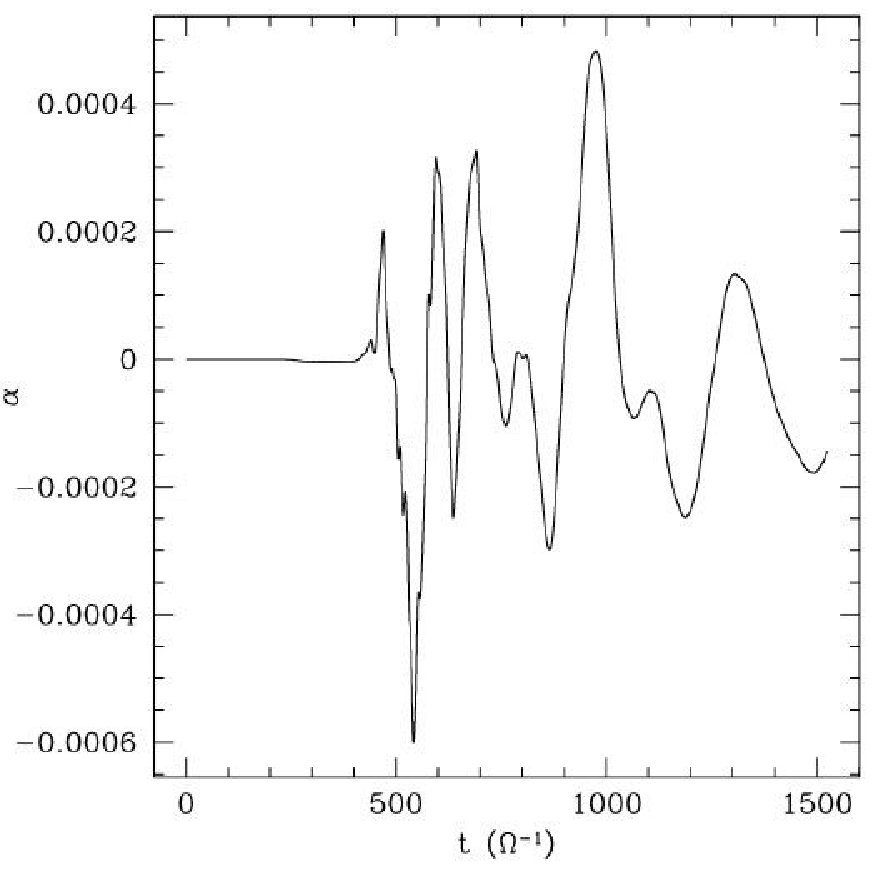}
\caption
{Evolution of the dimensionless angular momentum flux due to radial convection.}
\label{alpharun4}
\end{figure}

Consistent with results from numerical simulations of vertical convection
\citep{sb96,cab96}, the angular momentum transport associated with radial 
convection is {\it inwards}.  Figure~\ref{alpharun4} shows the evolution of 
the dimensionless angular momentum flux
\be
\alpha \equiv \frac{1}{L_x L_y \<P_0\>} \int \Sigma \delta v_x \delta v_y dx dy,
\ee
where $\<P_0\>$ is the radial average of the equilibrium pressure, for an 
extended version of Run 4.  Averaging over the last $1200 \Omega^{-1}$ yields 
$\alpha \sim -10^{-5}$.

There are two reasons for the larger error in the measured growth rate for
this run: 1) the equilibrium velocity gives rise to numerical diffusion due to
the motion of the fluid variables with respect to the grid; and 2) since the
growing modes are being advected in the azimuthal direction, the maximum growth
does not occur at the grid scale.  The latter effect can be seen in Figure~\ref{pap4f8};
several grid cells are required for a well-resolved wavelength.  In order to resolve
smaller wavelengths, we have repeated this run with $L_y = 6$, $3$ and $1.5$.  The
results are plotted in Figure~\ref{pap4f7} along with the results from the $L_y = 12$
run.  The measured growth rate for the $L_y = 1.5$ run is $0.0924$.

\begin{figure}[hp]
\centering
\includegraphics[width=4.in,clip]{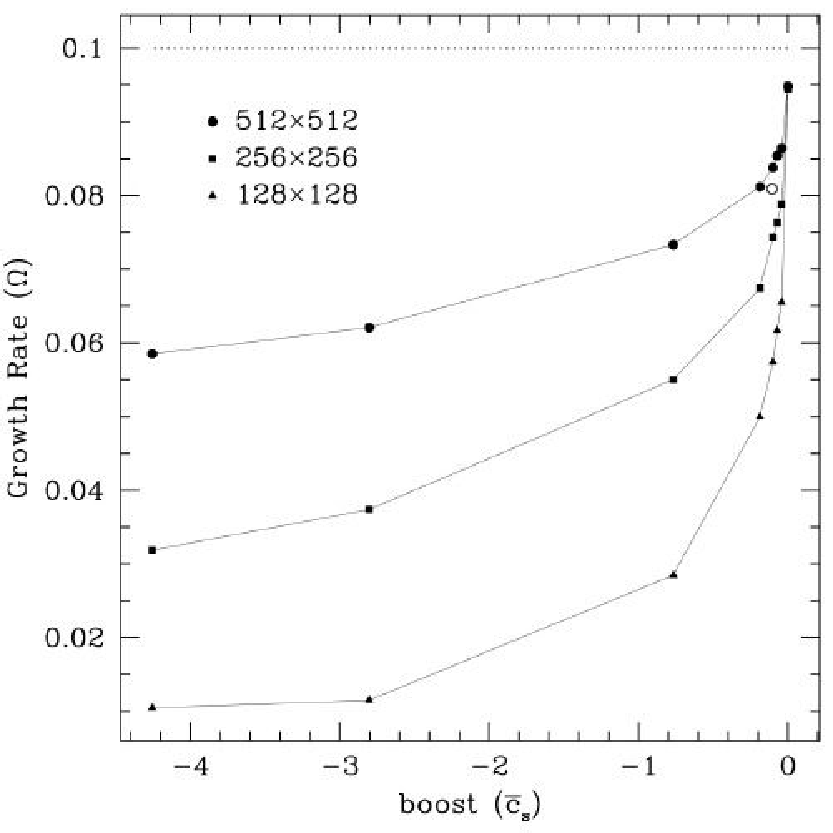}
\caption[Growth rates as a function of azimuthal boost.]
{Growth rates as a function of azimuthal boost in a series of runs with
an external potential and $N_{x,min}^2 = -0.01$.  The dotted line shows
the analytic growth rate from linear theory.  The open circle denotes the
growth rate that was measured in Run 4, with the boost corresponding to
the magnitude of the velocity at the minimum in $N_x^2$ for Run 3 ($q = 0$).
The largest boost magnitude corresponds to the velocity at the minimum in
$N_x^2$ for Run 10 ($q = 1.5$).}
\label{pap4f9}
\end{figure}

To quantify the effects of numerical diffusion, we have performed a series
of tests similar to Run 2 (external potential in a rotating frame) but with
an overall boost in the azimuthal direction.  Figure~\ref{pap4f9} shows measured growth
rates as a function of boost at three different numerical resolutions.  The
largest boost magnitude in this plot corresponds to the velocity at the
minimum in $N_x^2$ for a run with $q = 1.5$.

\subsection{Shearing Sheet}

\begin{figure}[hp]
\centering
\includegraphics[width=3.5in,clip]{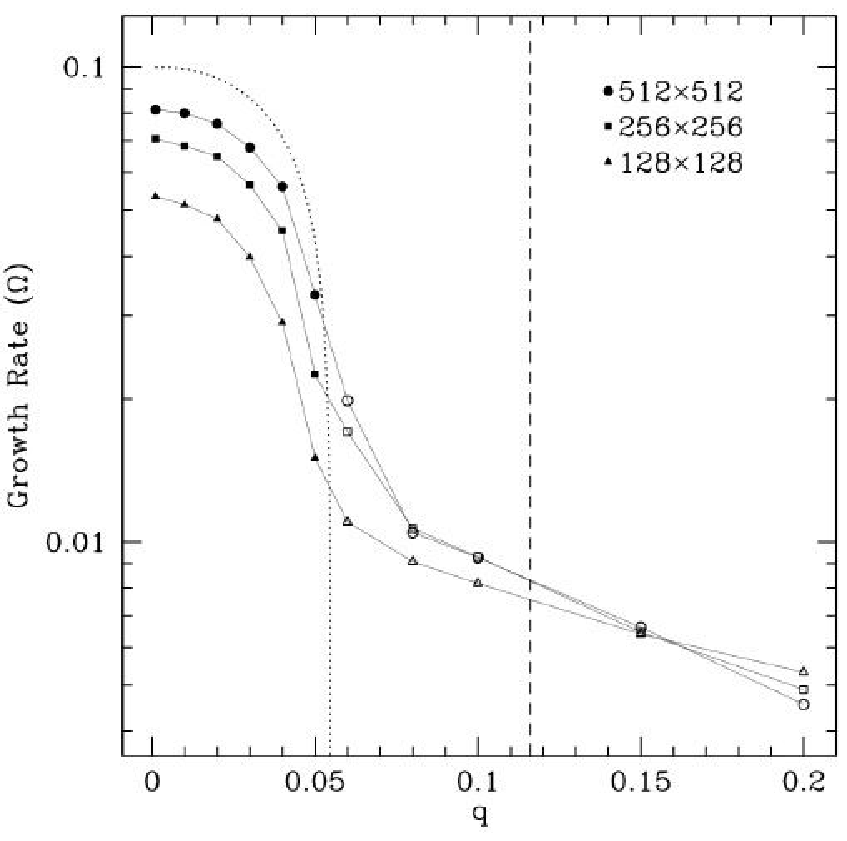}
\caption[Growth rates as a function of $q$.]
{Growth rates as a function of $q$ with $N_{x,min}^2 = -0.01$.
See the text for a discussion.}
\label{pap4f10}
\end{figure}

\begin{figure}[hp]
\centering
\includegraphics[width=5.in,clip]{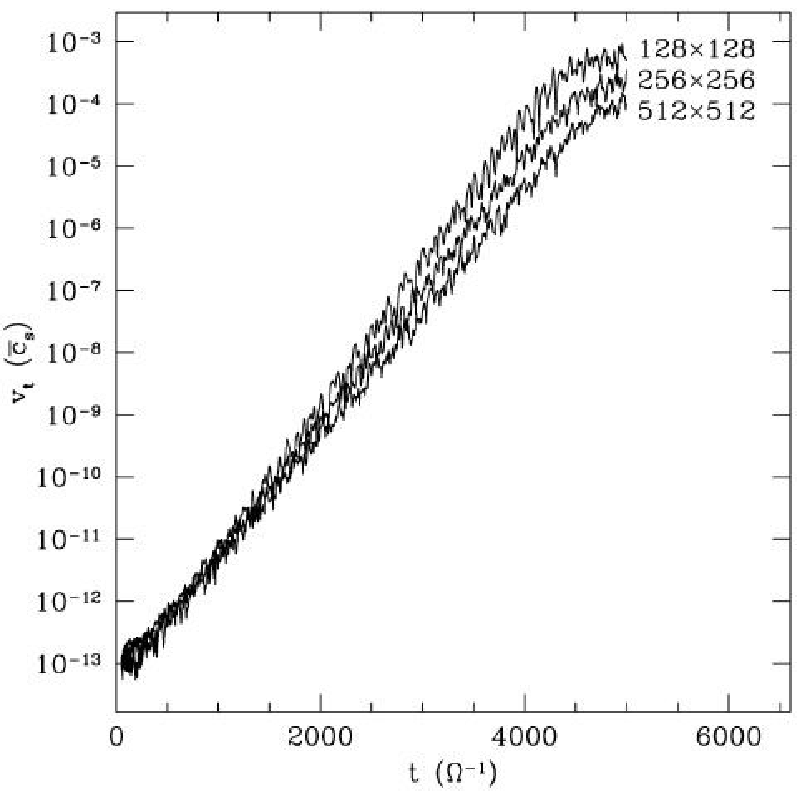}
\caption[Evolution of $v_t$ as a function of time for Run 7.]
{Evolution of $v_t$ as a function of time for Run 7 ($q = 0.2$ and $N_{x,min}^2 = -0.01$).}
\label{pap4f11}
\end{figure}

To investigate the effect of differential rotation upon the growth of this
instability, we have performed a series of simulations with nonzero $q$.
Intuitively, one expects the instability to be suppressed when the shear
rate is greater than the growth rate, i.e. for ${\rm Ri} \gtrsim -1$.
Figure~\ref{pap4f10} shows growth rates from a series of runs with $N_{x,min}^2
= -0.01$ and small, nonzero values of $q$ at three numerical resolutions.  This
figure clearly demonstrates our main result: convective instability is suppressed by
differential rotation.  The expected growth rate from linear theory ($\sqrt{|N_x^2|}$
at $\qe = 0$) is shown in Figure~\ref{pap4f10} as a dotted line.  If there is a
radial position where $\qe(x) = 0$ (i.e., ${\rm Ri} = -\infty$), $v_t$ at that
position looks similar to that of the previous runs (very little deviation from a
straight line); these measurements are indicated on the plot with solid points.
For $q \gtrsim 0.055$ there is no longer any point where $\qe(x) = 0$; in that case
$v_t$ was measured at the radial average between the minimum in $N_x^2(x)$ and the
minimum in $\qe(x)$, since this is where the maximum growth occured.  The data for
these measurements, which are indicated in Figure~\ref{pap4f10} with open points, is
not as clean as it is for the runs with ${\rm Ri} = -\infty$ (see Figure~\ref{pap4f11}).
All of the growth rate measurements in Figure~\ref{pap4f10} were obtained by a least-squares
fit of the data in the range $1 \times 10^{-9} < v_t/\bar{c}_s < 1 \times 10^{-5}$.
The dashed line in Figure~\ref{pap4f10} indicates the value of $q$ for which
${\rm Ri}_{min} = -1$.

Some of the growth in Figure~\ref{pap4f10} appears to be due to aliasing.  This is a
numerical effect in finite-difference codes that results in an artificial transfer of
power from trailing shwaves into leading shwaves as the shwave is lost at the grid scale.
One expects aliasing to occur approximately at intervals of
\be\label{ALIAS}
\Delta \tilde{\tau} = \frac{N_x}{n_y}\frac{L_y}{L_x},
\ee
where $n_y$ is the azimuthal shwave number.  This interval corresponds to $\Delta
\tilde{k}_x(t) = 2\pi/dx$, where $dx = L_x/N_x$ is the radial grid scale.  Based upon
expression (\ref{ALIAS}), aliasing effects should be more pronounced at lower numerical
resolution because the code has less time to evolve a shwave before the wavelength of the
shwave becomes smaller than the grid scale.  It can be seen from the far-right data point
in Figure~\ref{pap4f10} (Run 7 in Table~\ref{pap4t1}) that the measured growth rate
{\it decreases} with increasing resolution.  The evolution of $v_t$ for this run is
shown in Figure~\ref{pap4f11}.

\begin{figure}[hp]
\centering
\includegraphics[width=4.5in,clip]{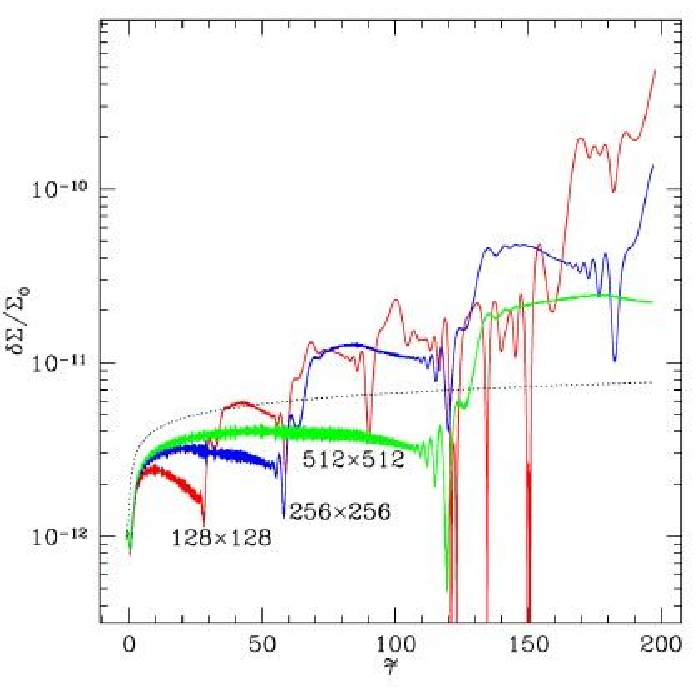}
\caption[Evolution of the density perturbation for Run 8.]
{Evolution of the density perturbation for a single shwave with $q = 0.2$, $N_{x,min}^2 =
-0.01$ and $L_y = L_x$ (Run 8).  The linear theory result is shown as a dotted line, along
with results at three numerical resolutions.
Aliasing occurs when $\tilde{k}_x(t) = 2\pi/dx$.  The overall growth, which is greater at
lower numerical resolution, requires $N_x^2 < 0$.}
\label{pap4f12}
\end{figure}

The effects of aliasing can be seen explicitly by evolving a single shwave, as was done for
our linear theory test (Figures~\ref{pap4f1a} and \ref{pap4f1b}).  Figure~\ref{pap4f12}
shows the evolution of the density perturbation for a single shwave using the same
parameters that were used for Run 7: $L_x = L_y = 12$, $N_{x,min}^2 = -0.01$ and $q = 0.2$.
The initial shwave vector used was $(k_{x0}, k_y) = (-8\pi/L_x, 8\pi/L_y)$.  This
corresponds to $n_y = 4$, and the expected aliasing interval (\ref{ALIAS}) is therefore
$\Delta \tilde{\tau} = N_x/4$.  Runs at three numerical resolutions are plotted in
Figure~\ref{pap4f12}, and the aliasing interval at each resolution is consistent with
expression (\ref{ALIAS}).  It is clear from Figure~\ref{pap4f12} that a lower resolution
results in a larger overall growth at the end of the run.  It also appears that the growth
seen in Figure~\ref{pap4f12} requires a negative entropy gradient.  We have performed this
same test with $N_x^2 > 0$, and while aliasing occurs at the same interval, there is no
overall growth in the perturbations.  This is likely due to the fact that the perturbations
decay asymptotically for $N_x^2 > 0$ (see expression [\ref{ASYMP}]).

\begin{figure}[hp]
\centering
\includegraphics[width=4.5in,clip]{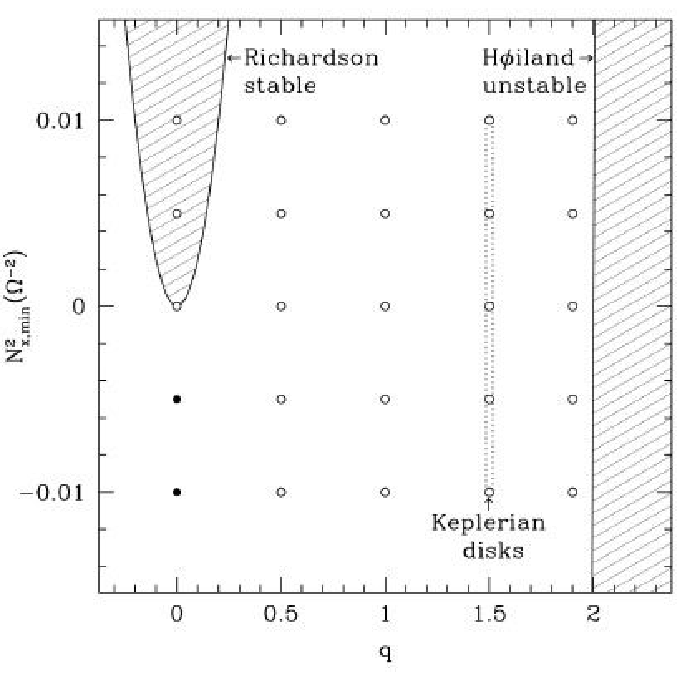}
\caption[Parameter space surveyed in a search for nonlinear instabilities.]
{Parameter space surveyed in a search for nonlinear instabilities.  Closed
(open) circles denote runs that were unstable (stable).  The
only instability found was convective instability for $\qe \simeq 0$
and $N_x^2 < 0$ (${\rm Ri} \rightarrow -\infty$).  (We do not include on
this plot the runs shown in Figure~\ref{pap4f10}.)}
\label{pap4f13}
\end{figure}

\begin{figure}[ht]
\centering
\includegraphics[width=4.in,clip]{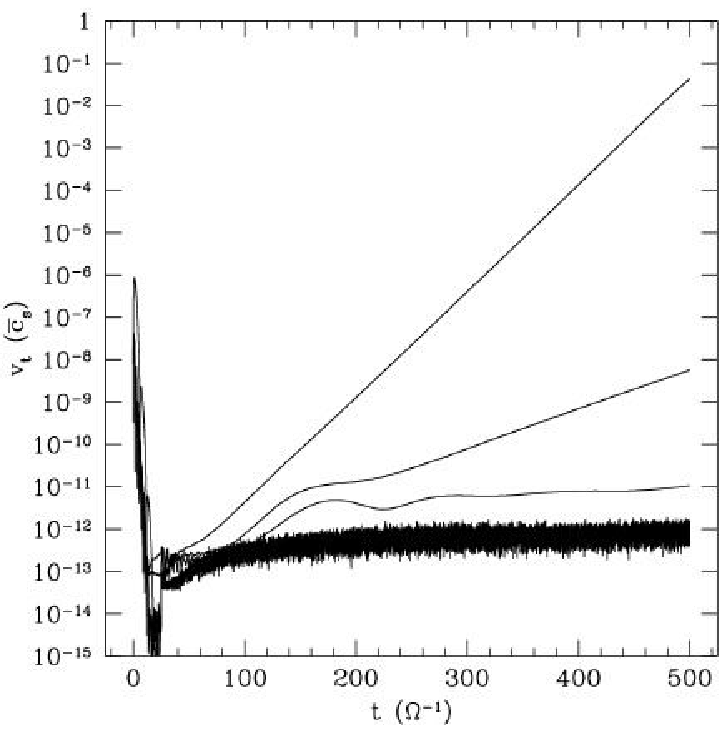}
\caption[Evolution of $v_t$ as a function of time for Run 10.]
{Evolution of $v_t$ as a function of time for Run 10 ($q = 1.5$,
$N_{x,min}^2 = -0.01$).  Also shown are runs with $q = 0$ and an overall
boost equivalent to the velocity at the minimum in $N_x^2$ for Run 10, for
$N_{x,min}^2 = -0.01$ (measured growth rate of $0.058$), $N_{x,min}^2 = -0.003$
(measured growth rate of $0.021$) and $N_{x,min}^2 = -0.001$ (measured growth
rate of $0.0025$).}
\label{pap4f14}
\end{figure}

Figure~\ref{pap4f13} summarizes the parameter space we have surveyed, indicating that
there is instability only for $\qe \simeq 0$ and $N_x^2 < 0$.  The numerical
resolution in all of these runs is $512 \times 512$.  Figure~\ref{pap4f14} shows the
evolution of the radial velocity in Run 10, a run with realistic
parameters for a disk with a nearly-Keplerian rotation profile and radial
gradients on the order of the disk radius: $q = 1.5$ and $N_{x,min}^2 =
-0.01$ (corresponding to ${\rm Ri} \simeq -0.004$).  Clearly no instability
is occurring on a dynamical timescale.  This plot is typical of all runs
for which the evolution was stable.  To give a sense for the minimum growth
rate that we are able to measure, we have also plotted in Figure~\ref{pap4f14}
the results from several unstable runs with $q = 0$ and a boost equivalent to the
velocity at the minimum in $N_x^2$ for Run 10.  It is difficult to measure a growth rate for
the smallest value of $N_{x,min}^2$, but it is clear that there is activity present in this
run which does not occur in the stable run.  Based upon Figure~\ref{pap4f14}, a
conservative estimate for the minimum growth rate that should be detectable in our
simulations is $0.0025 \Omega$.

\section{Implications}

Our results seem to indicate that nearly-Keplerian disks with weak radial
gradients are stable to local nonaxisymmetric disturbances, although we cannot
exclude instability at very high Reynolds number.  Figure~\ref{pap4f10}
demonstrates that convective instability, present when the shear is nearly zero,
is stabilized by differential rotation.  Perturbations simply do not have time to grow
before they are pulled apart by the shear.

An important implication of our results is that the instability claimed by \cite{kb03}
is {\it not} a linear or nonlinear local nonaxisymmetric instability.
Figure~\ref{pap4f12} suggests that the results of \cite{kb03} may be due,
at least in part, to aliasing.  They use a finite difference code at fairly
low numerical resolution ($\leq 128^2$), and growth is only observed in
runs with a negative entropy gradient.  Curvature effects and the effects of
boundary conditions, which may also play a role in their global results,
cannot be tested in our local model.

\end{spacing}

\chapter{Vortices in Thin, Compressible, Unmagnetized Disks}\label{paper3}

\begin{spacing}{1.5}

\section{Chapter Overview}

We consider the formation and evolution of vortices in a hydrodynamic
shearing-sheet model.  The evolution is done numerically using a version
of the ZEUS code.  Consistent with earlier results, an injected vorticity
field evolves into a set of long-lived vortices each of which has radial
extent comparable to the local scale height.  But we also find that the
resulting velocity field has positive shear stress $\<\Sigma \delta v_r
\delta v_\phi\>$.  This effect appears only at high resolution.  The
transport, which decays with time as $t^{-1/2}$, arises primarily because the vortices drive compressive
motions.  This result suggests a possible mechanism for angular momentum
transport in low-ionization disks, with two important caveats: a
mechanism must be found to inject vorticity into the disk, and the
vortices must not decay rapidly due to three-dimensional 
instabilities.\footnote{Submitted to ApJ.  Reproduction for this dissertation
is authorized by the copyright holder.}

\section{Introduction}

Astrophysical disks are common because the specific angular momentum of
the matter inside them is well-conserved.  They evolve because angular
momentum conservation is weakly compromised, either because of diffusion
of angular momentum within the disk or because of direct application of
external torques.

In astrophysical disks composed of a well-ionized plasma it is likely
that some, perhaps most, of the evolution is driven by diffusion of
angular momentum within the disk.  This view is certainly consistent
with observations of steadily accreting cataclysmic variable systems
like UX Ursa Majoris \citep{bap98,bap04}, whose radial
surface-brightness profile is consistent with steady accretion-flow
models in which the bulk of the accretion energy is dissipated within
the disk.

Angular momentum diffusion in well-ionized disks is likely driven by
magnetohydrodynamic (MHD) turbulence.  Analytic analyses, numerical
experiments, and laboratory evidence strongly suggest that
well-coupled plasmas in differentially-rotating flows are subject to
the magnetorotational instability (MRI; \citealt{bh91,bh98,bal03}).
But MHD turbulence is initiated by the MRI only so long as the
plasma is sufficiently ionized to couple to the magnetic field
\citep{kb04,des04}.  In disks around young stars, cataclysmic-variable 
and X-ray binary disks in quiescence, and possibly the outer parts of AGN 
disks, the plasma may be too neutral to support magnetic activity 
\citep{gm98,men00,sgbh00,mq01,ftb02}. This motivates interest in
non-MHD angular momentum transport mechanisms.

Within the last few years, a body of work has been developed suggesting
that vortices can be generated as a result of global hydrodynamic
instability \citep{haw87,bh88,haw90,llcn99,lflc00} or local hydrodynamic
instability \citep{kb03}, that vortices in disks may be long-lived
\citep{gl99,gl00,ur04,bm05}, and that these vortices may be
related to an outward flux of angular momentum \citep{lcwl01,bm05}.
If these claims can be verified then the consequences for low-ionization
disks would be profound.

Here we investigate the evolution of a disk that is given a large
initial vortical velocity perturbation.  Our study is done in the
context of a (two-dimensional) shearing-sheet model, which permits us to
resolve the dynamics to a degree that is not currently possible in a
global disk model.  Our model is also fully compressible, unlike previous
work using a local model \citep{ur04,bm05}.  The former assume incompressible
flow and the latter use the anelastic approximation (e.g., \citealt{gou69}),
which filters out the high-frequency acoustic waves.  We will show that
compressibility and acoustic waves play an essential part in the angular
momentum transport.

Our paper is organized as follows.  In \S2 we describe the model.  In
\S3 we describe the evolution of a fiducial, high-resolution model.
In \S4 we investigate the dependence of the results on model
parameters.  And in \S5 we describe implications, with an emphasis on
key open questions: are the vortices destroyed by three-dimensional
instabilities?; and do mechanisms exist that can inject vorticity into the
disk?

\section{Model}\label{pap3s2}

The shearing-sheet model is obtained via a rigorous expansion of the
two-dimensional hydrodynamic equations of motion to lowest order in
$H/R$, where $H = c_s/\Omega$ is the disk scale height ($c_s$ is the
isothermal sound speed and $\Omega$ is the local rotation frequency)
and $R$ is the local radius.  See \cite{ngg87} for a description. Adopting
a local Cartesian coordinate system where the $x$ axis is oriented
parallel to the radius vector and the $y$ axis points forward in
azimuth, the equations of motion become
\begin{equation}\label{EQUA1}
\dv{\Sigma}{t} + \Sigma \bld{\nabla} \cdot \bld{v} = 0,
\end{equation}
\begin{equation}\label{EQUA2}
\dv{\bld{v}}{t}
+ \frac{\bld{\nabla} P}{\Sigma} + 2\bld{\Omega}\times\bld{v} - 2q\Omega^2 x
\ex = 0,
\end{equation}
where $\Sigma$ and $P$ are the two-dimensional density and pressure,
$\bld{v}$ is the fluid velocity and $d/dt$ is the
Lagrangian derivative. The third and fourth terms in equation
(\ref{EQUA2}) represent the Coriolis and centrifugal forces in the local
model expansion, where $q = -(1/2) \, d\ln\Omega^2/d\ln r$ is
the shear parameter.  We will assume throughout that $q = 3/2$,
corresponding to a Keplerian shear profile.  We
close the above equations with an isothermal equation of state
\begin{equation}\label{EQUA3}
P = c_s^2 \Sigma,
\end{equation}
where $c_s$ is constant in time and space.

Equations (\ref{EQUA1}) through (\ref{EQUA3}) can be combined to show
that the vertical component of potential vorticity
\begin{equation}
\xi \equiv \frac{(\bld{\nabla} \times \bld{v} + 2\bld{\Omega})\cdot \ez}{\Sigma}
\end{equation}
is a constant of the motion; i.e., the potential vorticity of fluid elements
in two dimensions is conserved.

An equilibrium solution to the equations of motion is
\begin{equation}
\Sigma = \Sigma_0 = const.
\end{equation}
\begin{equation}
P = c_s^2 \Sigma_0 = const.
\end{equation}
\begin{equation}
v_x = 0
\end{equation}
\begin{equation}
v_y = -q \Omega x
\end{equation}
Thus the differential rotation of the disk makes an appearance in the
form of a linear shear.

We integrate the above equations using a version of the ZEUS code
\citep{sn92}.  ZEUS is a time-explicit, operator-split scheme on a
staggered mesh.  It uses artificial viscosity to capture shocks.  Our
computational domain is a rectangle of size $L_x \times L_y$ containing
$N_x \times N_y$ grid cells.  The numerical resolution is therefore $\Delta x
\times \Delta y = L_x/N_x \times L_y/N_y$.

Our code differs from the standard ZEUS algorithm in two respects.
First, we have implemented a version of the shearing-box boundary
conditions.  The model is then periodic in the $y$ direction;  the $x$
boundaries are initial joined in a periodic fashion, but they are
allowed to shear with respect to each other, becoming periodic again
when $t = n L_y/(q\Omega L_x)$, $n = 1,2,\ldots$.  A detailed
description of the boundary conditions is given in \cite{hgb95}.

Second, we treat advection by the mean flow $\bld{v}_0 = -q\Omega x \ey$
separately from advection by the perturbed flow $\delta \bld{v} \equiv
\bld{v} - \bld{v}_0$.  Mean-flow advection can be done by interpolation, using
the algorithm described in \cite{gam01}, which is similar to the FARGO
scheme \citep{mass00}.  This has the advantage that the timestep is not
limited by the mean flow velocity (it is $|\delta \bld{v}|$ rather than
$|\bld{v}|$ that enters the Courant condition).  This permits the use of a
timestep that is larger than the usual timestep by $\sim L_x/H$ if $L_x
\gg H$.  The shear-interpolation scheme also makes the algorithm more
nearly translation-invariant in the $x-y$ plane, thereby more
nearly embodying an important symmetry of the underlying equations.

\subsection{Initial Conditions}

Without a specific model for the process that is injecting the
vorticity, it is difficult to settle on a particular set of initial
conditions, or to know how these initial conditions ought to vary
when the size of the box is allowed to vary.  Our choice of initial
conditions is therefore somewhat arbitrary. We use a set of initial
(incompressive) velocity perturbations drawn from a Gaussian random
field.  The amplitude of the perturbations is characterized by $\sigma =
\<|\delta \bld{v}/c_s|^2\>^{1/2}$.  The power spectrum is $|\delta \bld{v}|^2
\sim k^{-8/3}$, corresponding to the energy spectrum ($E_k \sim k^{-5/3}$)
of a two-dimensional Kolmogorov inverse turbulent cascade, with
cutoffs at $k_{min} = (1/2) (2\pi/H)$ and $k_{max} = 32 k_{min}$\footnote{We
have compared our fiducial run to runs with a different range in $k$,
corresponding to vorticity injection either at scales $\sim H$ or scales
$\sim 0.1 H$.  The results are qualitatively the same.}.
The surface density is not perturbed. These initial conditions correspond
to a set of purely vortical perturbations.  The parameters for our fiducial
run are $L_x = L_y = 4H$ and $\sigma = 0.4$.

\subsection{Code Verification}

Although our basic algorithm has already been tested (see \citealt{gam01}),
we test the current version of our code by making a comparison with linear
theory.  Due to the underlying shear, small-amplitude perturbations in the
shearing sheet are naturally decomposed in terms of shearing waves
or {\it shwaves} (see Chapter~\ref{paper2}), Fourier
components in the ``co-shearing'' frame. These have time-dependent
wavenumber $\bld{k}(t) = k_x(t)\ex + k_y\ey$, where $k_x(t) = k_{x0}
+ q\Omega k_y t$ and $k_{x0}$ and $k_y$ are constant. The evolution of
a single Fourier component can be calculated by integrating an ordinary
differential equation for the amplitude of the shwave.  For purely vortical
(nonzero {\it potential} vorticity)
or non-vortical perturbations, the evolution can be obtained analytically. The
explicit expression for the amplitude of a vortical (incompressive) shwave is
\begin{equation}
\delta v_{xi} = \delta v_{x0} \frac{k_0^2}{k^2 } =
\delta v_{x0} \frac{1 + \tau_0^2}{1+\tau^2},
\end{equation}
where $k^2 = k_x^2 + k_y^2$, $\tau = q\Omega t + k_{x0}/k_y$ and a
subscript $0$ on a quantity indicates its value at $t = 0$.\footnote{This
solution is valid at all times only for short-wavelength vortical perturbations
($kH \gg 1$).} The amplitude of a non-vortical (compressive) shwave
satisfies the differential equation
\begin{equation}
\dv{^2 \delta v_{yc}}{t^2} + \left(c_s^2 k^2 + \Omega^2\right) \delta v_{yc} = 0,
\end{equation}
the solutions of which are parabolic cylinder functions. See Chapter~\ref{paper2}
for further details on the shwave solutions.

\begin{figure}[hp]
  \hfill
  \begin{minipage}[t]{1.\textwidth}
    \begin{center}
      \includegraphics[width=3.5in,clip]{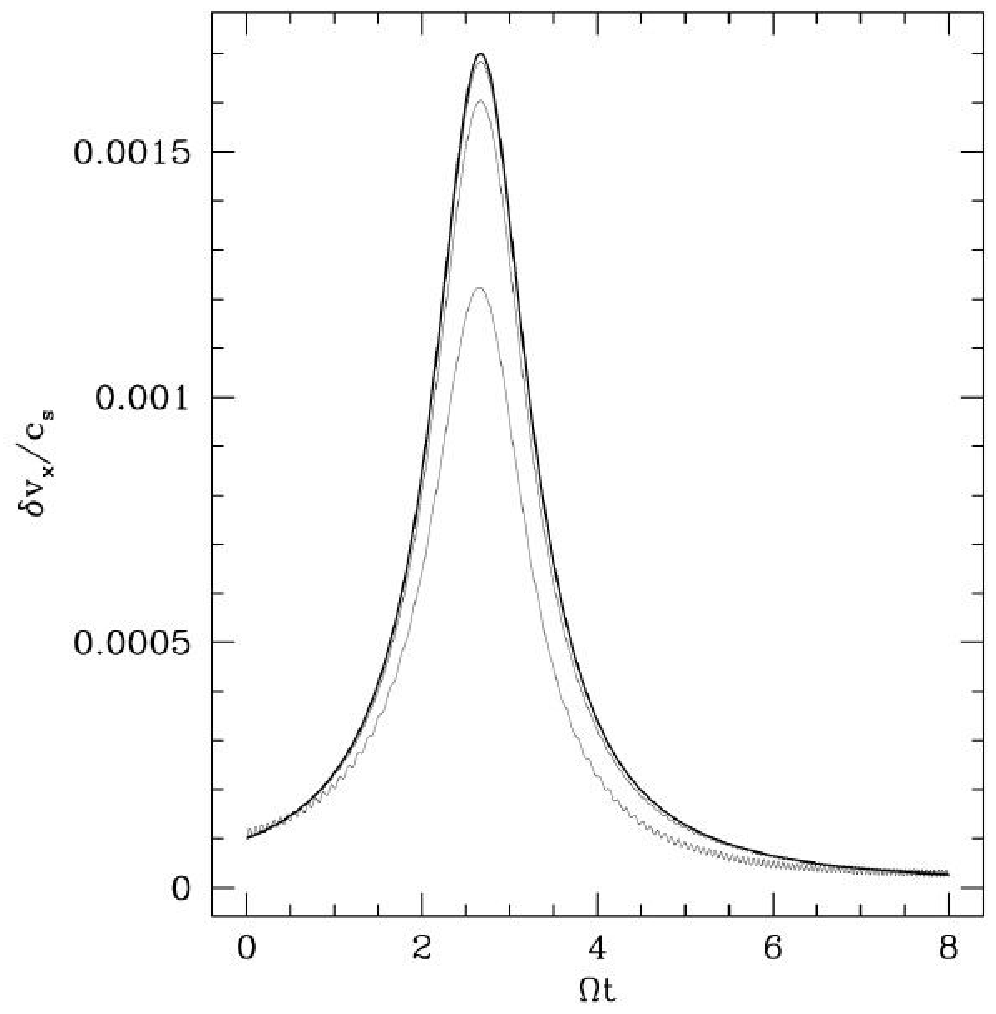}
      \caption[Evolution of the radial velocity amplitude for a vortical shwave.]
      {Evolution of the radial velocity amplitude for a vortical shwave.
      The heavy line is the analytic result, and the light lines
      are numerical results with (in order of increasing accuracy) $N_x = N_y
      = 32, 64, 128$ and $256$.}
      \label{pap3f1a}
    \end{center}
  \end{minipage}
  \hfill
  \begin{minipage}[b]{1.\textwidth}
    \begin{center}
      \includegraphics[width=3.5in,clip]{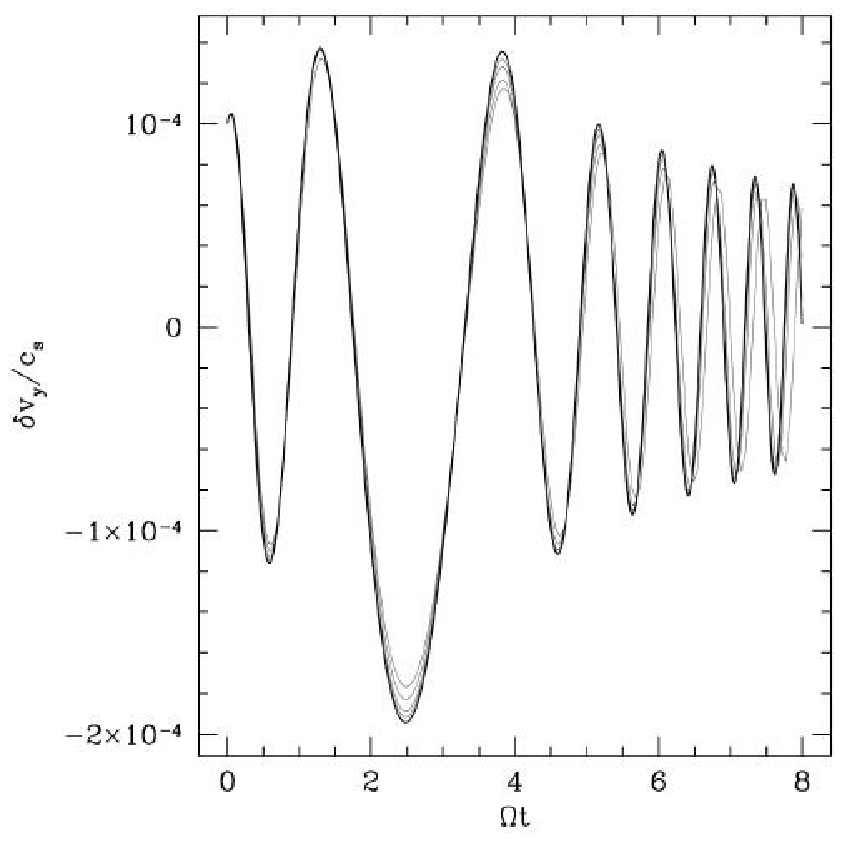}
      \caption[Evolution of the azimuthal velocity amplitude for a nonvortical shwave.]
      {Evolution of the azimuthal velocity amplitude for a nonvortical shwave.
      The heavy line is the analytic result, and the light lines
      are numerical results with (in order of increasing accuracy) $N_x = N_y
      = 32, 64, 128$ and $256$.}
      \label{pap3f1b}
    \end{center}
  \end{minipage}
  \hfill
\end{figure}

Figures~\ref{pap3f1a} and \ref{pap3f1b} compare the numerical evolution of both vortical and
compressive shwave amplitudes with their analytic solutions.  The initial
shwave vector ($k_{x0}, k_y$) is ($-16\pi/L_x, 4\pi/L_y$) for the
vortical shwave and ($-8\pi/L_x, 2\pi/L_y$) for the compressive shwave.
The other model parameters

\noindent
are the same as those in the fiducial run, except
that $L = 0.5H$ for the vortical-shwave evolution since $k_y H \gg 1$ is
required to prevent mixing between vortical and non-vortical shwaves
near $\tau = 0$.  The shwaves are well resolved until the radial
wavelength $\lambda_x = 4 \times \Delta x$, and the code is capable of
tracking both potential-vorticity and compressive perturbations with
high accuracy.

Without a specific model for the process that is injecting the
vorticity, it is difficult to settle on a particular set of initial
conditions, or to know how these initial conditions ought to vary
when the size of the box is allowed to vary.  Our choice of initial
conditions is therefore somewhat arbitrary. We use a set of initial
(incompressive) velocity perturbations drawn from a Gaussian random
field.  The amplitude of the perturbations is characterized by $\sigma =
\<|\delta \bld{v}|^2\>^{1/2}$.  The power spectrum is appropriate for two
dimensional Kolmogorov turbulence, $|\delta \bld{v}|^2 \sim k^{-8/3}$, with
cutoffs at $k_{min} = (1/2) (2\pi/H)$ and $k_{max} = 32 k_{min}$.
The surface density is not perturbed. These initial conditions correspond
to a set of purely vortical perturbations.

\section{Results}\label{pap3s3}

\begin{figure}[hp]
\centering
\includegraphics[width=6.in,clip]{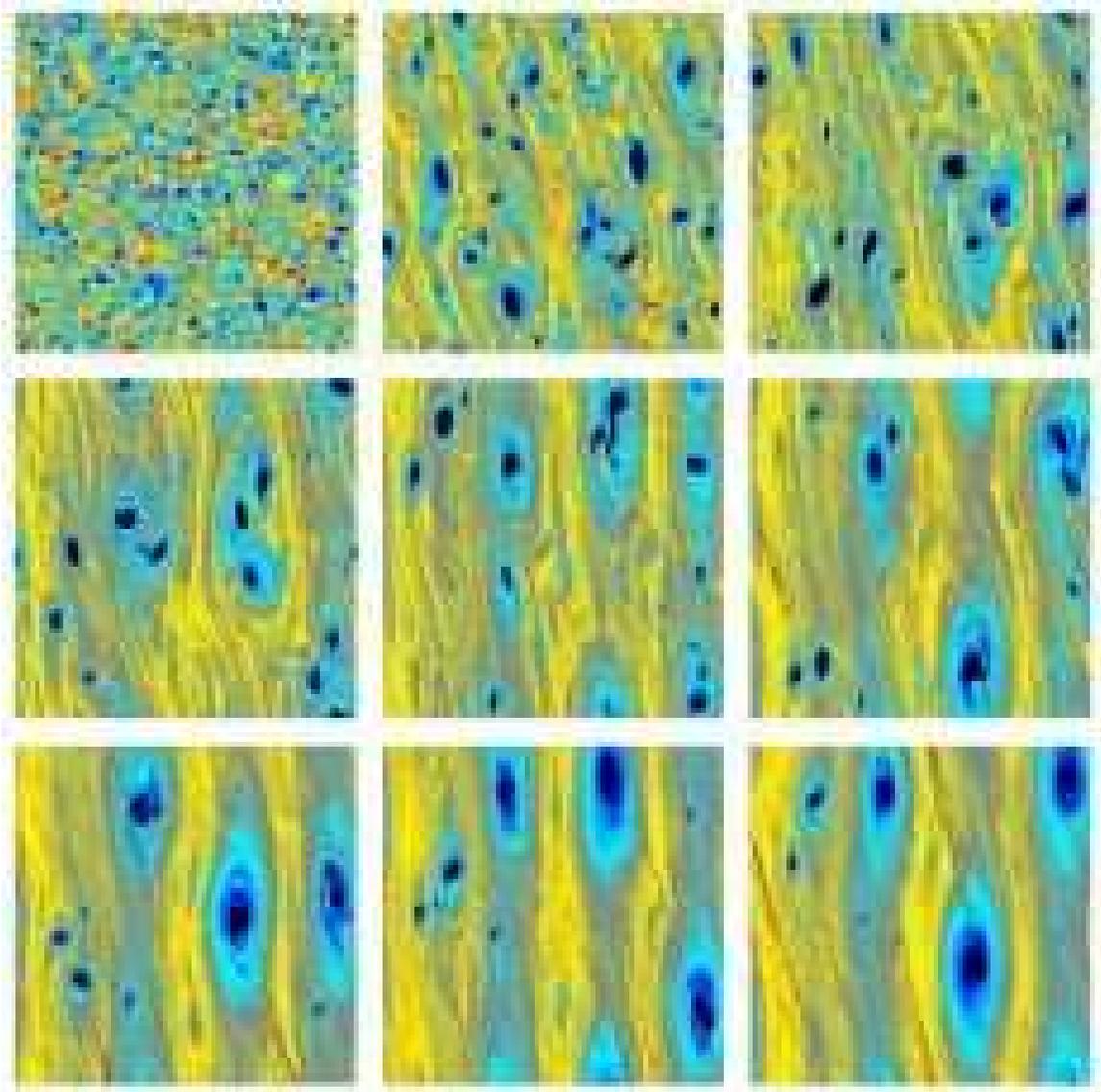}
\caption[Evolution of the potential vorticity in the fiducial run.]
{Panels show the evolution of the potential vorticity in the fiducial
run.  The size is $4 H \times 4 H$ and the numerical resolution is
$1024^2$.  The initial conditions are shown in the upper left corner,
and the other frames follow in lexicographic order at intervals of $22.2
\Omega^{-1}$.  Dark shardes (blue and black in the color version) indicate potential vorticity smaller than $\Omega/(2
\Sigma_0)$; light shades (yellow and red in the color version) show positive potential vorticity
perturbations.  Evidently only the ``anticyclonic'' (negative potential
vorticity perturbation) vortices survive.  Each vortex sheds sound
waves, which steepen into trailing shocks.}
\label{pap3f2}
\end{figure}

The evolution of the potential vorticity in our fiducial run is shown in
Figure~\ref{pap3f2}.  The snapshots are shown in lexicographic order beginning with
the initial conditions, which have equal positive and negative $\delta \xi$.

\begin{figure}[hp]
  \hfill
  \begin{minipage}[t]{1.\textwidth}
    \begin{center}
      \includegraphics[width=3.5in,clip]{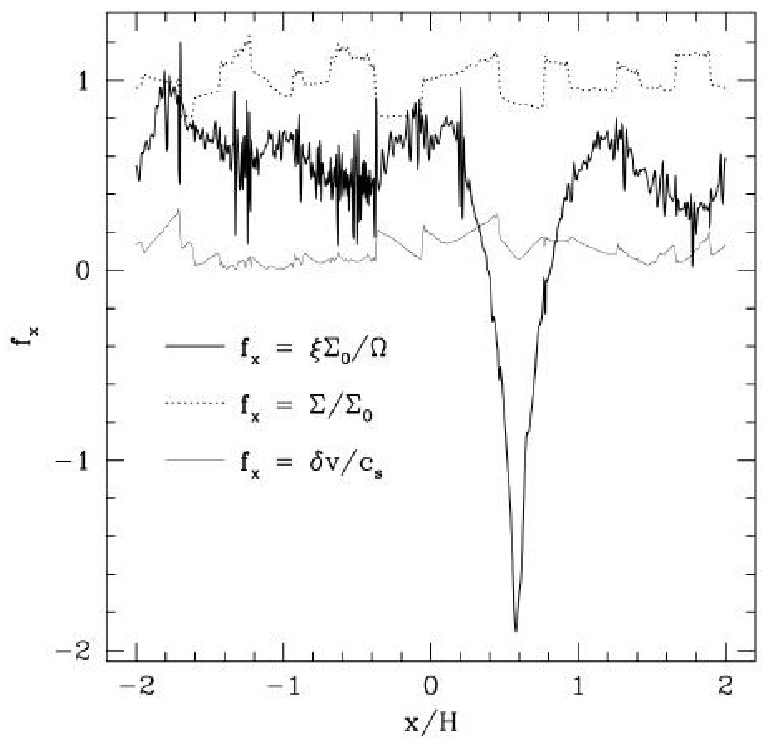}
      \caption[Radial slice of a vortex at the end of the fiducial run.]
      {Radial slice of a vortex at the end of the fiducial run.  The heavy line
      shows the potential vorticity, the light line shows the magnitude of the
      velocity and the dotted line shows the surface density.}
      \label{pap3f3a}
    \end{center}
  \end{minipage}
  \hfill
  \begin{minipage}[b]{1.\textwidth}
    \begin{center}
      \includegraphics[width=3.5in,clip]{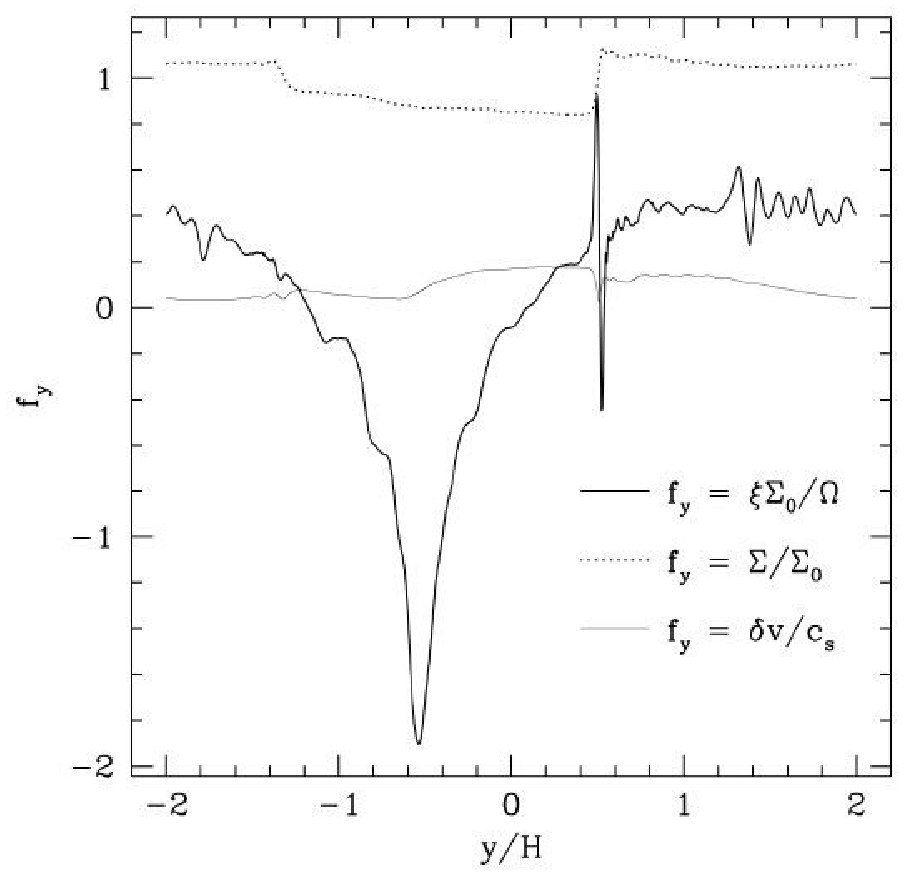}
      \caption[Azimuthal slice of a vortex at the end of the fiducial run.]
      {Azimuthal slice of a vortex at the end of the fiducial run.  The heavy line
      shows the potential vorticity, the light line shows the magnitude of the
      velocity and the dotted line shows the surface density.}
      \label{pap3f3b}
    \end{center}
  \end{minipage}
  \hfill
\end{figure}

One of the most remarkable features of the fiducial run evolution is the
appearance of comparatively stable, long-lived vortices.  These vortices
have negative $\delta \xi$ and are therefore dark in Figure~\ref{pap3f2}.  Similar
vortices have been seen by \cite{gl99,gl00}, \cite{lcwl01} and \cite{ur04}.  Cross
sections of one of the vortices at the end of the run are shown in 
Figures~\ref{pap3f3a} and \ref{pap3f3b}.
In our models the vortices are not associated with easily identifiable
features in the surface density, since the perturbed vorticity is not large
enough to require, through the equilibrium condition, an order unity
increase in the local pressure.

\begin{figure}[hp]
\centering
\includegraphics[width=6.in,clip]{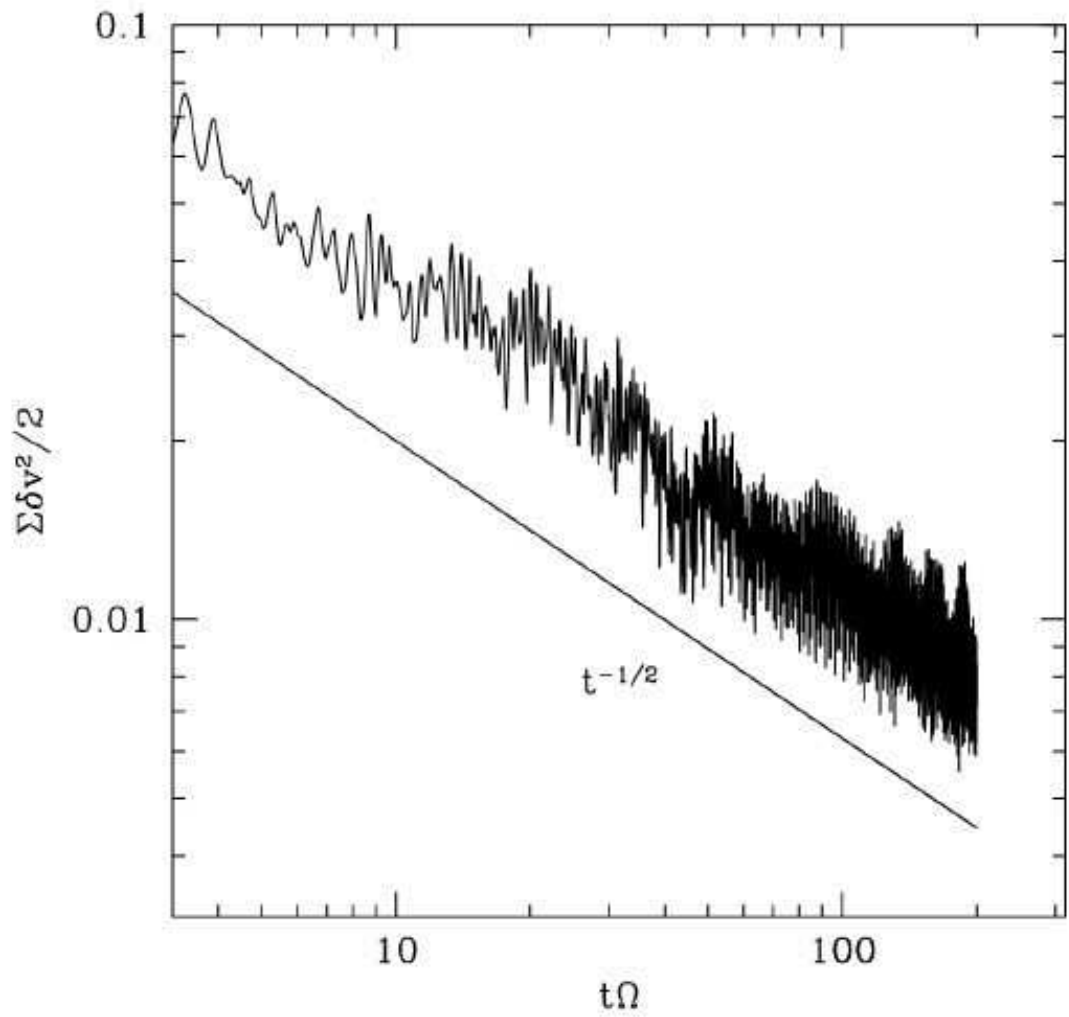}
\caption[Evolution of kinetic energy in time for the fiducial run.]
{Evolution of kinetic energy in time for the fiducial run, on a log-log
scale.  The solid line shows a $t^{-1/2}$ decay for comparison purposes.}
\label{pap3f4}
\end{figure}

While the vortices are long-lived, they do decay.  Figure~\ref{pap3f4} shows the
evolution of the perturbed (noncircular) kinetic energy
\begin{equation}
E_K \equiv \frac{1}{2} \Sigma (\delta v_x^2 + \delta v_y^2)
\end{equation}
in the fiducial run.  Evidently the kinetic energy decays approximately
as $t^{-1/2}$ (which is remarkable in that, if the vortices would
correspond to features in {\it luminosity} that decay as $t^{-1/2}$,
they could produce flicker noise; see \citealt{press78}).  Runs with half and
twice the resolution decay in the same fashion, but if the resolution is
reduced to $64^2$ the kinetic energy decays exponentially.  Resolution

\noindent
of at least 128 zones per scale height appears to be
required.

\begin{figure}[hp]
\centering
\includegraphics[width=5.25in,clip]{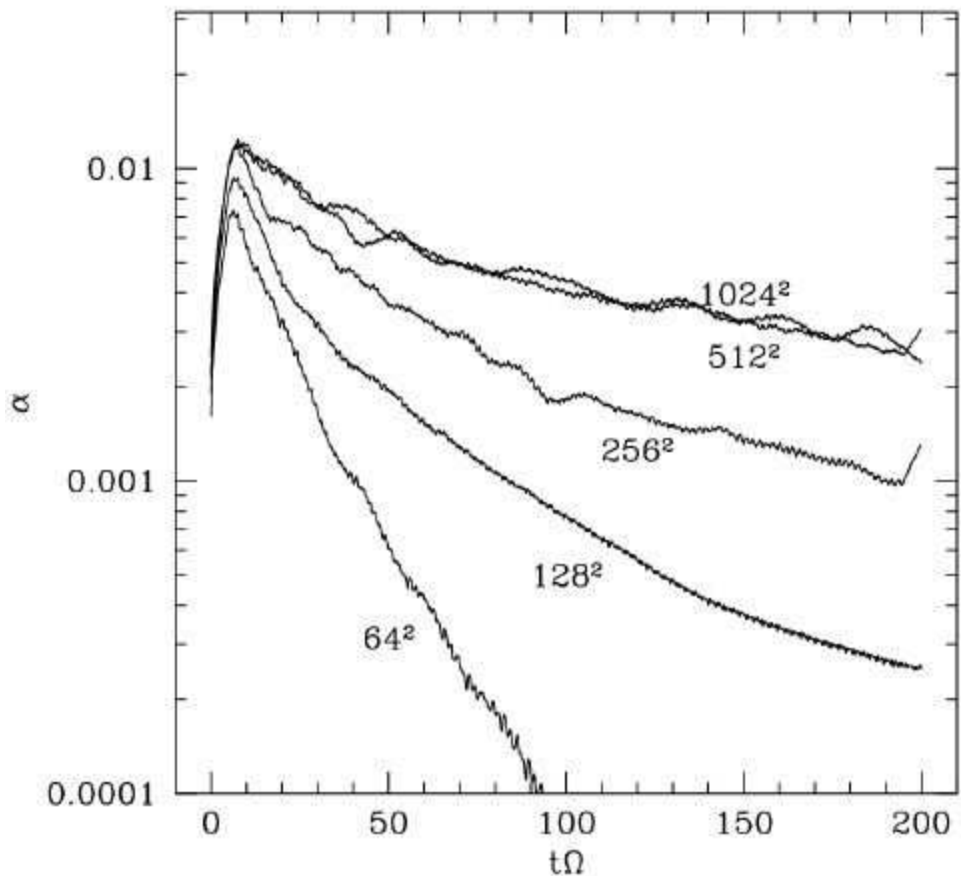}
\caption{Evolution of the shear stress $\alpha$ in the fiducial run and a set of
runs at lower resolutions.}
\label{pap3f5}
\end{figure}

What is even more remarkable is that the vortices are associated with an
outward angular momentum flux, due to the driving of compressive
motions by the vortices.  Figure~\ref{pap3f5} shows the evolution
of the dimensionless angular momentum flux
\begin{equation}
\alpha \equiv \frac{1}{L_x L_y\Sigma_0 c_s^21} \int \Sigma \delta v_x \delta v_y dx dy
\end{equation}
for models with a variety of resolutions.  The data has been boxcar
smoothed over an interval $\Delta t = 10 \Omega^{-1}$ to make the plot
readable.  Again, a resolution of at least $512^2$ appears to be
required for a converged measurement of the shear stress.  For the most
highly resolved models $\alpha$ evolves like the kinetic energy,
$\propto t^{-1/2}$.

Compressibility is crucial for development of the anglar momentum flux.
We have demonstrated this in two ways.  First, we have taken the
fiducial run and decomposed the velocity field into a compressive and an
incompressive part (i.e., into potential and solenoidal pieces in Fourier 
space) and measured the stress associated with each.  For a
set of snapshots taken from the last half of the fiducial run, the
average total $\alpha = 0.0036$; the incompressive component is
$\alpha_i = -0.0006$; the compressive component is $\alpha_c = 0.0032$.
The remaining alpha $\alpha_x = 0.00099$ is in cross-correlations
between the incompressive and compressive pieces of the velocity field.
As argued in \cite{bal00} and \cite{bal03}, both incompressive trailing
shwaves and incompressive turbulence tend to transport angular
momentum inward, whereas trailing compressive disturbances transport
angular momentum outward. Our negative (positive) value for $\alpha_i$
($\alpha_c$) is consistent with this.

Second, we have reduced the size of the model and reduced the amplitude
of the initial perturbation so that it scales with the shear velocity at
the edge of the model (constant ``intensity'' of the turbulence, in
Umurhan and Regev's parlance).  Thus the Mach number of the turbulence
is reduced in proportion to the size of the box.  We have compared four
models, with $L = (4, 2, 1, 0.5) H$ and $\sigma = (0.8, 0.4, 0.2, 0.1)
c_s$.  We would expect the lower Mach number models to have
smaller-amplitude compressive velocity fields and therefore, consistent
with the above results, smaller angular momentum flux $\alpha$.
Averaging over the second half of the simulation, we find $\alpha =
(0.0031, 0.0018, 7.2 \times 10^{-5}, -9.5 \times 10^{-7})$.

\begin{figure}[hp]
  \hfill
  \begin{minipage}[t]{1.\textwidth}
    \begin{center}
      \includegraphics[width=3.5in,clip]{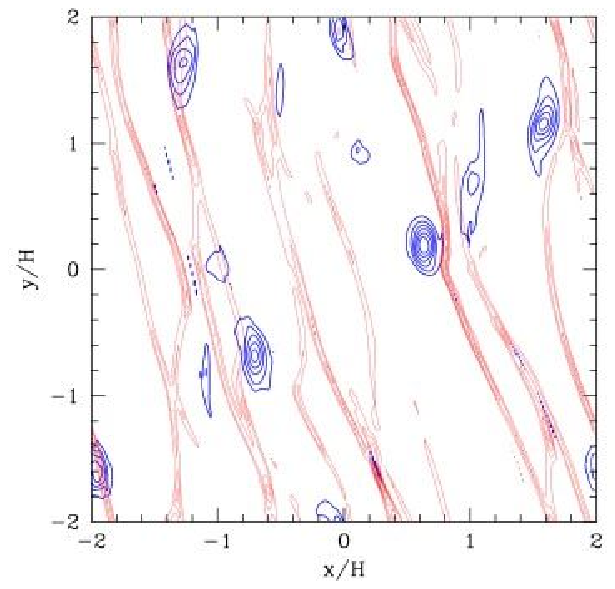}
      \caption[Snapshot of the velocity divergence superimposed on the potential
vorticity in a medium-resolution version of the fiducial run.]
      {Snapshot of the velocity divergence superimposed on the potential
vorticity in a medium-resolution ($256^2$) version of the fiducial run.
The thin (red in the color version) contours indicate negative divergence and are
associated with shocks. The thick (blue in the color version) contours indicate negative
potential vorticity.}
      \label{pap3f6}
    \end{center}
  \end{minipage}
  \hfill
  \begin{minipage}[b]{1.\textwidth}
    \begin{center}
      \includegraphics[width=3.5in,clip]{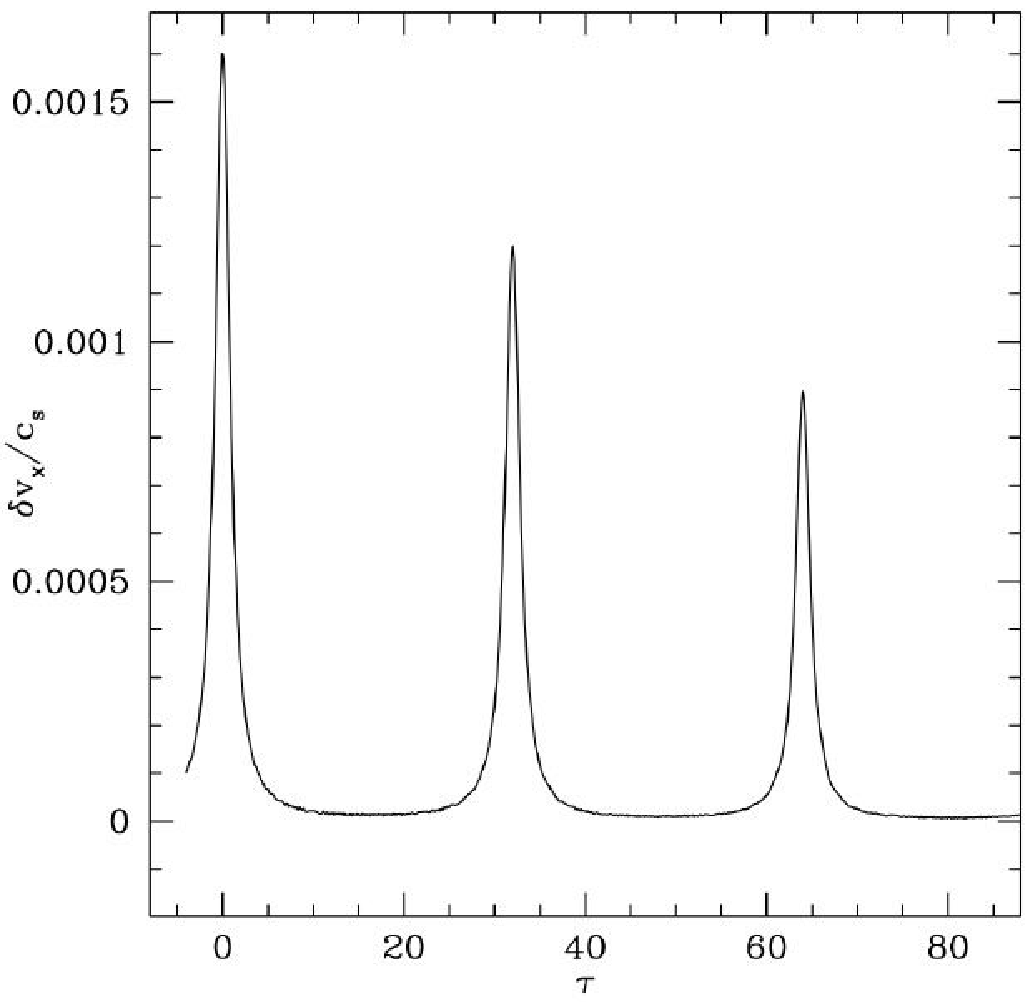}
      \caption[Evolution of a vortical shwave amplitude in a low-resolution run]
      {Evolution of a vortical shwave amplitude in a low-resolution ($64^2$)
run, in units of $\tau$. The initial shwave vector ($k_{x0},k_y$) is
($-16\pi/L_x, 4\pi/L_y$), corresponding to $\tau_0 = -4$. The interval
between successive peaks (a numerical effect due to aliasing) is $\tau =
N_x/n_y$, where $n_y = 2$ is the azimuthal wavenumber.}
      \label{pap3f7}
    \end{center}
  \end{minipage}
  \hfill
\end{figure}

An additional confirmation of our overall picture can be seen in
Figure~\ref{pap3f6}, in which we show a snapshot of the velocity divergence
superimposed on the potential vorticity for a medium-resolution
($256^2$) version of the fiducial run.\footnote{At higher resolutions,
shocks are generated earlier in the simulation from smaller vortices,
and it is more difficult to see the effect we are describing due to the
random nature of the vortices at this early stage.} The position of
the shocks with respect to the vortices in this figure is consistent
with our interpretation that the former are generated by the latter.

The smallest of our simulations ($L = 0.5H$) is nearly incompressible,
but we continue to observe $t^{-1/2}$ decay (least squares fit power law
is $-0.49$) at late times.  The reason that we see decay while
\cite{ur04} do not may be that: (1) the remaining compressibility in our
model causes added dissipation; (2) the numerical dissipation in our
code is larger than that of \cite{ur04}; (3) the code used by
\cite{ur04} could somehow be aliasing power from trailing shwaves to
leading shwaves (although they do explicitly discuss, and dismiss, this
possibility).

To highlight the dangers of aliasing for our finite-difference code, in
Figure~\ref{pap3f7} we show the evolution of a vortical shwave amplitude at low
resolution ($64^2$), in units of $\tau$. We use the same parameters as
those in our linear-theory test (Figures~\ref{pap3f1a} and \ref{pap3f1b}), 
for which the initial shwave
vector corresponds to $\tau_0 = -4$. The initially-leading shwave swings
into a trailing shwave, the radial wavelength is eventually lost near the
grid scale, and due to aliasing the code picks up the evolution of the
shwave again as a leading shwave.  Repeating this test at higher
resolutions indicates that successive swings from leading to trailing
occur at an interval of $\tau = N_x/n_y$, where $n_y = 2$ is the
azimuthal wavenumber of the shwave.  This is equivalent to $k_x(t) =
2\pi/\Delta x$. The decay of the successive linear solutions with time
is due to numerical diffusion.

Figure~\ref{pap3f7} suggests that it is easier to inject power into the simulation
due to aliasing rather than to remove power due to numerical diffusion.
We do not believe, however, that aliasing is affecting our
high-resolution results.  In addition, if we assume that the flow in our
simulations can be modeled as two-dimensional Kolmogorov turbulence,
then $\delta v_{rms} \sim \lambda^ {1/3}$, where $\delta v_{rms}$ is
the rms velocity variation across a scale $\lambda$.  The velocity due to the
mean shear at these scales is $\delta v_{shear} \sim q \Omega \lambda$,
and $\delta v_{rms}/\delta v_{shear} \sim \lambda^{-2/3}$. The velocities at
the smallest scales are thus dominated by turbulence rather than by the
mean shear. This conclusion is supported by the convergence of our
numerical results at high resolution.

\begin{figure}[hp]
  \hfill
  \begin{minipage}[t]{1.\textwidth}
    \begin{center}
      \includegraphics[width=3.5in,clip]{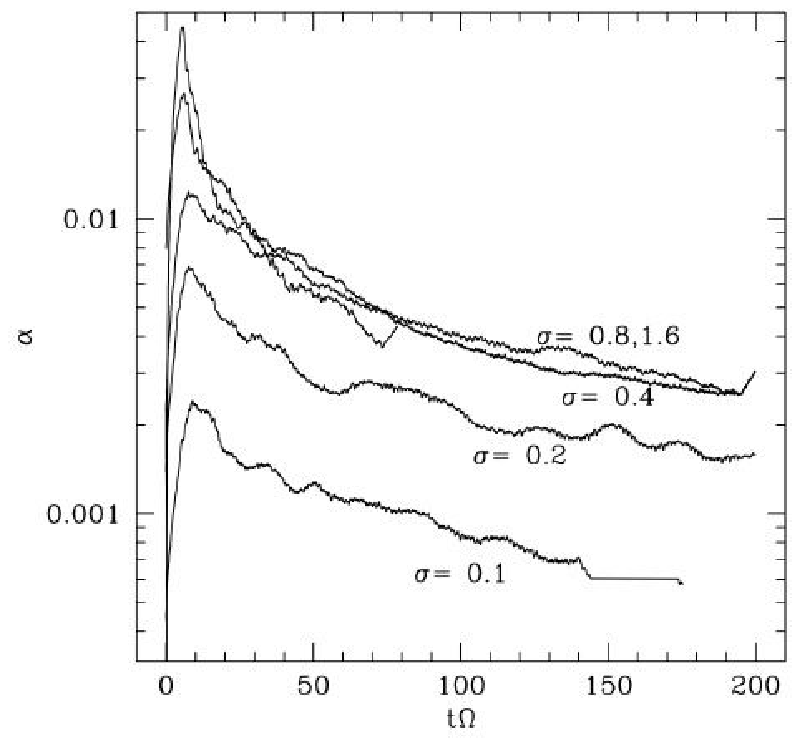}
      \caption[Evolution of the shear stress $\alpha$ in a set of runs at with varying
initial $\sigma$.]
      {Evolution of the shear stress $\alpha$ in a set of runs at with varying
initial $\sigma$.  Apparently for low values of $\sigma$ the shear
stress is reduced, but for initial Mach number near $1$ the stress
saturates.  All runs have $L = 4 H$.}
      \label{pap3f8}
    \end{center}
  \end{minipage}
  \hfill
  \begin{minipage}[b]{1.\textwidth}
    \begin{center}
      \includegraphics[width=3.5in,clip]{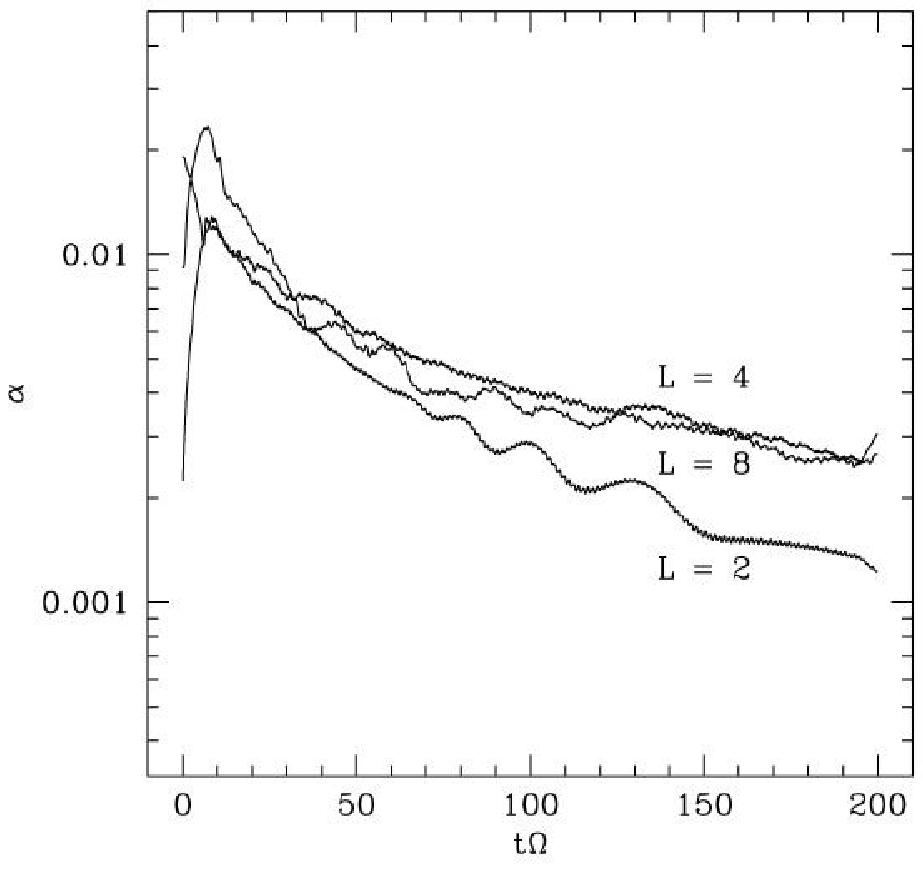}
      \caption[Evolution of a vortical shwave amplitude in a low-resolution run]
      {Evolution of the shear stress $\alpha$ in a set of runs at with varying
initial $L$, but the same initial Mach number $\sigma$.}
      \label{pap3f9}
    \end{center}
  \end{minipage}
  \hfill
\end{figure}

Our model contains two additional numerical parameters: the size $L$ and
the initial turbulence amplitude $\sigma$. Figure~\ref{pap3f8} shows the evolution
of $\alpha$ for several values of $\sigma$.  Evidently for small enough
values of $\sigma$ the $\alpha$ amplitude is reduced, but for near-sonic
initial Mach numbers the $\alpha$ amplitude saturates (or at least the
dependence on $\sigma$ is greatly weakened).  Figure~\ref{pap3f9} shows the
evolution for several values of $L$ but the same initial $\sigma$ and
the identical initial power spectrum. For large enough $L$ the shear
stress appears to be independent of $L$.

\begin{figure}[hp]
  \hfill
  \begin{minipage}[t]{1.\textwidth}
    \begin{center}
      \includegraphics[width=3.5in,clip]{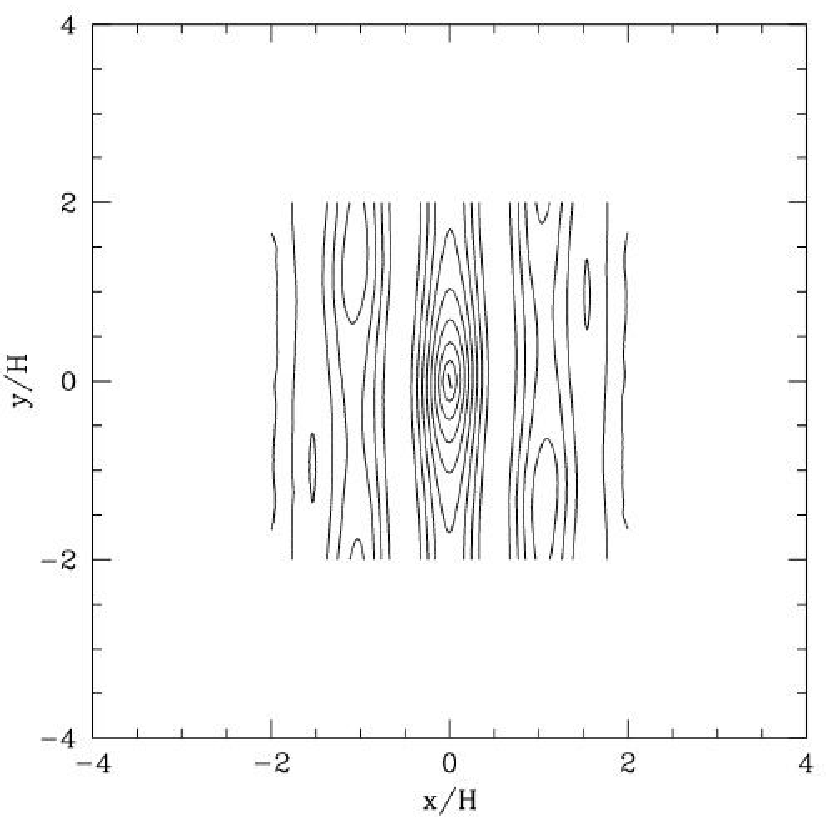}
\caption{Autocorrelation function of the potential vorticity $\xi$ for the fiducial
model with $L = 4H$.}
      \label{pap3f10a}
    \end{center}
  \end{minipage}
  \hfill
  \begin{minipage}[b]{1.\textwidth}
    \begin{center}
      \includegraphics[width=3.5in,clip]{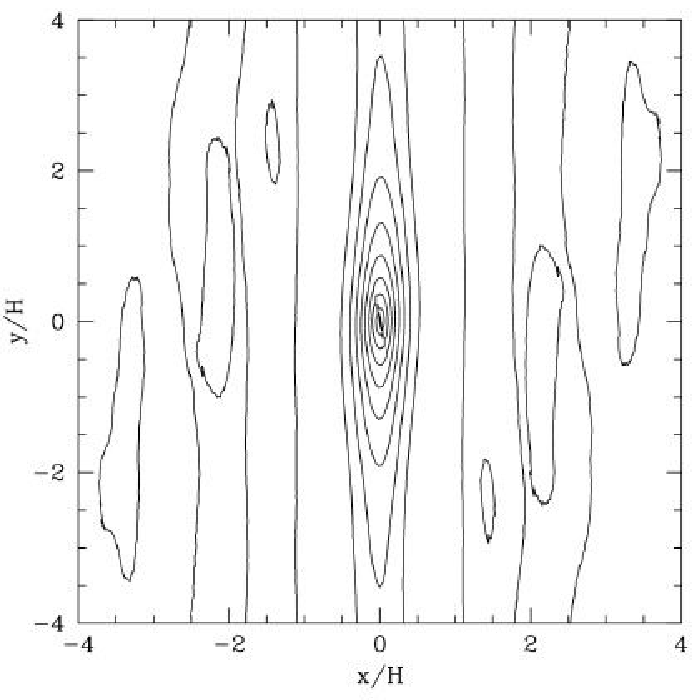}
\caption{Autocorrelation function of the potential vorticity $\xi$ for a model
with $L = 8H$.}
      \label{pap3f10b}
    \end{center}
  \end{minipage}
  \hfill
\end{figure}

Finally, we have studied the autocorrelation function of the potential
vorticity as a means of characterizing structure inside the flow.
Figures~\ref{pap3f10a} and \ref{pap3f10b} shows the autocorrelation function
measured in the fiducial
model and in an otherwise identical model with $L = 8H$.  Evidently
the potential vorticity is correlated over about one-half a scale height
in radius, independent of the size of the model.  This supports the idea
that compressive effects limit the size of the vortices, since the shear
flow becomes supersonic across a vortex of size $\sim H$ \citep{bs95,lcwl01}.

\section{Conclusion}\label{pap3s4}

The presence of long-lived vortices in weakly-ionized disks may be an
integral part of the angular momentum transport mechanism in these
systems. The key result we have shown here is that compressibility of
the flow is an extremely important factor in providing a significant,
positively-correlated average shear stress with its associated outward
transport of angular momentum. Previous results using a local model
have assumed incompressible flow and either report no angular momentum
transport \citep{ur04} or report a value ($\alpha \sim 10^{-5}$, \citealt{
bm05}) that is two orders of magnitude lower than what we find when we
include the effects of compressibility. Global simulations
\citep{gl99,gl00,lcwl01} have a difficult time accessing the high
resolution that we have shown is required for a significant shear
stress due to compressibility.

Our work leaves open the key question of what happens in three
dimensions. Our vortices, which have radial and azimuthal extent
$\lesssim H$, are inherently three-dimensional. Three-dimensional
vortices are susceptible to the elliptical instability \citep{ker02} and
are likely to be destroyed on a dynamical timescale. The fact that 
vortices persist in our two-dimensional simulations and not in the local 
(three-dimensional) shearing-box calculations of \cite{bhs96} is likely 
due to dimensionality. The recent numerical results of \cite{bm05} indicate 
that vortices near the disk midplane are quickly destroyed, whereas vortices 
survive if they are a couple of scale heights away from the midplane.  
Strong vertical stratification away from the midplane may enforce 
two-dimensional flow and allow the vortices that we consider here to survive.

The initial conditions in \cite{bm05} are analytic solutions for
two-dimensional vortices that are stacked into a three-dimensional
column. The stable, off-midplane vortices apparently arise due to the
breaking of internal gravity waves generated by the midplane vortices
before they become unstable. There is also an unidentified instability
that breaks a single off-midplane vortex into several vortices.  These
simulations leave open the question of whether stable off-midplane
vortices can be generated from a random set of initial vorticity 
perturbations rather than the special vortex solutions that are imposed.

Our work also leaves open the key question of what generates the initial
vorticity. One possibility is that material builds up at particular
radii in the disk, resulting in a global instability (e.g.
\citealt{pp84,pp85}) and a breakdown of the flow into vortices
\citep{lcwl01}. Another possibility for vortex generation in variable
systems is that the MHD turbulence, which likely operates during an
outburst but decays as the disk cools \citep{gm98}, leaves behind some
residual vorticity. The viability of such a mechanism could be tested
with non-ideal MHD simulations such as those of \cite{fs03} and
\cite{ss03}. Yet another possibility is that differential illumination of the disk
somehow produces vorticity.  Since the temperature of most circumstellar
disks is controlled by stellar illumination, small variations in
illumination could produce hot and cold spots in the disk that interact
to produce vortices. The final possibility that we consider is the generation
of vorticity via baroclinic instability, which is likely to operate in disks
whose vertical stratification is close to adiabatic \citep{ks86}. The
nonlinear outcome of this instability in planetary atmospheres is the
formation of vortices, although it is far from clear that the same outcome
will occur in disks. Finally, we note that a residual amount of vorticity 
can be generated from finite-amplitude compressive perturbations. We have 
performed a series of runs with zero initial vorticity and perturbation 
wavelengths on the order of the scale height, and the results are qualitatively similar to Figure~\ref{pap3f5} with the shear stress reduced by nearly two 
orders of magnitude.

\end{spacing}

\chapter{Summary and Outlook}\label{conclusion}

\begin{spacing}{1.5}

Angular momentum transport is key to the evolution of accretion disks.  In ionized disks,
momentum transport is likely to be mediated internally by MHD turbulence generated by the
MRI.  Despite the success of this local shear instability in elucidating the accretion
process in ionized disks, the complexity of its nonlinear outcome has raised a whole new
set of questions regarding its effects upon disk evolution.  The answers to most of these
questions will require the use of three-dimensional numerical simulations.  I discuss some
of the remaining open questions in Section~\ref{saos1} and propose some simple numerical
experiments that will attempt to answer them.

The mechanism driving accretion in weakly-ionized disks remains unclear.  I summarize the
main results of this dissertation in Section~\ref{saos2}, results which mostly argue against
a turbulent transport of angular momentum in these disks.  I also discuss some possible
directions to take in further pursuit of such a mechanism.

The fact that decades of research have not uncovered a robust turbulent transport mechanism
for weakly-ionized disks raises the possibility that at least some of these disks (or portions
of them) are stable and do not accrete in a steady state as the standard disk model assumes.
Proposals for exploring the implications of this possibility are discussed in
Section~\ref{saos3}.

Finally, while modeling turbulent shear stresses as an alpha viscosity has turned out to be
useful phenomenologically, representing disk turbulence as an alpha viscosity has its
limitations (see Section~\ref{dmts}), and any model results that depend upon an accurate
representation of this aspect of disk physics are therefore suspect.  The fundamental
understanding of turbulent shear stresses in disks that has begun to emerge in recent years
has opened up an exciting opportunity for developing physically-motivated disk models based
upon a first-principles treatment of disk turbulence.  I conclude in Section~\ref{saos4}
with a discussion of proposed research along these lines.

\section{Ionized Disks}\label{saos1}

The phenomenological approach to modeling turbulent transport in accretion disks has been
challenged by the recent improved physical understanding of that transport in ionized disks.
At the same time, there are important issues with regard to MHD turbulence and its 
ramifications for disk physics that must be understood before the standard disk model can be 
replaced with models based upon a more accurate representation of turbulent transport.  This 
section discusses a few of these issues and proposes ways in which they can be investigated.

\subsection{Transition from 2D to 3D MHD Turbulence}

An important assumption underlying standard disk theory is that the global and local disk 
scales are well separated; small-scale turbulence is assumed to average to a smoothly-varying 
flow on large scales.  The validity of this assumption can be tested by measuring the power 
spectrum of MHD turbulence in a series of three-dimensional shearing-box simulations with 
horizontal scales much larger than the vertical scale, looking for a transition between 
two-dimensional and three-dimensional behavior.  If there is a transition at some 
characteristic scale, presumably on the order of the disk scale height, this will confirm 
the standard picture as well as provide a guide for the scales at which the assumptions of
standard disk theory can be appropriately applied in global models.  If there is no transition 
to smoothly-varying two-dimensional flow, much of standard disk theory is invalid.

\subsection{Dynamics of MRI Turbulent Stresses}\label{dmts}

Turbulence is an extremely complex phenomenon, and the standard approach of modeling it as
a viscosity is clearly oversimplified.  For example, there are key dynamical properties of
MHD turbulence, such as the elastic properties that produce magnetic tension \citep{op03},
that cannot be modeled with a viscosity \citep{mg02}.  In addition, the standard disk model
assumes that the turbulent shear stress is isotropic, whereas MHD turbulence is inherently
anisotropic.  Any model results that depend upon an accurate representation of the turbulent
shear stress are therefore suspect.  \cite{ogl03} has developed an analytic model for MHD
turbulent stresses consisting of evolution equations for the turbulent stress tensor.  Such
a model could provide the basis for more realistic analytic and numerical studies of accretion
disks, as well as be useful for global disk simulations since it would allow one to represent
the underlying turbulence accurately over long time scales without having to resolve the
small-scale structure.  I will attempt to test this model against local numerical
simulations of MHD turbulence and seek to constrain the values of its free parameters.

\subsection{Interaction of Waves with Turbulence}\label{iwt}

Another key assumption underlying many disk studies is that waves will propagate through
turbulence relatively unhindered.  For example, models that have been developed to explain
quasi-periodic oscillations (QPOs) in binary systems often invoke characteristic oscillations
of the accretion disk as the source of the QPOs \citep{wso01}.  The implicit assumption in
these models is that the disk oscillations are negligibly affected by the disk turbulence.
As another example, analytic studies of warped accretion disks \citep{pp83,pl95} have shown
that warps will propagate as non-dispersive waves if the turbulence in the disk is
sufficiently small.  The turbulence in these studies is modeled as a shear viscosity, but
it is not clear how the turbulence will interact with warp propagation self-consistently.  
As a final example, the standard disk model assumes that thermal energy is both generated 
and radiated locally, which results in a temperature dependence with radius that is not 
always observed \citep{rww99}.  An alternative to this local dissipation process is a 
global process whereby waves are excited at one point in the disk and carry energy and 
angular momentum to another point in the disk before dissipating \citep{asl88}.  Again, 
wave propagation through the disk is assumed to proceed unhindered by the turbulence.

The general problem of the interaction of waves with turbulence has been well-studied
analytically, particularly in the context of solar oscillations (\citealt{gk88,gk90}; see
also \citealt{fg04}), but numerical studies along these lines are less advanced (see
\citealt{tea00} for one example).  I will seek to further our understanding in this area by
investigating, via numerical experiments, the effect of turbulence on waves propagating
through a disk.  A wave or superposition of wave modes calculated from linear theory will
be inserted into a magnetized turbulent numerical model, and the subsequent evolution
will be compared with the linear theory results.  The outcome of these experiments will
determine whether or not the neglect of wave-turbulence interactions is valid.  If there
is significant interaction, these experiments may provide a means for developing a
predictive theory of wave propagation in turbulent disks.

\section{Weakly-Ionized Disks}\label{saos2}

The prospects of discovering a turbulent transport mechanism in weakly-ionized accretion 
disks appear to be dim.  Since proving stability is much more difficult than demonstrating 
instability, however, the search for such a transport mechanism continues.  The most 
promising mechanism based upon the work discussed in this dissertation is the driving
of compressive motions by vortices, but even this faces serious difficulties due to the 
decaying angular momentum flux and the potential for the vortices to be unstable in three 
dimensions.  It would be useful to redo in three dimensions the simulations discussed in 
Chapter~\ref{paper3}, to determine the rate at which the vortices become unstable and the 
nonlinear outcome of such an instability.  Testing some of the vorticity-generation 
mechanisms discussed in \S\ref{pap3s4} is also a necessary next step in understanding the 
relevance of this overall picture for driving accretion in weakly-ionized disks.

The possibility of a bypass transition to turbulence due to the transient amplification of
linear disturbances followed by a nonlinear feedback from trailing shwaves (shearing waves) 
into leading shwaves
has not been fully explored.  The effects of aliasing shown in Figure~\ref{pap4f12} clearly
demonstrate that such a feedback mechanism can result in the overall growth of linear
disturbances into the nonlinear regime.  While the feedback in Figure~\ref{pap4f12} is
entirely numerical, the high-resolution requirements for tracking these shwaves (as can
be seen in Figure~\ref{pap4f1b}) implies that the nonlinear outcome of transient amplification
has not been tested at the resolutions we employ for the runs discussed in Chapter~\ref{paper4}.

\cite{vmw98} calculated three-shwave interactions in the unstratified (incompressible) shearing
sheet and found that there is feedback from trailing shwaves into leading shwaves for a small
subset of initial shwave vectors.  It would be of interest to revisit this calculation in the
stratified shearing sheet.  One key difference between the stratified and unstratified
shearing-sheet models is that in the latter case all linear perturbations decay after their transient
growth, whereas in the former case the density perturbation does not decay.  This implies that quasilinear interactions are more likely to take place.  
A comparison of Figures~\ref{pap3f7} and \ref{pap4f12} indicates that 
feedback due to aliasing in the unstratified sheet does not result in
any overall growth, whereas feedback in the stratified sheet does.

\section{Layered Disks}\label{saos3}

\begin{figure}[h]
\centering
\includegraphics[width=6.5in,clip]{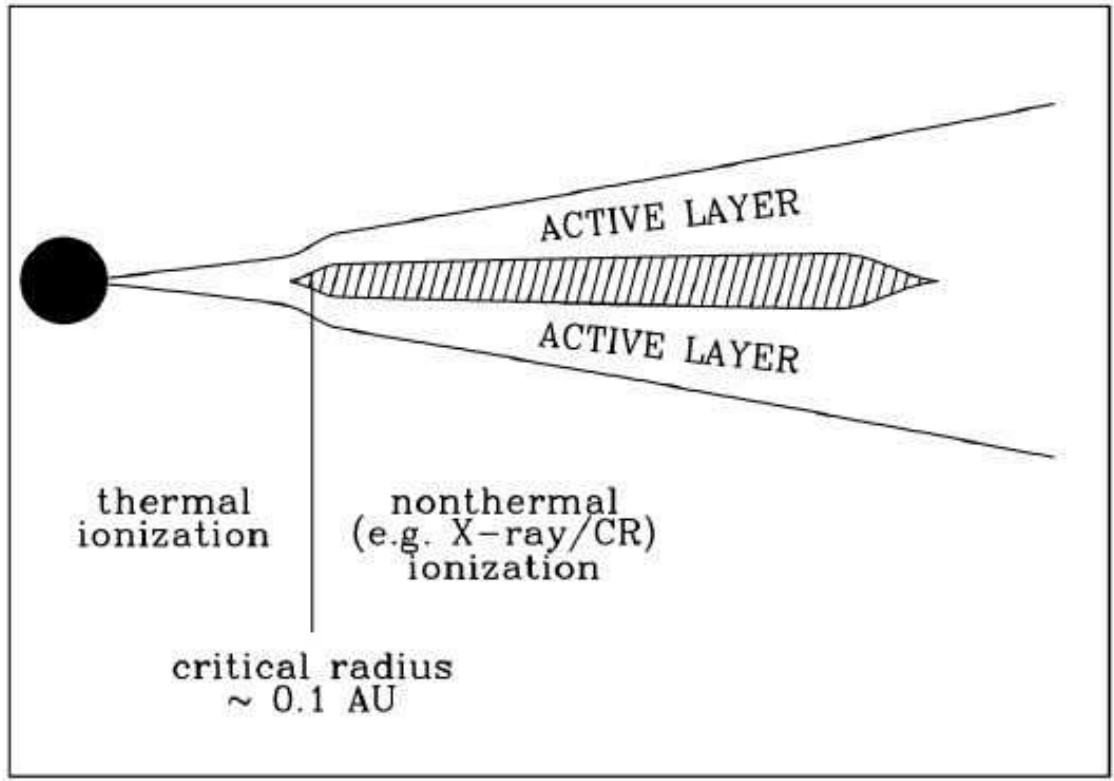}
\caption[Ionization structure of YSO disks.]{
Ionization structure of YSO disks.  The inner disk is coupled
to the field via thermal ionization.  At larger radii the surface layers
are coupled but the midplane (hashed) is inactive.  At still larger
radii the density is lower and nonthermal ionization provides effective
coupling throughout the disk.}
\label{conclf1}
\end{figure}

A more fruitful line of research may be to simply assume that a weakly-ionized flow is
stable.  Accretion in that case could proceed in surface layers that are ionized by
non-thermal radiative processes, with the mid-plane of the disk remaining inactive
(see Figure~\ref{conclf1} and \citealt{gam96}).  One of the few numerical studies of 
layered accretion has shown that there may be some wave transport from the active, 
MHD-turbulent zone to the inactive zone \citep{fs03}.  This work could be expanded upon 
with a more realistic treatment of the vertical structure and vertical boundary conditions, 
as well as a closer inspection of the stresses in the inactive zone, including an 
investigation of their dependence on simulation parameters.  Understanding these processes
would be useful for developing more sophisticated models of layered disks.

In addition, since the dynamics of the dust layer in protoplanetary disks (disks which are
likely to be weakly-ionized) have important implications for understanding planet formation,
a useful and natural extension to the research proposed here is to incorporate gas-dust
interactions in some of my calculations.  The effects of turbulence and wave transport on
the dust layer could be investigated by including dust particles of various sizes in numerical
simulations similar to those described in Section~\ref{iwt}. In addition, interactions could be
calculated self-consistently by incorporating dust dynamics into layered disk models using
a two-fluid approach.

\section{Advanced Physical Disk Models}\label{saos4}

The majority of analytic \citep{ss73,lbp74,prin81,har98} and numerical \citep{ia99,spb99,ia00,
ian00,mg02} accretion-disk studies incorporate the assumption of an alpha viscosity to model
disk turbulence.  As discussed in \S\ref{dmts}, such an assumption has its shortcomings.
A fundamental understanding of MHD turbulent stresses can be used to develop advanced disk
models that will enhance our ability to explain observations of accretion systems.

Numerical models that rely on an alpha viscosity could be improved upon by incorporating a more
sophisticated model for the turbulent stresses based upon simulations and theoretical studies
of turbulence, such as those described in Section~\ref{dmts}.  In addition to being useful
for modeling accretion systems, these models would provide a more reliable bridge between
local and global MHD simulations and possibly be a means for understanding the complex
results of global MHD  simulations at a more fundamental level.

Analytic models that incorporate more sophisticated turbulent-stress modeling could also be
developed in an effort to improve upon the standard disk model.  In addition, there may
be key analytic results that need to be revisited with an enhanced understanding of turbulent
stresses, such as the work on wave propagation in warped disks discussed in \S\ref{iwt}
\citep{pp83,pl95}.  All of these proposals represent potential progress towards a
first-principles understanding and modeling of accretion-disk systems.

\end{spacing}

\appendix

\chapter{The Boussinesq Approximation}\label{appendix}

We demonstrate here that the Boussinesq approximation to the linear
perturbation equations is formally equivalent to a short-wavelength,
low-frequency limit of the full set of linear equations. We perform the
demonstration for the stratified shearing-sheet model since the standard
shearing sheet is recovered in the limit of zero stratification.

Combining equations (\ref{LIN1a}) through (\ref{LIN5a}) into a single
equation for $\delta v_x$ yields the following differential equation,
fourth-order in time:
\be\label{DV4DT}
F_4 \delta v_x^{(4)} + F_3 \delta v_x^{(3)} + F_2\delta v_x^{(2)}  + F_1
\delta v_x^{(1)}  + F_0 \delta v_x = 0,
\ee
where
\be\label{COEF4}
F_4 = \tilde{k}_x^2 \left[(k_y^2 + k_z^2) \left(1-\frac{i}{\tilde{k}_x L_P}
\right)^2 + k_y^2\frac{2(\qe+1)(\qe+2)}{\tilde{k}_x^2 H^2}\right],
\ee
\be
F_3 = -2 \qe \Omega \tilde{k}_x k_y (k_y^2 + k_z^2) \left(1+\frac{i}
{\tilde{k}_x L_P}\right),
\ee
\begin{eqnarray}
F_2 = c_s^2 \tilde{k}_x^2 \left[\tilde{k}_x^2 \left(1 + \frac{1}{\tilde{k}
_x^2 L_P^2} + \frac{N_x^2 + \tilde{\kappa}^2}{\tilde{k}_x^2 H^2}\right) 
\left\{ (k_y^2 + k_z^2)\left(1-\frac{i}{\tilde{k}_x L_P}\right)^2 +  k_y^2
\frac{2(\qe + 1)(\qe + 2)}{\tilde{k}_x^2 H^2} \right\} \, + \right. \nonumber
\\ \left. (k_y^2 + k_z^2) \left\{(k_y^2 + k_z^2)\left(1-\frac{i}{\tilde{k}_x L_P}
\right)^2 + k_y^2 \frac{2(\qe + 1)(3\qe + 2)}{\tilde{k}_x^2 H^2}\right\} \right],
\end{eqnarray}
\begin{eqnarray}
F_1 = 4\qe \Omega c_s^2 k_y \tilde{k}_x^3\left[(k_y^2 + k_z^2)\left(1+\frac{i}{\tilde{k}_x L_P}\right) \left\{1 + \frac{i(3\qe - 2)}
{2\qe \tilde{k}_x L_P} + \frac{\tilde{\kappa}^2}{4\qe \tilde{k}_x^2 L_P^2} 
- \frac{N_x^2 + \tilde{\kappa}^2}{2\tilde{k}_x^2 H^2}\right\} \, + \right.
\nonumber \\ \left. 3k_y^2\frac{(\qe + 1)(\qe + 2)}{\tilde{k}_x^2 H^2}
\left(1 - \frac{2i}{3\qe \tilde{k}_x L_P}\right) \right],
\end{eqnarray}
\be\label{COEF0}
F_0 = c_s^2 \tilde{k}_x^2 \left[k_y^2 (N_x^2 + 2\qe^2 \Omega^2) +
k_z^2(N_x^2 + \tilde{\kappa}^2)\right] \left[(k_y^2 + k_z^2)\left(1-\frac{i}
{\tilde{k}_x L_P}\right)^2 + k_y^2 \frac{2(\qe+1)(3\qe+2)}
{\tilde{k}_x^2 H^2}\right].
\ee

The above expressions have been written to make the short-wavelength
limit more apparent: all but the leading-order terms in brackets are
proportional to factors of $(\tilde{k}_xL_P)^{-1}$ or
$(\tilde{k}_xH)^{-1}$. Notice also that since one expects $H/L_P \ll 1$ for
Keplerian disks with modest radial gradients,
\be
\frac{1}{\tilde{k}_x L_P} = \frac{1}{k_y H\tilde{\tau}}\frac{H}{L_P} 
\ll \frac{1}{k_y H\tilde{\tau}},
\ee
(for $\tilde{\tau} \neq 0$) and therefore the short-wavelength limit is sufficient.
One needs to be careful in taking this limit, however, since $\tilde{k}_x = k_y
\tilde{\tau}$ goes through zero as a shwave goes from leading to trailing. The
approximation is rigorously valid only for $\tilde{\tau} \neq 0$, but we have 
numerically integrated the full set of linear equations (equations (\ref{LIN1a})
through (\ref{LIN5a}) with $k_z = 0$) and found good agreement with the 
Boussinesq solutions described in \S 4.3 for all $\tilde{\tau}$ at sufficiently
short wavelengths.\footnote{One must start with a set of initial conditions 
consistent with equations (\ref{LIN2}), (\ref{LIN3}), (\ref{LIN1}) and
(\ref{LIN5}) in order to accurately track the incompressive-shwave solutions. 
In addition, suppression of the high-frequency compressive-shwave solutions
near $\tilde{\tau} = 0$ requires $k_y L_P \gtrsim 200$, which for $H/L_P = 
0.1$ implies $H k_y \gtrsim 20$.}

With these assumptions in mind, to leading order in $(H k_y)^{-1}$ equation
(\ref{DV4DT}) becomes
\begin{eqnarray}\label{EQVX}
\tilde{k}_x \delta v_x^{(4)} - 2 \qe \Omega k_y \delta v_x^{(3)} +
c_s^2 \tilde{k}_x \tilde{k}^2 \delta v_x^{(2)}  + 4\qe \Omega c_s^2
\tilde{k}_x^2 k_y \delta v_x^{(1)} \, + \nonumber \\ c_s^2 \tilde{k}_x
\left[k_y^2(N_x^2 + 2\qe^2 \Omega^2) + k_z^2(N_x^2 + \tilde{\kappa}^2)
\right]\delta v_x = 0.
\end{eqnarray}
If we assume $\partial_t \ll c_s k_y$, the two highest-order time derivatives
are of lower order and can be neglected (thereby eliminating the compressive
shwaves) and we have
\be\label{DV2DT}
\tilde{k}^2 \ddot{\delta v_x} + 4 \qe \Omega \tilde{k}_x k_y\dot{\delta v_x}
\left[k_y^2(N_x^2 + 2\qe^2 \Omega^2) + k_z^2(N_x^2 +\tilde{\kappa}^2)\right]
\delta v_x = 0.
\ee
This is equivalent to equation (\ref{BOUSSVX}).

Notice also that the assumption $\partial_t \sim O(c_s k_y)$ applied to
equation (\ref{EQVX}) yields
\be
\delta v_x^{(4)} + c_s^2 \tilde{k}^2 \delta v_x^{(2)} = 0
\ee
to leading order in $(H k_y)^{-1}$. This equation is of the same form
as the short-wavelength limit of equation (\ref{SOUND}) for the
compressive shwaves in the unstratified shearing sheet, confirming our
claim that short-wavelength compressive shwaves are unchanged at
leading order by stratification.

\backmatter

\bibliography{thesis}

\begin{thebibliography}{188}
\expandafter\ifx\csname natexlab\endcsname\relax\def\natexlab#1{#1}\fi
\expandafter\ifx\csname url\endcsname\relax
  \def\url#1{{\tt #1}}\fi

\bibitem[{Abramowitz} and {Stegun}(1972)]{as72}
M.~{Abramowitz} and I.~A. {Stegun}.
\newblock {\em {Handbook of Mathematical Functions}}.
\newblock Handbook of Mathematical Functions, New York: Dover, 1972, 1972.

\bibitem[{Adams} and {Lin}(1993)]{al93}
F.~C. {Adams} and D.~N.~C. {Lin}.
\newblock {Transport processes and the evolution of disks}.
\newblock In {\em Protostars and Planets III}, pages 721--748, 1993.

\bibitem[{Adams} et~al.(1988){Adams}, {Shu}, and {Lada}]{asl88}
F.~C. {Adams}, F.~H. {Shu}, and C.~J. {Lada}.
\newblock {The disks of T Tauri stars with flat infrared spectra}.
\newblock {\em \apj}, 326:\penalty0 865--883, March 1988.

\bibitem[{Afshordi} et~al.(2004){Afshordi}, {Mukhopadhyay}, and
  {Narayan}]{amn04}
N.~{Afshordi}, B.~{Mukhopadhyay}, and R.~{Narayan}.
\newblock {Bypass to Turbulence in Hydrodynamic Accretion: Lagrangian Analysis
  of Energy Growth}.
\newblock {\em ArXiv Astrophysics e-prints}, December 2004.

\bibitem[{Arlt} and {R{\" u}diger}(2001)]{ar01}
R.~{Arlt} and G.~{R{\" u}diger}.
\newblock {Global accretion disk simulations of magneto-rotational
  instability}.
\newblock {\em \aap}, 374:\penalty0 1035--1048, August 2001.

\bibitem[{Armitage}(1998)]{arm98}
P.~J. {Armitage}.
\newblock {Turbulence and Angular Momentum Transport in Global Accretion Disk
  Simulation}.
\newblock {\em \apjl}, 501:\penalty0 L189, July 1998.

\bibitem[{Baggett} and {Trefethen}(1997)]{bt97}
J.~S. {Baggett} and L.~N. {Trefethen}.
\newblock {Low-dimensional models of subcritical transition to turbulence}.
\newblock {\em Physics of Fluids}, 9:\penalty0 1043--1053, April 1997.

\bibitem[{Balbus}(1988)]{bal88}
S.~A. {Balbus}.
\newblock {Local interstellar gasdynamical stability and substructure in spiral
  arms}.
\newblock {\em \apj}, 324:\penalty0 60--74, January 1988.

\bibitem[{Balbus}(2000)]{bal00}
S.~A. {Balbus}.
\newblock {Stability, Instability, and ``Backward'' Transport in Stratified
  Fluids}.
\newblock {\em \apj}, 534:\penalty0 420--427, May 2000.

\bibitem[{Balbus}(2003)]{bal03}
S.~A. {Balbus}.
\newblock {Enhanced Angular Momentum Transport in Accretion Disks}.
\newblock {\em \araa}, 41:\penalty0 555--597, 2003.

\bibitem[{Balbus}(2004)]{bal04}
S.~A. {Balbus}.
\newblock {Nonlinear Scale Invariance in Local Disk Flows}.
\newblock {\em ArXiv Astrophysics e-prints}, August 2004.

\bibitem[{Balbus} and {Hawley}(1991)]{bh91}
S.~A. {Balbus} and J.~F. {Hawley}.
\newblock {A powerful local shear instability in weakly magnetized disks. I -
  Linear analysis. II - Nonlinear evolution}.
\newblock {\em \apj}, 376:\penalty0 214--233, July 1991.

\bibitem[{Balbus} and {Hawley}(1992)]{bh92}
S.~A. {Balbus} and J.~F. {Hawley}.
\newblock {Is the Oort A-value a universal growth rate limit for accretion disk
  shear instabilities?}
\newblock {\em \apj}, 392:\penalty0 662--666, June 1992.

\bibitem[{Balbus} and {Hawley}(1998)]{bh98}
S.~A. {Balbus} and J.~F. {Hawley}.
\newblock {Instability, turbulence, and enhanced transport in accretion disks}.
\newblock {\em Reviews of Modern Physics}, 70:\penalty0 1--53, January 1998.

\bibitem[{Balbus} et~al.(1996){Balbus}, {Hawley}, and {Stone}]{bhs96}
S.~A. {Balbus}, J.~F. {Hawley}, and J.~M. {Stone}.
\newblock {Nonlinear Stability, Hydrodynamical Turbulence, and Transport in
  Disks}.
\newblock {\em \apj}, 467:\penalty0 76, August 1996.

\bibitem[{Balbus} and {Terquem}(2001)]{bt01}
S.~A. {Balbus} and C.~{Terquem}.
\newblock {Linear Analysis of the Hall Effect in Protostellar Disks}.
\newblock {\em \apj}, 552:\penalty0 235--247, May 2001.

\bibitem[{Baptista}(2004)]{bap04}
R.~{Baptista}.
\newblock {What can we learn from accretion disc eclipse mapping experiments?}
\newblock {\em Astronomische Nachrichten}, 325:\penalty0 181--184, 2004.

\bibitem[{Baptista} et~al.(1998){Baptista}, {Horne}, {Wade}, {Hubeny}, {Long},
  and {Rutten}]{bap98}
R.~{Baptista}, K.~{Horne}, R.~A. {Wade}, I.~{Hubeny}, K.~S. {Long}, and
  R.~G.~M. {Rutten}.
\newblock {HST spatially resolved spectra of the accretion disc and gas stream
  of the nova-like variable UX Ursae Majoris}.
\newblock {\em \mnras}, 298:\penalty0 1079--1091, August 1998.

\bibitem[{Barge} and {Sommeria}(1995)]{bs95}
P.~{Barge} and J.~{Sommeria}.
\newblock {Did planet formation begin inside persistent gaseous vortices?}
\newblock {\em \aap}, 295:\penalty0 L1--L4, March 1995.

\bibitem[{Barranco} and {Marcus}(2005)]{bm05}
J.~A. {Barranco} and P.~S. {Marcus}.
\newblock {Three-Dimensional Vortices in Stratified Protoplanetary Disks}.
\newblock {\em ArXiv Astrophysics e-prints}, January 2005.

\bibitem[{Bayly} et~al.(1988){Bayly}, {Orszag}, and {Herbert}]{bo88}
B.~J. {Bayly}, S.~A. {Orszag}, and T.~{Herbert}.
\newblock {Instability mechanisms in shear-flow transition}.
\newblock {\em Annual Review of Fluid Mechanics}, 20:\penalty0 359--391, 1988.

\bibitem[{Bell} and {Lin}(1994)]{bl94}
K.~R. {Bell} and D.~N.~C. {Lin}.
\newblock {Using FU Orionis outbursts to constrain self-regulated protostellar
  disk models}.
\newblock {\em \apj}, 427:\penalty0 987--1004, June 1994.

\bibitem[{Binney} and {Tremaine}(1987)]{bt87}
J.~{Binney} and S.~{Tremaine}.
\newblock {\em {Galactic dynamics}}.
\newblock Princeton, NJ, Princeton University Press, 1987, 747 p., 1987.

\bibitem[{Blaes}(1987)]{bla87}
O.~M. {Blaes}.
\newblock {Stabilization of non-axisymmetric instabilities in a rotating flow
  by accretion on to a central black hole}.
\newblock {\em \mnras}, 227:\penalty0 975--992, August 1987.

\bibitem[{Blaes} and {Balbus}(1994)]{bb94}
O.~M. {Blaes} and S.~A. {Balbus}.
\newblock {Local shear instabilities in weakly ionized, weakly magnetized
  disks}.
\newblock {\em \apj}, 421:\penalty0 163--177, January 1994.

\bibitem[{Blaes} and {Hawley}(1988)]{bh88}
O.~M. {Blaes} and J.~F. {Hawley}.
\newblock {Nonaxisymmetric disk instabilities - A linear and nonlinear
  synthesis}.
\newblock {\em \apj}, 326:\penalty0 277--291, March 1988.

\bibitem[{Blandford} and {Payne}(1982)]{bp82}
R.~D. {Blandford} and D.~G. {Payne}.
\newblock {Hydromagnetic flows from accretion discs and the production of radio
  jets}.
\newblock {\em \mnras}, 199:\penalty0 883--903, June 1982.

\bibitem[{Boss}(1997)]{boss97}
A.~P. {Boss}.
\newblock {Giant planet formation by gravitational instability.}
\newblock {\em Science}, 276:\penalty0 1836--1839, 1997.

\bibitem[{Boss}(1998)]{boss98}
A.~P. {Boss}.
\newblock {Evolution of the Solar Nebula. IV. Giant Gaseous Protoplanet
  Formation}.
\newblock {\em \apj}, 503:\penalty0 923, August 1998.

\bibitem[{Boss}(2002)]{boss02}
A.~P. {Boss}.
\newblock {Evolution of the Solar Nebula. V. Disk Instabilities with Varied
  Thermodynamics}.
\newblock {\em \apj}, 576:\penalty0 462--472, September 2002.

\bibitem[{Burrows} et~al.(1996){Burrows}, {Stapelfeldt}, {Watson}, {Krist},
  {Ballester}, {Clarke}, {Crisp}, {Gallagher}, {Griffiths}, {Hester},
  {Hoessel}, {Holtzman}, {Mould}, {Scowen}, {Trauger}, and {Westphal}]{burr96}
C.~J. {Burrows}, K.~R. {Stapelfeldt}, A.~M. {Watson}, J.~E. {Krist}, G.~E.
  {Ballester}, J.~T. {Clarke}, D.~{Crisp}, J.~S. {Gallagher}, R.~E.
  {Griffiths}, J.~J. {Hester}, J.~G. {Hoessel}, J.~A. {Holtzman}, J.~R.
  {Mould}, P.~A. {Scowen}, J.~T. {Trauger}, and J.~A. {Westphal}.
\newblock {Hubble Space Telescope Observations of the Disk and Jet of HH 30}.
\newblock {\em \apj}, 473:\penalty0 437--+, December 1996.

\bibitem[{Cabot}(1984)]{cab84}
W.~{Cabot}.
\newblock {The nonaxisymmetric baroclinic instability in thin accretion disks}.
\newblock {\em \apj}, 277:\penalty0 806--812, February 1984.

\bibitem[{Cabot}(1996)]{cab96}
W.~{Cabot}.
\newblock {Numerical Simulations of Circumstellar Disk Convection}.
\newblock {\em \apj}, 465:\penalty0 874, July 1996.

\bibitem[{Cameron}(1978)]{cam78}
A.~G.~W. {Cameron}.
\newblock {Physics of the primitive solar accretion disk}.
\newblock {\em Moon and Planets}, 18:\penalty0 5--40, February 1978.

\bibitem[{Cassen}(1993)]{cass93}
P.~{Cassen}.
\newblock {Why convective heat transport in the solar nebula was inefficient}.
\newblock In {\em Lunar and Planetary Institute Conference Abstracts}, pages
  261--262, 1993.

\bibitem[{Chagelishvili} et~al.(1997){Chagelishvili}, {Tevzadze}, {Bodo}, and
  {Moiseev}]{chag97}
G.~D. {Chagelishvili}, A.~G. {Tevzadze}, G.~{Bodo}, and S.~S. {Moiseev}.
\newblock {Linear Mechanism of Wave Emergence from Vortices in Smooth Shear
  Flows}.
\newblock {\em Physical Review Letters}, 79:\penalty0 3178--3181, October 1997.

\bibitem[{Chagelishvili} et~al.(2003){Chagelishvili}, {Zahn}, {Tevzadze}, and
  {Lominadze}]{cztl03}
G.~D. {Chagelishvili}, J.-P. {Zahn}, A.~G. {Tevzadze}, and J.~G. {Lominadze}.
\newblock {On hydrodynamic shear turbulence in Keplerian disks: Via transient
  growth to bypass transition}.
\newblock {\em \aap}, 402:\penalty0 401--407, May 2003.

\bibitem[{Chimonas}(1970)]{chi70}
G.~{Chimonas}.
\newblock {The extension of the Miles-Howard theorem to compressible fluids}.
\newblock {\em J. Fluid Mech.}, 43:\penalty0 833--836, 1970.

\bibitem[{Desch}(2004)]{des04}
S.~J. {Desch}.
\newblock {Linear Analysis of the Magnetorotational Instability, Including
  Ambipolar Diffusion, with Application to Protoplanetary Disks}.
\newblock {\em \apj}, 608:\penalty0 509--525, June 2004.

\bibitem[{Drazin} and {Reid}(1981)]{dr81}
P.~G. {Drazin} and W.~H. {Reid}.
\newblock {Hydrodynamic stability}.
\newblock {\em NASA STI/Recon Technical Report A}, 82:\penalty0 17950, 1981.

\bibitem[{Dubrulle}(1993)]{dub93}
B.~{Dubrulle}.
\newblock {Differential rotation as a source of angular momentum transfer in
  the solar nebula}.
\newblock {\em Icarus}, 106:\penalty0 59, November 1993.

\bibitem[{Dubrulle} and {Knobloch}(1992)]{dk92}
B.~{Dubrulle} and E.~{Knobloch}.
\newblock {On the local stability of accretion disks}.
\newblock {\em \aap}, 256:\penalty0 673--678, March 1992.

\bibitem[{Dubrulle} and {Knobloch}(1993)]{dk93}
B.~{Dubrulle} and E.~{Knobloch}.
\newblock {On instabilities in magnetized accretion disks}.
\newblock {\em \aap}, 274:\penalty0 667, July 1993.

\bibitem[{Durisen} et~al.(2001){Durisen}, {Mejia}, {Pickett}, and
  {Hartquist}]{dmph01}
R.~H. {Durisen}, A.~C. {Mejia}, B.~K. {Pickett}, and T.~W. {Hartquist}.
\newblock {Gravitational Instabilities in the Disks of Massive Protostars as an
  Explanation for Linear Distributions of Methanol Masers}.
\newblock {\em \apjl}, 563:\penalty0 L157--L160, December 2001.

\bibitem[{Eliassen} et~al.(1953){Eliassen}, {H{\o}iland}, and {Riis}]{ehr53}
A.~{Eliassen}, E.~{H{\o}iland}, and E.~{Riis}.
\newblock {Two-Dimensional Perturbation of a Flow with Constant Shear of a
  Stratified Fluid}.
\newblock {\em Institute for Weather and Climate Research, Nowegian Academy of
  Sciences and Letters, Publ. no. 1}, 1, 1953.

\bibitem[{Farmer} and {Goldreich}(2004)]{fg04}
A.~J. {Farmer} and P.~{Goldreich}.
\newblock {Wave Damping by Magnetohydrodynamic Turbulence and Its Effect on
  Cosmic-Ray Propagation in the Interstellar Medium}.
\newblock {\em \apj}, 604:\penalty0 671--674, April 2004.

\bibitem[{Farrell} and {Ioannou}(1993)]{fi93}
B.~F. {Farrell} and P.~J. {Ioannou}.
\newblock {Optimal excitation of three-dimensional perturbations in viscous
  constant shear flow}.
\newblock {\em Physics of Fluids}, 5:\penalty0 1390--1400, June 1993.

\bibitem[{Fleming} and {Stone}(2003)]{fs03}
T.~{Fleming} and J.~M. {Stone}.
\newblock {Local Magnetohydrodynamic Models of Layered Accretion Disks}.
\newblock {\em \apj}, 585:\penalty0 908--920, March 2003.

\bibitem[{Fleming} et~al.(2000){Fleming}, {Stone}, and {Hawley}]{fsh00}
T.~P. {Fleming}, J.~M. {Stone}, and J.~F. {Hawley}.
\newblock {The Effect of Resistivity on the Nonlinear Stage of the
  Magnetorotational Instability in Accretion Disks}.
\newblock {\em \apj}, 530:\penalty0 464--477, February 2000.

\bibitem[{Frank} et~al.(1981){Frank}, {King}, {Sherrington}, {Jameson}, and
  {Axon}]{fea81}
J.~{Frank}, A.~R. {King}, M.~R. {Sherrington}, R.~F. {Jameson}, and D.~J.
  {Axon}.
\newblock {Infrared and optical light curves of UX Ursae Majoris and U
  Geminorum}.
\newblock {\em \mnras}, 195:\penalty0 505--516, May 1981.

\bibitem[{Fromang} et~al.(2002){Fromang}, {Terquem}, and {Balbus}]{ftb02}
S.~{Fromang}, C.~{Terquem}, and S.~A. {Balbus}.
\newblock {The ionization fraction in {$\alpha$} models of protoplanetary
  discs}.
\newblock {\em \mnras}, 329:\penalty0 18--28, January 2002.

\bibitem[{Gammie}(1996)]{gam96}
C.~F. {Gammie}.
\newblock {Layered Accretion in T Tauri Disks}.
\newblock {\em \apj}, 457:\penalty0 355, January 1996.

\bibitem[{Gammie}(1999)]{gam99}
C.~F. {Gammie}.
\newblock {Instabilities in Circumstellar Discs}.
\newblock In {\em ASP Conf. Ser. 160: Astrophysical Discs - an EC Summer
  School}, page 122, April 1999.

\bibitem[{Gammie}(2001)]{gam01}
C.~F. {Gammie}.
\newblock {Nonlinear Outcome of Gravitational Instability in Cooling, Gaseous
  Disks}.
\newblock {\em \apj}, 553:\penalty0 174--183, May 2001.

\bibitem[{Gammie} and {Menou}(1998)]{gm98}
C.~F. {Gammie} and K.~{Menou}.
\newblock {On the Origin of Episodic Accretion in Dwarf Novae}.
\newblock {\em \apjl}, 492:\penalty0 L75, January 1998.

\bibitem[{Gammie} et~al.(1999){Gammie}, {Narayan}, and {Blandford}]{gnb99}
C.~F. {Gammie}, R.~{Narayan}, and R.~{Blandford}.
\newblock {What Is the Accretion Rate in NGC 4258?}
\newblock {\em \apj}, 516:\penalty0 177--186, May 1999.

\bibitem[{Garaud} and {Lin}(2004)]{gl04}
P.~{Garaud} and D.~N.~C. {Lin}.
\newblock {On the Evolution and Stability of a Protoplanetary Disk Dust Layer}.
\newblock {\em \apj}, 608:\penalty0 1050--1075, June 2004.

\bibitem[{Garaud} and {Ogilvie}(2005)]{go05}
P.~{Garaud} and G.~I. {Ogilvie}.
\newblock {A model for the nonlinear dynamics of turbulent shear flows}.
\newblock {\em ArXiv Astrophysics e-prints}, March 2005.

\bibitem[{Godon} and {Livio}(1999)]{gl99}
P.~{Godon} and M.~{Livio}.
\newblock {Vortices in Protoplanetary Disks}.
\newblock {\em \apj}, 523:\penalty0 350--356, September 1999.

\bibitem[{Godon} and {Livio}(2000)]{gl00}
P.~{Godon} and M.~{Livio}.
\newblock {The Formation and Role of Vortices in Protoplanetary Disks}.
\newblock {\em \apj}, 537:\penalty0 396--404, July 2000.

\bibitem[{Goldreich} et~al.(1986){Goldreich}, {Goodman}, and {Narayan}]{ggn86}
P.~{Goldreich}, J.~{Goodman}, and R.~{Narayan}.
\newblock {The stability of accretion tori. I - Long-wavelength modes of
  slender tori}.
\newblock {\em \mnras}, 221:\penalty0 339--364, July 1986.

\bibitem[{Goldreich} and {Kumar}(1988)]{gk88}
P.~{Goldreich} and P.~{Kumar}.
\newblock {The interaction of acoustic radiation with turbulence}.
\newblock {\em \apj}, 326:\penalty0 462--478, March 1988.

\bibitem[{Goldreich} and {Kumar}(1990)]{gk90}
P.~{Goldreich} and P.~{Kumar}.
\newblock {Wave generation by turbulent convection}.
\newblock {\em \apj}, 363:\penalty0 694--704, November 1990.

\bibitem[{Goldreich} and {Lynden-Bell}(1965)]{glb65}
P.~{Goldreich} and D.~{Lynden-Bell}.
\newblock {II. Spiral arms as sheared gravitational instabilities}.
\newblock {\em \mnras}, 130:\penalty0 125, 1965.

\bibitem[{Goldreich} and {Tremaine}(1978)]{gt78}
P.~{Goldreich} and S.~{Tremaine}.
\newblock {The excitation and evolution of density waves}.
\newblock {\em \apj}, 222:\penalty0 850--858, June 1978.

\bibitem[{Goodman}(2003)]{good03}
J.~{Goodman}.
\newblock {Self-gravity and quasi-stellar object discs}.
\newblock {\em \mnras}, 339:\penalty0 937--948, March 2003.

\bibitem[{Goodman} et~al.(1987){Goodman}, {Narayan}, and {Goldreich}]{gng87}
J.~{Goodman}, R.~{Narayan}, and P.~{Goldreich}.
\newblock {The stability of accretion tori. II - Non-linear evolution to
  discrete planets}.
\newblock {\em \mnras}, 225:\penalty0 695--711, April 1987.

\bibitem[{Goodman} and {Rafikov}(2001)]{gr01}
J.~{Goodman} and R.~R. {Rafikov}.
\newblock {Planetary Torques as the Viscosity of Protoplanetary Disks}.
\newblock {\em \apj}, 552:\penalty0 793--802, May 2001.

\bibitem[{Gough}(1969)]{gou69}
D.~O. {Gough}.
\newblock {The Anelastic Approximation for Thermal Convection}.
\newblock {\em Journal of Atmospheric Sciences}, 26:\penalty0 448--456, 1969.

\bibitem[{Grossmann}(2000)]{gross00}
S.~{Grossmann}.
\newblock {The onset of shear flow turbulence}.
\newblock {\em Reviews of Modern Physics}, 72:\penalty0 603--618, April 2000.

\bibitem[{Gullbring} et~al.(1998){Gullbring}, {Hartmann}, {Briceno}, and
  {Calvet}]{ghbc98}
E.~{Gullbring}, L.~{Hartmann}, C.~{Briceno}, and N.~{Calvet}.
\newblock {Disk Accretion Rates for T Tauri Stars}.
\newblock {\em \apj}, 492:\penalty0 323, January 1998.

\bibitem[{Hartmann}(1998)]{har98}
L.~{Hartmann}.
\newblock {\em {Accretion processes in star formation}}.
\newblock Accretion processes in star formation / Lee Hartmann.~Cambridge, UK ;
  New York : Cambridge University Press, 1998.~(Cambridge astrophysics series ;
  32) ISBN 0521435072., 1998.

\bibitem[{Hawley}(1987)]{haw87}
J.~F. {Hawley}.
\newblock {Non-linear evolution of a non-axisymmetric disc instability}.
\newblock {\em \mnras}, 225:\penalty0 677--694, April 1987.

\bibitem[{Hawley}(1990)]{haw90}
J.~F. {Hawley}.
\newblock {Nonaxisymmetric instabilities in a slender torus - Two- and
  three-dimensional simulations}.
\newblock {\em \apj}, 356:\penalty0 580--590, June 1990.

\bibitem[{Hawley}(1991)]{haw91}
J.~F. {Hawley}.
\newblock {Three-dimensional simulations of black hole tori}.
\newblock {\em \apj}, 381:\penalty0 496--507, November 1991.

\bibitem[{Hawley}(2000)]{haw00}
J.~F. {Hawley}.
\newblock {Global Magnetohydrodynamical Simulations of Accretion Tori}.
\newblock {\em \apj}, 528:\penalty0 462--479, January 2000.

\bibitem[{Hawley} et~al.(1999){Hawley}, {Balbus}, and {Winters}]{hbw99}
J.~F. {Hawley}, S.~A. {Balbus}, and W.~F. {Winters}.
\newblock {Local Hydrodynamic Stability of Accretion Disks}.
\newblock {\em \apj}, 518:\penalty0 394--404, June 1999.

\bibitem[{Hawley} et~al.(1995){Hawley}, {Gammie}, and {Balbus}]{hgb95}
J.~F. {Hawley}, C.~F. {Gammie}, and S.~A. {Balbus}.
\newblock {Local Three-dimensional Magnetohydrodynamic Simulations of Accretion
  Disks}.
\newblock {\em \apj}, 440:\penalty0 742, February 1995.

\bibitem[{Hawley} et~al.(1996){Hawley}, {Gammie}, and {Balbus}]{hgb96}
J.~F. {Hawley}, C.~F. {Gammie}, and S.~A. {Balbus}.
\newblock {Local Three-dimensional Simulations of an Accretion Disk
  Hydromagnetic Dynamo}.
\newblock {\em \apj}, 464:\penalty0 690, June 1996.

\bibitem[{Hersant} et~al.(2005){Hersant}, {Dubrulle}, and {Hur{\' e}}]{hdh05}
F.~{Hersant}, B.~{Dubrulle}, and J.-M. {Hur{\' e}}.
\newblock {Turbulence in circumstellar disks}.
\newblock {\em \aap}, 429:\penalty0 531--542, January 2005.

\bibitem[{Hessman}(1991)]{hess91}
F.~V. {Hessman}.
\newblock {The long-term evolution of FU Orionis variables}.
\newblock {\em \aap}, 246:\penalty0 137--145, June 1991.

\bibitem[{Houghton}(2002)]{hou02}
J.~T. {Houghton}.
\newblock {\em {The physics of atmospheres}}.
\newblock The physics of atmospheres, 3rd ed.~by John Houghton.~Cambridge, UK:
  Cambridge University Press, 2002 xv, 320 p.~ISBN \#0521011221, 2002.

\bibitem[{Howard}(1961)]{how61}
L.~N. {Howard}.
\newblock {Note on a paper of John W. Miles}.
\newblock {\em J. Fluid Mech.}, 10:\penalty0 509--512, 1961.

\bibitem[{Hubeny}(1990)]{hub90}
I.~{Hubeny}.
\newblock {Vertical structure of accretion disks - A simplified analytical
  model}.
\newblock {\em \apj}, 351:\penalty0 632--641, March 1990.

\bibitem[{Igumenshchev} and {Abramowicz}(1999)]{ia99}
I.~V. {Igumenshchev} and M.~A. {Abramowicz}.
\newblock {Rotating accretion flows around black holes: convection and
  variability}.
\newblock {\em \mnras}, 303:\penalty0 309--320, February 1999.

\bibitem[{Igumenshchev} and {Abramowicz}(2000)]{ia00}
I.~V. {Igumenshchev} and M.~A. {Abramowicz}.
\newblock {Two-dimensional Models of Hydrodynamical Accretion Flows into Black
  Holes}.
\newblock {\em \apjs}, 130:\penalty0 463--484, October 2000.

\bibitem[{Igumenshchev} et~al.(2000){Igumenshchev}, {Abramowicz}, and
  {Narayan}]{ian00}
I.~V. {Igumenshchev}, M.~A. {Abramowicz}, and R.~{Narayan}.
\newblock {Numerical Simulations of Convective Accretion Flows in Three
  Dimensions}.
\newblock {\em \apjl}, 537:\penalty0 L27--L30, July 2000.

\bibitem[{Ioannou} and {Kakouris}(2001)]{ik01}
P.~J. {Ioannou} and A.~{Kakouris}.
\newblock {Stochastic Dynamics of Keplerian Accretion Disks}.
\newblock {\em \apj}, 550:\penalty0 931--943, April 2001.

\bibitem[{Jang-Condell} and {Sasselov}(2003)]{js03}
H.~{Jang-Condell} and D.~D. {Sasselov}.
\newblock {Radiative Transfer on Perturbations in Protoplanetary Disks}.
\newblock {\em \apj}, 593:\penalty0 1116--1123, August 2003.

\bibitem[{Jeffreys}(1930)]{jeff30}
H.~{Jeffreys}.
\newblock {Note on a paper of John W. Miles}.
\newblock {\em Proc. Cambridge Phil. Soc.}, 26:\penalty0 170, 1930.

\bibitem[{Jin}(1996)]{jin96}
L.~{Jin}.
\newblock {Damping of the Shear Instability in Magnetized Disks by Ohmic
  Diffusion}.
\newblock {\em \apj}, 457:\penalty0 798, February 1996.

\bibitem[{Julian} and {Toomre}(1966)]{jt66}
W.~H. {Julian} and A.~{Toomre}.
\newblock {Non-Axisymmetric Responses of Differentially Rotating Disks of
  Stars}.
\newblock {\em \apj}, 146:\penalty0 810, December 1966.

\bibitem[{Kerswell}(2002)]{ker02}
R.~R. {Kerswell}.
\newblock {Elliptical instability}.
\newblock {\em Annual Review of Fluid Mechanics}, 34:\penalty0 83--113, 2002.

\bibitem[{Klahr}(2004)]{klr04}
H.~{Klahr}.
\newblock {The Global Baroclinic Instability in Accretion Disks. II. Local
  Linear Analysis}.
\newblock {\em \apj}, 606:\penalty0 1070--1082, May 2004.

\bibitem[{Klahr} and {Bodenheimer}(2003)]{kb03}
H.~H. {Klahr} and P.~{Bodenheimer}.
\newblock {Turbulence in Accretion Disks: Vorticity Generation and Angular
  Momentum Transport via the Global Baroclinic Instability}.
\newblock {\em \apj}, 582:\penalty0 869--892, January 2003.

\bibitem[{Knobloch} and {Spruit}(1986)]{ks86}
E.~{Knobloch} and H.~C. {Spruit}.
\newblock {Baroclinic waves in a vertically stratified thin accretion disk}.
\newblock {\em \aap}, 166:\penalty0 359--367, September 1986.

\bibitem[{K{\"o}nigl} and {Pudritz}(2000)]{kp00}
A.~{K{\"o}nigl} and R.~E. {Pudritz}.
\newblock {Disk Winds and the Accretion-Outflow Connection}.
\newblock {\em Protostars and Planets IV}, page 759, May 2000.

\bibitem[{Kunz} and {Balbus}(2004)]{kb04}
M.~W. {Kunz} and S.~A. {Balbus}.
\newblock {Ambipolar diffusion in the magnetorotational instability}.
\newblock {\em \mnras}, 348:\penalty0 355--360, February 2004.

\bibitem[{Larson}(1984)]{lar84}
R.~B. {Larson}.
\newblock {Gravitational torques and star formation}.
\newblock {\em \mnras}, 206:\penalty0 197--207, January 1984.

\bibitem[{Larson}(1989)]{lars89}
R.~B. {Larson}.
\newblock {The evolution of protostellar disks}.
\newblock In {\em The Formation and Evolution of Planetary Systems}, pages
  31--48, 1989.

\bibitem[{Lasota} et~al.(1996){Lasota}, {Abramowicz}, {Chen}, {Krolik},
  {Narayan}, and {Yi}]{las96}
J.-P. {Lasota}, M.~A. {Abramowicz}, X.~{Chen}, J.~{Krolik}, R.~{Narayan}, and
  I.~{Yi}.
\newblock {Is the Accretion Flow in NGC 4258 Advection Dominated?}
\newblock {\em \apj}, 462:\penalty0 142, May 1996.

\bibitem[{Laughlin} and {Rozyczka}(1996)]{lr96}
G.~{Laughlin} and M.~{Rozyczka}.
\newblock {The Effect of Gravitational Instabilities on Protostellar Disks}.
\newblock {\em \apj}, 456:\penalty0 279, January 1996.

\bibitem[{Li} et~al.(2001){Li}, {Colgate}, {Wendroff}, and {Liska}]{lcwl01}
H.~{Li}, S.~A. {Colgate}, B.~{Wendroff}, and R.~{Liska}.
\newblock {Rossby Wave Instability of Thin Accretion Disks. III. Nonlinear
  Simulations}.
\newblock {\em \apj}, 551:\penalty0 874--896, April 2001.

\bibitem[{Li} et~al.(2000){Li}, {Finn}, {Lovelace}, and {Colgate}]{lflc00}
H.~{Li}, J.~M. {Finn}, R.~V.~E. {Lovelace}, and S.~A. {Colgate}.
\newblock {Rossby Wave Instability of Thin Accretion Disks. II. Detailed Linear
  Theory}.
\newblock {\em \apj}, 533:\penalty0 1023--1034, April 2000.

\bibitem[{Lin} and {Papaloizou}(1980)]{lp80}
D.~N.~C. {Lin} and J.~{Papaloizou}.
\newblock {On the structure and evolution of the primordial solar nebula}.
\newblock {\em \mnras}, 191:\penalty0 37--48, April 1980.

\bibitem[{Livio} and {Spruit}(1991)]{ls91}
M.~{Livio} and H.~C. {Spruit}.
\newblock {On the mechanism of angular momentum transport in accretion disks}.
\newblock {\em \aap}, 252:\penalty0 189--192, December 1991.

\bibitem[{Lodato} and {Bertin}(2003)]{lb03}
G.~{Lodato} and G.~{Bertin}.
\newblock {Non-Keplerian rotation in the nucleus of NGC 1068: Evidence for a
  massive accretion disk?}
\newblock {\em \aap}, 398:\penalty0 517--524, February 2003.

\bibitem[{Longaretti}(2002)]{long02}
P.~{Longaretti}.
\newblock {On the Phenomenology of Hydrodynamic Shear Turbulence}.
\newblock {\em \apj}, 576:\penalty0 587--598, September 2002.

\bibitem[{Lovelace} et~al.(1999){Lovelace}, {Li}, {Colgate}, and
  {Nelson}]{llcn99}
R.~V.~E. {Lovelace}, H.~{Li}, S.~A. {Colgate}, and A.~F. {Nelson}.
\newblock {Rossby Wave Instability of Keplerian Accretion Disks}.
\newblock {\em \apj}, 513:\penalty0 805--810, March 1999.

\bibitem[{Lynden-Bell} and {Pringle}(1974)]{lbp74}
D.~{Lynden-Bell} and J.~E. {Pringle}.
\newblock {The evolution of viscous discs and the origin of the nebular
  variables.}
\newblock {\em \mnras}, 168:\penalty0 603--637, September 1974.

\bibitem[{Machida} et~al.(2000){Machida}, {Hayashi}, and {Matsumoto}]{mhm00}
M.~{Machida}, M.~R. {Hayashi}, and R.~{Matsumoto}.
\newblock {Global Simulations of Differentially Rotating Magnetized Disks:
  Formation of Low-{$\beta$} Filaments and Structured Coronae}.
\newblock {\em \apjl}, 532:\penalty0 L67--L70, March 2000.

\bibitem[{Marcus} and {Press}(1977)]{mp77}
P.~S. {Marcus} and W.~H. {Press}.
\newblock {On Green's functions for small disturbances of plane Couette flow}.
\newblock {\em Journal of Fluid Mechanics}, 79:\penalty0 525--534, 1977.

\bibitem[{Masset}(2000)]{mass00}
F.~{Masset}.
\newblock {FARGO: A fast eulerian transport algorithm for differentially
  rotating disks}.
\newblock {\em \aaps}, 141:\penalty0 165--173, January 2000.

\bibitem[{Matsumoto} and {Tajima}(1995)]{mt95}
R.~{Matsumoto} and T.~{Tajima}.
\newblock {Magnetic viscosity by localized shear flow instability in magnetized
  accretion disks}.
\newblock {\em \apj}, 445:\penalty0 767--779, June 1995.

\bibitem[{Matsuzaki} et~al.(1997){Matsuzaki}, {Matsumoto}, {Tajima}, and
  {Shibata}]{mmts97}
T.~{Matsuzaki}, R.~{Matsumoto}, T.~{Tajima}, and K.~{Shibata}.
\newblock {Three Dimensional MHD Simulations of Parker Instability in
  Differentially Rotating Disk}.
\newblock In {\em Astronomical Society of the Pacific Conference Series}, page
  766, 1997.

\bibitem[{Mayer} et~al.(2002){Mayer}, {Quinn}, {Wadsley}, and {Stadel}]{mqws02}
L.~{Mayer}, T.~{Quinn}, J.~{Wadsley}, and J.~{Stadel}.
\newblock {Formation of Giant Planets by Fragmentation of Protoplanetary
  Disks}.
\newblock {\em Science}, 298:\penalty0 1756--1759, November 2002.

\bibitem[{McKinney} and {Gammie}(2002)]{mg02}
J.~C. {McKinney} and C.~F. {Gammie}.
\newblock {Numerical Models of Viscous Accretion Flows near Black Holes}.
\newblock {\em \apj}, 573:\penalty0 728--737, July 2002.

\bibitem[{Menou}(2000)]{men00}
K.~{Menou}.
\newblock {Viscosity Mechanisms in Accretion Disks}.
\newblock {\em Science}, 288:\penalty0 2022--2024, June 2000.

\bibitem[{Menou} and {Quataert}(2001)]{mq01}
K.~{Menou} and E.~{Quataert}.
\newblock {Ionization, Magnetorotational, and Gravitational Instabilities in
  Thin Accretion Disks Around Supermassive Black Holes}.
\newblock {\em \apj}, 552:\penalty0 204--208, May 2001.

\bibitem[{Miles}(1961)]{jwm61}
J.~W. {Miles}.
\newblock {On the stability of heterogeneous shear flows}.
\newblock {\em J. Fluid Mech.}, 10:\penalty0 496--508, 1961.

\bibitem[{Miyoshi} et~al.(1995){Miyoshi}, {Moran}, {Herrnstein}, {Greenhill},
  {Nakai}, {Diamond}, and {Inoue}]{miy95}
M.~{Miyoshi}, J.~{Moran}, J.~{Herrnstein}, L.~{Greenhill}, N.~{Nakai},
  P.~{Diamond}, and M.~{Inoue}.
\newblock {Evidence for a Black-Hole from High Rotation Velocities in a
  Sub-Parsec Region of NGC4258}.
\newblock {\em \nat}, 373:\penalty0 127, January 1995.

\bibitem[{Mukhopadhyay} et~al.(2004){Mukhopadhyay}, {Afshordi}, and
  {Narayan}]{man04}
B.~{Mukhopadhyay}, N.~{Afshordi}, and R.~{Narayan}.
\newblock {Bypass to Turbulence in Hydrodynamic Accretion Disks: An Eigenvalue
  Analysis}.
\newblock {\em ArXiv Astrophysics e-prints}, December 2004.

\bibitem[{Mukhopadhyay} et~al.(2005){Mukhopadhyay}, {Afshordi}, and
  {Narayan}]{man05}
B.~{Mukhopadhyay}, N.~{Afshordi}, and R.~{Narayan}.
\newblock {Hydrodynamic Turbulence in Accretion Disks}.
\newblock {\em ArXiv Astrophysics e-prints}, January 2005.

\bibitem[{Muzerolle} et~al.(2000){Muzerolle}, {Calvet}, {Brice{\~ n}o},
  {Hartmann}, and {Hillenbrand}]{mea00}
J.~{Muzerolle}, N.~{Calvet}, C.~{Brice{\~ n}o}, L.~{Hartmann}, and
  L.~{Hillenbrand}.
\newblock {Disk Accretion in the 10 MYR Old T Tauri Stars TW Hydrae and Hen
  3-600A}.
\newblock {\em \apjl}, 535:\penalty0 L47--L50, May 2000.

\bibitem[{Narayan} et~al.(1987){Narayan}, {Goldreich}, and {Goodman}]{ngg87}
R.~{Narayan}, P.~{Goldreich}, and J.~{Goodman}.
\newblock {Physics of modes in a differentially rotating system - Analysis of
  the shearing sheet}.
\newblock {\em \mnras}, 228:\penalty0 1--41, September 1987.

\bibitem[{Narita} et~al.(1994){Narita}, {Kiguchi}, and {Hayashi}]{nkh94}
S.~{Narita}, M.~{Kiguchi}, and C.~{Hayashi}.
\newblock {The structure and evolution of thin viscous disks. 1: Non-steady
  accretion and excretion}.
\newblock {\em \pasj}, 46:\penalty0 575--587, December 1994.

\bibitem[{Nelson} et~al.(2000){Nelson}, {Benz}, and {Ruzmaikina}]{nbr00}
A.~F. {Nelson}, W.~{Benz}, and T.~V. {Ruzmaikina}.
\newblock {Dynamics of Circumstellar Disks. II. Heating and Cooling}.
\newblock {\em \apj}, 529:\penalty0 357--390, January 2000.

\bibitem[{Ogilvie}(2003)]{ogl03}
G.~I. {Ogilvie}.
\newblock {On the dynamics of magnetorotational turbulent stresses}.
\newblock {\em \mnras}, 340:\penalty0 969--982, April 2003.

\bibitem[{Ogilvie} and {Proctor}(2003)]{op03}
G.~I. {Ogilvie} and M.~R.~E. {Proctor}.
\newblock {On the relation between viscoelastic and magnetohydrodynamic flows
  and their instabilities}.
\newblock {\em Journal of Fluid Mechanics}, 476:\penalty0 389--409, February
  2003.

\bibitem[{Ostriker} et~al.(1992){Ostriker}, {Shu}, and {Adams}]{osa92}
E.~C. {Ostriker}, F.~H. {Shu}, and F.~C. {Adams}.
\newblock {Near-resonant excitation and propagation of eccentric density waves
  by external forcing}.
\newblock {\em \apj}, 399:\penalty0 192--212, November 1992.

\bibitem[{Owen} et~al.(1999){Owen}, {Mahaffy}, {Niemann}, {Atreya}, {Donahue},
  {Bar-Nun}, and {de Pater}]{owen99}
T.~{Owen}, P.~{Mahaffy}, H.~B. {Niemann}, S.~{Atreya}, T.~{Donahue},
  A.~{Bar-Nun}, and I.~{de Pater}.
\newblock {A low-temperature origin for the planetesimals that formed Jupiter}.
\newblock {\em \nat}, 402:\penalty0 269--270, November 1999.

\bibitem[{Paczy{\' n}ski}(1978)]{pac78}
B.~{Paczy{\' n}ski}.
\newblock {A model of selfgravitating accretion disk}.
\newblock {\em Acta Astronomica}, 28:\penalty0 91--109, 1978.

\bibitem[{Paczy{\' n}ski}(1983)]{pac83}
B.~{Paczy{\' n}ski}.
\newblock {Mass of large Magellanic Cloud X-3}.
\newblock {\em \apjl}, 273:\penalty0 L81--L84, October 1983.

\bibitem[{Papaloizou} and {Lin}(1995)]{pl95}
J.~C.~B. {Papaloizou} and D.~N.~C. {Lin}.
\newblock {0n the dynamics of warped accretion disks}.
\newblock {\em \apj}, 438:\penalty0 841--851, January 1995.

\bibitem[{Papaloizou} and {Pringle}(1983)]{pp83}
J.~C.~B. {Papaloizou} and J.~E. {Pringle}.
\newblock {The time-dependence of non-planar accretion discs}.
\newblock {\em \mnras}, 202:\penalty0 1181--1194, March 1983.

\bibitem[{Papaloizou} and {Pringle}(1984)]{pp84}
J.~C.~B. {Papaloizou} and J.~E. {Pringle}.
\newblock {The dynamical stability of differentially rotating discs with
  constant specific angular momentum}.
\newblock {\em \mnras}, 208:\penalty0 721--750, June 1984.

\bibitem[{Papaloizou} and {Pringle}(1985)]{pp85}
J.~C.~B. {Papaloizou} and J.~E. {Pringle}.
\newblock {The dynamical stability of differentially rotating discs. II}.
\newblock {\em \mnras}, 213:\penalty0 799--820, April 1985.

\bibitem[{Papaloizou} and {Pringle}(1987)]{pp87}
J.~C.~B. {Papaloizou} and J.~E. {Pringle}.
\newblock {The dynamical stability of differentially rotating discs. III}.
\newblock {\em \mnras}, 225:\penalty0 267--283, March 1987.

\bibitem[{Pedlosky}(1979)]{ped87}
J.~{Pedlosky}.
\newblock {\em {Geophysical Fluid Dynamics}}.
\newblock New York: Springer-Verlag, 1979, 1979.

\bibitem[{Pickett} et~al.(2000){Pickett}, {Cassen}, {Durisen}, and
  {Link}]{pcdl00}
B.~K. {Pickett}, P.~{Cassen}, R.~H. {Durisen}, and R.~{Link}.
\newblock {The Effects of Thermal Energetics on Three-dimensional Hydrodynamic
  Instabilities in Massive Protostellar Disks. II. High-Resolution and
  Adiabatic Evolutions}.
\newblock {\em \apj}, 529:\penalty0 1034--1053, February 2000.

\bibitem[{Press}(1978)]{press78}
W.~H. {Press}.
\newblock {Flicker noises in astronomy and elsewhere}.
\newblock {\em Comments on Astrophysics}, 7:\penalty0 103--119, 1978.

\bibitem[{Pringle}(1981)]{prin81}
J.~E. {Pringle}.
\newblock {Accretion discs in astrophysics}.
\newblock {\em \araa}, 19:\penalty0 137--162, 1981.

\bibitem[{Rempfer}(2003)]{rem03}
D.~{Rempfer}.
\newblock {Low-Dimensional Modeling and Numerical Simulation of Transition in
  Simple Shear Flows}.
\newblock {\em Annual Review of Fluid Mechanics}, 35:\penalty0 229--265, 2003.

\bibitem[{Rice} et~al.(2003){Rice}, {Armitage}, {Bate}, and {Bonnell}]{rabb03}
W.~K.~M. {Rice}, P.~J. {Armitage}, M.~R. {Bate}, and I.~A. {Bonnell}.
\newblock {The effect of cooling on the global stability of self-gravitating
  protoplanetary discs}.
\newblock {\em \mnras}, 339:\penalty0 1025--1030, March 2003.

\bibitem[{Richard}(2001{\natexlab{a}})]{rich01}
D.~{Richard}.
\newblock {Instabilit{\' e}s Hydrodynamiques dans les Ecoulements en Rotation
  Diff{\' e}rentielle}.
\newblock {\em Ph.D.~Thesis}, December 2001{\natexlab{a}}.

\bibitem[{Richard}(2001{\natexlab{b}})]{rddz01}
D.~{Richard}.
\newblock {Subcritical Instabilities of Astrophysical Interest in
  Couette-Taylor System}.
\newblock {\em Proc. 12th Couette-Taylor Workshop, Evanston, IL USA},
  2001{\natexlab{b}}.

\bibitem[{Richard}(2003)]{rich03}
D.~{Richard}.
\newblock {On non-linear hydrodynamic instability and enhanced transport in
  differentially rotating flows}.
\newblock {\em \aap}, 408:\penalty0 409--414, September 2003.

\bibitem[{Richard} and {Davis}(2004)]{rd04}
D.~{Richard} and S.~S. {Davis}.
\newblock {A note on transition, turbulent length scales and transport in
  differentially rotating flows}.
\newblock {\em \aap}, 416:\penalty0 825--827, March 2004.

\bibitem[{Richard} and {Zahn}(1999)]{rz99}
D.~{Richard} and J.~{Zahn}.
\newblock {Turbulence in differentially rotating flows. What can be learned
  from the Couette-Taylor experiment}.
\newblock {\em \aap}, 347:\penalty0 734--738, July 1999.

\bibitem[{Robinson} et~al.(1999){Robinson}, {Wood}, and {Wade}]{rww99}
E.~L. {Robinson}, J.~H. {Wood}, and R.~A. {Wade}.
\newblock {Application of Realistic Model Atmospheres to Eclipse Maps of
  Accretion Disks: The Effective Temperature and Flare of the Disk in the Dwarf
  Nova Z Chamaeleontis}.
\newblock {\em \apj}, 514:\penalty0 952--958, April 1999.

\bibitem[{Ruden} and {Pollack}(1991)]{rp91}
S.~P. {Ruden} and J.~B. {Pollack}.
\newblock {The dynamical evolution of the protosolar nebula}.
\newblock {\em \apj}, 375:\penalty0 740--760, July 1991.

\bibitem[{Ryu} and {Goodman}(1992)]{rg92}
D.~{Ryu} and J.~{Goodman}.
\newblock {Convective instability in differentially rotating disks}.
\newblock {\em \apj}, 388:\penalty0 438--450, April 1992.

\bibitem[{Salmeron}(2004)]{sal04}
R.~{Salmeron}.
\newblock {\em Ph.D.~Thesis, University of Sydney}, 2004.

\bibitem[{Salmeron} and {Wardle}(2003)]{sw03}
R.~{Salmeron} and M.~{Wardle}.
\newblock {Magnetorotational instability in stratified, weakly ionized
  accretion discs}.
\newblock {\em \mnras}, 345:\penalty0 992--1008, November 2003.

\bibitem[{Sano} and {Stone}(2002{\natexlab{a}})]{ss02a}
T.~{Sano} and J.~M. {Stone}.
\newblock {The Effect of the Hall Term on the Nonlinear Evolution of the
  Magnetorotational Instability. I. Local Axisymmetric Simulations}.
\newblock {\em \apj}, 570:\penalty0 314--328, May 2002{\natexlab{a}}.

\bibitem[{Sano} and {Stone}(2002{\natexlab{b}})]{ss02b}
T.~{Sano} and J.~M. {Stone}.
\newblock {The Effect of the Hall Term on the Nonlinear Evolution of the
  Magnetorotational Instability. II. Saturation Level and Critical Magnetic
  Reynolds Number}.
\newblock {\em \apj}, 577:\penalty0 534--553, September 2002{\natexlab{b}}.

\bibitem[{Sano} and {Stone}(2003)]{ss03}
T.~{Sano} and J.~M. {Stone}.
\newblock {A Local One-Zone Model of MagnetoHydroDynamic Turbulence in Dwarf
  Nova Disks}.
\newblock {\em \apj}, 586:\penalty0 1297--1304, April 2003.

\bibitem[{Sari} and {Goldreich}(2004)]{sg04}
R.~{Sari} and P.~{Goldreich}.
\newblock {Planet-Disk Symbiosis}.
\newblock {\em \apjl}, 606:\penalty0 L77--L80, May 2004.

\bibitem[{Shakura} and {Sunyaev}(1973)]{ss73}
N.~I. {Shakura} and R.~A. {Sunyaev}.
\newblock {Black holes in binary systems. Observational appearance.}
\newblock {\em \aap}, 24:\penalty0 337--355, 1973.

\bibitem[{Shepherd}(1985)]{shep85}
T.~G. {Shepherd}.
\newblock {Time Development of Small Disturbances to Plane Couette Flow.}
\newblock {\em Journal of Atmospheric Sciences}, 42:\penalty0 1868--1872,
  September 1985.

\bibitem[{Shlosman} et~al.(1990){Shlosman}, {Begelman}, and {Frank}]{sbf90}
I.~{Shlosman}, M.~C. {Begelman}, and J.~{Frank}.
\newblock {The fuelling of active galactic nuclei}.
\newblock {\em \nat}, 345:\penalty0 679--686, June 1990.

\bibitem[{Shu} et~al.(1994){Shu}, {Najita}, {Ostriker}, {Wilkin}, {Ruden}, and
  {Lizano}]{shu94}
F.~{Shu}, J.~{Najita}, E.~{Ostriker}, F.~{Wilkin}, S.~{Ruden}, and S.~{Lizano}.
\newblock {Magnetocentrifugally driven flows from young stars and disks. 1: A
  generalized model}.
\newblock {\em \apj}, 429:\penalty0 781--796, July 1994.

\bibitem[{Shu} et~al.(2000){Shu}, {Najita}, {Shang}, and {Li}]{shu00}
F.~H. {Shu}, J.~R. {Najita}, H.~{Shang}, and Z.-Y. {Li}.
\newblock {X-Winds Theory and Observations}.
\newblock {\em Protostars and Planets IV}, page 789, May 2000.

\bibitem[{Spiegel} and {Veronis}(1960)]{sv60}
E.~A. {Spiegel} and G.~{Veronis}.
\newblock {On the Boussinesq Approximation for a Compressible Fluid.}
\newblock {\em \apj}, 131:\penalty0 442, March 1960.

\bibitem[{Spruit} et~al.(1995){Spruit}, {Stehle}, and {Papaloizou}]{ssp95}
H.~C. {Spruit}, R.~{Stehle}, and J.~C.~B. {Papaloizou}.
\newblock {Interchange instability in and accretion disc with a poloidal
  magnetic field}.
\newblock {\em \mnras}, 275:\penalty0 1223--1231, August 1995.

\bibitem[{Stepinski}(1997)]{step97}
T.~F. {Stepinski}.
\newblock {Modeling the evolutionary history of the solar nebula}.
\newblock In {\em Lunar and Planetary Institute Conference Abstracts}, page
  1373, March 1997.

\bibitem[{Sterzik} and {Morfill}(1994)]{sm94}
M.~F. {Sterzik} and G.~E. {Morfill}.
\newblock {Evolution of protoplanetary disks with condensation and
  coagulation}.
\newblock {\em Icarus}, 111:\penalty0 536--546, October 1994.

\bibitem[{Stone} and {Balbus}(1996)]{sb96}
J.~M. {Stone} and S.~A. {Balbus}.
\newblock {Angular Momentum Transport in Accretion Disks via Convection}.
\newblock {\em \apj}, 464:\penalty0 364, June 1996.

\bibitem[{Stone} et~al.(2000){Stone}, {Gammie}, {Balbus}, and {Hawley}]{sgbh00}
J.~M. {Stone}, C.~F. {Gammie}, S.~A. {Balbus}, and J.~F. {Hawley}.
\newblock {Transport Processes in Protostellar Disks}.
\newblock {\em Protostars and Planets IV}, page 589, May 2000.

\bibitem[{Stone} and {Norman}(1992)]{sn92}
J.~M. {Stone} and M.~L. {Norman}.
\newblock {ZEUS-2D: A radiation magnetohydrodynamics code for astrophysical
  flows in two space dimensions. I - The hydrodynamic algorithms and tests.}
\newblock {\em \apjs}, 80:\penalty0 753--790, June 1992.

\bibitem[{Stone} et~al.(1999){Stone}, {Pringle}, and {Begelman}]{spb99}
J.~M. {Stone}, J.~E. {Pringle}, and M.~C. {Begelman}.
\newblock {Hydrodynamical non-radiative accretion flows in two dimensions}.
\newblock {\em \mnras}, 310:\penalty0 1002--1016, December 1999.

\bibitem[{Toomre}(1964)]{toom64}
A.~{Toomre}.
\newblock {On the gravitational stability of a disk of stars}.
\newblock {\em \apj}, 139:\penalty0 1217--1238, May 1964.

\bibitem[{Toomre}(1969)]{toom69}
A.~{Toomre}.
\newblock {Group Velocity of Spiral Waves in Galactic Disks}.
\newblock {\em \apj}, 158:\penalty0 899, December 1969.

\bibitem[{Torkelsson} et~al.(2000){Torkelsson}, {Ogilvie}, {Brandenburg},
  {Pringle}, {Nordlund}, and {Stein}]{tea00}
U.~{Torkelsson}, G.~I. {Ogilvie}, A.~{Brandenburg}, J.~E. {Pringle},
  {\AA}.~{Nordlund}, and R.~F. {Stein}.
\newblock {The response of a turbulent accretion disc to an imposed epicyclic
  shearing motion}.
\newblock {\em \mnras}, 318:\penalty0 47--57, October 2000.

\bibitem[{Umurhan} et~al.(2005){Umurhan}, {Nemirovsky}, {Regev}, and
  {Shaviv}]{unrs05}
O.~M. {Umurhan}, A.~{Nemirovsky}, O.~{Regev}, and G.~{Shaviv}.
\newblock {Hydrodynamical stability of thin accretion discs: transient growth
  of global axisymmetric perturbations}.
\newblock {\em ArXiv Astrophysics e-prints}, February 2005.

\bibitem[{Umurhan} and {Regev}(2004)]{ur04}
O.~M. {Umurhan} and O.~{Regev}.
\newblock {Hydrodynamic stability of rotationally supported flows: Linear and
  nonlinear 2D shearing box results}.
\newblock {\em \aap}, 427:\penalty0 855--872, December 2004.

\bibitem[{van Paradijs}(1996)]{vanp96}
J.~{van Paradijs}.
\newblock {On the Accretion Instability in Soft X-Ray Transients}.
\newblock {\em \apjl}, 464:\penalty0 L139, June 1996.

\bibitem[{Vanneste} et~al.(1998){Vanneste}, {Morrison}, and {Warn}]{vmw98}
J.~{Vanneste}, P.~J. {Morrison}, and T.~{Warn}.
\newblock {Strong echo effect and nonlinear transient growth in shear flows}.
\newblock {\em Physics of Fluids}, 10:\penalty0 1398--1404, June 1998.

\bibitem[{Vrtilek} et~al.(1991){Vrtilek}, {Penninx}, {Raymond}, {Verbunt},
  {Hertz}, {Wood}, {Lewin}, and {Mitsuda}]{vea91}
S.~D. {Vrtilek}, W.~{Penninx}, J.~C. {Raymond}, F.~{Verbunt}, P.~{Hertz},
  K.~{Wood}, W.~H.~G. {Lewin}, and K.~{Mitsuda}.
\newblock {Observations of Scorpius X-1 with IUE - Ultraviolet results from a
  multiwavelength campaign}.
\newblock {\em \apj}, 376:\penalty0 278--288, July 1991.

\bibitem[{Wagoner} et~al.(2001){Wagoner}, {Silbergleit}, and
  {Ortega-Rodr{\'{\i}}guez}]{wso01}
R.~V. {Wagoner}, A.~S. {Silbergleit}, and M.~{Ortega-Rodr{\'{\i}}guez}.
\newblock {``Stable'' Quasi-periodic Oscillations and Black Hole Properties
  from Diskoseismology}.
\newblock {\em \apjl}, 559:\penalty0 L25--L28, September 2001.

\bibitem[{Waleffe}(1997)]{wal97}
F.~{Waleffe}.
\newblock {On a self-sustaining process in shear flows}.
\newblock {\em Physics of Fluids}, 9:\penalty0 883--900, April 1997.

\bibitem[{Wardle}(1999)]{war99}
M.~{Wardle}.
\newblock {The Balbus-Hawley instability in weakly ionized discs}.
\newblock {\em \mnras}, 307:\penalty0 849--856, August 1999.

\bibitem[{Wardle} and {K{\"o}nigl}(1993)]{wk93}
M.~{Wardle} and A.~{K{\"o}nigl}.
\newblock {The structure of protostellar accretion disks and the origin of
  bipolar flows}.
\newblock {\em \apj}, 410:\penalty0 218--238, June 1993.

\bibitem[{Warner}(1995)]{warn95}
B.~{Warner}.
\newblock {\em {Cataclysmic variable stars}}.
\newblock Cambridge Astrophysics Series, Cambridge, New York: Cambridge
  University Press, |c1995, 1995.

\bibitem[{Watson} and {Wallin}(1994)]{ww94}
W.~D. {Watson} and B.~K. {Wallin}.
\newblock {Evidence from masers for a rapidly rotating disk at the nucleus of
  NGC 4258}.
\newblock {\em \apjl}, 432:\penalty0 L35--L38, September 1994.

\bibitem[{Wilner} et~al.(2003){Wilner}, {Bourke}, {Wright}, {J{\o}rgensen},
  {van Dishoeck}, and {Wong}]{wea03}
D.~J. {Wilner}, T.~L. {Bourke}, C.~M. {Wright}, J.~K. {J{\o}rgensen}, E.~F.
  {van Dishoeck}, and T.~{Wong}.
\newblock {Disks around the Young Stars TW Hydrae and HD 100546 Imaged at 3.4
  Millimeters with the Australia Telescope Compact Array}.
\newblock {\em \apj}, 596:\penalty0 597--602, October 2003.

\bibitem[{Yecko}(2004)]{yeck04}
P.~A. {Yecko}.
\newblock {Accretion disk instability revisited. Transient dynamics of rotating
  shear flow}.
\newblock {\em \aap}, 425:\penalty0 385--393, October 2004.

\bibitem[{Zahn}(1991)]{zahn91}
J.~P. {Zahn}.
\newblock {On the Nature of Disk Viscosity.}
\newblock In {\em IAU Colloq. 129: The 6th Institute d'Astrophysique de Paris
  (IAP) Meeting: Structure and Emission Properties of Accretion Disks},
  page~87, 1991.

\end{thebibliography}

\chapter{Curriculum Vitae}

\setlength{\parskip}{0ex} \setlength{\parindent}{0ex}

\section*{\centering Bryan Mark Johnson}

\subsection*{ADDRESS}

Loomis Laboratory of Physics

University of Illinois at Urbana-Champaign

1110 West Green Street

Urbana, IL 61801

Phone: (217) 333-2327

e-mail: bmjohnso@uiuc.edu

\subsection*{EDUCATION}

Ph.D. in Physics, University of Illinois at Urbana-Champaign, 2005 (Advisor: Gammie).

Dissertation: {\it Turbulent Angular Momentum Transport in Weakly-Ionized Accretion Disks}.

B.S. Electrical Engineering, LeTourneau University, 1996 (Advisor: Knoop).

Senior thesis: {\it System Measurement with White Noise}.

\subsection*{EXPERIENCE}

Teaching Assistant, University of Illinois at Urbana-Champaign, 2000-2002.

Systems Integration Engineer, Northrop Grumman Corporation, 1997-2000.

\subsection*{HONORS AND AWARDS}

Drickamer Research Fellowship, University of Illinois, 2004: {\it to recognize a graduate student
who has demonstrated significant ability in research, the most prestigious prize given by
the Department of Physics to a graduate student}.

R.G. LeTourneau Award for Outstanding Senior Engineering Student, LeTourneau University, 1996.

Gold Key Honor Society, LeTourneau University, 1996: {\it represents the highest honor given
by LeTourneau University to outstanding senior students}.

Dean's Scholarship, LeTourneau University, 1994.

Northrop Corporation Engineering Scholarship, Harper College, 1993.

Square D Engineering Scholarship, Harper College, 1992.

President's Scholarship, Crown College, 1990.

\end{document}